\documentclass[11pt]{article}
\usepackage{graphicx}
\usepackage{amssymb}
\usepackage{amsmath}
\usepackage{graphicx}
\usepackage{amstext}
\usepackage{leftidx}
\usepackage{esint}
\usepackage[utf8]{inputenc}
\usepackage{color}
\usepackage{cite}

\numberwithin{equation}{section}

\def\beq{\begin{eqnarray}}    
\def\eeq{\end{eqnarray}}      

\newcommand{\rL}{\rho_\Lambda}

\newcommand{\CC}{\Lambda}

\newcommand{\rv}{\rho_{\rm vac}}
\newcommand{\rveff}{\rho_{\rm vac}^{\rm eff}}
\newcommand{\Pv}{P_{\rm vac}}
\newcommand{\rvo}{\rho^0_{\rm vac}}

\newcommand{\wm}{\omega_m}

\newcommand{\nueff}{\nu_{\rm eff}}

\newcommand{\bk}{{\bf k}}
\newcommand{\mpl}{m_{\rm Pl}}
\newcommand{\MPl}{{\cal M}_{\rm Pl}}
\newcommand{\bv}{b_{\rm vac}}
\newcommand{\wv}{w_{\rm vac}}

\newcommand{\be}{\begin{equation}}
\newcommand{\ee}{\end{equation}}

\newcommand{\cH}{\mathcal{H}}

\newcommand{\txi}{\tilde{\xi}}

\newcommand{\astar}{a_{*}}

\newcommand{\rI}{\rho_I}

\newcommand{\tHI}{\tilde{H}_I}
\newcommand{\GB}{\mathfrak{G}}
\newcommand{\tal}{\tilde{\alpha}}

\usepackage{multido}
\makeatletter
\ams@newcommand{\vardot}[2]{%
  {\mathop{#2\kern0pt}\limits^{\vbox to-1.4\ex@{\kern-\tw@\ex@
   \hbox{\normalfont\multido{}{#1}{.}}\vss}}}}
\makeatother

\begin{document}



 \hyphenation{nu-cleo-syn-the-sis u-sing si-mu-la-te ma-king par-ti-cu-lar-ly
cos-mo-lo-gy know-led-ge e-vi-den-ce stu-dies be-ha-vi-or streng-thens
res-pec-ti-ve-ly appro-xi-ma-te-ly gra-vi-ty sca-ling ha-ving
ge-ne-ra-li-zed re-gu-la-ri-za-tion mo-del mo-dels po-wers ex-cee-din-gly ho-we-ver me-tric pa-ra-me-ter  vacu-um per-for-ming appro-xi-ma-tion li-te-ra-tu-re pro-pa-ga-tor}




\begin{center}
{\bf \Large  Renormalizing the vacuum energy in cosmological spacetime: implications for  the cosmological constant problem} \vskip 2mm

 \vskip 8mm

\textbf{Cristian Moreno-Pulido and  Joan Sol\`a Peracaula}

\vskip 0.5cm
Departament de F\'isica Qu\`antica i Astrof\'isica, \\
and \\ Institute of Cosmos Sciences,\\ Universitat de Barcelona, \\
Av. Diagonal 647, E-08028 Barcelona, Catalonia, Spain

\vskip0.5cm

\vskip0.4cm

E-mails:  cristian.moreno@fqa.ub.edu, sola@fqa.ub.edu

 \vskip2mm

\end{center}
\vskip 15mm

\begin{quotation}
\noindent {\large\it \underline{Abstract}}.
The renormalization of the vacuum energy in quantum field theory (QFT) is usually plagued with theoretical conundrums related not only with the renormalization procedure itself, but also with the fact that the final result leads usually to very large (finite) contributions incompatible with the measured value of $\Lambda$ in cosmology. As a consequence, one is  bound to extreme fine-tuning of the parameters and so to sheer unnaturalness of the result and of the entire approach.  We may however  get over this adversity using  a different perspective. Herein, we compute the zero-point energy (ZPE) for a nonminimally coupled (massive) scalar field in FLRW spacetime using the off-shell adiabatic renormalization technique employed in previous work. The on-shell renormalized result first appears at sixth adiabatic order, so the calculation is rather cumbersome. The general off-shell result yields a smooth function $\rho_{\rm vac}(H)$ made out of powers of the Hubble rate and/or of its time derivatives involving different (even) adiabatic orders $\sim H^N$ ($N=0, 2,4,6,...)$, i.e.  it leads, remarkably enough,  to the running vacuum model (RVM) structure. We have verified the same result from the effective action formalism and used it to find the $\beta$-function of the running quantum vacuum. No undesired contributions $\sim m^4$ from particle masses appear and hence no fine-tuning of the parameters is needed in $\rho_{\rm vac}(H)$. Furthermore, we find that the higher power $\sim H^6$ could naturally drive RVM-inflation in the early universe. Our calculation also elucidates in detail the equation of state of the quantum vacuum: it proves to be not exactly $-1$ and is moderately dynamical. The form of $\rho_{\rm vac}(H)$ at low energies is also characteristic of the RVM and consists of an additive term  (the so-called `cosmological constant') together with a small dynamical component $\sim \nu H^2$ ($|\nu|\ll1$). Finally, we predict a slow ($\sim\ln H$) running of Newton's gravitational coupling $G(H)$. The physical outcome of our semiclassical QFT calculation is revealing: today's cosmic vacuum and the gravitational strength should be both mildly dynamical.

\end{quotation}
\vskip 5mm

\newpage

\tableofcontents

\newpage


\vspace{1cm}

\section{Executive Summary}

 After the general introduction in Sec.\,\ref{intro}, wherein we try to motivate and contextualize the framework of this calculation, in Sections \ref{sec:EMT} and \ref{eq:RegZPE} we define our quantum field theoretical model under study, which consists of a neutral scalar field nonminimally coupled to gravity and with no self-interactions. We assume a  spatially flat Friedmann-Lema\^\i tre-Robertson-Walker (FLRW) background and solve for the mode functions of the scalar field  using the WKB approximation up to $4th$ and $6th$  adiabatic orders.   This enables us to compute, in  Sections  \ref{sec:RenormEMT} and \ref{sec:RenormalizedVED},  the vacuum expectation value of the energy-momentum tensor (EMT) within QFT in curved spacetime up to the same orders.  For renormalization purposes,  however,  the calculation up to $4th$ order  suffices. We employ  an off-shell generalization of the usual adiabatic renormalization procedure to compute the zero-point energy (ZPE) of the quantum fluctuations and show that the scaling evolution of the vacuum energy density (VED), $\rv$,  is  free from quartic powers of the masses.  We discuss the absence of  fine-tuning.   Remarkably, we find that $\rv$ carries a dynamical  component $\sim H^2$ which is characteristic of the running vacuum model (RVM) at low energies.  In Section \ref{sec:Trace} we  compute the trace of the EMT  up to $6th$  adiabatic order, which will be used to extract the quantum vacuum pressure at the same adiabaticity order.  In passing we verify, as a useful cross-check, that our results correctly reproduce the trace anomaly. The  vacuum pressure is used in Section \ref{sec:EoSvacuum} to find out the equation of state (EoS) of the quantum vacuum. We discover  that it is not exactly equal to $-1$, in contrast to the usual situation. Owing to the quantum effects,  the EoS  becomes moderately dynamical and can mimic quintessence in the current universe. Our $6th$ order calculation is instrumental to unveil a generalized form of the  RVM at high energies with potential implications for the physics of the very early universe. Indeed, in Sec.\,\ref{sec:RVMInflation}  we show that the RVM amounts to a new mechanism of inflation triggered by the higher order term $\sim H^6$.  Next, with the purpose of bolt securing the renormalization results for the EMT that have been obtained from direct calculation of the expansion modes in the previous sections, we recompute them anew  in  Section \ref{HeatKernel} within the effective action formalism using the heat-kernel expansion of the propagator with the DeWitt-Schwinger technique.  We use the effective action to compute  the running couplings, in particular the $\beta$-function  and  renormalization group equation (RGE) for the vacuum energy density (VED) itself, $\rv(H)$,  showing it to be consistent with the absence of quartic mass scales in the running.  To our knowledge, this is the first time that the dynamical VED  is derived from first principles.
 Finally, in Sec.\,\ref{sec:RenormalizedFriedmann} we study  Friedmann's equation in the presence of the  running $\rv(H)$ and observe that the gravitational coupling  $G$ is also a running quantity, although evolving only logarithmically with the expansion rate: $G=G( \ln H)$.  We compute the local conservation law for $\rv$ and verify (as a robust check of our calculation) that  it only depends on the $4th$ adiabatic terms (all of the $6th$ order effects cancel nontrivially in it).  {In Sec. \ref {sec:PhenoImplications} we provide some discussion on the possible phenomenological implications of the RVM}. Our conclusions are delivered in Section \ref{sec:conclusions}.  Four appendices at the end furnish specific technical details and complementary materials, such as our conventions and other useful formulas which are referred to from the main text.

\newpage

\section{Introduction}\label{intro}

The vacuum energy in cosmology is a most subtle concept which has challenged theoretical physicists and cosmologists for many decades, specially with the advent of Quantum Theory. The problem stems from the interpretation of the cosmological constant (CC) term,  $\CC$,  in Einstein's equations as a term being connected with the notion of vacuum energy density (VED), $\rv$, a fundamental concept in quantum field theory (QFT).   The precise connection with the current value  is $\rvo=\CC/(8\pi G_N)$,  where $G_N$ is the locally measured Newton's constant.
The CC term is $104$ years old and was introduced by Einstein in order to make the universe static\,\cite{Einstein1917}.  This is possible for $\CC>0$, but even then the equilibrium is unstable and hence the proposed solution is not admissible (as it was proven later on by Eddington\,\cite{Eddington1930}).  It goes without saying that such a static solution was soon ruled out also by Hubble's discovery of the expansion of the universe.  But even so the $\CC$-term remained unscathed, because the most important achievement of Einstein at this point was actually quite another, namely the fact that the tensorial structure $\CC\,g_{\mu\nu}$ is  another fundamental ingredient to be added as an indispensable  completion of  the gravitational field equations that he had found just two years before.  Such a  structure is indeed fully allowed by the general covariance inherent to General Relativity (GR).
There is, therefore, no reason to withdraw such a term only because it cannot serve a particular phenomenological purpose.  Once we realize it is there,  only a supreme theoretical reason can overrule its existence. But such reason has not been found. In fact, quite the opposite: the same $\CC$-term that was later  (groundlessly) rejected  by Einstein  in 1931\,\cite{Einstein1931} on account of  Hubble's finding of the expansion of the universe,  and which was resuscitated shortly afterwards by   Lema\^\i tre in 1934\,\cite{Lemaitre1934}  only to be   dismissed anew by others,  is still fully alive and kicking with us.  After a long checkered history,  currently  it has been  rescued from oblivion and promoted to be the simplest possible explanation for the observed acceleration of the universe. Such a groundbreaking discovery, made almost a quarter of century ago from the luminosity measurements of distant supernovae\,\cite{SNIa}, is fully borne out at present after collecting a lot more of supernovae data\,\cite{Scolnic2018} and from precision cosmology measurements of the anisotropies of the  cosmic microwave background (CMB)\,\cite{PlanckCollab}.  Thus, beyond reasonable doubt,  the $\CC$-term  is there after all, and we have every reason to want it be there, even if its introduction was categorized by Einstein himself as the ``biggest blunder'' he ever made in his life\,\cite{Gamow1970}.  However, as a famous proverb wisely says,  once the genie is let out of the bottle there is no turning back!

In spite of its current phenomenological success, the $\CC$-term and the very notion of vacuum energy  keeps on being the most challenging  theoretically ingredient of GR and maybe the most daring problem in the realm of Fundamental Physics\,\cite{Weinberg89,Witten2000}.  The reason for that is the preposterous disagreement between QFT and GR when we consider the notion of quantum vacuum.  The connection between vacuum energy and $\CC$ had already been glimpsed by Lema\^\i tre in 1934\,\cite{Lemaitre1934} who proposed for the first time the famous EoS of  vacuum  $\Pv=-\rv$ together with the aforementioned  VED relation $\rv=\CC/(8\pi G_N)$, although accompanied with a peculiar association of $\CC>0$ with  a negative energy density of vacuum (sic), and  still without hinting  at any  relationship with the quantum theory at this point.  Such a  connection first appeared with the work of Gliner\cite{Gliner966} and the more elaborated considerations by Zeldovich\,\cite{Zeldovich1967a,Zeldovich1967b} in 1967,  just fifty years after the $\CC$-term was first introduced by Einstein.  Retaking and promoting  old Planck's \textit{Nullpunktsenergie} to the QFT arena, he noted that the zero-point energy (ZPE) caused by the  vacuum fluctuations of all quantized (massive) fields induce a value of $\rv$  of order  of the quartic power of the mass of such a particle ($\sim \hbar m^4$). This is usually many orders of magnitude above the observational value,  $\rvo\sim 10^{-47}$ GeV$^4$.   Notice that the ZPE effects are of pure quantum origin since they are associated to vacuum-to-vacuum diagrams.  At one loop they are all proportional to $\hbar$.  The fact that we usually set $\hbar=1$ in natural units should not mislead us. Following Zeldovich, an estimate  of the ZPE contribution from the top quark should be of order   $m_t^4\sim  10^9$ GeV$^4$, which is 56 orders of magnitude larger than $\rvo$.  But even the electron, whose ZPE is of order $m_e^4\sim 6\times 10^{-14}$ GeV$^4$,  generates a discordance of `only' $33-34$ orders of magnitude. Except for a tiny neutrino mass in the meV range, all of the particles of the SM afford devastatingly large contributions to the ZPE which are many orders of magnitude away from the measured value of $\rvo$.  This is of course the famous Cosmological Constant Problem (CCP)\,\cite{Weinberg89,Witten2000}, on which so many efforts have been invested.  Yet, it remains constantly debated and is perhaps the biggest and toughest enigma faced by modern theoretical physics over time\,\cite{Sahni2000,PeeblesRatra2003,Padmanabhan2003,Copeland2006,DEBook,JMartin2012,JSPRev2013,JSPRev2022}.  Many of these attempts involved scalar fields  in different ways, such as quintessence and phantom dark energy (DE), see e.g.\,\cite{PeeblesRatra2003} for a review. But other ideas can also be contemplated and will be revived here on a more solid base,  cf.\cite{JSPRev2013,JSPRev2022,SPRev2014,JSPRev2015} and references therein.  Part of the CCP is that the  pure quantum effects  $\sim \hbar m^4$ pointed out by Zeldovich lead to intolerable fine-tuning among the parameters.  For this reason even the Higgs boson discovery at CERN\,\cite{HiggsDiscovery2012}, which was celebrated almost a decade ago as the crowning event in the history of particle physics and the most decisive endorsement to the Standard Model (SM) of the strong and electroweak interactions, becomes deeply involved in the gory details of this story.  This is both because of the huge $\sim M_H^4\sim  10^8$ GeV$^4$   contribution to the ZPE owing to its large mass ($M_H\sim 125$ GeV), and also because of its induced contribution to the VED from the ground state or vacuum expectation value (VEV) of the Higgs potential, which is of order  $<V_{H}>\sim M_H^2 v^2\sim 10^9$ GeV$^4$ and negative (where  $v\sim 246$ GeV is the  Higgs VEV). Taken separately or together, the two contributions (which are independent and uncorrelated, in principle)  amount to an outrageous mismatch of 55-56 orders of magnitude with respect to the measured cosmic vacuum energy density associated to the CC, viz.  $\rvo\sim 10^{-47}$  GeV$^4$,  not to mention  the quantum contributions to the effective potential\,\cite{JSPRev2013,JSPRev2022}.  Other vacuum effects, such as the QCD quark and gluon condensates, are smaller but still go more than $43$ orders of magnitude astray from observation, although these strong interacting contributions have been disputed in the literature, see e.g.\,\cite{Brodsky2011,Brodsky2022}.  The CCP has been looked upon  as being a reflex of the in-depth clash between elementary particle physics and GR at a most fundamental level.  If so, one should expect that  among the virtues of the promised  `third cosmological paradigm' -- which is foreseen to come to rescue  (hopefully) in the near  future\,\cite{Turner2021} -- will be  to throw shining light on the resolution of such a longstanding conundrum of modern theoretical physics and cosmology.

The above situation shows that even if $\CC$ has the virtue of maximal simplicity and is a core ingredient of the `concordance'  $\Lambda$CDM model of cosmology\,\cite{Peebles1984,Peebles1993,KraussTurner1995,OstrikerSteinhardt1995}, we do not have an explanation for it based on  first principles.  Not surprisingly, theoretical physicists and cosmologists have invented all kinds of ersatz concepts somehow replacing the notion of vacuum energy in an attempt  to escape from such a phenomenal `cul de sac'. But there is no escape if the ZPE strictly follows Zeldovich's interpretation.  As it should be obvious by now,  the CCP is much more than just the wavering phenomenological use of the CC term  in cosmology and astrophysics over the years\cite{ORaifeartaigh2018}, it is a profound fundamental theoretical problem.  If the $\sim m^4$ contributions  are there and are alive for all the (massive)  fields, we have to understand how to eschew or at least how to renormalize away their  implications.  On top of it we have the vacuum contributions emerging from every single mechanism of spontaneous symmetry breaking (SSB), starting from the   Brout-Englert-Higgs  mechanism\,\cite{BEH-Mechanism64}  of the  SM of particle physics.  We may change the name of this nightmare, we may call it CCP, quintessence  or DE  riddle, but the issue of the value of $\CC$ and the abhorrent fine-tuning problem being implied stay both fully upright. It all suggests that we might be doing something which is not quite right  in our  endeavour at giving a sense to the energy density of the quantum vacuum, even after the manyfold efforts devoted to this important subject and after devising all its poor DE substitutes\footnote{We point out that sophisticated dynamical adjustment mechanisms can be concocted  in QFT to dodge  fine-tuning issues\,\cite{Florian2010}, but unfortunately lacking at the moment of a connection with known fundamental theories.}.

The first inkling of the problem at a fundamental level comes from a simple observation. The concordance $\CC$CDM model  is formulated in  the FLRW context  and the latter is deeply ingrained in the heart of the General Relativity (GR) paradigm.  However, one of the most important drawbacks of GR  is that it is a non-renormalizable theory.  This can be considered a serious theoretical objection for GR to be the ultimate theory of gravity, and hence it adversely impacts on the $\CC$CDM status too.  GR cannot properly describe the short distance effects of gravity (the ultraviolet regime, UV), only the large distance effects (or infrared regime, IR).  As a consequence,  GR cannot furnish by itself a framework for quantizing gravity (the spacetime metric field)  along with the rest of the elementary  interactions  (assuming of course that gravity is amenable to be  quantized on conventional grounds). In this sense, a first (rougher but effective) approach is to treat gravity as a classical (external or background) field and quantize the matter fields only. This is the essence of the semiclassical approach, namely  the point of view of QFT in curved spacetime, see e.g.\cite{BirrellDavies82,ParkerToms09,Fulling89,MukhanovWinitzki07} for textbook  materials on the subject.  Herein it will be our approach too, and despite its limitations it will still allow us to compute meaningful quantum effects of matter on top of that classical gravitational background.

On the other hand, winds of change also blow on the phenomenological side. In the last  few years, issues of more practical nature  hint at the prospect that the concordance model of cosmology might not be such an immaculate model bordering perfection, not even at the phenomenological level.  The $\CC$CDM is known to be currently in tension in different observational fronts, e.g. with the structure formation data (the so-called $\sigma_8$-tension) and most significantly  with the local value of the Hubble parameter $H_0$ as compared to its determination from the CMB,  cf.  e.g.\,\cite{TensionsLCDM,IntertwinedH0,PerivoSkara2021,ValentinoReviewTensions,Intertwined8,TensionsJSP2018,WhitePaper2022}
and quoted references for a review of these tensions.  The possibility that these (persisting) discrepancies may be a serious symptom of physics beyond the $\CC$CDM model remains perfectly sound\,\cite{Riess2019}. As it has been shown in different works,  models mimicking in different ways a time-evolving  $\CC$ (or  dynamical  DE) could help in alleviating these problems\,\cite{ApJL2015,ApJ2017,RVMphenoOlder1,Elae2015,RVMphenoOlder2,RVMpheno1,RVMpheno2,PericoTamayo2017,CQG2017,ApJL2019,Mehdi2019,Tsiapi2019,BD2020,Singh2021,Manos2021,Yu2022,deCruzPerez:2023wzd}, see especially the most recent analysis \cite{EPL2021}.  In point of fact, a wide assort of different proposals aiming at curing the current challenges for the $\CC$CDM are available in the literature,  cf. \cite{Valentino1, Valentino2,PerivoSkara2021} and the long list of references therein. Many of these models are just phenomenological and are designed \textit{ad hoc} for this particular purpose, and hence lacking a formal support  from QFT or string theory.  Here we would like to show that the methods of QFT in curved spacetime can give a hand and can be applied to the context of the cosmological FLRW metric to find out  a  possible clue on these problems.

In our previous work \cite{CristianJoan2020}, we found  that such a QFT approach can indeed be used to derive a particular form of dynamical vacuum energy density called in the literature `running vacuum model' (RVM) -- cf. \cite{JSPRev2013,JSPRev2022,JSPRev2014,JSPRev2015} along with \cite{AdriaPhD2017,JavierPhD2021} containing ample bibliography --, in which the effective vacuum density $\rv$ (feeded by the quantum effects of the quantized matter fields)   appears as an expansion in powers of the Hubble function and its time derivatives, i.e. $\rv=\rv(H,\dot{H}, \ddot{H},...)$.  Such a VED  form has been tested previously on strict phenomenological grounds and  has implications both for the early universe (inflation) as well as for the current (slowly accelerated)  universe.  The first hints and additional support on such possible dynamical character of the VED in the QFT context actually appeared early on along the idea of  $\CC$  being a running quantity following a  renormalization group equation\,\cite{ShapSol1}, which was linked later on with a possible action functional\,\cite{Fossil2008} and other implications\,\cite{ShapSol2} --  for a review see  \cite{JSPRev2013,JSPRev2022,JSPRev2014,JSPRev2015}.  A variety of studies  bear relation in different ways with these investigations, see  e.g. \cite{Babic2005,Antoniadis2007,Maggiore2011,Maggiore2011b,Bilic2011,Bilic2012,Domazet2012,Antipin2017,BennieW2013,KohriMatsui2017,FerreiroNavarroSalas,FerreiroNavarroSalas2,NS2021,Pavlovic2021,Firouzjahi2022,Mottola2022} and corresponding references. However, it was only recently, in the aforementioned work \cite{CristianJoan2020},  where an explicit form of the RVM vacuum density could be derived within the QFT context.  Here we shall revisit and reinforce such a derivation  by  extending these calculations up to a point enabling us to investigate for the first time the  EoS of the quantum vacuum and show that it is not exactly equal to $-1$. Interestingly enough, we will also argue that the mentioned quantum effects  (in the extended form computed here, which accounts up to the  $\sim H^6$ contributions) can produce inflation in the early universe without invoking the existence of an  inflaton field mechanism.  This was not discussed in Ref.\cite{CristianJoan2020},  since only  renormalization issues were addressed in it and hence it was sufficient to consider the adiabatic expansion up to $4th$ order.  The more complete results involve the nontrivial  calculation of  a plethora of (finite) $6th$-order effects. These are now disclosed in the current comprehensive presentation.  The low energy form of the RVM includes the terms $H^2$ and $\dot{H}$ only (on top of a dominant constant term) and they mimic quintessence, without need of introducing an explicit quintessence field. In addition, very recently it has been shown that the RVM  can help mitigating in a very significant way the aforementioned $\sigma_8$ and $H_0$ tensions\,\cite{EPL2021}.

The leitmotif of the calculation put forward in the current study is the  renormalization of the stress-energy-momentum tensor using the adiabatic regularization prescription (ARP)\,\cite{BirrellDavies82,ParkerToms09} and  to explore the  implications for the cosmic evolution. The method is based on  the WKB expansion of the field modes  in the FLRW background.  By using an off-shell variant of this subtraction method, fully along the lines of the $4th$-order calculation presented in \cite{CristianJoan2020}, we find that the evolution of the  properly renormalized VED  does not depend on the unwanted contributions proportional to the fourth power of the particle masses ($\sim m^4$) and hence it is free from fine-tuning.   The extension to the next adiabaticity order ($6th$) is a rather cumbersome ulterior step in this calculational procedure. But it is a necessary step, as only in this way we are able to compute the renormalized on-shell ZPE of a quantized scalar field of mass $m$ nonminimally coupled to gravity. The final result turns out to be not only free of undesirable $\sim m^4$ contributions, it appears entirely harmless for the fate of the present universe (in contrast to the traditional outcome of more conventional calculations within simpler renormalization schemes).  To strengthen the robustness of our approach, we have cross-checked these results also in the effective action formalism using the heat-kernel expansion of the propagator.  It is noteworthy that our results have implications both for the current and for the early universe. As indicated,  apart from predicting a mild dynamics of the VED today that can  mimic quintessence,  the full result  up to   $\sim H^6$  order has the power to trigger inflation in the early universe. In this respect we  point out the resemblance (albeit not at all identification) of this mechanism with a  stringy version of the RVM-inflationary mechanism which has recently been developed\,\cite{BMS2020,NickJoan2020,NickJoan2021}, see also ~\cite{NickPhiloTrans,NickMG16,NickTorsion} for more details and interesting spin-off possibilities.  Whether our approach may contribute in some way to resolve the CCP in the future is difficult to say,  but the kind of results obtained here are  unprecedented in the literature (to the best of our knowledge).  Even though overcoming fine tuning issues should have an  impact on the CCP,  we should not forget that the latter  is a multifarious theoretical challenge that involves not only the  fine-tuning ordeal, but also calls for an eventual understanding of the value itself of the measured CC and its physical origin, which should gather contributions not only from the ZPE of all the quantum matter fields but also from the different SSB  mechanisms.

There is, of course, still a long trail to  solve the quantum vacuum problem.  In this paper, we have just tried to pave the way within  a simplified framework. Specifically, since the calculation of the ZPE  in QFT requires  renormalization,  it must be a scale-dependent quantity.  Using an appropriate renormalization scheme  based on adiabatic regularization we have produced a  renormalized  ZPE in the cosmological spacetime  free of fine-tuning oddities, and then used it to investigated a variety of implications for the current as well as for the early universe. What we have found  is that the VED evolves with the expansion and that it depends only on the Hubble rate and its time derivatives.  Obviously, much more  work will be needed to substantiate the reach of these considerations. Renormalization theory, however,  stops at this precise point since its job is to relate renormalized quantities at different scales;  it is  not its business to predict the value of the VED at some particular point, for example the VED at present.  The latter may be taken as an experimental input, which can then be used in our approach  as a boundary condition to predict the evolution of the VED at other expansion history epochs. What we have found is that such an evolution should be free from unnatural adjustments among the parameters of the theory.

\section{Vacuum energy for a scalar field in the FLRW background}\label{sec:EMT}

In this paper, we are addressing  the QFT calculation of the VED  of a quantized free scalar field nonminimally coupled to FLRW spacetime.   The corresponding VED  involves the cosmological term $\CC$ (or, more precisely, the parameter $\rL$)  in the Einstein-Hilbert (EH) action and the zero-point energy (ZPE) of the quantum matter  fluctuations.  We should emphasize from the beginning that we do not address here the issues of quantum gravity (QG) and the  functional integration over metrics, with or without the $\CC$ term, see e.g. \cite{ChristensenDuff1,ChristensenDuff2,ChristensenDuff3}.  While these are potentially important matters,  in the current context  gravity is treated as a classical background field and hence the QG considerations in connection to quantizing the metric lie out of the scope of our semiclassical approach.  We take a scalar field only as the simplest representative  of the kind of results and difficulties we expect to encounter when dealing with generic quantized fields in a curved background.  Calculational details for  fermions or vector bosons do not change in any essential way our approach to computing the  vacuum fluctuations in the gravitational context \cite{JCS2022}.   Needless to say,  the QFT calculation of vacuum-to-vacuum diagrams  implies to perform renormalization since we meet UV-divergent integrals. The usual procedures to account for the regularization and renormalization of divergent quantities in QFT, such as e.g. the minimal subtraction (MS) scheme\,\cite{BolliniGiambiagi72,Hooft73}, do not yield a sensible answer  for the VED. Despite producing finite results, these are useless and incongruent with the physical facts.  They  lead to a quartic dependence on the mass of the fields ($\sim m^4$) and this enforces a very serious (in fact incommensurable) fine-tuning among the parameters, this being so both in Minkowski and in curved spacetime\,\footnote{See e.g. \cite{JSPRev2013,JSPRev2022}  for  a pedagogical  introduction to the ZPE calculation in flat and curved spacetime. }.  In this paper, we  forgo making use of such an unsuccessful method and rather adhere to the adiabatic renormalization procedure (ARP)\,\cite{BirrellDavies82,ParkerToms09,Fulling89}.  However, we adopt a more recent variant of this method, which we used extensively in our previous paper\,\cite{CristianJoan2020} in order to renormalize the ZPE in a more meaningful way and to connect the results with the general framework of the running vacuum model (RVM) mentioned in the introduction\footnote{The method was introduced for the study of the running couplings in curved spacetime in \cite{FerreiroNavarroSalas}, although it was not applied to the VED nor to the study of the running of this quantity throughout the cosmic evolution. This was done for the first time in\,\cite{CristianJoan2020}. We shall comment on some differences with respect to our approach later on.}.   In this section, we shall summarize the approach followed in\,\cite{CristianJoan2020}  in view of  fixing the notation and also to prepare the ground to extend the  results previously obtained in that reference to higher adiabatic orders. Such an ascension in the adiabatic order  is mandatory if we are interested to compute the on-shell renormalized zero-point energy (ZPE) of a scalar field in FLRW spacetime.  Recall that in the momentum subtraction scheme,  the renormalized Green's functions and running couplings are obtained by subtracting their values at a renormalization point $p^2=M^2$ (space-like in our metric, which becomes an Euclidean point after Wick rotation) or at the time-like one $p^2=-M^2$  (depending  on the kinematical region involved). Here the situation is similar,  but since  for vacuum diagrams we do not have external momenta,  we renormalize the ZPE by subtracting (to its on-shell value) the corresponding value computed to $4th$ adiabatic order but at an arbitrary mass scale $M$.  This suffices both  to eliminate the divergent terms in the first four adiabatic orders, which are the only ones that can be divergent in the renormalization of the EMT\,\cite{CristianJoan2020}  and to relate the ZPE  at different scales.  However, if we wish to compute the zero-point energy at  the value of the particle mass $m$, then the renormalization is to be performed on-shell, i.e. at $M=m$. In such a case it is evident that a subtraction at  $4th$-order would give a vanishing result. Therefore, since general covariance of the effective action  requires that the vacuum energy can only  be expanded in even adiabatic orders, the leading contribution to the on-shell renormalized ZPE must appear at the sixth order of adiabaticity.  The main aim of this work is to perform such a nontrivial calculation and extract some interesting consequences.  But prior to undertaking this rather heavy task we prepare our basic framework, starting with the classical action, energy-momentum tensor and field equations.

\subsection{Classical field equations for a scalar field nonminimally coupled to gravity}\label{Sec:FieldEquations}

We start from the EH action for gravity plus matter:
\begin{equation}\label{eq:EH}
S_{\rm EH+m}=  S_{\rm EH}+S_{\rm m}=\frac{1}{16\pi G}\int d^4 x \sqrt{-g}\, R  -  \int d^4 x \sqrt{-g}\, \rL + S_{\rm m}\,.
\end{equation}
The (constant) term $\rL$ has dimension of energy density and is usually called the vacuum energy density. However, we will not call it that way here since it is not yet the physical vacuum energy density, $\rv$,  as we shall see. The term $\rL$ is at this point just a bare parameter of the EH action, as $G$ itself.  We prefer not to introduce special notations by now. The physical values will be identified  only after renormalizing the bare theory\,\footnote{If we would write  $\rL=\CC/(8\pi G_N)$  the parameter $\CC$  could not be interpreted as a physical cosmological constant, but just as the bare cosmological term.  As mentioned in the introduction, the quantity that can be associated with the physically measured cosmological constant, $\CC$ , is  defined through $\rv=\Lambda/(8\pi G_N)$,  where $\rv$  and $G_N$ are the physical quantities. The latter, however, can only be identified after properly renormalizing the QFT calculation. }. Varying the part involving  the $\rL$ term  yields
\begin{equation}\label{eq:SrL}
\delta S_{\CC}= -\int d^4 x \,\delta \sqrt{-g}\, \rL =  -\frac{1}{2}\int  d^4 x\, \sqrt{-g}\, \left(- \rL\, g_{\mu\nu}\right)\,\delta g^{\mu\nu}\,.
\end{equation}
Together with the variation of the EH and matter terms of \eqref{eq:EH}, the  gravitational field equations can be put in the convenient form
\begin{equation} \label{FieldEq2}
\frac{1}{8\pi G}G_{\mu \nu}=-\rho_\Lambda g_{\mu \nu}+T_{\mu \nu}^{\rm m}\,,
\end{equation}
where $G_{\mu\nu}=R_{\mu\nu}-(1/2) g_{\mu\nu} R$  is the usual Einstein tensor and  $ T_{\mu \nu}^{\rm m}$  is the  stress-energy-momentum tensor, or just energy-momentum tensor (EMT for short) of matter\,\footnote{Our conventions and other formulas  of interest for this calculation are collected in the Appendix \ref{sec:appendixA1}.}:
\begin{equation}\label{eq:deltaTmunu2}
  T_{\mu \nu}^{\rm m}=-\frac{2}{\sqrt{-g}}\frac{\delta S_{\rm m}}{\delta g^{\mu\nu}}\,.
\end{equation}
 For simplicity we will assume that there is only one (matter) quantum field contribution to the EMT  on the right hand side of \eqref{FieldEq2} in the form of  a real scalar field, $\phi$, with mass $m$. Such a contribution will be denoted $T_{\mu \nu}^{\phi}$.   We neglect for the moment the incoherent matter contributions  from dust and radiation.  They can be added a posteriori  without altering the pure  QFT aspects on  which we wish to focus right now.  We assume that $\phi$ is nonminimally coupled to gravity and with no potential $V(\phi)$.  Thus, the part of the action involving $\phi$ reads
\begin{equation}\label{eq:Sphi}
  S[\phi]=-\int d^4x \sqrt{-g}\left(\frac{1}{2}g^{\mu \nu}\partial_{\mu} \phi \partial_{\nu} \phi+\frac{1}{2}(m^2+\xi R)\phi^2 \right)\,.
\end{equation}
Here, $\xi$ is the nonminimal coupling of $\phi$ to gravity.  For $\xi=1/6$, the massless ($m=0$)  classical action is conformally invariant in $4$ spacetime dimensions.
Since no classical potential for $\phi$ is present  in our analysis  we need not consider  the  quantum corrections and corresponding renormalization of  the effective potential.  Under these conditions $\xi$ is not necessary for renormalizing the theory. Even so, by keeping $\xi\neq 0$ we can obtain more general results, which will be particularly useful for the connection with the RVM framework.  Furthermore, it allows us to perform a nontrivial test of our calculations by reproducing the conformal anomaly for the quantum corrected action.  In general, the presence of a nonminimal coupling is expected in a variety of contexts, e.g. in extended gravity theories\,\cite{Sotiriou2010,Capozziello2008,Capozziello2011,CapozzielloFaraoni2011} and in models of Higgs-induced inflation\,\cite{Barvinsky2008}.  However,  as previously indicated, even in the absence of  $V(\phi)$  the presence of $\xi$ can be very useful.  From hereon in, we will exclusively target the adiabatic renormalization of the  ZPE  of $\phi$, which in itself  is already quite involved in curved spacetime.

The classical  energy-momentum tensor can be obtained by computing the functional derivative with respect to the metric, Eq.\,\eqref{eq:deltaTmunu2}. For the specific action \eqref{eq:Sphi}, we obtain
\begin{equation}\label{EMTScalarField}
\begin{split}
T_{\mu \nu}^{\phi}=&-\frac{2}{\sqrt{-g}}\frac{\delta S[\phi]}{\delta g^{\mu\nu}}= (1-2\xi) \partial_\mu \phi \partial_\nu\phi+\left(2\xi-\frac{1}{2} \right)g_{\mu \nu}\partial^\sigma \phi \partial_\sigma\phi\\
& -2\xi \phi \nabla_\mu \nabla_\nu \phi+2\xi g_{\mu \nu }\phi \Box \phi +\xi G_{\mu \nu}\phi^2-\frac{1}{2}m^2 g_{\mu \nu} \phi^2.
\end{split}
\end{equation}
For $\xi=0$ we recover the trivial result for the free and minimally coupled scalar field.

The field $\phi$ obeys the  Klein-Gordon (KG) equation in curved spacetime, which follows from varying the action \eqref{eq:Sphi} with respect to $\phi$:
\be\label{eq:KG}
(\Box-m^2-\xi R)\phi=0\,,
\ee
where $\Box\phi=g^{\mu\nu}\nabla_\mu\nabla_\nu\phi=(-g)^{-1/2}\partial_\mu\left(\sqrt{-g}\, g^{\mu\nu}\partial_\nu\phi\right)$  is the standard box operator in curved spacetime. As it is well-known, the time and space variables in the KG equation can be separated, i.e. placed in the form $\phi(t,x)\sim \sum_k\psi_k(x)\phi_k(t)$ (the sum usually being in the continuum limit)  provided the metric is   conformally static\,\cite{Fulling89},  which means $ds^2=-dt^2+a^2(t)\gamma_{ij}(x)dx^idx^j$, where $a(t)$ is the scale factor and $\gamma_{ij}$ is the metric of any three-dimensional Riemannian manifold as its basic spatial section. The spacetime metric can then be put in the form   $ds^2= C(\tau)(-d\tau^2+\gamma_{ij}(x)dx^idx^j)$, where the conformal scale factor $C(\tau)=a^2(\tau)$  is a function of the conformal time $\tau$. The latter is connected to the cosmic time through $\tau=\int dt/a$. The separability condition  certainly holds for any  FLRW metric.  In the following, however,  we  will focus  on the spatially flat three-dimensional case, $\gamma_{ij}=\delta_{ij}$.  The FLRW line element  is then conformally static and even  conformally flat,  and can be written   $ds^2=a^2(\tau)\eta_{\mu\nu}dx^\mu dx^\nu$, where  $\eta_{\mu\nu}={\rm diag} (-1, +1, +1, +1)$ is the Minkowski  metric in our conventions (cf.  Appendix \ref{sec:appendixA1}).   The derivative with respect to the conformal time will be denoted   $^\prime\equiv d/d\tau$ and thus the  Hubble rate in conformal time  reads $\mathcal{H}(\tau)\equiv a^\prime /a$.  Since  $dt=a d\tau$,  the relation between the Hubble rate in cosmic and conformal times is $\mathcal{H}(\tau)=a  H(t)$, where  $H(t)=\dot{a}/a$ (with $\dot{}\equiv d/dt$) is the usual Hubble rate.

In conformally flat coordinates, the KG equation \eqref{eq:KG} reads explicitly as follows:
\begin{equation}\label{eq:KGexplicit}
 \phi''+2\cH\phi'-\nabla^2\phi+a^2(m^2+\xi R)\phi=0\,,
\end{equation}
where  $\Box\phi=-a^{-2}\left(\phi''+2\cH\phi-\nabla^2\phi\right)$  (with $\cH=a'/a$) and the Ricci scalar reads  $R=6a^{\prime\prime}/a^3$ (cf.\,Appendix \ref{sec:appendixA1}) .
The separability condition  in these coordinates, namely the factorization  $\phi(\tau,x)\sim \int d^3k \ A_{\bf k}\psi_k({\bf x})\phi_k(\tau)+cc$, can  be implemented with $\psi_k(x)=e^{i{\bf k\cdot x}}$, but in contradistinction to the Minkowskian case we cannot take  $\phi_k(\tau)=e^{\pm i\omega_k \tau}$ since the frequencies of the modes are no longer constant.  The precise form of the  modes $\phi_k(\tau)$  in the curved spacetime case are determined by the KG equation.  In fact, starting from the Fourier expansion with separated space and time variables
\begin{equation}\label{FourierModes}
\phi(\tau,{\bf x})=\int d^3{k} \left[ A_\bk  u_k(\tau,{\bf x})+A_\bk^\ast  u_k^*(\tau,{\bf x}) \right]=\int\frac{d^3{k}}{(2\pi)^{3/2}} \left[ A_\bk e^{i{\bf k\cdot x}} \phi_k(\tau)+A_\bk^\ast e^{-i{\bf k\cdot x}} \phi_k^*(\tau) \right]
\end{equation}
($A_\bk $  and  $A_\bk^\ast$  being the Fourier coefficients, treated still classically at this point) and substituting it into \eqref{eq:KGexplicit} we find that  the mode functions $\phi_k(\tau)$ are determined by the nontrivial differential equation
\begin{equation}\label{eq:KGFourier}
 \phi_k''+2\cH\phi'_k+\left(\omega_k^2(m)+a^2\xi R\right)\phi_k=0\,.
\end{equation}
 Because   $\omega_k^2(m)\equiv k^2+a^2m^2 $, the mode functions depend only on the modulus $k\equiv|\bk|$ of the  momenta, where   $k$ is  the comoving momentum and $\tilde{k}=k/a$ the physical one. The frequencies are seen to be functions of the time-evolving scale factor $a=a(\tau)$. This is the first unmistakable  sign that a particle interpretation will become hard in this context, or in other words, it amounts to the phenomenon  of particle creation in a time-dependent gravitational field\,\cite{Parker1966,Parker1968,Parker1969} -- for a review see e.g.  \cite{DeWitt1975,ParkerCargese1978,Ford1997}.  If we perform the change of field mode variable  $\phi_k=\varphi_k/a$   the above equation simplifies to a more amenable one in which the damping term is absent:
\begin{equation}\label{eq:KGFourier}
\varphi_k^{\prime \prime}+\left(\omega_k^2(m)+a^2\,(\xi-1/6)R)\right)\varphi_k=0\,.
\end{equation}
Despite it being the equation of an  harmonic oscillator,  it has a time-dependent frequency and cannot be solved analytically except in a few cases. For example,  for conformally invariant matter, i.e. for  massless scalar field  ($m=0$)  and conformal  coupling  ($\xi=1/6$),  the above equation boils down to the form $\varphi_k^{\prime \prime}+k^2\varphi_k=0$, whose positive- and negative-energy solutions are just  $e^{- ik\tau}$  and $e^{+ ik\tau}$, respectively. These are the very same solutions as in the massless Minkowskian case (for which $R=0$), which is ultimately the reason why no particles are created in the quantized version of the theory (in which  $A_\bk $  and $A_\bk^\dagger$ -- the latter replacing  $A_\bk^\ast$ --  become the annihilation and creation operators) in the conformally invariant case.  In the massless case with minimal coupling ($\xi=0$)  Eq.\,\eqref{eq:KGFourier} takes on  the form
\be
\varphi_k''+(k^2-a^2R/6)\varphi_k=0\,.
\ee
In the radiation epoch ($a\propto\tau$, thus  $R=6a^{\prime\prime}/a^3=0$) we find once more  the trivial modes $\varphi_k(\tau)=e^{\pm ik\tau}$.  On the other hand, both in the de Sitter ($a=-1/(H\tau)$,  $H=$const.) and matter-dominated ($a\propto\tau^2$) epochs we have $a^2R=12/\tau^2$, which leads to
$\varphi_k''+(k^2-2/\tau^2)\varphi_k=0\,.$
This equation is nontrivial but  admits an exact (positive-energy) solution in terms of  Bessel  functions. In the de Sitter case ($\tau<0$) one may impose the Bunch-Davies vacuum limit $\sim e^{-ik|\tau|}$ in the far remote past ($\tau\to-\infty$)\footnote{For a given mode $k$ this condition  insures $k|\tau|\gg1$ and hence the modes can be thought of as being essentially insensitive to curvature effects, since $a^2R=12/\tau^2\to 0$ in this limit.  In this way we are free to fix the convenient initial condition  $\phi_k(\tau)\sim  e^{ik\tau}=e^{-ik|\tau|}$ in the remote past. }  and  one finds  the solution in terms of  Bessel/Hankel functions: $\varphi(\tau)\propto\sqrt{k|\tau|}\left(J_{3/2}(k|\tau|)-iJ_{-3/2}(k|\tau|)\right) =\sqrt{k|\tau|}H_{3/2}^{(2)}(k|\tau|)$. Because of the half-integer order of these functions in this case, it leads to a close analytic form:  $\varphi_k(\tau)\propto (1-i/(k|\tau|))e^{-ik|\tau|}$. The same solution is valid for the matter-dominated era (for which $\tau>0$).    The corresponding solutions for $\phi_k$ are  of course $\phi_k(\tau)=\varphi_k(\tau)/a(\tau)$ for each relevant epoch. For $m\neq0$ and/or $\xi\neq 1/6$ a solution in terms of modified Bessel functions is also possible in the de Sitter epoch. In general, however, there is no analytic solution of \eqref{eq:KGFourier} for the whole cosmic expansion history of the universe up to the current DE epoch.  Therefore,  we are generally led  to search for a WKB (Wentzel-Kramers-Brillouin)  expansion of the solution.  But before doing that let us take up the quantization of the scalar field $\phi$, since we are mainly interested in computing the vacuum fluctuations.

 \subsection{Quantum fluctuations and expansion modes}\label{sec:AdiabaticVacuum}

The previous equations are classical. To account for the quantum fluctuations of the field $\phi$  we must consider the expansion of the field around its background value $\phi_b$:
\begin{equation}
\phi(\tau,x)=\phi_b(\tau)+\delta\phi (\tau,x). \label{ExpansionField}
\end{equation}
We wish to compute the  vacuum expectation value (VEV)  of the EMT of $\phi$, i.e. $\langle T_{\mu \nu}^\phi \rangle\equiv \langle  0 |T_{\mu \nu}^\phi |0 \rangle$.  But for this we need to define
a useful vacuum state for the QFT in a curved background,  $|0 \rangle$,  called the adiabatic vacuum\,\cite{Bunch1980}. The VEV of the field is identified with the background value,  $\langle 0 | \phi (\tau, x) | 0\rangle=\phi_b (\tau)$, whereas the VEV of the fluctuation is zero:  $\langle  \delta\phi  \rangle\equiv \langle 0 | \delta\phi | 0\rangle =0$. Not so, of course, the VEV of the bilinear products of fluctuations, e.g. $\langle \delta\phi^2 \rangle\neq0$. These and other bilinear VEV's will be responsible for the zero-point energy (ZPE) of the field. For an appropriate definition of the ZPE,  we  first decompose  $\langle T_{\mu \nu}^\phi \rangle=\langle T_{\mu \nu}^{\phi_b} \rangle+\langle T_{\mu \nu}^{\delta\phi}\rangle$, where
$\langle T_{\mu \nu}^{\phi_{b}} \rangle =T_{\mu \nu}^{\phi_{b}} $
is the  contribution  from the classical background part,  whilst $\langle T_{\mu \nu}^{\delta\phi}\rangle\equiv \langle 0 | T_{\mu \nu}^{\delta\phi}| 0\rangle$  is  the genuine vacuum contribution from the field fluctuations $\delta\phi$.  Because  $\rho_\Lambda$  is also part of the vacuum action  \eqref{eq:EH}, the full vacuum contribution is  not only the expectation value of the EMT but the sum of these two terms in the form prescribed on the \textit{r.h.s.} of Eq.\,\eqref{FieldEq2}:
\begin{equation}\label{EMTvacuum}
\langle T_{\mu \nu}^{\rm vac} \rangle=-\rho_\Lambda g_{\mu \nu}+\langle T_{\mu \nu}^{\delta \phi}\rangle\,.
\end{equation}
This is an important point, which will play a role in our calculation:  the total vacuum part  receives contributions from both the cosmological term as well as from the quantum fluctuations  of the field. However, since these quantities are formally UV-divergent, the physical vacuum contribution can only be identified upon suitable regularization and renormalization of our calculation. Rather than using minimal subtraction, as it has been customary in addressing the vacuum problem in QFT, we will use the adiabatic method along the lines of our previous work\,\cite{CristianJoan2020}.

We start by noting that the classical and  quantum parts of the  field (\ref{ExpansionField}) obey the  curved spacetime KG equation  \eqref{eq:KGFourier} separately.  Similarly for $\varphi=\varphi_b+\delta\varphi$  (where $\phi=\varphi/a$).  Denoting the frequency modes of the fluctuating part  $\delta\varphi$ by   $h_k(\tau)$, we can write
\begin{equation}\label{FourierModesFluc}
\delta \varphi(\tau,{\bf x})=\int \frac{d^3{k}}{(2\pi)^{3/2}} \left[ A_\bk e^{i{\bf k\cdot x}} h_k(\tau)+A_\bk^\dagger e^{-i{\bf k\cdot x}} h_k^*(\tau) \right]\,.
\end{equation}
Here  $A_\bk$ and  $A_\bk^\dagger $ are no longer the classical Fourier coefficients but are now promoted to be (time-independent) annihilation and creation quantum operators, which satisfy the commutation relations
\begin{equation}\label{CommutationRelation}
[A_\bk, A_\bk'^\dagger]=\delta({\bf k}-{\bf k'}), \qquad [A_\bk,A_ \bk']=0.
\end{equation}
The frequency modes of the fluctuations, $h_k(\tau)$, observe the same  differential equation \eqref{eq:KGFourier}:
\begin{equation}\label{eq:ODEmodefunctions}
h_k^{\prime \prime}+\Omega_k^2(\tau) h_k=0\ \ \ \ \ \ \ \ \ \ \ \Omega_k^2(\tau) \equiv\omega_k^2(m)+a^2\, (\xi-1/6)R\,.
\end{equation}
Except in the simple cases mentioned above, the solution of that equation can only be found approximately through a recursive self-consistent iteration.  The starting point is the phase integral
\begin{equation}\label{eq:phaseIntegral}
h_k(\tau)=\frac{1}{\sqrt{2W_k(\tau)}}\exp\left(-i\int^\tau W_k(\tilde{\tau})d\tilde{\tau} \right)\,.
\end{equation}
The normalization factor satisfies the Wronskian condition
$ h_k^{} h_k^{*\prime}- h_k^*h_k^\prime=i$, 
which preserves  the standard equal-time commutation relations.
The effective frequency function $W_k$ in the above ansatz follows  from inserting \eqref{eq:phaseIntegral} into \eqref{eq:ODEmodefunctions}, with the result
\begin{equation} \label{WKBIteration}
W_k^2(\tau)=\Omega_k^2(\tau) -\frac{1}{2}\frac{W_k^{\prime \prime}}{W_k}+\frac{3}{4}\left( \frac{W_k^\prime}{W_k}\right)^2\,.
\end{equation}
This (non-linear) equation can be solved using the WKB expansion, or Carlini-Liouville–Green approximation\,\cite{Fulling89}. For a sufficiently differentiable function $\Omega(\tau)$, such an expansion takes the form
\be\label{eq:WKBexpansion}
W_k(\tau)=\Omega_k\left[1+\delta_2(\tau)\Omega_k^{-2}+\delta_4(\tau)\Omega_k^{-4}+...\right]\,.
\ee
The leading term holds when the time variation of the frequency $W_k(\tau)$ is supposed to be very small as compared  to $k$.  In that case, the derivative terms on the \textit{r.h.s} of \eqref{eq:WKBexpansion}  can be neglected and the phase integral \eqref{eq:phaseIntegral} with $W_k(\tau)\simeq\Omega_k(\tau)$ furnishes a sufficient approximation.  The remaining terms of \eqref{eq:WKBexpansion} improve the accuracy and can be computed by iterating the procedure in what is known as the adiabatic expansion. The implementation in the gravitational context is well-known since long\,\,\cite{ParkerToms09,Fulling89}.  The WKB approximation is applicable for large $k$, hence  short wave lengths (as e.g. in geometrical Optics),  and weak gravitational fields.  In our case such a regime is appropriate to study the short-distance behavior of the theory, i.e. the UV-divergences and the renormalization procedure.   Because the general mode functions $h_k(\tau)$ are not the canonical  $\varphi_k(\tau)=e^{\pm i\omega_k\tau}$ anymore,  particles with definite frequencies cannot be strictly defined in a curved background.  Yet an approximate Fock space interpretation is still feasible if the  vacuum is defined as the quantum state which is annihilated by all the operators $A_{\bf k}$ of the above Fourier expansion. This defines the adiabatic vacuum\,\cite{Bunch1980}, see also \,\cite{BirrellDavies82,ParkerToms09,Fulling89,MukhanovWinitzki07}\footnote{A simple physical example in hydrodynamics  is that of small amplitude
adiabatic acoustic waves in an otherwise homobaric fluid (i.e.  whose unperturbed pressure is constant). If the equilibrium/background state (the counterpart of the adiabatic ``vacuum'' in QFT) varies only little over the characteristic lengthscale $\lambda=1/k$  of variation of the wave,  then  a ``wave-like solution'' can be found through the WKB method for the pressure perturbation $\delta p$ \,\cite{Gough2007}.}.  Our VEV's actually refer to that adiabatic vacuum. The Bunch-Davies vacuum mentioned above was a particular form of adiabatic vacuum for the case of the de Sitter space.  In the general case and in the absence of a clear-cut particle interpretation,  a more physical approach to the vacuum effects of the expanding universe can be obtained by computing  the vacuum part of the EMT of the scalar field in the cosmologically expanding background.  To accomplish this task, we need to insert the above Fourier expansions in \eqref{EMTScalarField} and compute the VEV in Fourier space, hence integrating over all modes,  $\int\frac{d^3k}{(2\pi)^3}(...) $.  In the process we must use the  expansion \eqref{eq:WKBexpansion} in order to compute the explicit form of the modes \eqref{eq:phaseIntegral},  and this  yields UV-divergent integrals up to fourth adiabatic order.  Therefore, at this point we need to renormalize the VEV of the EMT (sometimes referred to here as `vacuum EMT')  by appropriately subtracting the first four  (UV-divergent) adiabatic orders. The  orders higher than $4$  decay sufficiently quick at large momentum $k$ (short-distances)  so as to make the corresponding integrals convergent. This is a reflex  of the Appelquist-Carazzone decoupling theorem\,\cite{AppelquistCarazzone75}.

Sometimes the following notation is used in the literature. Let $T$ be a dimensionless parameter and let us replace $\Omega_k\to T\Omega_k$. Then the above series \eqref{eq:WKBexpansion} can be regarded as an expansion in $T^{-2}$ for $T\rightarrow\infty$.  The power of $T^{-1}$ defines the adiabaticity order. Upon rescaling, this is equivalent to replace $a(\tau)\to a(\tau/T)$. Then the derivatives of $\Omega_k(a(\tau))$ with respect to $\tau$ all go to $0$ for  $T\rightarrow\infty$.  The number of derivatives coincides with the power of $T^{-1}$, i.e. the adiabaticity order.  This shows that the condition of validity of the expansion is that $\Omega_k(a)$ varies slowly in time, whence the name adiabatic expansion.  In practice we need not keep $T$ explicitly, it suffices to count the number of time derivatives of the various terms of the expansion. But the power $T^{-N}$ of a given term serves as a practical book-keeping device to identify the $Nth$ adiabaticity order of such a term.

\subsection{WKB expansion of the mode functions: systematics of the procedure}\label{sec:WKB}

 In the gravitational context, the  WKB approximation  is organized through the mentioned adiabatic orders and constitutes the basis for the  adiabatic regularization procedure (ARP)\,\cite{ParkerFulling1974,FullingParker1974,BunchParker1979} and the definition of the adiabatic vacuum \cite{Bunch1980} in QFT in curved spacetime.  For a review, see e.g.  the classic books \cite{BirrellDavies82,ParkerToms09}.  In the present study we make a systematic use of it by extending our previous work\,\cite{CristianJoan2020} up to $6th$ adiabatic order, i.e. ${\cal O}(T^{-6})$.  Notice that the adiabatic expansion is an asymptotic expansion, and therefore going to higher and higher orders (which becomes extremely cumbersome in practice) does not necessarily imply a degree of better convergence of the series.  Expanding up to $6th$ adiabatic order is already cumbersome, but it is feasible and actually necessary for the study of the on-shell renormalized theory and other properties of the quantum vacuum, as we shall see.   As noted above, the counting of adiabatic orders follows in most cases the number of time derivatives, so it goes as follows:  $k^2$ and $a$ are taken of adiabatic order $0$;  $a^\prime$ and $\mathcal{H}=a'/a$  of adiabatic order 1;  $a^{\prime \prime},a^{\prime 2},\mathcal{H}^\prime$ and $\mathcal{H}^2$ as well as $R$, are of adiabatic order $2$. Each additional derivative increases the adiabatic order  by one unit.

The form \eqref{eq:WKBexpansion} of the  WKB expansion was useful to discuss the meaning of the adiabatic expansion.  Let us now rewrite it in a way in which we collect the different adiabatic orders:
\begin{equation}\label{WKB}
W_k=\omega_k^{(0)}+\omega_k^{(2)}+\omega_k^{(4)}+\omega_k^{(6)}+\cdots,
\end{equation}
where the various $\omega_k^{(j)}$ are corrections of adiabatic order $j$.  The non-appearance of odd adiabatic orders can be explained by the general covariance, which forbid tensors with an odd number of derivatives of the scale factor in the effective action and  gravitational field equations.  The  $\omega_k^{(j)}$ can be expressed in terms of $\Omega_k(\tau)$ and its time derivatives.  However,  since $\Omega_k(\tau)$ in our case adopts the explicit form indicated in \eqref{eq:ODEmodefunctions}, with $R$ being of adiabatic order $2$,  to insure that the adiabaticity order is preserved it suffices that the derivatives in the terms on the \textit{r.h.s.} of  \eqref{WKBIteration}  are performed on  $\omega_k(\tau) $ only.  We will see this feature in the formulas given below.

Following \cite{CristianJoan2020} we adhere to an off-shell prescription whereby the frequency $\omega_k$ of a given mode  is defined not at the mass $m$ of the particle but at an arbitrary mass scale $M$ generally different from the physical mass\footnote{We distinguish $M$  from 't Hooft's mass unit $\mu$ in dimensional regularization (DR), which will be used together with $M$ in Sec.\,\ref{HeatKernel}  to regularize the UV divergences of the effective action. The parameter $\mu$ is unphysical and is used in the MS scheme with DR to define the renormalization point.  We should stress, however,  that we do not use such a  scheme at all in our calculation, even if we make some (optional) use of DR in certain parts.   In our physical results, $\mu$  always cancels  out and the final renormalized quantities depend on $M$ only. }:
\be\label{eq:omegaM}
\omega_k\equiv\omega_k(\tau, M)\equiv \sqrt{k^2+a^2(\tau) M^2}\,.
\ee
At the moment we will use just the notation $\omega_k$ to indicate such an off-shell value, and when necessary we will distinguish it from the on-shell one using the forms $\omega_k(M)$ and  $\omega_k(m)$, both being of course functions of $\tau$ (which we will omit to simplify notation).  At this stage the method coincides with the proposal  made in \cite{FerreiroNavarroSalas} (see, however, later on in  Sec.\,\ref{sec:RunningCouplings}).

Upon working out  the second and  fourth terms of \eqref{WKB}, we find
\begin{equation}
\begin{split}
\omega_k^{(0)}&= \omega_k\,,\\
\omega_k^{(2)}&= \frac{a^2 \Delta^2}{2\omega_k}+\frac{a^2 R}{2\omega_k}(\xi-1/6)-\frac{\omega_k^{\prime \prime}}{4\omega_k^2}+\frac{3\omega_k^{\prime 2}}{8\omega_k^3}\,,\\
\omega_k^{(4)}&=-\frac{1}{2\omega_k}\left(\omega_k^{(2)}\right)^2+\frac{\omega_k^{(2)}\omega_k^{\prime \prime}}{4\omega_k^3}-\frac{\omega_k^{(2)\prime\prime}}{4\omega_k^2}-\frac{3\omega_k^{(2)}\omega_k^{\prime 2}}{4\omega_k^4}+\frac{3\omega_k^\prime \omega_k^{(2)\prime}}{4\omega_k^3}\,.
\end{split}\label{WKBexpansions1}
\end{equation}
As in \cite{CristianJoan2020}, the quadratic mass differences  $\Delta^2\equiv m^2-M^2$  must be counted as being of adiabatic order 2 since they appear in the WKB expansion along with other terms of the same adiabatic order\footnote{In Sec.\,\ref{HeatKernel} and Appendix \ref{sec:appendixB}, we provide an alternative justification of this method using the heat-kernel expansion. }.  The on-shell result is recovered for $M=m$, for which  $\Delta = 0$ and corresponds to the usual ARP procedure\,\cite{BirrellDavies82,ParkerToms09}.  We now go one step further to our previous study\,\cite{CristianJoan2020}  by extending  the adiabatic  expansion up to the next nonvanishing  order, i.e. the $6th$-order, which is a rather  bulky contribution. It will be essential to cover different aspects of the current work.  After some computational effort, one finds (with the help of Mathematica\,\cite{Mathematica})\footnote{We are not aware that this result has been previously reported in the literature.}:
\begin{equation}\label{WKBexpansion6thOrder}
\begin{split}
\omega_k^{(6)}&=\frac{\omega_k^{\prime\prime}\omega_k^{(4)}}{4\omega_k^3}-\frac{\omega_k^{\prime\prime}\left(\omega_k^{(2)}\right)^2}{4\omega_k^4}+\frac{\left(\omega_k^{(2)}\right)^{\prime\prime}\omega_k^{(2)}}{4\omega_k^3}-\frac{\left(\omega_k^{(4)}\right)^{\prime\prime}}{4\omega_k^2} -\frac{3 \left(\omega_k^\prime\right)^2\omega_k^{(4)}}{4\omega_k^4}+\frac{9\left(\omega_k^\prime\right)^2\left(\omega_k^{(2)}\right)^2}{8\omega_k^5}\\
&+\frac{3\left(\left(\omega_k^{(2)}\right)^\prime\right)^2}{8\omega_k^3}
 +\frac{3\omega_k^\prime \left(\omega_k^{(4)}\right)^\prime}{4\omega_k^3}-\frac{3\omega_k^\prime \left(\omega_k^{(2)}\right)^\prime\omega_k^{(2)}}{2\omega_k^4}-\frac{\omega_k^{(2)}\omega_k^{(4)}}{\omega_k}\,.
\end{split}
\end{equation}
It is easily checked with the mentioned book-keeping device that the above terms $\omega_k^{(2)}$, $\omega_k^{(4)}$ and $\omega_k^{(6)}$ (and each component piece in them) are indeed of ${\cal O}(T^{-2})$, ${\cal O}(T^{-4})$  and ${\cal O}(T^{-6})$ , respectively.
For simplicity we have presented here the above results in terms of the previous orders and their derivatives with respect to the conformal time.
The first two derivatives of $\omega_k$  read
\begin{equation}\label{omegak0}
\omega_k^\prime=a^2\mathcal{H}\frac{M^2}{\omega_k}, \qquad\omega_k^{\prime \prime}=2a^2\mathcal{H}^2\frac{M^2}{\omega_k}+a^2\mathcal{H}^\prime \frac{M^2}{\omega_k}-a^4\mathcal{H}^2\frac{M^4}{\omega_k^3}\,.
\end{equation}
From these elementary differentiations one can then compute the more laborious derivatives appearing in the above expressions, such as $ \left(\omega_k^{(2)}\right)^\prime, \left(\omega_k^{(2)}\right)^{\prime\prime}, \left(\omega_k^{(4)}\right)^\prime, \left(\omega_k^{(4)}\right)^{\prime\prime}$, etc.
The explicit form with all of the terms after computing the various derivatives  and expanding the products and powers of the different terms leads to a rather formidable output. We  refrain from quoting it here, but of course it will be used for the computation of the EMT up to ${\cal O}(T^{-6})$.  One can see immediately that the adiabatic expansion becomes an expansion in powers of $\mathcal{H}$ and its time derivatives.  This is a noticeable property which will be of paramount importance for our considerations. Notice that if the final formulas for the physical quantity (in our case the ZPE) are written in terms of the ordinary Hubble function, $H(t)$, no factor of $a$ can remain.  All the terms with $n$ cosmic time derivatives of the scale factor in different ways are of adiabatic order $N$.  For example, for $N=4$ one can have, in principle, $5$ possible combinations:  $H^4, \dot{H}^2,\, H^2 \dot{H},\vardot{3}{H}$  and  $ H\ddot{H}$, all of them being ${\cal O}(T^{-4})$;  and for $N=6$  we can have $11$ structures of order  ${\cal O}(T^{-6})$, to wit:  $H^6, H^4\dot{H}, \dot{H}^3, H^3\ddot{H}, H^2\vardot{3}{H}, \dot{H}\vardot{3}{H},\ddot{H}^2, H\vardot{4}{H}, H^2\dot{H}^2, H\dot{H}\ddot{H}$ and $\vardot{5}{H}$.  We shall find explicitly all  the actual terms. Somewhat unexpectedly, though, we will find that terms of a given order in the list do not show up in the final result. So the correct adiabaticity order is a necessary but not a sufficient condition to appear in the final result.

\section{Adiabatic expansion of the ZPE up to $4th$  and  $6th$ orders}\label{eq:RegZPE}

We have now all the necessary ingredients  to compute the ZPE associated to the quantum vacuum fluctuations in curved spacetime with FLRW metric.  We closely follow the presentation of \cite{CristianJoan2020}, but we will take into account that  the field modes will be expanded up to $6th$ order, not just up to $4th$ order.  The starting procedure is the same,  we can insert the decomposition (\ref{ExpansionField}) of the quantum field $\phi$ in the  EMT as given in Eq.\,\eqref{EMTScalarField} and select only the fluctuating parts  $\delta\phi$ decomposed as  in (\ref{ExpansionField}).  However, we are interested just on the contribution from the fluctuations, so we pick out the quadratic fluctuations in  $\delta\phi$ only since, as previously indicated, we have zero VEV for the fluctuation itself.  By the same token,  the crossed terms with the background part $\phi_b$ and the fluctuation $\delta\phi$ vanish.   The ZPE is obtained from just the  $00th$-component,  so we find
\begin{equation}\label{EMTInTermsOfDeltaPhi}
\langle T_{00}^{\delta \phi}\rangle =\left\langle \frac{1}{2}\left(\delta\phi^{\prime}\right)^2+\left(\frac{1}{2}-2\xi\right)\left(\nabla\delta \phi\right)^2+6\xi\mathcal{H}\delta \phi \delta \phi^\prime
-2\xi\delta\phi\,\nabla^2\delta\phi+3\xi\mathcal{H}^2\delta\phi^2+\frac{a^2m^2}{2}(\delta\phi)^2 \right\rangle\,.
\end{equation}
Notice that $\delta\phi'$, the fluctuation of the differentiated field (with respect to conformal time), is equal to the derivative of the fluctuating field, i.e. $\delta\phi^{\prime}\equiv\delta\partial_0\phi= \partial_0\delta\phi=(\delta\phi)'$. Next, we substitute the Fourier expansion of $\delta\phi=\delta\varphi/a$, as given in \eqref{FourierModesFluc},  into Eq.\,\eqref{EMTInTermsOfDeltaPhi} and use the commutation relations \eqref{CommutationRelation}. At the same time we  symmetrize  the operator field products $\delta\phi \delta\phi^\prime$ with respect to the creation and annihilation operators.  We present the final result in Fourier space, and hence  we integrate  $\int\frac{d^3k}{(2\pi)^3}(...) $ over solid angles:
\begin{equation}\label{T00}
\begin{split}
\langle T_{00}^{\delta \phi}\rangle=&\frac{1}{4\pi^2 a^2}\int dk k^2 \left[ \left|h_k^\prime\right|^2+(\omega_k^2+a^2\Delta^2)\left|h_k\right|^2\right.\\
&+\left.\left(\xi-\frac{1}{6}\right)\left(-6\mathcal{H}^2\left|h_k\right|^2+6\mathcal{H}\left(h_k^\prime h_k^{*}+h_k^{*\prime}h_k\right)\right)\right]\,,
\end{split}
\end{equation}
where the remaining integrals are over  $k=|\bk|$.
Using now the WKB approximations  (\ref{WKBexpansions1}) and  (\ref{WKBexpansion6thOrder}), we expand the various terms of the above integral consistently up to $6th$ order. After some tedious calculations, we find
\begin{equation}\label{eq:exphk2}
\begin{split}
&|h_k|^2=\frac{1}{2W_k}=\frac{1}{2\omega_k}-\frac{\omega_k^{(2)}}{2\omega_k^2}-\frac{\omega_k^{(4)}}{2\omega_k^2}-\frac{\omega_k^{(6)}}{2\omega_k^2}+\frac{\left(\omega_k^{(2)}\right)^2}{2\omega_k^3}
+\frac{\omega_k^{(2)}\omega_k^{(4)}}{\omega_k^3}-\frac{\left(\omega_k^{(2)}\right)^3}{2\omega_k^4}\,,\\
\end{split}
\end{equation}\label{eq:exphk2b}
\begin{equation}
\begin{split}
\left| h_k^\prime \right|^2&=\frac{\left(W_k^\prime\right)^2}{8W_k^3}+\frac{W_k}{2}=\frac{\omega_k}{2}\left(1 +\frac{\omega_k^{(2)}}{\omega_k} + \frac{\omega_k^{(4)}}{\omega_k} +\frac{\omega_k^{(6)}}{\omega_k}\right) +\frac{\left(\omega_k^\prime\right)^2}{8\omega_k^3}\left(1 -\frac{3\omega_k^{(2)}}{\omega_k} -  \frac{3\omega_k^{(4)}}{\omega_k} + 6\frac{\left(\omega_k^{(2)}\right)^2}{\omega_k^2}\right)\\
&+\frac{\left(\left(\omega_k^{(2)}\right)^\prime\right)^2}{8\omega_k^3}
+ \frac{\left(\omega_k^{(2)}\right)^\prime\omega_k^\prime}{4\omega_k^3}\left(1 -\frac{3\omega_k^{(2)}}{\omega_k}\right) +\frac{\left(\omega_k^{(4)}\right)^\prime\omega_k^\prime}{4\omega_k^3}\,,
\end{split}
\end{equation}
\begin{equation}\label{eq:exphk2c}
\begin{split}
\begin{split}
 h_k^\prime h_k^*+\left(h_k^*\right)^\prime h_k=& -\frac{W_k^\prime}{2W_k^2}=-\frac{\omega_k^\prime}{2\omega_k^2}\left(1 - \frac{2\omega_k^{(2)}}{\omega_k} -\frac{2\omega_k^{(4)}}{\omega_k} + \frac{3\left(\omega_k^{(2)}\right)^2}{\omega_k^2}\right)\\
&-\frac{\left(\omega_k^{(2)}\right)^\prime}{2\omega_k^2}\left(1 - \frac{2\omega_k^{(2)}}{\omega_k} \right)  -\frac{\left(\omega_k^{(4)}\right)^\prime}{2\omega_k^2}\,.
\end{split}
\end{split}
\end{equation}
In view of these explicit results it is obvious that the VEV \eqref{T00} is UV-divergent, specifically the integrals   $\int dk k^2  \left|h_k^\prime\right|^2$ and $\int dk k^2  \omega_k^2\left|h_k\right|^2$  in it are both quartically divergent, $\int dk k^2  \left|h_k\right|^2$ is quadratically divergent and $\int dk k^2  \left(h_k^\prime h_k^{*}+h_k^{*\prime}h_k\right)$ is  logarithmically divergent. No terms can be left in the EMT being linear in $\cal H$,  nor any odd power of it, as they would violate the covariance of the result.  Only even powers of $\cH$ can remain in the final result (strictly speaking, terms with an even number of derivatives of the scale factor), as we shall further  reconfirm below.

 With the help of \eqref{omegak0} these expressions can be made more explicit before being inserted in \eqref{T00}, but it demands a significant amount of algebra.
 The result can be conveniently split into the various contribution up to $6th$ adiabatic order  (plus higher orders, if necessary, but not in our case):
 \begin{equation}\label{eq:T00expansion}
 T_{00}^{\delta \phi} =T_{00}^{\delta \phi (0)}+T_{00}^{\delta \phi (2)} +T_{00}^{\delta \phi (4)} +T_{00}^{\delta \phi (6)} +... =T_{00}^{\delta \phi (0-4)} +T_{00}^{\delta \phi (6)} +...
 \end{equation}
 where for convenience we have collected the contribution from the terms up to $4th$ adiabatic order in the expression
 $T_{00}^{\delta \phi (0-4)}\equiv  T_{00}^{\delta \phi (0)}+T_{00}^{\delta \phi (2)} +T_{00}^{\delta \phi (4)}$.

 Explicitly\,\cite{CristianJoan2020}:
\begin{equation*}
\begin{split}
\langle T_{00}^{\delta \phi} \rangle^{ (0-4)} & =\frac{1}{8\pi^2 a^2}\int dk k^2 \left[ 2\omega_k+\frac{a^4M^4 \mathcal{H}^2}{4\omega_k^5}-\frac{a^4 M^4}{16 \omega_k^7}(2\mathcal{H}^{\prime\prime}\mathcal{H}-\mathcal{H}^{\prime 2}+8 \mathcal{H}^\prime \mathcal{H}^2+4\mathcal{H}^4)\right.\\
&+\frac{7a^6 M^6}{8 \omega_k^9}(\mathcal{H}^\prime \mathcal{H}^2+2\mathcal{H}^4) -\frac{105 a^8 M^8 \mathcal{H}^4}{64 \omega_k^{11}}\\
&+\left(\xi-\frac{1}{6}\right)\left(-\frac{6\mathcal{H}^2}{\omega_k}-\frac{6 a^2 M^2\mathcal{H}^2}{\omega_k^3}+\frac{a^2 M^2}{2\omega_k^5}(6\mathcal{H}^{\prime \prime}\mathcal{H}-3\mathcal{H}^{\prime 2}+12\mathcal{H}^\prime \mathcal{H}^2)\right. \\
& \left. -\frac{a^4 M^4}{8\omega_k^7}(120 \mathcal{H}^\prime \mathcal{H}^2 +210 \mathcal{H}^4)+\frac{105a^6 M^6 \mathcal{H}^4}{4\omega_k^9}\right)\\
\end{split}
\end{equation*}
\begin{equation}\label{EMTFluctuations}
\begin{split}
&+\left. \left(\xi-\frac{1}{6}\right)^2\left(-\frac{1}{4\omega_k^3}(72\mathcal{H}^{\prime\prime}\mathcal{H}-36\mathcal{H}^{\prime 2}-108\mathcal{H}^4)+\frac{54a^2M^2}{\omega_k^5}(\mathcal{H}^\prime \mathcal{H}^2+\mathcal{H}^4) \right)
\right]\\
&+\frac{1}{8\pi^2 a^2} \int dk k^2 \left[  \frac{a^2\Delta^2}{\omega_k} -\frac{a^4 \Delta^4}{4\omega_k^3}+\frac{a^4 \mathcal{H}^2 M^2 \Delta^2}{2\omega_k^5}-\frac{5}{8}\frac{a^6\mathcal{H}^2 M^4\Delta^2}{\omega_k^7} \right.\\
& \left. +\left( \xi-\frac{1}{6} \right) \left(-\frac{3a^2\Delta^2 \mathcal{H}^2}{\omega_k^3}+\frac{9a^4 M^2 \Delta^2 \mathcal{H}^2}{\omega_k^5}\right)\right]\,.
\end{split}
\end{equation}
As expected, only even powers of $\cal H$ remain in the final result.
We point out the appearance of the  $\Delta$-dependent terms in the last two rows, which contribute at second ($\Delta^2$) and fourth ($\Delta^4$) adiabatic order, according to the mentioned rule.  Mind that $k$ in the above formulas  is the comoving momentum, whereas the physical momentum is $\tilde{k}=k/a$.

Suppose we  fix the scale $M$ at the physical mass of the particle ($M=m$), so that the $\Delta$-terms vanish. The first two adiabatic orders  $T_{00}^{\delta \phi (0-2)}\equiv  T_{00}^{\delta \phi (0)}+T_{00}^{\delta \phi (2)}$  can be easily identified:
\begin{equation}\label{eq:T002}
  \left.\langle T_{00}^{\delta \phi}\rangle^{ (0-2)}\right|_{M=m}  =\frac{1}{8\pi^2 a^2}\int dk k^2 \left[ 2\omega_k(m)+\frac{a^4m^4 \mathcal{H}^2}{4\omega_k^5(m)}
-\left(\xi-\frac{1}{6}\right)\left(\frac{6\mathcal{H}^2}{\omega_k(m)}+\frac{6 a^2 m^2\mathcal{H}^2}{\omega_k^3(m)}\right)\right]\,,
\end{equation}
where   $\omega_k(m)\equiv \sqrt{k^2+a^2 m^2}$. Let us next project  the UV-divergent terms  of this formula only  and assume that the nonminimal coupling to gravity is absent ($\xi=0$).  We are then left with
\begin{equation}\label{eq:LowOrderZPE1}
  \left.\langle T_{00}^{\delta \phi}\rangle^{ (0-2)}\right|_{(M=m, \xi=0)}^{\rm UV}=\frac{1}{8\pi^2 a^2}\int dk k^2\left( 2\omega_k(m)+\frac{\cH^2}{\omega_k(m)}+\frac{a^2m^2 \cH^2}{\omega_k^3(m)}\right)\,.
\end{equation}
Finally, the  Minkowskian spacetime result is obtained for $a=1$ ($\mathcal{H}=0)$:
\begin{equation}\label{eq:Minkoski}
 \left.\langle T_{00}^{\delta \phi}\rangle^{\rm Mink}\right|_{(M=m, \xi=0)}=\frac{1}{4\pi^2}\int dk k^2 \omega_k =  \int\frac{d^3k}{(2\pi)^3}\,\left(\frac12\,\hbar\,\omega_k\right)\,,
\end{equation}
where $\hbar$ has been restored only in the trailing term for a better identification of the result. The last quantity  is the vacuum energy density  of the quantum fluctuations in flat spacetime, i.e. the  ZPE in Minkowski spacetime. It is of course the traditional contribution found in usual calculations. It is quartically UV-divergent.  Usual attempts to regularize and renormalize this result by e.g.  cancelling the corresponding UV-divergence against the bare $\rL$ term  in the action \eqref{eq:EH} within the context of  a simple cutoff method or  appealing to the Pauli-Villars, more formal, procedure; or even using the MS scheme and related ones,  leads in all these cases to the well-known ugly fine-tuning  problem inherent to the CCP, see Sec.\,\ref{sec:VEDMSS} for a summarized discussion.  We will certainly not proceed in this way  here. We seek out (and will find) an alternative way  for renormalizing the above result \eqref{EMTFluctuations} in  its full general form\,\footnote{Let us  note that Supersymmetry is not sufficiently helpful for solving the CCP since the cancellation of quartic divergences (warranted e.g. in the Wess-Zumino model\,\cite{Zumino1975}, cf. also \cite{Barvinsky2018}) does not guarantee the cancellation of the subleading ones, e.g. the quadratic divergences\,\cite{Bilic2011}.  The quadratic parts are of the form $\Lambda_c^2 H^2$ (where $\Lambda_c$ can serve as a UV cutoff). See \cite{Maggiore2011,Maggiore2011b,Visser2018,Donoghue2021} for  a discussion in nonsupersymmetric contexts.  For $\Lambda_c$ around the Planck mass,  it can be phenomenological acceptable only  if  $\Lambda_c^2 H^2$  carries a small coefficient, as it was noticed much earlier in \cite{ShapSol1,Fossil2008}.  The current calculation and the companion one\,\cite{CristianJoan2020} substantiate these results for the first time in a rigorous QFT context, see Sec.\,\ref{sec:RenormalizedVED}. }.
The previous simpler formulas, including the more complicated ones  are  UV divergent and require appropriate regularization and renormalization.  The result \,(\ref{EMTFluctuations}) constitutes the WKB approximation up to $4th$ adiabatic order.  It is enough to encompass all the UV-divergences that appear in the WKB expansion of the ZPE.  However, we need to continue such an expansion one more step since we want to compute the on-shell value of the ZPE and, as it will be clear in the next section,  the effort  is necessary.

Thus, we now move on to the calculation of the $6th$-order contribution,  $T_{00}^{\delta \phi (6)}$,  which is more cumbersome than the contributions up to  $4th$-order, Eq.\,\eqref{EMTFluctuations} .  We will quote the expression only at the on-shell point $M=m$ (so all of  the terms proportional to $\Delta$ vanish in this case). There is no need to compute $\langle T_{00}^{\delta \phi }\rangle^{(6)}$ at an arbitrary scale $M$ since no subtraction is needed for a contribution which is fully convergent, piece by piece.  In Fourier space, it reads as follows:
%
\begin{equation*}
\begin{split}
& \langle T_{00}^{\delta \phi} \rangle^{\rm (6)}(m)=\frac{1}{4\pi^2 a^2}\int dk k^2\left[ \frac{a^4 m^4}{128\omega_k^9}\Bigg(16\mathcal{H}^6 +96 \mathcal{H}^4 \mathcal{H}^\prime+84 \mathcal{H}^2\left(\mathcal{H}^\prime\right)^2-12 \left(\mathcal{H}^\prime\right)^3\right.\\
&+56\mathcal{H}^3\mathcal{H}^{\prime\prime}+36\mathcal{H}\mathcal{H}^\prime\mathcal{H}^{\prime\prime}+\left(\mathcal{H}^{\prime\prime}\right)^2+16\mathcal{H}^2 \mathcal{H}^{\prime\prime\prime}-2 \mathcal{H}^\prime \mathcal{H}^{\prime\prime\prime}+2\mathcal{H}\mathcal{H}^{\prime\prime\prime\prime}\Bigg)\\
&+\frac{a^6m^6}{32\omega_k^{11}}\Bigg( -152 \mathcal{H}^6 -396 \mathcal{H}^4\mathcal{H}^\prime -114 \mathcal{H}^2\left(\mathcal{H}^\prime\right)^2 +5 \left(\mathcal{H}^\prime\right)^3-102 \mathcal{H}^3\mathcal{H}^{\prime\prime}-15 \mathcal{H}\mathcal{H}^\prime \mathcal{H}^{\prime\prime}-9 \mathcal{H}^2 \mathcal{H}^{\prime\prime\prime} \Bigg)\\
&\left.+\frac{33a^8m^8}{256\omega_k^{13}}\left(212  \mathcal{H}^6 + 264\mathcal{H}^4 \mathcal{H}^\prime+27  \mathcal{H}^2 \left(\mathcal{H}^\prime\right)^2+26  \mathcal{H}^3 \mathcal{H}^{\prime\prime}\right) -\frac{3003a^{10}m^{10}}{128 \omega_k^{15}}\left(2 \mathcal{H}^6 + \mathcal{H}^4 \mathcal{H}^\prime\right)+\frac{25025 a^{12}\mathcal{H}^6 m^{12}}{1024 \omega_k^{17}}\right]
\end{split}
\end{equation*}
\begin{equation}
\begin{split}
&+\frac{\left(\xi-\frac{1}{6}\right)}{4\pi^2 a^2}\int dk k^2 \Bigg[\frac{3a^2m^2}{16\omega_k^7}\Bigg(-32\mathcal{H}^4\mathcal{H}^\prime -32 \mathcal{H}^2 \left(\mathcal{H}^\prime\right)^2 +8 \left(\mathcal{H}^\prime\right)^3-24\mathcal{H}^3\mathcal{H}^{\prime\prime}-24 \mathcal{H}\mathcal{H}^\prime \mathcal{H}^{\prime\prime} - \left(\mathcal{H}^{\prime\prime}\right)^2\nonumber\\
&-8  \mathcal{H}^2\mathcal{H}^{\prime\prime\prime}+2\mathcal{H}^\prime\mathcal{H}^{\prime\prime\prime}-2  \mathcal{H}\mathcal{H}^{\prime\prime\prime\prime}\Bigg)\\
&+\frac{21a^4m^4}{32\omega_k^9}\Bigg(76  \mathcal{H}^6 +232 \mathcal{H}^4 \mathcal{H}^\prime +73  \mathcal{H}^2\left(\mathcal{H}^\prime\right)^2-4\left(\mathcal{H}^\prime\right)^3+74 \mathcal{H}^3 \mathcal{H}^{\prime\prime}+12 \mathcal{H}\mathcal{H}^\prime\mathcal{H}^{\prime\prime}+8 \mathcal{H}^2 \mathcal{H}^{\prime\prime\prime}\Bigg)\\
\end{split}
\end{equation}
\begin{equation}\label{Tsixthhorder}
\begin{split}
&-\frac{63a^6m^6}{16\omega_k^{11}}\Bigg(92 \mathcal{H}^6 +123 \mathcal{H}^4\mathcal{H}^\prime+13  \mathcal{H}^2 \left(\mathcal{H}^\prime\right)^2 +14 \mathcal{H}^3 \mathcal{H}^{\prime\prime}\Bigg)+\frac{693a^8m^8}{128\omega_k^{13}}\left(123 \mathcal{H}^6+64  \mathcal{H}^4 \mathcal{H}^\prime\right)-\frac{45045a^{10}m^{10}\mathcal{H}^6}{128\omega_k^{15}}\Bigg]\\
&+\frac{\left(\xi-\frac{1}{6}\right)^2}{4\pi^2 a^2}\int dk k^2 \Bigg[\frac{9}{8\omega_k^5}\Bigg(-4\mathcal{H}^2\left(\mathcal{H}^\prime\right)^2-4\left(\mathcal{H}^\prime\right)^3-8\mathcal{H}^3\mathcal{H}^{\prime\prime}+12 \mathcal{H}\mathcal{H}^\prime \mathcal{H}^{\prime\prime}+ \left(\mathcal{H}^{\prime\prime}\right)^2-2 \mathcal{H}^\prime \mathcal{H}^{\prime\prime\prime}+2 \mathcal{H}\mathcal{H}^{\prime\prime\prime\prime}\Bigg)\\
&+\frac{45a^2 m^2}{4\omega_k^7}\Bigg(-8  \mathcal{H}^4 \mathcal{H}^\prime -9  \mathcal{H}^2 \left(\mathcal{H}^\prime\right)^2 +  \left(\mathcal{H}^\prime \right)^3-8  \mathcal{H}^3 \mathcal{H}^{\prime\prime}-3 \mathcal{H}\mathcal{H}^\prime\mathcal{H}^{\prime\prime}-2 \mathcal{H}^2 \mathcal{H}^{\prime\prime\prime}\Bigg)\\
&+\frac{315a^4m^4}{16\omega_k^{9}}\left(13 \mathcal{H}^6 +32 \mathcal{H}^4 \mathcal{H}^\prime +7  \mathcal{H}^2 \left(\mathcal{H}^\prime\right)^2+6  \mathcal{H}^3\mathcal{H}^{\prime\prime}\right)-\frac{2835a^6m^6}{8 \omega_k^{11}}\mathcal{H}^4\left(\mathcal{H}^2+\mathcal{H}^\prime\right)\Bigg]\\
&+\frac{\left(\xi-\frac{1}{6}\right)^3}{4\pi^2 a^2}\int dk k^2 \Bigg[\frac{9}{2\omega_k^5}\left(-15\mathcal{H}^6+9\mathcal{H}^2\left(\mathcal{H}^\prime\right)^2-6\left(\mathcal{H}^\prime\right)^3+18\mathcal{H}^3\mathcal{H}^{\prime\prime}
+18\mathcal{H}\mathcal{H}^\prime\mathcal{H}^{\prime\prime}\right)\nonumber\\
&-\frac{405a^2m^2}{2\omega_k^7}\left(\mathcal{H}^6+2\mathcal{H}^4 \mathcal{H}^\prime +\mathcal{H}^2\left(\mathcal{H}^\prime\right)^2\right)\Bigg]\,.\hspace{8cm} (4.11)
\end{split}
\end{equation}
Some of the integrals in \eqref{EMTFluctuations} are UV-divergent, as we have seen, and others are convergent.  The integrals in \eqref{Tsixthhorder}, instead,  are all convergent. One may compute/regularize every single integral in these formulas (convergent or divergent)  using the master formula for DR in Appendix \ref{sec:appendixA2}.


\section{Renormalization of the energy-momentum tensor  in curved spacetime}\label{sec:RenormEMT}

We may compare the evolving vacuum energy density (VED) of cosmological spacetime with a Casimir device wherein the parallel plates slowly move apart (“expand”)\,\cite{JSPRev2014}.
While the  total vacuum energy density cannot be measured, the `differential'  effect associated to the presence of the plates, and then also to their increasing separation with time, it can. Similarly,  in the expanding FLRW spacetime there is a genuine nonvanishing spacetime curvature, $R$, as compared to Minkowskian spacetime  and such a curvature is changing with the expansion. The VED must vary accordingly and we naturally expect that  there is a contribution proportional to $R$, hence to $H^2$ and $\dot{H}$  (plus  higher derivative (HD) effects  $R^2$, $R^{\mu\nu}R_{\mu\nu}$, etc. in the early universe).  Both spacetimes, Minkowski and FLRW,  are obviously similar at short distances, in the sense that the curved background is locally flat.  However, the short distance singularities are not really  identical since the curvature carries additional ones related to the nontrivial geometric structures.

More formally, the energy-momentum tensor (EMT), Eq.\,\eqref{EMTScalarField}, is a quadratic functional of the field $\phi(x)$. However, in the context of  QFT,  $\phi(x)$ is an operator-valued distribution and hence terms like $g_{\mu\nu} m^2\phi^2(x)$, $ \partial_\mu \phi(x) \partial_\nu\phi(x)$,  etc. in EMT  are  not well defined at a given point $x$ since a square of a distribution is not generally defined. This is ultimately the source of the UV-divergences of QFT in configuration space.  For this reason it is advisable to consider  a bilinear functional replacing the original  EMT, which we may denote as  $T_{\mu\nu}(x,x')=T_{\mu\nu}(\phi(x)\phi(x'))$, where a point-splitting has been operated in order to avoid the UV-divergence\,\cite{BirrellDavies82,ParkerToms09,Fulling89}.  The coincidence point limit in configuration space amounts, of course, to the UV-limit in momentum space.  In practice, we need the VEV of that bilinear functional and the point splitting regularization of $\langle T_{\mu\nu}(x,x')\rangle$  is carried out through a (differential)  operator ${D}_{\mu\nu}$ acting on an appropriate  two-point (Green's, Hadamard's, etc.)  function $G(x,x')$ as follows: $\langle T_{\mu\nu}(x,y)\rangle={D}_{\mu\nu} G(x,y)$.  The operator ${ D}_{\mu\nu}$ can be easily identified from the terms involved in \eqref{EMTScalarField} and is usually expressed in a symmetrized form.  For instance, the VEV of the first term on the \textit{r.h.s.} of \eqref{EMTScalarField} is treated as
\be\label{eq:pointsplit}
 (1-2\xi)\langle \nabla_\mu \phi \nabla_\nu\phi\rangle\longrightarrow (1-2\xi)\frac12\left(\nabla_\mu\nabla_{\nu'}+\nabla_{\mu'}\nabla_{\nu}\right)  G(x,x')\,,
\ee
where the derivatives with primed indices are assumed to act on $x'$ and those without primes on $x$.  In the jargon of QFT, this part would be regularization.
The renormalization of the EMT is then performed by subtracting the vacuum expectation value  through the  coincidence limit $x'\to x$.  In the simplest case of Minkowski space, with Minkowskian vacuum $|0\rangle^{\rm Mink}$,  it would be natural to define the renormalized EMT operator as
\begin{equation}\label{eq:RenormEMTpointsplit}
T_{\mu\nu}(x)= \lim\limits_{x'\to x} \left[T_{\mu\nu} (x,x')-\langle  0|T_{\mu\nu}(x,x')|0\rangle^{\rm Mink}\right]\,.
\end{equation}
since in this case the VEV of the renormalized EMT is expected to be zero for sound physical reasons.  In our  Casimir example, the short-distance behavior in the region between the plates is the same as that outside the plates and the limit gives a finite result.  However, as warned above, curved spacetime induces new types of infinities as compared to Minkowskian spacetime.  The latter, however, are still there and  may still carry the core of the quantum vacuum problem if the Minkowskian result is not renormalized to zero (cf. Sec.\,\ref{sec:VEDMSS}).

 The generalization of \eqref{eq:RenormEMTpointsplit} in curved spacetime is more delicate, but under appropriate conditions  it is natural to use a similar definition where we replace the Minkowskian vacuum $|0\rangle^{\rm Mink}$ with  the adiabatic vacuum, simply denoted $|0\rangle$ as we have been doing in the previous sections.  We may define  the renormalized EMT operator performing a suitable subtraction, but in this case we should not presume a zero result for the VEV of the renormalized EMT. We would rather extract the nonvanishing renormalized vacuum energy density and pressure in curved spacetime as a function of the background itself,  in such a way that when the background is Minkowskian we ought to recover the previous vanishing VEV.  There are, however, some additional specifications  to handle correctly the UV-divergences.  Moreover, we wish to provide an off-shell definition enabling us to explore the VED at different scales. Thereby we define the renormalized EMT operator in  $n$-dimensional curved spacetime (with $n-1$ spatial dimensions)  up to adiabatic order $N\geq n$  through the following off-shell subtraction prescription (which we shall refer to also as the off-shell ARP):
 \begin{equation}\label{eq:RenormEMToperator}
 \begin{split}
T^{(0-N)}_{\mu\nu}(x)_{\rm ren}(M) =   T^{(0-N)}_{\mu\nu}(x)(m) - \langle  0| T^{(0-n)}_{\mu\nu}(x)|0\rangle (M)\,.
\end{split}
\end{equation}
In this equation, $ T^{(0-N)}_{\mu\nu}(x)(M)$ refers to the computation of the renormalized ETM  to adiabatic order $N\geq n$  at the scale $M$ (not necessarily equal to the on-shell mass value $m$), whereas $ \langle  0| T^{(0-n)}_{\mu\nu}(x)|0\rangle (M)$ is the VEV of the EMT computed up to adiabatic order $n$ (the dimension of spacetime, i.e. $n=4$ in our context).  The on-shell value is just $T^{(0-N)}_{\mu\nu}(x)(m)$, of course.  The subtraction is, therefore,  performed upon  that on-shell  value. By virtue of general covariance, the adiabatic orders involved in the EMT must be even ($N=0,2,4,6,...$).  However,  irrespective of the adiabaticity order $N$ at which the on-shell value  is computed, the subtracted quantity at the scale $M$,  i.e. $T^{(0-n)}_{\mu\nu}(x)(M)$, must include just the first $\frac{n}{2}+1$ (nonvanishing) even orders $N=0,2,4\cdots n$, as these are the only ones  which are UV-divergent (in $n$ spacetime dimensions). In $n=4$, this means that $T^{(0-4)}_{\mu\nu}(x)(M)$ must contain the first three even adiabatic orders $N=0,2,4$.

We have mentioned  point-splitting regularization\,\cite{DeWitt1975}, see also\,\cite{Christensen1976,Christensen1978,BunchDavies1978},  because it  illustrates very clearly the origin of the UV-divergences in QFT computations and because it is in general a consistent, and  physically justified, covariant procedure to define the rnormalized EMT.   Proceeding in this way,  however,  can be rather cumbersome. Indeed, in the general case one has to start from the adiabatic expansion of the Green function $G(x,x')$ and the structure of divergences is not apparent until the mode integral has been performed.  Examples are well described in the literature \,\cite{BirrellDavies82,ParkerToms09,Fulling89}.  Fortunately, however, the ARP procedure defined above  for renormalizing the EMT and other local quantities can be shown to be equivalent to the point-splitting procedure\,\cite{Birrell1978}.  In particular, when the field equation can be solved by separation of variables,  as in the case under study, one can resort to a simpler method of renormalization which is to perform a mode by mode subtraction process under the integral sign using the adiabatic expansion of the modes, Eq.\,\eqref{FourierModes}\,\cite{Fulling89}.  In Sec.\,\ref{Sec:FieldEquations} we have seen that the properly normalized form of these  modes is
 \begin{equation}\label{eq:NormalizedModes}
 u_k(\tau,{\bf x})=(2\pi)^{-3/2}\, a^{-1}(\tau)\,e^{i{\bf k}\cdot{\bf x}}\,\varphi_k(\tau)\,,
 \end{equation}
in which the space and time variables are separated and the time-evolving part $\varphi_k(\tau)$  obeys the nontrivial Eq.\,\eqref{eq:KGFourier}.  Similarly for the equations satisfied by the  fluctuating parts, \eqref{FourierModesFluc} and \eqref{eq:ODEmodefunctions}.
When the field modes can be expressed in separated form it is possible to arrange for the explicit cancellation of UV-divergences before the mode integral is computed. The advantage is clear since the arrangement of terms can be made inside the subtracted integrand such that no UV-divergence is present and the integral appears manifestly convergent \textit{ab initio}.

The VEV of the  $00th$-component of  $ T^{(0-N)}_{\mu\nu}(x)(m)$ in \eqref{eq:RenormEMToperator} is precisely given by Eq.\,\eqref{EMTInTermsOfDeltaPhi}  in our case.
Thus, the renormalized  vacuum EMT up to ${\cal O}(T^{-N})$  in $n=4$ spacetime dimensions reads
 \begin{equation}\label{eq:RenormEMTAdiabatic}
\begin{split}
\langle  0| T^{(0-N)}_{\mu\nu}(x)|0\rangle_{\rm ren}(M) &=   \langle  0|T^{(0-N)}_{\mu\nu}(x)|0\rangle (m) - \langle  0| T^{(0-4)}_{\mu\nu}(x)|0\rangle (M) \,,
\end{split}
\end{equation}
where it is supposed, of course,  that the mode expansion has been performed to ${\cal O}(T^{-N})$.
Since the  EMT structure is made of quadratic expressions of the fields, they are  expanded at that order  in terms of the above mentioned modes \eqref{eq:NormalizedModes} and the creation and annihilation operators, and finally one can move to momentum space by integrating $\int d^3k (...) $ the result.  The detailed computational results of this procedure  have already been given in Sec.\,\ref{eq:RegZPE}. Here we  just discuss the formal procedure and  furnish the practical recipe \eqref{eq:RenormEMTAdiabatic}, which is necessary  to achieve a renormalized finite result using such a mode by mode subtraction at any order.

\subsection{Off-shell renormalization of the EMT}\label{Sec:RenEMToffshell}
 It goes without saying that to call Eq.\,\eqref{eq:RenormEMToperator}  `renormalized' EMT and \eqref{eq:RenormEMTAdiabatic} the renormalized vacuum EMT  is almost unnecessary since the  mode by mode subtraction in the integrand makes the integral manifestly finite.  The ARP procedure (based on the adiabatic expansion) defines automatically the renormalized quantity. However,  as mentioned above, while in the usual adiabatic regularization method\,\cite{BirrellDavies82,ParkerToms09,Fulling89} the subtraction is always performed on-shell,  here  we shall  instead perform the subtraction off-shell, i.e. at a scale $M$ which is generally different from the mass of the particle. This enables  us to test the scale dependence of the renormalized result \eqref{eq:RenormEMTAdiabatic}.
 The explicit calculations which we provided in \cite{CristianJoan2020} concerning the regularization of the ZPE of a nonminimally coupled scalar field  can serve as a very practical illustration of this procedure.  In the present paper we  continue and extend those calculations for the same scalar field model and hence we adopt such a renormalization framework, although we will summarize it below for the sake of convenience of the reader.  No new divergences appear in the present calculation as compared to that of\,\cite{CristianJoan2020}. We shall reconfirm these results here using the effective action approach (cf. Sec. \ref{HeatKernel})  and  will extend the computation by determining the (difficult) $6th$ order adiabatic term in the mode expansion along with the corresponding contribution to the density and pressure of the vacuum. The new terms are perfectly finite, but fairly cumbersome, as we have seen in previous sections (see also  our Appendix \ref{sec:appendixC}).  We will use these  results to extract physical consequences as to the scale dependence of the ZPE and its on-shell value, as well as to derive the equation of state of the quantum vacuum.

But before that, let us stress that the consistency of such a renormalization method has been explicitly verified in our previous work, where we have shown that it is equivalent to the renormalization of coupling constants in Einstein’s equations, see Appendix B of Ref.\,\cite{CristianJoan2020}.  If dimensional regularization (DR)  is used, the needed counterterms to cancel the poles can be generated from the basic three parameters  $G^{-1}$,  $\rho_\Lambda$  and $\alpha$  appearing in the generalized form of Einstein's equations (compare with the original form \eqref{FieldEq2}):
\begin{equation}\label{eq:MEEs}
\frac{1}{8\pi G} G_{\mu \nu}+\rho_\Lambda g_{\mu \nu}+\alpha\ \leftidx{^{(1)}}{\!H}_{\mu \nu}= T_{\mu \nu}\,.
\end{equation}
Here  $\leftidx{^{(1)}}{\!H}_{\mu \nu}$ is the HD tensor  which appears from the metric variation of  the $R^2$-term in the  higher derivative vacuum action for FLRW spacetime.  We remind the reader that  the  term emerging from  the variation of the square of the Ricci tensor, called  $\leftidx{^{(2)}}{\!H}_{\mu \nu}$, is not necessary in our case since it is not independent of $\leftidx{^{(1)}}{\!H}_{\mu \nu}$  for  FLRW spacetimes (confer Appendix \ref{sec:appendixA1}).
All three  couplings  $G^{-1}$,  $\rho_\Lambda$ and $\alpha$  are necessary to generate the counterterms that cancel all the divergences in the regularized EMT:
\begin{equation}\label{eq:splitcounters}
\begin{split}
G^{-1}&= G^{-1}(M)+\delta_\epsilon G^{-1},\\
\rho_\Lambda&=\rho_\Lambda(M)+\delta_\epsilon\rho_\Lambda,\\
\alpha&=\alpha(M)+\delta_\epsilon \alpha.
\end{split}
\end{equation}
The counterterms are denoted with the subscript $\epsilon$ to emphasize that they  depend on the regulator $\epsilon$ and become infinite for $\epsilon\to 0$ (see below and Appendix \ref{sec:appendixA2}).  The subscript is also useful to distinguishing this notation from other quantities introduced in Sec.\,\ref{sec:RenormalizedVED}  which bare some notational resemblance.
The specific forms of the three counterterms mentioned above is\,\cite{CristianJoan2020}:
\begin{equation}\label{eq:counters}
\begin{split}
  \delta_\epsilon G^{-1} &=-\frac{m^2}{2\pi}\,\left(\xi-\frac{1}{6}\right)\, D_\epsilon\,,\\
  \delta_\epsilon\rho_\Lambda &=+\frac{m^4}{64\pi^2}\, D_\epsilon\,,\\
  \delta_\epsilon \alpha &= -\frac{1}{32\pi^2}\,\left(\xi-\frac{1}{6}\right)^2\, D_\epsilon\,,
  \end{split}
\end{equation}
with
\begin{equation}\label{eq:Depsilon}
  D_\epsilon\equiv\frac{1}{\epsilon}-\gamma_E+\ln 4\pi=\frac{2}{4-n}-\gamma_E+\ln 4\pi\,.
\end{equation}
The pole is at $n-1=3$ space  (resp. $n=4$ spacetime) dimensions, where $\epsilon=0$.
No more counterterms are needed in the present calculation.  In particular, we do not need the nonminimal coupling $\xi$ to generate an additional counterterm for the free scalar field theory that we are addressing here. Even so it is useful to keep a nonvanishing value of  $\xi$ in the action \eqref{eq:Sphi} for the general reasons explained in Sec.\,\ref{Sec:FieldEquations} and for more specific ones that we will consider in the coming chapters.
Overall the results obtained using the counterterm method and renormalization of constants in the generalized Einstein's equations  is identical to that of performing the mode by mode subtraction directly in the integrand until evincing the convergent nature of the integrals. We do not furnish more details here on the counterterm method since they are given in full in Ref.\,\,\cite{CristianJoan2020}.  We have just reminded  the reader that in these cases  the two procedures are equivalent.

Following \,\cite{CristianJoan2020},  after  computing  the  adiabatic WKB  expansion of the integrand of the divergent integrals  a subtraction is carried out at an arbitrary scale $M$, i.e. we apply the off-shell ARP\,\eqref{eq:RenormEMTAdiabatic}.
Taking into account that in $4$ spacetime dimensions the only adiabatic orders that are divergent in the case of the EMT are the first four ones, the subtraction at the scale $M$ is performed only up to the fourth adiabatic order. The on-shell value of the EMT can be computed of course at any order,  all terms beyond  $4th$-order being  finite.
Let us apply this procedure to the UV-divergent ZPE  as given by  Eq.\,\eqref{EMTFluctuations}.  The renormalized $00th$-component of the EMT in this context therefore reads\,\footnote{Equation \eqref{EMTRenormalized} implies a subtraction between two UV-divergent integrals. In the cases under consideration,  the two integrals can be combined  into a single one in which the integrands are subtracted and the overall integration becomes convergent, as shown right next.  But one may equally regularize  the UV-divergences of both integrals  through e.g.  DR, and then one can check the cancellation of the UV-divergent parts, see Appendix B of \cite{CristianJoan2020}  for more details on this procedure.}
\begin{eqnarray}\label{EMTRenormalized}
\langle T_{00}^{\delta \phi}\rangle_{\rm ren}(M)&=&\langle T_{00}^{\delta \phi}\rangle(m)-\langle T_{00}^{\delta \phi}\rangle^{(0-4)}(M)\,.
\end{eqnarray}
This subtraction prescription is, of course,  equally valid for any component of the EMT, as it is obvious from Eq.\,\eqref{eq:RenormEMTAdiabatic}.  In the above equation and hereafter we omit the adiabaticity order $N$ up to which the EMT is computed. In our context, the spacetime dimension is $n=4$ and hence it is understood that $N\geq 4$.  The value $N=4$ is the minimum one which is necessary to perform the renormalization of the EMT,  but for some applications we will consider also up to $N=6$, as it is obvious from the calculations already presented in Sec.\ref{eq:RegZPE}.

To ease the presentation of the explicit result,  it proves convenient to recover at least in part  the more explicit notation \eqref{eq:omegaM} in order to distinguish explicitly  between the off-shell energy mode $\omega_k(M)=\sqrt{k^2+a^2 M^2}$  (formerly denoted just as $\omega_k$) and the on-shell one  $\omega_k(m)=\sqrt{k^2+a^2 m^2}$.  With this notation, calculations lead to the following result up to fourth adiabatic order\,\cite{CristianJoan2020}:
\begin{equation}\label{Renormalized}
\begin{split}
&\langle T_{00}^{\delta \phi}\rangle^{\rm (0-4)}_{\rm ren}(M)=\frac{1}{8\pi^2 a^2}\int dk k^2 \left[ 2 \left(\omega_k (m)- \omega_k (M)\right)-\frac{a^2 \Delta^2}{\omega_k (M)}+\frac{a^4 \Delta^4}{{4\omega^3_k (M)}}\right]\\
&-\left(\xi-\frac{1}{6}\right) \frac{6\mathcal{H}^2}{8\pi^2 a^2}\int dk k^2 \left[\frac{1}{\omega_k(m)}-\frac{1}{\omega_k (M)}-\frac{a^2 M^2}{{\omega^3_k (M)}}-\frac{a^2 \Delta^2}{2{\omega^3_k (M)}}+\frac{a^2 m^2}{{\omega^3_k (m)}} \right]\\
&-\left(\xi-\frac{1}{6}\right)^2 \frac{9\left(2 \mathcal{H}^{\prime \prime}\mathcal{H}-\mathcal{H}^{\prime 2}-3 \mathcal{H}^{4}\right)}{8\pi^2 a^2}\int dk k^2 \left[ \frac{1}{{\omega^3_k (m)}}-\frac{1}{{\omega^3_k (M)}}\right]
-\left(\xi-\frac{1}{6}\right)\frac{3\Delta^2  \mathcal{H}^2}{8\pi^2}
\end{split}
\end{equation}
\begin{equation}
\begin{split} \label{RenormalizedExplicit2}
&=\frac{a^2}{128\pi^2 }\left(-M^4+4m^2M^2-3m^4+2m^4 \ln \frac{m^2}{M^2}\right)\\
&-\left(\xi-\frac{1}{6}\right)\frac{3 \mathcal{H}^2 }{16 \pi^2 }\left(m^2-M^2-m^2\ln \frac{m^2}{M^2} \right)+\left(\xi-\frac{1}{6}\right)^2 \frac{9\left(2  \mathcal{H}^{\prime \prime} \mathcal{H}- \mathcal{H}^{\prime 2}- 3  \mathcal{H}^{4}\right)}{16\pi^2 a^2}\ln \frac{m^2}{M^2}\,.
\end{split}
\end{equation}
Even though some of the individual terms in the integrand of \eqref{Renormalized} look formally UV-divergent, one can check upon careful inspection that the overall integral is not, and this explains why the final result \eqref{RenormalizedExplicit2} is perfectly finite\footnote{For instance, the expression under square brackets in the first line of \eqref{Renormalized} can be written:
$$2 (\omega_k (m)- \omega_k (M))-\frac{a^2 \Delta^2}{\omega_k (M)}+\frac{a^4 \Delta^4}{{4\omega^3_k (M)}}= \Delta^6 a^6 \frac{\omega_k (m)+3\omega_k (M)}{4\omega^3_k (M)(\omega_k (m)+\omega_k (M))^3}\sim \Delta^6 a^6  \frac{1}{k^5}\ \ \textrm{ as }k\rightarrow \infty\,.$$ }.  As noted, the same result can be obtained  from the counterterm procedure,  see Appendix B of \cite{CristianJoan2020}. The counterterms take the precise form \eqref{eq:splitcounters}, which only depends on the physical mass $m$ of the particle, not on the arbitrary scale $M$. Hence they cancel in the subtraction \eqref{EMTRenormalized}.  However, the counterterms  can also be used  to cancel the poles and write down the generalized Einstein's equations \eqref{eq:MEEs} fully in terms of  finite, renormalized, quantities at the scale $M$, as we shall do in the next section.
Let us emphasize that the expression \eqref{RenormalizedExplicit2} is not yet the renormalized vacuum energy density, it is only the renormalized ZPE.

\subsection{The full  ZPE up to  $6th$ adiabatic order}\label{sec:ZPE6th}

The explicit form of the  $6th$ adiabatic order is obtained by computing the integrals in the expression\,\eqref{Tsixthhorder}, which is one of the main  objectives of this work.   In the absence of the $6th$-order terms, the  $4th$-order result that we have obtained, Eq.\,\eqref{RenormalizedExplicit2},  vanishes on-shell (i.e. for $M=m$), as it should be expected  from the definition itself, Eq.\,(\ref{EMTRenormalized}).  As a matter of fact, this is the reason why we need to include the next nonvanishing adiabatic order  so as to get the first nonvanishing contribution to the on-shell value of the ZPE. The  higher order finite effects must satisfy the Appelquist-Carazzone decoupling theorem\,\cite{AppelquistCarazzone75} since they must be suppressed  for  large values of the physical mass $m$  of the quantum field.  We may now compute these finite  contributions.  On using the master integral formulas given in Appendix \ref{sec:appendixA2}, the  final renormalized result computed up to $6th$-order  is:
\begin{equation}
\begin{split}\label{renormalized6th}
&\langle T_{00}^{\delta \phi}\rangle^{\rm (0-6)}_{\rm ren}(M)=\frac{a^2}{128\pi^2}\left(-M^4+4m^2M^2-3m^4+2m^4\ln \frac{m^2}{M^2}\right)-\left(\xi-\frac{1}{6}\right)\frac{3 \mathcal{H}^2 }{16 \pi^2 }\left(m^2-M^2-m^2\ln \frac{m^2}{M^2} \right)\\
&+\left(\xi-\frac{1}{6}\right)^2 \frac{9\left(2  \mathcal{H}^{\prime \prime} \mathcal{H}- \mathcal{H}^{\prime 2}- 3  \mathcal{H}^{4}\right)}{16\pi^2 a^2}\ln \frac{m^2}{M^2}+\frac{1}{20160 \pi^2 a^4m^2}\left(4\left(\mathcal{H}^\prime\right)^3-24\mathcal{H}^3\mathcal{H}^{\prime\prime}-6\mathcal{H}^\prime\mathcal{H}^{\prime\prime\prime}+96\mathcal{H}^4\mathcal{H}^\prime\right.\\
&\left.-12\mathcal{H}\mathcal{H}^\prime\mathcal{H}^{\prime\prime}+6\mathcal{H}\mathcal{H}^{\prime\prime\prime\prime}+
3\left(\mathcal{H}^{\prime\prime}\right)^2-12\mathcal{H}^2\left(\mathcal{H}^\prime\right)^2-24\mathcal{H}^2\mathcal{H}^{\prime\prime\prime}\right)\\
&+\left(\xi-\frac{1}{6}\right)\frac{1}{160\pi^2a^4m^2}\left(12\mathcal{H}^3\mathcal{H}^{\prime\prime}-\left(\mathcal{H}^{\prime\prime}\right)^2-40\mathcal{H}^4\mathcal{H}^\prime+2\mathcal{H}^\prime\mathcal{H}^{\prime\prime\prime}-2\mathcal{H}\mathcal{H}^{\prime\prime\prime\prime}+10\mathcal{H}^2\left(\mathcal{H}^{\prime}\right)^2+8\mathcal{H}^2\mathcal{H}^{\prime\prime\prime}\right)\\
&+\left(\xi-\frac{1}{6}\right)^2\frac{3}{32\pi^2 a^4 m^2}\left(4\mathcal{H}^6+48\mathcal{H}^4\mathcal{H}^\prime-12\mathcal{H}^2\left(\mathcal{H}^\prime\right)^2-16\mathcal{H}^3\mathcal{H}^{\prime\prime}+\left(\mathcal{H}^{\prime\prime}\right)^2-8\mathcal{H}^2\mathcal{H}^{\prime\prime\prime}-2\mathcal{H}^\prime\mathcal{H}^{\prime\prime\prime}+2\mathcal{H}\mathcal{H}^{\prime\prime\prime\prime}\right)\\
&-\left(\xi-\frac{1}{6}\right)^3\frac{9}{8\pi^2a^4m^2}\left(\mathcal{H}^2+\mathcal{H}^\prime\right)\left(11\mathcal{H}^4+\mathcal{H}^2\mathcal{H}^\prime+
2\left(\mathcal{H}^\prime\right)^2-6\mathcal{H}\mathcal{H}^{\prime\prime}\right)\,.
\end{split}
\end{equation}

The first two lines of this expression embody just the  $4th$-order renormalized result \eqref{RenormalizedExplicit2}.  We can easily convince ourselves that the remaining  terms of \eqref{renormalized6th} are of $6th$ adiabatic order,  i.e. ${\cal O}(T^{-6})$, where we are using  the notation introduced in Sec.\,\ref{sec:AdiabaticVacuum} for the adiabaticity order.   Moreover, they are all suppressed by two powers of the particle's mass  $m$, i.e. they fall off as $\sim 1/m^2$, or to be more precise as  ${\cal O}(T^{-6})/m^2$.  Thus, as formerly announced, the $6th$-order terms satisfy the Appelquist-Carazzone decoupling theorem for large $m$\,\cite{AppelquistCarazzone75}\footnote{An alternative way to express this decoupling result for large $m$  is to say that in the opposite limit ($m\to 0$) the higher order adiabatic terms beyond $N>4$ (all of them of even order owing to covariance,  $N=6,8,...$) are infrared divergent for $m\to 0$ in four dimensions. This is a well-known behavior expected from the effective action \cite{BirrellDavies82},  which e.g.  can be immediately appraised in the explicit form of the $6th$ order adiabatic integral \eqref{Tsixthhorder}. Although we are not affected by  IR effects  ($m$ is in our case  very large), the IR limit of ARP must be treated with care\,\cite{IR-ARP}.}.  The  next adiabatic order would be the  $8th$ one. These terms also fulfill the decoupling theorem and are further  suppressed as  ${\cal O}(T^{-8})/m^4$. We shall not be concerned with them.

We should also note that the $6th$-order terms do not depend on the arbitrary mass scale $M$, but only on the mass of the particle, $m$.  The reason is that $M$ enters only the terms up to adiabatic order $4$, which are the only ones which are subtracted (because they are the only ones which are originally UV-divergent), as it is obvious from the definition \eqref{EMTRenormalized}. As a consequence, the on-shell value of \eqref{renormalized6th} is now nonvanishing and it is exclusively determined by the higher order adiabatic  terms beyond ${\cal O}(T^{-4})$. It is convenient to express the result in terms of the ordinary Hubble rate defined in terms of the cosmic time, $H(t)$, with  $\mathcal{H}(\tau)=a  H(t)$. We may now use  the conversion relations between the derivatives of $\mathcal{H}$ with respect to the conformal time and the derivatives of $H$ with respect to the cosmic time (see Appendix \ref{sec:appendixA1}).  After some algebra we find the following expression for  the renormalized on-shell value ($M=m$) at 6th adiabatic order, i.e.  at  ${\cal O}(T^{-6})$ (which, we stress again,  is the first and hence the leading  nonvanishing order in the on-shell case):
\begin{equation*}
\begin{split}
&\langle T_{00}^{\delta \phi}\rangle^{(6)}_{\rm ren}(m)=
\frac{a^2}{20160 \pi^2 m^2}\left(-8H^6-36H^4\dot{H}-20\dot{H}^3+42H^3\ddot{H}+3\ddot{H}^2-6\dot{H}\vardot{3}{H}\right.\\
&\phantom{XXXXXXXXXXXXXXXXXXXXXX}\left.+84H^2\dot{H}^2+36H^2\vardot{3}{H}+60H\dot{H}\ddot{H}+6H\vardot{4}{H}\right)\\
\end{split}
\end{equation*}
\begin{equation}
\begin{split}\label{renormalizedONSHELL6th}
&+\left(\xi-\frac{1}{6}\right)\frac{a^2}{160\pi^2m^2}\left(2H^6+12H^4\dot{H}+8\dot{H}^3-14H^3 \ddot{H}-\ddot{H}^2+2\dot{H}\ddot{H}-34H^2\dot{H}^2\right.\\
&\phantom{XXXXXXXXXXXXXXXXXXXXXX}\left.-12H^2\vardot{3}{H}-24H\dot{H}\ddot{H}-2H\vardot{4}{H}\right)\\
&+\left(\xi-\frac{1}{6}\right)^2\frac{3 a^2}{32\pi^2 m^2}\left(-24H^4\dot{H}-8\dot{H}^3+10H^3\ddot{H}+\ddot{H}^2-2\dot{H}\vardot{3}{H}+32H^2\dot{H}^2\right.\\
&\phantom{XXXXXXXXXXXXXXXXXXXXXX}\left.+12H^2\vardot{3}{H}+24H\dot{H}\ddot{H}+2H\vardot{4}{H}\right)\\
&-\left(\xi-\frac{1}{6}\right)^3\frac{9 a^2}{8\pi^2 m^2}\left(2H^2+\dot{H}\right)\left(2H^4-19H^2\dot{H}+2\dot{H}^2-6H\ddot{H}\right)\,.
\end{split}
\end{equation}

\section{Renormalized vacuum energy density and running vacuum}\label{sec:RenormalizedVED}

The renormalized expression for the vacuum fluctuations, $\langle T_{\mu\nu}^{\delta \phi}\rangle_{\rm ren}(M)$, is not yet the final one for the renormalized VED. As indicated in \eqref{EMTvacuum}, the latter is obtained upon including the contribution from the $\rL$-term in the Einstein-Hilbert action \eqref{eq:EH}.  This term is initially a bare quantity, but we take its renormalized value at the same  scale $M$, Eq.\,\eqref{eq:splitcounters}.  Therefore, the renormalized vacuum  EMT at the scale $M$ is given by
\begin{equation}\label{RenEMTvacuum}
\langle T_{\mu\nu}^{\rm vac}\rangle_{\rm ren}(M)=-\rho_\Lambda (M) g_{\mu \nu}+\langle T_{\mu \nu}^{\delta \phi}\rangle_{\rm ren}(M)\,.
\end{equation}
For  the  considerations in this section we will use only the  renormalized expressions up to $4th$ adiabatic order, since these suffice to discuss the renormalization of the EMT.

 \subsection{Renormalizing the  vacuum energy density in FLRW spacetime}\label{sec:RenormalizedZPE}
The renormalized VED obtains from extracting the $00th$-component of the expression \eqref{RenEMTvacuum}:
\begin{equation}\label{RenVDE}
\rv(M)= \frac{\langle T_{00}^{\rm vac}\rangle_{\rm ren}(M)}{a^2}=\rho_\Lambda (M)+\frac{\langle T_{00}^{\delta \phi}\rangle_{\rm ren}(M)}{a^2}\,,
\end{equation}
where we have used the fact that $g_{00}=-a^2$ in the conformal metric that we are using.  The above equation stems from treating the vacuum as a perfect fluid, namely with an EMT of the form
\begin{equation}\label{eq:VaccumIdealFluid}
\langle T_{\mu\nu}^{{\rm vac}}\rangle=\Pv g_{\mu \nu}+\left(\rv+\Pv\right)u_\mu u_\nu\,,
\end{equation}
where $u^\mu$ is the $4$-velocity ($u^\mu u_\mu=-1$). In conformal coordinates in  the comoving cosmological (FLRW)  frame, $u^\mu=(1/a,0,0,0)$ and hence  $u_\mu=(-a,0,0,0)$. Taking the $00th$-component of  \eqref{eq:VaccumIdealFluid},  the relation  $\langle T_{00}^{{\rm vac}}\rangle=-a^2 \Pv+\left(\rv+\Pv\right) a^2=a^2\rho_{\rm vac}$ follows, irrespective of $\Pv$.  Whence
\begin{equation}\label{eq:Defrhovac}
\rv(M)= \frac{\langle T_{00}^{\rm vac}\rangle_{\rm ren}(M)}{a^2}\,.
\end{equation}
Finally, inserting the $00th$-component of \eqref{RenEMTvacuum} into \eqref{eq:Defrhovac} we obtain Eq.\,\eqref{RenVDE}, as desired.

Notice that we distinguish between VED and ZPE. The latter is generated from the vacuum fluctuations of the fields whereas the former combines the ZPE and the parameter $\rL$ in the action.  When they are both renormalized quantities, the sum \eqref{RenVDE} provides the physically measurable quantity at the scale $M$. In a symbolic way, we can write
\begin{equation}\label{eq:Symbolic}
  {\rm VED}=\rL+{\rm ZPE}\,.
\end{equation}
More explicitly, we can write it out on taking cognizance of  the important result presented in Eq.\,\eqref{RenormalizedExplicit2}:
\begin{equation}\label{RenVDEexplicit}
\begin{split}
\rv(M)&= \rho_\Lambda (M)+\frac{1}{128\pi^2 }\left(-M^4+4m^2M^2-3m^4+2m^4 \ln \frac{m^2}{M^2}\right)\\
&+\left(\xi-\frac{1}{6}\right)\frac{3 \mathcal{H}^2 }{16 \pi^2 a^2}\left(M^2-m^2+m^2\ln \frac{m^2}{M^2} \right)+\left(\xi-\frac{1}{6}\right)^2 \frac{9\left(2  \mathcal{H}^{\prime \prime} \mathcal{H}- \mathcal{H}^{\prime 2}- 3  \mathcal{H}^{4}\right)}{16\pi^2 a^4}\ln \frac{m^2}{M^2}+\cdots
\end{split}
\end{equation}
Here the dots denote that we have not written higher order adiabatic orders beyond the 4th one. It is important to remark that these terms do not depend on the renormalization parameter $M$. The simplified notation  $\rho_{\rm vac}(M)$  should not obscure the fact that the VED in FLRW spacetime is dynamical, as  it rests on the expansion  rate of the universe and its derivatives, apart from the scale $M$,
i.e.  $\rv(M)\equiv\rv(M,\cH, \cH',\cH'',...)$.  We note that $M$ itself is dynamical in cosmology since we will associate  $M$  with a  cosmological variable.  Thus, the dynamical character of the VED  enters both through the explicit dependence in the Hubble rate and also implicitly through $M$ (cf. Sec. \ref{sec:RVMcurrentUniverse}).

It should be clear that  $\rho_{\rm vac}(M)$ cannot be computed  from the above expression, even if  the expansion rate is known at a given time  since we do not know the value of the renormalized parameter $\rL(M)$ in the action.  As usual, we need some experimental input, we will comment more on this fact in Sec. \ref{sec:RVMcurrentUniverse}.
We should not forget that the main aim of renormalization theory  is not so much to predict the value of a quantity, for example the VED,  at a certain scale (and time, in cosmology) but to relate it at different scales or renormalization points,  and hence to account for its evolution with the scale $M$.  We next compute the difference of VED values between the two scales $M$ and $M_0$.   We find
\begin{equation}\label{eq:VEDscalesMandM0}
\begin{split}
&\rv(M)-\rv(M_0)=\frac{\langle T_{00}^{\rm vac}\rangle_{\rm ren}(M)-\langle T_{00}^{\rm vac}\rangle_{\rm ren}(M_0)}{a^2}\\
&=\rL(M)-\rL(M_0)+\frac{\langle T_{00}^{\delta \phi}\rangle_{\rm ren}(M)-\langle T_{00}^{\delta \phi}\rangle_{\rm ren}(M_0)}{a^2}\,,
\end{split}
\end{equation}
where
\begin{equation}\label{RenT00vacuumMM0}
\begin{split}
&\langle T_{00}^{\delta \phi}\rangle_{\rm ren}(M)-\langle T_{00}^{\delta \phi}\rangle_{\rm ren}(M_0)=-\frac{a^2}{128\pi^2}\left(M^4-M_0^{4}-4m^2(M^2-M_0^{2})+2m^4\ln  \frac{M^{2}}{M_0^2}\right)\\
&+\left(\xi-\frac16\right)\frac{3\cH^2}{16\pi^2}\,\left(M^2 - M_0^{2} -m^2\ln \frac{M^{2}}{M_0^2}\right)+\left(\xi-\frac16\right)^2\frac{9}{16 \pi^2 a^2}\left(\mathcal{H}^{\prime 2}-2\mathcal{H}^{\prime \prime}\mathcal{H}+3 \mathcal{H}^4 \right)\ln \frac{M^2}{M_0^{2}}\,.\\
\end{split}
\end{equation}
 To account for this difference we  have just used the  $4th$-order form  \eqref{RenormalizedExplicit2} since, as noticed,  the $6th$ adiabatic contribution  does not carry along  any new dependency on the scale $M$  (it only adds up new contributions which depend on the mass $m$); and hence all the higher order adiabatic effects beyond  $4th$-order cancel out when one is just relating scales rather than computing the result at a given scale.  The same subtraction can be performed using the generalized Einstein's equations \eqref{eq:MEEs}. We take now these equations in vacuo and in terms of the renormalized couplings at the scale $M$.  For convenience we write them  as follows:
 \begin{equation}\label{eq:EqsVac}
\MPl^2(M) G_{\mu \nu}+\rho_\Lambda (M) g_{\mu \nu}+\alpha(M)\ \leftidx{^{(1)}}{\!H}_{\mu\nu}= \langle T_{\mu\nu}^{\delta \phi}\rangle_{\rm ren}(M)\,.
\end{equation}
Apart from $\rL(M)$ and $\alpha(M)$ we have defined  another renormalized coupling at the scale $M$,
\begin{equation}\label{eq:RenCouplings}
\MPl^2(M)=\frac{G^{-1}(M)}{8\pi}\,,
\end{equation}
which is nothing but the  reduced Planck mass squared at that scale. Its relation with the ordinary Planck mass is $\MPl (M)=\mpl (M) /\sqrt{8\pi}$.  As we shall further discuss  in what follows (see also  Appendix \ref{sec:appendixAbis}), the setting $M=H$ is the most appropriate one  to make contact between  the renormalized value of a parameter and its physical value  at the epoch $H$.  In accordance with this prescription, the measured local value of gravity, $G_N$,  is obtained when  $M$ is set to the current value of the Hubble parameter, i.e.  $G(H_0)\equiv G_N=1/\mpl^2$, where  $\mpl\simeq 1.2\times10^{19}$ GeV.  Performing the subtraction of the $00th$-component of \eqref{eq:EqsVac} at the two scales $M$ and $M_0$, we find\,\footnote{Let us note that  the relations in which we perform subtractions at two scales are obviously unaffected by the background contributions to the EMT since these are independent of $M$, and hence cancel in the subtraction.}:
\begin{equation}\label{RenT00vacuumMM0Eeqs}
\begin{split}
&\langle T_{00}^{\delta \phi}\rangle_{\rm ren}(M)-\langle T_{00}^{\delta \phi}\rangle_{\rm ren}(M_0)=- a^2\left(\rL(M)-\rL(M_0)\right)\\
&\phantom{XXXXXXXXXXXXXX}+\left(\MPl^2(M)-\MPl^2(M_0)\right)G_{00}+\left(\alpha(M)-\alpha(M_0)\right)\leftidx{^{(1)}}{\!H}_{00}\,,
\end{split}
\end{equation}
where in the first line we have used once more that  $g_{00}=-a^2$.  Comparison between \eqref{RenT00vacuumMM0} and \eqref{RenT00vacuumMM0Eeqs} yields the important relation
\begin{equation}\label{SubtractionrL}
\delta\rL(m,M,M_0)\equiv\rL(M)-\rL(M_0)=\frac{1}{128\pi^2}\left(M^4-M_0^{4}-4m^2(M^2-M_0^{2})+2m^4\ln  \frac{M^{2}}{M_0^2}\right)\,,
\end{equation}
and  upon using the known form of $G_{00}$ and  $\leftidx{^{(1)}}{\!H}_{00}$ in the conformal metric  (given in Appendix \ref{sec:appendixA1}) we collect also the two relations:
\begin{equation} \label{SubtractionMPl}
\delta\MPl^2(m,M,M_0)\equiv\MPl^2(M)-\MPl^2(M_0)=\left(\xi-\frac{1}{6}\right)\frac{1}{16\pi^2}\left[M^2 - M_0^{2} -m^2\ln \frac{M^{2}}{M_0^2}\right]
\end{equation}
and\,\footnote{The scale shifts quoted in equations \eqref{SubtractionrL}-\eqref{Subtractionalpha} are finite quantities in our renormalization scheme and should not be confused with counterterms, such as those in \eqref{eq:splitcounters}-\eqref{eq:counters}.  Strictly speaking, we do not need counterterms in the ARP since we perform a subtraction of UV-divergent quantities at two scales, and this renders a finite result.  }
\begin{equation} \label{Subtractionalpha}
\delta\alpha(M,M_0)\equiv\alpha(M)-\alpha(M_0)= -\frac{1}{32\pi^2}\left(\xi-\frac{1}{6}\right)^2 \ln \frac{M^2}{M_0^{2}}\,.
\end{equation}
These relations are important not only because they furnish the scaling laws of the couplings $\MPl^2(M)$ and $\alpha(M)$ in the modified Einstein's equations,  but also because they help to properly identify the various terms in Eq.\,\eqref{RenT00vacuumMM0Eeqs}.  In particular they contribute to isolate the shift of the renormalized vacuum parameter  $\rL(M)$, Eq.\,\eqref{SubtractionrL}.   Using \eqref{SubtractionrL}  we can  rewrite \eqref{RenT00vacuumMM0} in the following  form
\begin{equation}\label{RenT00vacuumMM0rhos}
\begin{split}
&\frac{\langle T_{00}^{\delta \phi}\rangle_{\rm ren}(M)-\langle T_{00}^{\delta \phi}\rangle_{\rm ren}(M_0)}{a^2}=-\delta\rL(m,M,M_0)
+\left(\xi-\frac16\right)\frac{3\cH^2}{16\pi^2 a^2}\,\left(M^2 - M_0^{2} -m^2\ln \frac{M^{2}}{M_0^2}\right)\\
&\phantom{XXXXXXXXXXXXXXXX}+\left(\xi-\frac16\right)^2\frac{9}{16 \pi^2 a^4}\left(\mathcal{H}^{\prime 2}-2\mathcal{H}^{\prime \prime}\mathcal{H}+3 \mathcal{H}^4 \right)\ln \frac{M^2}{M_0^{2}}\,.\\
\end{split}
\end{equation}

Finally, on combining  equations\,\eqref{RenT00vacuumMM0rhos} and \eqref{eq:VEDscalesMandM0} we see that the expression $\delta\rL(m,M,M_0)$ exactly cancels and we are left with
\begin{equation}\label{eq:VEDscalesMandM0Final}
\begin{split}
&\rv(M)-\rv(M_0)=\left(\xi-\frac16\right)\frac{3\cH^2}{16\pi^2 a^2}\,\left(M^2 - M_0^{2} -m^2\ln \frac{M^{2}}{M_0^2}\right)\\
&\phantom{XXXXXXXXXXXXXXXX}+\left(\xi-\frac16\right)^2\frac{9}{16 \pi^2 a^4}\left(\mathcal{H}^{\prime 2}-2\mathcal{H}^{\prime \prime}\mathcal{H}+3 \mathcal{H}^4 \right)\ln \frac{M^2}{M_0^{2}}\\\
&=\left(\xi-\frac{1}{6}\right) \frac{3H^2 }{16\pi^2}\,\left[M^2 - M_0^{2}-m^2\ln \frac{M^{2}}{M_0^2}\right]\\
&\phantom{XXXXXXXXXXXXXXXX} +\left( \xi-\frac{1}{6}\right)^2\frac{9}{16\pi^2}\, \left(\dot{H}^2 - 2 H\ddot{H} - 6 H^2 \dot{H} \right)\ln \frac{M^2}{M_0^2}\,,\\
\end{split}
\end{equation}
where in the second equality we have written the result in terms of the Hubble rate in the ordinary cosmic time, and for this reason it does not depend explicitly on the scale factor.
The exact cancellation of the quantity \eqref{SubtractionrL}  in Eq.\,\eqref{eq:VEDscalesMandM0Final} is very important since such a term  is precisely the potentially conflicting quantity carrying all of the awkward quartic powers of the mass scales.  If this term would survive, it would recreate the traditional (ugly) fine-tuning conundrum between the values of the VED at the two scales $M$ and $M_0$.  Its remarkable cancellation in our renormalization setup, however, shows that the values of $\rv(M)$ and $\rv(M_0)$ differ only by a small quantity proportional to $H^2$ and another which is of ${\cal O}(H^4)$, both small in the current universe (the latter being utterly irrelevant for the entire FLRW regime).  Thus, no fine-tuning is needed to relate  $\rv(M)$ and $\rv(M_0)$ in the present renormalization framework.

The above considerations show that the generalized Einstein's equations \eqref{eq:EqsVac} can be written in terms of the full vacuum energy-momentum tensor \eqref{RenEMTvacuum}:
\begin{equation}\label{eq:EqsVac2}
\MPl^2 (M) G_{\mu \nu}+\alpha(M)\ \leftidx{^{(1)}}{\!H}_{\mu \nu}= \langle T_{\mu\nu}^{\rm vac}\rangle_{\rm ren}(M)\,.
\end{equation}
As a result the contribution \eqref{SubtractionrL} cancels automatically whenever we compare the renormalized vacuum EMT,  $ \langle T_{\mu\nu}^{\rm vac}\rangle_{\rm ren}(M)$, at two scales.   The formal quantity $\rL(M)-\rL(M_0)$ indeed never shows up physically  and hence only the difference of VED's at the two renormalization points $M$ and $M_0$  remains, Eq.\,\eqref{eq:VEDscalesMandM0Final}. Such a physically measurable quantity is a smooth function $\sim m^2 H^2$ of the cosmic evolution.   With no  quartic mass terms being involved in the physical measurements, there is no need of fine-tuning in this renormalization setup. This is also true in the special case of  Minkowskian spacetime, where neither  $\rL(M)$ nor the  ZPE can be measured in an isolated way, just the sum, which in this case is exactly zero (see next  section).  The foregoing considerations show that, in the context of the running vacuum model,  the dark energy that we observe is just the (non-constant) vacuum energy density predicted within QFT in FLRW spacetime,  which remains naturally of order $H^2$ at all times without fine tuning.  At any  cosmic time  $t$ characterized by $H(t)$  there is a (different)   `CC' term $\CC(H)=8\pi G_N\rv(H)$ acting (approximately)  as a cosmological constant for a long period around that  time, but there is no true CC valid  at all times!

\subsection{Vanishing vacuum energy density in Minkowskian spacetime}\label{sec:VEDMinkowski}

We note that Eq.\,\eqref{eq:VEDscalesMandM0Final}  can be thought of as being the result of subtracting  the Minkowskian  form of the ZPE in  FLRW spacetime\, \cite{CristianJoan2020}.  Let us evaluate
\begin{equation}\label{eq:RenEMTSubtractMink}
\begin{split}
&\rv(M)\equiv\frac{\langle T_{00}^{\delta \phi}\rangle_{\rm ren}(M )}{a^2}-\left[\frac{\langle T_{00}^{\delta \phi}\rangle_{\rm ren}(M )}{a^2}\right]^{\rm Mink}
=\frac{\langle T_{00}^{\delta \phi}\rangle_{\rm ren}(M )}{a^2}-\langle T_{00}^{\delta \phi}\rangle_{\rm ren}^{\rm Mink}(M)\,,
\end{split}
\end{equation}
where in Minkowski spacetime $a=1$, $\cH=H=0$.  It follows that equation \eqref{eq:EqsVac} boils down to just  $\rho_\Lambda (M) \eta_{\mu \nu}= \langle T_{\mu\nu}^{\delta \phi}\rangle_{\rm ren}^{\rm Mink}(M)$, or $\langle T_{00}^{\delta \phi}\rangle_{\rm ren}^{\rm Mink}(M)=-\rL(M)$. Thus, the above expression becomes
\begin{equation}\label{RenVDE2}
\rv(M)=\frac{\langle T_{00}^{\delta \phi}\rangle_{\rm ren}(M)}{a^2}+\rho_\Lambda (M)= \frac{\langle T_{00}^{\rm vac}\rangle_{\rm ren}(M)}{a^2}\,,
\end{equation}
which is precisely our starting formula for the renormalized VED, Eq.\,\eqref{RenVDE}.  Retaking from this point the subtraction procedure \eqref{eq:VEDscalesMandM0}, we reach once more the final result \eqref{eq:VEDscalesMandM0Final}.

As previously noted, the fact that the correct renormalized result can be viewed as subtracting the  Minkowskian contribution can be traced to an analogy with the Casimir effect. Namely, one expects that if we compute the expression for  $\langle T_{\mu\nu}^{\rm vac}\rangle$  in Minkowskian spacetime and  subtract it from its equivalent in curved spacetime the result should depend only on the  curvature of the latter and hence evolve only mildly with the cosmic evolution through a function of the Hubble rate (which is the key term providing the departure of the FLRW background from Minkowskian spacetime). Such a function is exactly given by the \textit{r.h.s.} of Eq.\,\eqref{eq:VEDscalesMandM0Final}.

In Minkowski space we should expect zero vacuum energy, as in such a case we can apply the normal ordering of the quantum operators in the canonical formalism.  In our context we encounter the same result. To start with, the renormalized ZPE in Minkowski space  is the  value of $\langle T_{00}^{\delta \phi}\rangle_{\rm ren}(M)$, given by Eq.\,\eqref{RenormalizedExplicit2} for $a=1$ and $\cH=0$:
\begin{equation}\label{eq:ZPEMinkowski}
\begin{split}
\langle T_{00}^{\delta \phi}\rangle_{\rm ren}^{\rm Mink}(M)=\frac{1}{128\pi^2 }\left(-M^4+4m^2M^2-3m^4+2m^4 \ln \frac{m^2}{M^2}\right)\,.
\end{split}
\end{equation}
However, this quantity is purely formal and does not appear in physical results since  it cancels exactly against $\rL(M)$.  Indeed, from \eqref{RenVDE2} we confirm that the VED in Minkowskian spacetime is exactly zero in our renormalization setup:
\begin{equation}\label{eq:VEDMinkowski}
\rv^{\rm Mink}=\left[\frac{\langle T_{00}^{\delta \phi}\rangle_{\rm ren}(M )}{a^2}\right]^{\rm Mink}+\rL(M)=\langle T_{00}^{\delta \phi}\rangle_{\rm ren}^{\rm Mink}(M)+\rL(M)=-\rL(M)+\rL(M)=0\,,
\end{equation}
 for all scales $M$.  In hindsight, this can be viewed as the practical implementation of the setting \eqref{eq:RenormEMTpointsplit}. Therefore, for Minkowski spacetime
\begin{equation} \label{eq:rLMMinkowski}
\begin{split}
\rL(M)=-\frac{1}{128\pi^2 }\left(-M^4+4m^2M^2-3m^4+2m^4 \ln \frac{m^2}{M^2}\right)\,.
\end{split}
\end{equation}
The two quantities $\rL(M)$ and $\langle T_{00}^{\delta \phi}\rangle_{\rm ren}^{\rm Mink}(M)$  both carry quartic dependencies on the mass scales, but they exactly conspire to sum up to zero, and hence the detailed structure of these formal quantities plays no role in the physically measured quantities.  In contrast to the Minkowski case, in curved spacetime such quantities cannot be isolated  since the sum is not zero, see Eq.\,\eqref{RenVDEexplicit}, but it yields a smooth quantity mildly evolving with the cosmological evolution.  For $a\to 1$  ($\cH\to 0$) the \textit{l.h.s} of \eqref{RenVDEexplicit} goes to zero, as we have seen, and hence the  \textit{r.h.s}  goes to zero too.   At this point we retrieve the Minkowskian space result \eqref{eq:VEDMinkowski} from the curved spacetime case \eqref{RenVDEexplicit}.  But at any intermediate stage of this limit we cannot determine $\rL(M)$  separately from the ZPE, only the sum is physically relevant and it defines the dynamical vacuum energy density in curved spacetime.  Recall that the renormalization point itself $M$ is dynamical in curved spacetime.  As it was previously indicated,  the scale setting prescription  $M=H$  is an appropriate ansatz for testing  the cosmological evolution of the VED at different stages of the expansion history (cf. Sec. \,\ref{sec:RVMcurrentUniverse} and Appendix \ref{sec:appendixAbis1}).   It is remarkable that the VED of the expanding universe  despite it being currently very small (of order $\rvo$)   is dynamical and such dynamics could be measured since it  is of  ${\cal O}(H^2)$.  The VED  is exactly zero only in Minkowski spacetime, where the vacuum energy plays no cosmological role.  In actual fact, the scale $M$ becomes in this case a purely formal quantity devoid of any physical meaning,  much the same as the artificial mass unit $\mu$ employed in DR (as discussed in the next section). There is no dynamics of gravity in Minkowski space and therefore nothing  can physically run with $M$  (or $\mu$).  In contrast, in FLRW spacetime the gravitational field is dynamical and hence the prescription $M=H$  is physically meaningful and  enables exploring the running of the vacuum energy density with the cosmic expansion (cf. and Appendix \ref{sec:appendixAbis1} for a thorough discussion). This is actually the original point of view of the RVM from the renormalization group approach\,\cite{JSPRev2013,JSPRev2022} and is also the main result advanced in our previous work \cite{CristianJoan2020}.

\subsection{Minimal Subtraction scheme and the fine-tuning problem}\label{sec:VEDMSS}

The former are certainly properties  we should expect from a correct, physically meaningful, renormalization of the vacuum energy density, in contrast to other formal treatments in the literature in the framework of different renormalization schemes. In particular, the absence of fine-tuning among the different terms is much welcome as well as the vanishing value of the VED in Minkowskian spacetime, which  in the simplified  notation \eqref{eq:Symbolic}  introduced above  reads $\rL+$ZPE$=0$.  This condition can be thought of as a necessary condition for the physical renormalization prescription for the quantum vacuum energy, cf Eq.\,\eqref{eq:RenormEMTpointsplit}, and is encoded in the general form  \eqref{RenVDE}.   Many other approaches and renormalization prescriptions have been advocated to deal with the vacuum energy,  see\,\cite{ShapSol1,Fossil2008,ShapSol2,Babic2005,Maggiore2011,Maggiore2011b,Bilic2011,Bilic2012,Domazet2012,BennieW2013,Antipin2017,KohriMatsui2017,FerreiroNavarroSalas,FerreiroNavarroSalas2,IR-ARP,Birrell1978,Akhmedov2002,Sirlin2003,Visser2018,Donoghue2021,Christensen1976,Christensen1978,BunchDavies1978}, for instance.  Arguably, the simplest treatments are those based on the Minimal Subtraction (MS) scheme\,\cite{BolliniGiambiagi72,Hooft73} (cf.\,\cite{Manohar97,Sterman93,Brown92,Collins84} for further explanations and  practical applications). Its use was soon  extended to QFT in curved spacetime\,\cite{Bunch1979}.  But in this context, simplicity does not necessarily mean adequacy to the physical purposes, and in fact the MS scheme does not lead to a physically acceptable approach to the renormalization of the VED, cf.\,\cite{ShapSol2} and references therein. Let us briefly summarize the situation of the fine-tuning problem in the MS scheme (we refer the reader e.g. to \,\cite{JSPRev2013,JSPRev2022} for more details). It will suffice to focus on Minkowskian spacetime for this consideration.

In flat spacetime ($a=1$, $\cH=0$)  the  ZPE  \eqref{EMTFluctuations}  shrinks  just to the compact form \eqref{eq:Minkoski}, where we shall continue with $\hbar=1$ in natural units.  Using  dimensional regularization in Minkowskian  $n$-spacetime (with $n-1$ spatial dimensions),  a simple calculation with the notation and formulas of Appendix \ref{sec:appendixA2} leads to the  result\,\cite{JSPRev2013,JSPRev2022}
\begin{equation}\label{eq:MinkoskiMSS}
\begin{split}
 \langle T_{00}^{\delta \phi}\rangle^{\rm Mink}(m)&= \int\frac{\mu^{2\epsilon}d^{n-1}k}{(2\pi)^{n-1}}\,\left(\frac12\,\omega_k(m)\right)=\frac{1}{2}\, I_{n-1}(p=-1,Q=m)\\
  &= \frac{m^4}{4(4\pi)^2}\,\left(-D_\epsilon+ \ln\frac{m^2}{\mu^2}-\frac32\right)\,,
  \end{split}
\end{equation}
where $D_\epsilon$ contains the pole at $n-1=3$ (i.e. at $\epsilon=0$, or equivalently at $n=4$ spacetime dimensions) as  given by Eq.\,\eqref{eq:Depsilon}.  It is natural to  assume that the VED in Minkowskian space is given by a similar  equation to \eqref{RenVDE}, but with the bare quantities at this point since \eqref{eq:MinkoskiMSS} is divergent, i.e. $\rv^{\rm Mink}=\rL+\langle T_{00}^{\delta \phi}\rangle^{\rm Mink}$.  We next split the bare term $\rL$ into the renormalized quantity $\rL(\mu)$  plus the counterterm, as shown  in Eq.\,\eqref{eq:splitcounters}. As we know, in the MS scheme  the running scale is usually represented by means of the arbitrary 't Hooft's mass unit $\mu$,  which displays dimensions away from $n=4$ and keeps control of dimensional analysis.
Using the MS scheme   (or its variant $\overline{\rm MS}$\,\cite{Bardeen1978,Manohar97}) to deal with the UV-divergences is very tempting for  we can choose the counterterm $\delta\rL$ in the  form given in Eq.\,\eqref{eq:counters}, which precisely cancels the pole in \eqref{eq:MinkoskiMSS} and this allows to define the renormalized ZPE in Minkowski space,  $\langle T_{00}^{\delta \phi}\rangle^{\rm Mink}_{\rm Ren}(\mu)$.   One  may then be tempted to interpret that the MS-renormalized VED in Minkowskian spacetime is the finite expression
\begin{equation}\label{eq:VEDMSS}
\rv^{\rm Mink}=\rL(\mu)+\langle T_{00}^{\delta \phi}\rangle^{\rm Mink}_{\rm Ren}(\mu)=\rL(\mu)+\frac{m^4}{4(4\pi)^2}\,\left( \ln\frac{m^2}{\mu^2}-\frac32\right)\,.
\end{equation}
However, in spite of  its formal simplicity the  above formula  leads to the usual  fine-tuning nightmare associated to the CCP,  which is brought about by the fact that the renormalized ZPE  is  proportional to the quartic power of the mass of the particle $\sim m^4$. As a result, the MS-renormalized term $\rL(\mu)$ must be fine tuned in a preposterous way against the renormalized  ZPE contribution so as to get a VED value in a reasonable phenomenological range, see \cite{JSPRev2013,JSPRev2022} for a detailed  exposition of the fine-tuning problem.  Since $\mu$ was introduced on mere dimensional grounds there is no special physical meaning to be ascribed  to it.  To set $\mu=H$ does not make much sense here since the above formula applies to Minkowskian spacetime, where $H=0$. Besides, any attempt to make sense of the above VED formula  (using  DR or  Pauli-Villars regularization, for example) leads to nowhere. No matter what physical quantity is chosen for $\mu$ or the type and number of fields involved, the numerical results are completely astrayed\,\cite{Visser2018,KoksmaProkopec2011}.  Let us stress once more that one should \textit{not} aim at a prediction of the value of the VED at present,  as this is out of the scope of renormalization theory.  Equation \eqref{eq:VEDMSS}  is certainly not the VED neither in flat nor in curved spacetime.  Such an expression is unphysical, it just  describes the mathematical running of the parameter $\rL(\mu)$ and the renormalized ZPE with $\mu$  in  such a way that their sum remains equal to the original bare (hence  RG-invariant) parameter $\rL$ in the action. There is not an inch of  physics in it  since $\mu$ cannot be related to any quantity of cosmological interest;  we reiterate that \eqref{eq:VEDMSS} was derived in Minkowki space, where we have seen that the VED is just zero. In Sec.\,\ref{Sec:RGE-VED}  we come back to this point,  after discussing  the running couplings in curved spacetime.

Nothing of this sort occurs in our renormalization scheme, where the VED is given by \eqref{RenVDEexplicit}.  To start with, the  value of that expression in Minkowski space is exactly zero, as we have shown above, in stark contrast  with  the MS  formula \eqref{eq:VEDMSS}.  Furthermore, as long as we hold on to the aforementioned prescription for Minkowskian spacetime,  the implication on the corresponding calculation for curved spacetime  is that the VED  is no longer zero but a mildly dynamical quantity, which evolves smoothly (without fine-tuning) from one scale to another throughout  the cosmic evolution following the `running law' \eqref{eq:VEDscalesMandM0Final}\,\footnote{If one naively extends the MS renormalization to curved spacetime, the fine-tuning problem persists unmodified, see\,\cite{JSPRev2013,JSPRev2022} for a summarized account. The two  fine-tuning-generating pieces on the \text{r.h.s} of \eqref{eq:VEDMSS} remain  exactly as they are.  The curved background only adds purely geometric terms  ${\cal O}(R, R^2, R^{\mu\nu}R_{\mu\nu},...)$   and the essence of the fine-tuning problem embodied in the Minkowskian spacetime  replicates identically in curved spacetime in such scheme,  see\,\cite{Bunch1979} for more technical details.  Renormalization of the VED   \textit{ à la}  MS   seems to be completely hopeless in cosmology. }.
Thus, while we do not aim  at a  prediction of the value of the VED at present from pure renormalization theory,  a prediction is made of its value at some scale, given its value e.g. at present. Such an evolution is governed  by the Hubble flow and a  quadratic (not quartic) dependence on the mass scale, which is made extremely smooth since it is  accompanied by $H^2$ and thereby evolving as $\sim m^2 H^2$.
In what follows we take up what are the implications for the late time universe and in particular for our present time.

\

\subsection{Running vacuum in the current universe}\label{sec:RVMcurrentUniverse}

While  for the current universe  ($H=H_0$)  we may neglect all terms of order  ${\cal O}(H^4)$ (which comprise also  $\dot{H}^2, H\ddot{H}$ and  $H^2 \dot{H}$)  the piece proportional to $H^2$ in \eqref{eq:VEDscalesMandM0Final}  may be significant in the present universe. It entails that the vacuum energy density is dynamical and such a dynamics is amenable to being measured, as we have previously shown in \cite{CristianJoan2020}. This property leads to the notion of `running VED'. By running we mean that the VED is not static but changing with the  cosmic evolution.  A good tracking of that evolution in the FLRW context is provided by the Hubble rate $H$.  At any given cosmic time characterized by the Hubble rate, the choice  $M=H$ may be taken as parameterizing the scaling evolution of the VED, see  \cite{JSPRev2013,JSPRev2022,JSPRev2014,JSPRev2015} and references therein for the old connection with the renormalization group arguments\,\cite{ShapSol1,Fossil2008}.   If  $\rv$ is known at some reference scale $M_0$ associated to the epoch $H_0$ we can use the  relation Eq.\,\eqref{RenVDEexplicit} to compute the value of $\rv$ at another scale $M$ associated to some other epoch $H$.  Since these scales may represent different stages of the cosmic evolution,  the idea of running vacuum could be a viable  framework for the possible time variation of the so-called fundamental constants of nature\,\cite{FritzschSola}.

If $M_0$  is taken to be the current Hubble parameter,  then  $\rv(H_0)=\rvo$  can be  identified as being the presently observed  value of the  VED  at $H=H_0$.   We may relate $\rvo$ with the value  $\rv(M=H)$ at another scale in the past corresponding to the cosmic epoch $H$,  which we typically select within the accessible FLRW cosmic history.
Using  Eq.\,\eqref{RenVDEexplicit},  the connection between the two values of the  VED  can be written as follows (cf. Appendix \ref{sec:appendixAbis1} for details and notation):
\begin{equation}\label{eq:RVM2}
\rv(H)\simeq \rvo+\frac{3\nueff}{8\pi}\,(H^2-H_0^2)\,\mpl^2=\rvo+\frac{3\nueff}{\kappa^2}\,(H^2-H_0^2)\,,
\end{equation}
where $\kappa^2=8\pi G_N$.   Here we have defined the `running parameter' $\nueff$, which is approximately given by
\begin{equation}\label{eq:nueffAprox}
\nueff\simeq\frac{1}{2\pi}\,\left(\xi-\frac{1}{6}\right)\,\frac{m^2}{\mpl^2}\ln\frac{m^2}{H_0^2}\,.
\end{equation}
The more detailed treatment in Appendix \ref{sec:appendixAbis1} shows that $\nueff$  is actually a slowly changing (logarithmic) function of the Hubble rate.  But, to a fairly good approximation, $\nueff$ can be taken essentially  as the constant value given above for values of $H$ corresponding to the relatively recent universe\footnote{In \cite{CristianJoan2020} a formula similar to that in \eqref{eq:RVM2} was derived, making  however reference  to a Grand Unified Theory (GUT) scale $M_X$. The result is not essentially different if $m$ is assumed very large, of order of $M_X$.}.
As it was foreseen from the beginning, the structure of the RVM vacuum does not necessarily require  the nonminimal coupling of  matter to the external gravitational field.  We can see from \eqref{eq:nueffAprox} that  $\xi=0$  does \textit{not } imply $\nueff=0$.  The vanishing of $\nueff$ and hence of  the dynamical $\sim H^2$ part of \eqref{eq:RVM2} is obtained only for conformal coupling: $\xi=1/6$.  If the scalar field pertains to a typical GUT, i.e. $m\sim M_X\sim 10^{16}$ GeV,  the ratio $m^2/\mpl^2\sim 10^{-6}$ remains sizeable. Taking into account that the parameter $\xi$ can be, in principle,  arbitrary  and that the multiplicity of states in a GUT is usually high, the value of $\nueff$ can actually be much larger.
There are, however, also the  fermionic contributions to $\nueff$. These are obviously   independent of  $\xi$,  but we shall not tackle them  here\,\footnote{A detailed account of the fermionic quantum effects on the RVM vacuum structure will be provided in a separate publication\,\cite{JCS2022}.}.  An accurate determination of  $\nueff$  can only be found by fitting the RVM to the overall cosmological data, as it has been done in detail e.g. in \cite{RVMpheno1,RVMpheno2}, and lately in \cite{EPL2021}.   The phenomenological results show that $\nueff$ is  positive and can be of order $10^{-3}$. A reecent analysis of Big Bang nucleosynthesis (BBN) constraints points to the same order of magnitude, although is not sensitive to the sign of $\nueff$\,\cite{Manos2021}.

The following comments are in order.  From Eq.\eqref{eq:RVM2},  we see that for  $\nueff>0$ the vacuum can be conceived as decaying into matter since the vacuum energy density is larger in the past (where $H>H_0$); whereas if $\nueff<0$ the opposite occurs.  The former  situation, however, is more natural from a thermodynamical point of view, for if the vacuum decays into matter  one can show that the Second Law of  Thermodynamics is satisfied by the general RVM, see \cite{Yu2020} for a detailed discussion.   Moreover, for $\nueff>0$  the RVM effectively behaves as quintessence since the vacuum energy density decreases with time.  For $\nueff<0$ the behavior is that of phantom DE.  One may also interpret here that $G$ is changing with time owing to vacuum decay. Both possibilities have been mooted  within the RVM in Ref.\,\cite{FritzschSola}, see also Sec.\,\ref{sec:RenormalizedFriedmann}. Recall that we expect  $|\nueff|\ll1$  from  \eqref{eq:nueffAprox}, whereby we cannot hope for observing dramatic deviations from the standard $\CC$CDM model. This would actually not be welcome, given the considerable success of the concordance cosmology.  But the fact that the existing analyses point to $\nueff= {\cal O}(10^{-3})$ suggests that the effects are not necessarily negligible, and in fact they can be helpful to cure or relieve some of the existing tensions in the context of the $\CC$CDM model. This has been shown in actual RVM fits and also in the framework of alternative cosmological models which  mimic the RVM behavior, see e.g.\,\cite{Mehdi2019,EPL2021} and \cite{ApJL2019,BD2020}.

\section{Trace of the vacuum EMT  in curved spacetime} \label{sec:Trace}
As we will need to compute the pressure and equation of state of  the  quantum vacuum, it is helpful to compute the trace of the vacuum part of the  EMT. We start by computing the trace of the classical EMT, which we denote  $T^{\rm cl.}\equiv T^{\mu}_{\ \mu}$.  Using  \eqref{EMTScalarField}, it can be expressed as
\begin{eqnarray}\label{eq:ClassicTrace}
T^{\rm cl.}
&=&\left(6\xi-1\right)\ \nabla^\mu\phi \nabla_\mu\phi+6\xi\phi \Box \phi-\xi R\phi^2-2m^2\phi^2\nonumber\\
&=&\left(6\xi-1\right) \nabla^\mu\phi \nabla_\mu\phi+\left(6\xi-1\right)\phi\Box \phi-m^2\phi^2\,,
\end{eqnarray}
where in the last step we have used the equation of motion \eqref{eq:KG}.  This last form is  useful since it makes transparent that the trace is null in the conformal limit ($m=0$ and $\xi=1/6$), as it should be (in four spacetime dimensions).   An alternative form which  will be more helpful for our purposes and still makes apparent the previous property,  is  obtained by trading $\phi\Box\phi$  for  $R\phi^2$ as follows:
\begin{eqnarray}\label{eq:ClassicTrace2}
T^{\rm cl.}
&=&\left(6\xi-1 \right)\nabla^\mu  \phi\nabla_\mu \phi +2(3\xi-1)m^2\phi^2+ \xi \left(6\xi-1\right) R \phi^2\nonumber\\
&=&\left(6\xi-1 \right) g^{\mu\nu}\nabla_\mu\phi\nabla_\mu \phi +2(3\xi-1)m^2\phi^2+6\left(\xi-\frac{1}{6} \right)^2R \phi^2+\left(\xi-\frac{1}{6} \right)R \phi^2\,,
\end{eqnarray}
where the last rearrangement is just for convenience.
On examining  the quantum fluctuations \eqref{ExpansionField} about the background field, we note that  the  vacuum expectation value (VEV) of  the trace $T_{\mu \nu}^{\phi}$  with the quantum field $\phi$, can only comprise terms quadratic (or, more rigorously, bilinear under the coincidence limit) on its fluctuations $\delta\phi$.  Denoting  by   $\langle T^{\delta \phi} \rangle\equiv  \langle 0| T^{\delta \phi}|0\rangle$ such a result, we find
\begin{equation}\label{eq:QuantumTrace}
\langle T^{\delta \phi} \rangle=\left\langle \left(6\xi-1 \right)g^{\mu\nu}\nabla_\mu\delta \phi\nabla_\mu \delta \phi +2(3\xi-1)m^2\delta \phi^2+6\left(\xi-\frac{1}{6} \right)^2R\delta \phi^2+\left(\xi-\frac{1}{6} \right)R\delta \phi^2\right\rangle\,.
\end{equation}
This result for the vacuum trace (i.e. the vacuum expectation value of the trace)  adopts the same form as \eqref{eq:ClassicTrace2}, with $\phi$ replaced by its fluctuating part $\delta\phi$.

\subsection{Trace calculation up to $4th$ and  $6th$ adiabatic  orders}

We may now explicitly compute  the VEV of the trace, i.e. Eq.\,\eqref{eq:QuantumTrace},  using the Fourier decompositions of the field fluctuation $\delta\phi$  in the mode functions $h_k(\tau)$ and also utilizing the commutation relations between the  creation and annihilation operators, i.e. we proceed  along the lines we already followed in Sec.\,\ref{sec:AdiabaticVacuum} with the components of the EMT. The result can be  presented in two steps as follows:
\begin{equation}\label{eq:TraceVEV}
\begin{split}
&\langle T^{\delta \phi} \rangle=
-\frac{\left(6\xi-1\right)}{a^2}\left(\frac{\mathcal{H}^2}{(2\pi)^3 a^2}\int d^3 k |h_k|^2+\frac{1}{(2\pi)^3 a^2}\int d^3 k |{h}_k^\prime|^2-\frac{\mathcal{H}}{(2\pi)^3 a^2}\int d^3k \left(h_k h_k^{\prime*}+h_k^{\prime} h_k^* \right) \right)\\
&+\frac{\left(6\xi-1\right)}{a^2}\frac{1}{(2\pi)^3 a^2}\int d^3k k^2 |h_k|^2+2(3\xi-1)m^2\frac{1}{(2\pi)^3 a^2}\int d^3k |h_k|^2+6\left(\xi-\frac{1}{6} \right)^2 R\frac{1}{(2\pi)^3 a^2}\int d^3k |h_k|^2\\
&+\left(\xi-\frac{1}{6} \right) R\frac{1}{(2\pi)^3 a^2}\int d^3k |h_k|^2\\
&=\frac{1}{(2\pi)^3 a^2}\int d^3 k\left(-(6\xi-1) \frac{\mathcal{H}^2}{a^2}+(6\xi-1) \frac{k^2}{a^2}+2(3\xi-1)m^2+6\left(\xi-\frac{1}{6} \right)^2 R+\left(\xi-\frac{1}{6} \right) R \right) |h_k|^2\\
&-\frac{(6\xi-1)}{a^2}\frac{1}{(2\pi)^3 a^2}\int d^3k |h_k^\prime|^2+\frac{\left(6\xi-1 \right)}{a^2}\frac{\mathcal{H}}{(2\pi)^3a^2}\int d^3k \left(h_k h_k^{\prime*}+h_k^{\prime} h_k^* \right)\,.
\end{split}
\end{equation}
The first equality makes it clearer the structure of the result. For instance,  using the fact that   $g^{\mu\nu}\nabla_\mu\delta \phi\nabla_\mu \delta \phi =-a^{-2}\left((\delta\phi')^2-\nabla^2\delta\phi\right)$ and taking into account that the expansion of $\delta\phi'=(\delta\phi)'$ involves the calculation of $(d/d\tau)(h_k(\tau)/a)=(h'_k-\cH h_k)/a)$  -- as can be seen from Eq.\eqref{FourierModesFluc} (with $\delta\phi=\delta\varphi/a$) --,  it is easy to understand the origin of the first line of  Eq.\,\eqref{eq:TraceVEV}, and similarly with the other terms.
Up to this point this result is generic and no approximation has been performed (apart from using the adiabatic vacuum, on which the creation and annihilation operators act upon).   We must now expand the above VEV with respect to such vacuum state up to the $6th$-order.  To this aim we employ the   $6th$-order adiabatic expansions of the mode functions given in equations \eqref{eq:exphk2}-\eqref{eq:exphk2c} in combination with the relations \eqref{omegak0}.  On substituting them in the above formula the result is a rather lengthy expression, which is given in full in Appendix \ref{sec:Full6thOrder}.  We will use this result at due time.

\subsection{Trace anomaly}\label{sec:TraceAnomaly}
 As we know,  at the classical level  the trace of the EMT vanishes in the massless ($m=0$)  conformal limit ($\xi=1/6$). Indeed, using equation \eqref{eq:ClassicTrace2} it is  obvious that for ($\xi=1/6$) we get the result
\begin{equation}\label{eq:ConformalClassicalLimit}
T^{\rm cl.}=-m^2\phi^2 \,.
\end{equation}
Moreover,  it is evident that $\lim\limits_{m\rightarrow 0}  T^{\rm cl.}=0$, and hence the classical trace of the EMT vanishes in the massless conformal limit, a well-known result which corresponds to the Noether identity following from the conformal invariance of the theory in that limit. However, if we move to the part of the trace  inherent to the quantum fluctuations, we find from \eqref{eq:QuantumTrace} that in the conformal limit takes on the form $\langle T^{\delta \phi} \rangle=-m^2\langle \delta\phi^2 \rangle$, still consistent with \eqref{eq:ConformalClassicalLimit} since equations \eqref{eq:QuantumTrace}  and \eqref{eq:ClassicTrace2} are formally identical, as we noted.  We may naturally  wonder if $\lim\limits_{m\rightarrow 0} \langle T^{\delta \phi} \rangle=0$ holds good too.  The naive answer is yes, but the correct (and also well-known) answer is no.  This is the origin of the famous trace anomaly (also called conformal anomaly), see e.g. \cite{BirrellDavies82} and references therein.
 Basically, what happens is that  the quantum fluctuation $\langle \delta\phi^2 \rangle$  involves terms $\sim 1/m^2$, which make the limit nonvanishing and independent of $m$.   Setting  $\xi=1/6$ in Eq.\,\eqref{eq:TraceVEV} we find the reduced result:
\begin{equation}\label{eq:TraceVEVconformal1}
\begin{split}
&\left.\langle T^{\delta \phi} \rangle\right|_{\xi=1/6}=
-\frac{m^2}{(2\pi)^3 a^2}\int d^3k |h_k|^2=-\frac{m^2}{2\pi^2 a^2}\int dk k^2 |h_k|^2\,.
\end{split}
\end{equation}
The explicit form of this integral, however, is nontrivial. Notice that only the $4th$ adiabatic order terms contribute to it for $m\to 0$ since the $6th$ adiabatic order must decouple for $m\to\infty$, so it cannot be independent of $m$. In  the Appendix\,\ref{sec:TraceAnomalyComp} we furnish the details of this calculation  for $m\to 0$ and show that it leads to  the standard form of the trace anomaly.  The latter is the result associated to the finite part of the effective action. Since the above calculation comes up from the  unrenormalized part of the EMT, the trace anomaly is just minus the above result since the vacuum trace   of the total EMT derived from the full effective action must be zero in the massless conformally coupled limit\,\cite{BirrellDavies82}.   The latter can be computed first in the conformal metric and subsequently expressed in a covariant form, with the final result (cf. Appendix\,\ref{sec:TraceAnomalyComp}):
\begin{equation}\label{eq:TraceAnomaly}
\left.\lim\limits_{m\to 0} \langle T^{\delta \phi}\rangle\right|_{(\xi=1/6,M=m )}^{\rm anomaly}  =\frac{1}{480\pi^2 a^4}\left(4\mathcal{H}^2\mathcal{H}^\prime-\mathcal{H}^{\prime\prime\prime}\right)=\frac{1}{2880\pi^2}\left[R^{\mu\nu}R_{\mu\nu}-\frac{1}{3} R^2+\Box R\right]\,.
\end{equation}
The last equation coincides with the standard covariant formulation of the anomaly, except that the square of the Weyl tensor  does not show up here (while it appears for more general backgrounds\,\cite{BirrellDavies82})   since the FLRW spacetime is conformal to Minkowski spacetime (i.e. it is a conformally flat spacetime)  and hence the Weyl tensor vanishes identically.

\subsection{Trace renormalization}

It is important to realize that the vacuum trace \eqref{eq:TraceVEV} is UV-divergent and therefore needs renormalization. For instance,  similar to the situation with Eq.\,\eqref{T00}, the integrals in the first line of \eqref{eq:TraceVEV} are quadratically, quartically and logarithmically UV-divergent, respectively, cf. Eqs. \eqref{eq:exphk2}-\eqref{eq:exphk2c}.
In our framework, as outlined in Sec.\,\ref{sec:RenormEMT}, renormalization amounts to  subtract the same expression at an arbitrary mass scale $M$ (but computed only up to $4th$ adiabatic order):
\begin{equation}\label{eq:TraceEMTsubtracted}
\langle T^{\delta \phi} \rangle_{\rm ren}(M)=\langle T^{\delta \phi} \rangle (m)-\langle T^{\delta \phi} \rangle^{(0-4)}  (M)\,.
\end{equation}
This is of course the same subtraction prescription that we have already  followed  with the components of the EMT, see \eqref{EMTRenormalized}.  Computational details are lengthy and the full expression is given in  Appendix \ref{sec:Full6thOrder}.  Fortunately, the  final form of the renormalized trace of the vacuum  EMT, exact up to 6th adiabatic order,  can be cast in a relatively compact form as follows:
%
\begin{equation}
\begin{split}
&\langle T^{\delta \phi}\rangle^{{\rm (0-6)}}_{\rm ren}(M)
=\frac{1}{32\pi^2}\left(3m^4-4m^2M^2+M^4-2m^2\ln \frac{m^2}{M^2}\right)\\
&+\frac{3\left(\xi-\frac{1}{6}\right)}{8\pi^2}\left(m^2-M^2-m^2\ln \frac{m^2}{M^2}\right)\left(2H^2+\dot{H}\right)\\
&-\frac{9}{8\pi^2}\left(\xi-\frac{1}{6}\right)^2\left(12H^2 \dot{H}+4\dot{H}^2+7H\ddot{H}+\vardot{3}{H}\right)\ln \frac{m^2}{M^2}\\
&+\frac{1}{10080\pi^2 m^2}\left(16H^6+96H^4\dot{H}-44\dot{H}^3-66H^3\ddot{H}-54\dot{H}\vardot{3}{H}-96H^2\dot{H}^2\right.\\
\end{split}
\end{equation}
\begin{equation}
\begin{split}
&\phantom{XXXXXXXXX}\left.-93H^2\vardot{3}{H}-267H\dot{H}\ddot{H}-30H\vardot{4}{H}-36\ddot{H}^2-3\vardot{5}{H}\right)\\
&-\frac{\left(\xi-\frac{1}{6}\right)}{80\pi^2 m^2}\left(4H^6+30H^4\dot{H}-44H^2\dot{H}^2-18\dot{H}^3-22H^3\ddot{H}\right.\\
&\phantom{XXXXXXXXX}\left.-103H\dot{H}\ddot{H}-14\ddot{H}^2-31H^2\vardot{3}{H}-20\dot{H}\vardot{3}{H}-10H \vardot{4}{H}-\vardot{5}{H}\right)\nonumber\\
\end{split}
\end{equation}
\begin{equation}\label{eq:TraceIntegrated}
\begin{split}
&+\frac{3\left(\xi-\frac{1}{6}\right)^2}{16\pi^2 m^2}\left( 48H^4\dot{H}-16\dot{H}^3-8H^3\ddot{H}-14\ddot{H}^2\right.\\
&\phantom{XXXXXXXXX}\left.-20\dot{H}\vardot{3}{H}-16H^2\dot{H}^2-29H^2\vardot{3}{H}-95H\dot{H}\ddot{H}-10H\vardot{4}{H}-\vardot{5}{H} \right)\\
&-\frac{9\left(\xi-\frac{1}{6}\right)^3}{4\pi^2   m^2}\left(-8H^6+60H^4\dot{H}+11\dot{H}^3+42H^3\ddot{H}\right.\\
&\phantom{XXXXXXXXX}\left.+45H\dot{H}\ddot{H}+3\ddot{H}^2+3\dot{H}\vardot{3}{H}+102H^2 \dot{H}^2+6H^2\vardot{3}{H}\right)\,.\\
\end{split}
\end{equation}
We have  used once more the conversion relations between the derivatives of $\mathcal{H}$ with respect to the conformal time and the derivatives of $H$ with respect to the cosmic time (see Appendix \ref{sec:appendixA1}) so as to express the final result in terms of $H=H(t)$. Notice that the first three lines of the above expression comprise the terms  up to the 4th adiabatic order while the remaining lines stand for  the complete $6th$-order contributions. The subsequent contribution would be of  adiabatic order $8th$, which we are not interested in here.


\section{Equation of state of the quantum vacuum}\label{sec:EoSvacuum}

We should not presume that the equation of state (EoS) of the quantum vacuum is exactly $\Pv=-\rv$, as we must first carefully evaluate the quantum effects.  Obviously the EoS cannot depart too much from the traditional one, but we will see that it is not exactly $-1$.  The vacuum pressure is defined in a way similar to the vacuum energy density \eqref{RenVDE}. Assuming the vacuum to be an homogeneous and isotropic  medium (it should preserve the Cosmological Principle) we may define the pressure  using any diagonal $ii$-component of the renormalized vacuum stress tensor.

\subsection{Quasi-vacuum EoS}\label{sec:Quasivacuum}
Adopting once more the perfect fluid form\,\eqref{eq:VaccumIdealFluid} for the vacuum EMT,  we may  infer the  expression for the vacuum pressure  by  following the same logic as for the vacuum energy density \eqref{RenVDE}.  We start taking the $11th$-component,  $T_{11}^{\rm vac}$, of  the mentioned EMT. As we said, any $ii$th-component would do equally well owing to isotropy, and we find  $T_{11}^{\rm vac}=a^2\Pv$ in the conformal metric.  Mind that since we are using again the comoving cosmological frame there is no contribution from the $4$-velocity part.  We subsequently equate this result to the  $11th$-component of \eqref{EMTvacuum},  $\langle T_{11}^{\rm vac} \rangle=-\rho_\Lambda g_{11}+\langle T_{11}^{\delta \phi}\rangle=-\rL a^2+\langle T_{11}^{\delta \phi}\rangle$.  Thus, the renormalized vacuum pressure  at the scale $M$ is given by
\begin{equation}\label{eq:VacuumPressureDef}
\Pv(M)\equiv   \frac{\langle T_{11}^{\rm vac}\rangle_{\rm ren}(M)}{a^2}= -\rho_\Lambda (M)+ \frac{\langle T_{11}^{\delta \phi} \rangle_{\rm ren}(M)}{a^2}\,,
\end{equation}
which looks similar to the renormalized VED, Eq.\,\eqref{RenVDE}, up to a sign in the $\rL$ term.  This sign points to the expected EoS for the vacuum, but we need to proceed carefully before unveiling the final result. Having computed the $00th$-component of the EMT and its trace in the previous sections,  the isotropy condition enables us to compute the $11th$-component of the EMT simply by means of the relation
\begin{equation}\label{eq:VacuumT11}
\frac{\langle T_{11}^{\delta \phi} \rangle_{\rm ren}(M)}{a^2}=\frac{1}{3}\left(\langle T^{\delta \phi} \rangle_{\rm ren}(M)+\frac{\langle T_{00}^{\delta \phi} \rangle_{\rm ren}(M)}{a^2}\right)\,.
\end{equation}
Using now our definition \eqref{RenVDE} of VED, we can eliminate $\rL(M)$ in favor of $\rv(M)$ in the above equations, and we find
\begin{equation}\label{eq:VacuumPressureDef2}
\Pv(M)=-\rv(M)+\frac{1}{3}\left( \langle T^{\delta \phi} \rangle_{\rm ren}(M)+4\frac{\langle T_{00}^{\delta \phi} \rangle_{\rm ren}(M)}{a^2}\right)\,.
\end{equation}
This equation clearly shows that the EoS of the quantum vacuum is not exactly $-1$, and the departure from this value can be obtained from the previously computed expressions.  We can provide a rather precise result by including terms up to $6th$ adiabatic order\,\footnote{The reader may carefully track the calculation and observe that there is once more an exact cancellation of the quartic mass scales in the sum of the  two terms in parenthesis on the \textit{r.h.s.} of Eq.\,\eqref{eq:VacuumPressureDef2}. To check this one has to use equations \eqref{RenormalizedExplicit2} and \eqref{eq:TraceIntegrated}.  It follows that the scaling evolution of the vacuum pressure is also free from quartic mass dependencies. This is of course reassuring and shows the consistency of our calculation.}:
\begin{equation*}
\begin{split}
\Pv(M)=&-\rv(M)+\frac{\left(\xi-\frac{1}{6}\right)}{8\pi^2}\dot{H}\left(m^2-M^2-m^2\ln\frac{m^2}{M^2}\right)\\
&-\frac{3}{8\pi^2}\left(\xi-\frac{1}{6}\right)^2\left(6\dot{H}^2+3H\ddot{H}+\vardot{3}{H}\right)\ln \frac{m^2}{M^2}\\
&+\frac{1}{10080\pi^2 m^2}\left(8H^4\dot{H}-28\dot{H}^3+6H^3\ddot{H}-10\ddot{H}^2-22\dot{H}\vardot{3}{H}\right.\\
&\left.+24H^2 \dot{H}^2-7H^2 \vardot{3}{H}-49H \dot{H}\ddot{H}-6H \vardot{4}{H}-\vardot{5}{H}\right)\\
&+\frac{\left(\xi-\frac{1}{6}\right)}{240\pi^2 m^2}\left(-6H^4\dot{H}+34\dot{H}^3-6H^3\ddot{H}+12\ddot{H}^2+24\dot{H}\vardot{3}{H}-24H^2\dot{H}^2\right.\\
&\left.+7H^2\vardot{3}{H}+55H\dot{H}\ddot{H}+6H\vardot{4}{H}+\vardot{5}{H}\right)\\
\end{split}
\end{equation*}
\begin{equation}\label{eq:VacuumPressureFull}
\begin{split}
&-\frac{\left(\xi-\frac{1}{6}\right)^2}{16\pi^2 m^2}\left(32\dot{H}^3-12H^3\ddot{H}+12\ddot{H}^2+24\dot{H}\vardot{3}{H}-48H^2\dot{H}^2\right.\\
&\left.+5H^2\vardot{3}{H}+47H\dot{H}\ddot{H}+6H\vardot{4}{H}+\vardot{5}{H}\right)\\
&+\frac{9\left(\xi-\frac{1}{6}\right)^3}{4\pi^2m^2}\left(4H^4\dot{H}-5\dot{H}^3-6H^3\ddot{H}-11H\dot{H}\ddot{H}-\ddot{H}^2\right.\\
&\left.-\dot{H}\vardot{3}{H}-24H^2\dot{H}^2-2H^2\vardot{3}{H}\right)+\cdots
\end{split}
\end{equation}
where  \dots represent the $8th$-order contributions and above, which we shall not consider at all. The obtained expression takes  the generic  form
\begin{equation}\label{eq:VacuumPressureFullsplit}
\Pv(M)=-\rv(M)+f_2(M,\dot{H})+ f_4(M,H,\dot{H},...,\vardot{3}{H})+f_6(\dot{H},...,\vardot{5}{H})+\cdots\,,
\end{equation}
in which $f_2$, $f_4$ and $f_6$ involve ${\cal O}(T^{-2})$,  ${\cal O}(T^{-4})$ and ${\cal O}(T^{-6})$  adiabatic contributions. They represent a small correction to the canonical relation $\Pv(M)=-\rv(M)$ for the vacuum EoS and therefore, strictly speaking,  make the quantum vacuum a quasi-vacuum state.
 Notice that the adiabatic contributions $f_i$  are specific effects on  the pressure not present in the vacuum energy density, which in its own also contains contributions to all these orders.
The specific  ${\cal O}(T^{-6})$  terms shown above for the pressure are essential if we want to compute the EoS on-shell since  $f_2=0$  and $f_4=0 $  for $M=m$, as it is obvious from the first two lines of \eqref{eq:VacuumPressureFull}.  Therefore, at leading order,  the  on-shell  value of the vacuum EoS  is
\begin{equation}\label{eq:onshellEoS}
 \Pv(m)=-\rv(m)+f_6=-\rho_\Lambda (m)-\frac{\langle T_{00}^{\delta \phi}\rangle^{(6)}_{\rm ren}(m)}{a^2}+f_6\,,
 \end{equation}
where $\langle T_{00}^{\delta \phi}\rangle^{(6)}_{\rm ren}(m)$  is given by Eq.\,\eqref{renormalizedONSHELL6th}.  The EoS ``parameter'' therefore reads
\begin{equation*}
\begin{split}
\wv(m)=\frac{\Pv(m)}{\rv(m)}&=-1+\frac{f_6(m)}{\rv(m)}\\
&=-1+\frac{1}{10080\pi^2  m^2 \rv(m)}\left(8H^4\dot{H}-28\dot{H}^3+6H^3\ddot{H}-10\ddot{H}^2-22\dot{H}\vardot{3}{H}\right.\\
&\left.+24H^2 \dot{H}^2-7H^2 \vardot{3}{H}-49H \dot{H}\ddot{H}-6H \vardot{4}{H}-\vardot{5}{H}\right)\\
&+\frac{\left(\xi-\frac{1}{6}\right)}{240\pi^2 m^2 \rv(m) }\left(-6H^4\dot{H}+34\dot{H}^3-6H^3\ddot{H}+12\ddot{H}^2\right.\\
&\left.+24\dot{H}\vardot{3}{H}-24H^2\dot{H}^2+7H^2\vardot{3}{H}+55H\dot{H}\ddot{H}+6H\vardot{4}{H}+\vardot{5}{H}\right)\\
&-\frac{3\left(\xi-\frac{1}{6}\right)^2}{48\pi^2 m^2\rv(m) }\left(32\dot{H}^3-12H^3\ddot{H}+12\ddot{H}^2\right.\\
\end{split}
\end{equation*}
\begin{equation}
\begin{split}
&\left.+24\dot{H}\vardot{3}{H}-48H^2\dot{H}^2+5H^2\vardot{3}{H}+47H\dot{H}\ddot{H}+6H\vardot{4}{H}+\vardot{5}{H}\right)\\
&+\frac{9\left(\xi-\frac{1}{6}\right)^3}{4\pi^2 m^2 \rv(m)}\left(4H^4\dot{H}-5\dot{H}^3-6H^3\ddot{H}-11H\dot{H}\ddot{H}-\ddot{H}^2\right.\\
&\left.-\dot{H}\vardot{3}{H}-24H^2\dot{H}^2-2H^2\vardot{3}{H}\right)\,.\\
\end{split}
\end{equation}
Quite obviously $w_{\rm vac}(m)=w_{\rm vac}(m,H, \dot{H}, \ddot{H},,...) $ is actually a function of $H$ and its derivatives.  These terms can  be relevant at the very early stages of the cosmological evolution. However, even during  the short inflationary period deviations from the vacuum EoS $w_{\rm vac}=-1$ are tiny since all the terms that can trigger a departure  depend on derivatives of $H$, but $H$ remains essentially constant during inflation.  We discuss RVM-inflation in Sec.\,\ref{sec:RVMInflation}.  At this point the important result that we have just obtained must be emphasized in a twofold manner, quantitatively and qualitatively. First,  quantitatively, we have just proven that the EoS of the quantum vacuum is essentially the expected one,  i.e.  close to $w_{\rm vac}=-1$, and hence this is no longer an assumption or imposition; second,  qualitatively, we have found that  the quantum vacuum is not a  static state but  is  dynamical: it changes very slowly at present but it could well have been a powerful driving force in the past.  Both of these conclusions are perfectly reasonable for a primeval vacuum which might have been highly ``creative'' in the past but became much more tempered at present. Even so, the effect at present may not be entirely negligible, as it will be discussed in the next two sections. In particular, in  Sec.\,\ref{sec:EoSLateUniverse} we discuss the late time EoS of the quantum vacuum.

\subsection{Generalized RVM at low energies}\label{sec:Generalized RVM}

The result \eqref{eq:VacuumPressureFull}, derived in the previous section, reveals an interesting new feature.  Among the various  ${\cal O}(T^{-2})$,  ${\cal O}(T^{-4})$ and  ${\cal O}(T^{-6}) $ terms that we have collected on its \textit{r.h.s.} (all of which are contributions to the vacuum pressure beyond those entering the VED), the  ${\cal O}(T^{-2})$ one is particularly  worth noticing, namely the term
\begin{equation}\label{eq:f2}
\begin{split}
f_2(M,\dot{H})=\frac{\left(\xi-\frac{1}{6}\right)}{8\pi^2}\dot{H}\left(m^2-M^2-m^2\ln\frac{m^2}{M^2}\right)\,.
\end{split}
\end{equation}
This term can have implications on the vacuum dynamics at low energy since $\dot{H}$ is of the same order as $H^2$.   To see this, let us write down the two ordinary Friedmann's equations for flat three-dimensional space and in the presence of a dominant matter component and vacuum energy:
\begin{equation}\label{eq:FriedmanEqs}
\begin{split}
&3H^2= 8\pi G (\rho_m+\rv)\,, \\
&2\dot{H}+3H^2=- 8\pi G (p_m+\Pv)\,,
\end{split}
\end{equation}
where $\rho_m$ and $p_m$ are the density and pressure of the dominant matter component (relativistic or non-relativistic). From these two equations one can derive the differential equation that is satisfied by the Hubble rate:
\begin{equation}\label{eq:DiffH1}
\dot{H}+\frac32\,(1+\wm)\,H^2=4\pi\,G \left(w_m\rv-\Pv\right)\,,
\end{equation}
where $w_m=p_m/\rho_m$ is the EoS of the dominant matter component.  For the  present universe,  we have $w_m=0$  and  the above equation reduces to
\begin{equation}\label{eq:DiffH}
\dot{H}+\frac32\,H^2=-4\pi\,G \Pv\equiv 4\pi\,G \rveff\,.
\end{equation}
Here we have defined an effective vacuum pressure  $\rveff=-\Pv$ as if the EoS of the quantum vacuum would be exactly $-1$ . However, as we know, this does not imply $\rv=-\Pv$.   We use $\rveff$ only to mimic the situation in the $\CC$CDM, but in reality the quantum vacuum contributes with a term that produces a departure of the EoS from the usual value. From Eq.\,\eqref{eq:VacuumPressureFullsplit} we have the dominant contribution \eqref{eq:f2} at the scale $M$, which is of second adiabatic order, and can still be sizeable in the current universe.  In fact, we have two pieces of second adiabatic order on the \textit{r.h.s.} of \eqref{eq:VacuumPressureFullsplit}, one contained in  $\rv$ and the other given by $f_2$, which when combined lead to
\begin{equation}\label{eq:VacuumPressureCurrentU}
\begin{split}
\rveff(M)&=-\Pv(M)=\rv(M)-f_2(\dot{H})+ ...\\
\thickapprox & \left(\xi-\frac{1}{6}\right)\frac{3 {H}^2 }{16 \pi^2}\left(M^2-m^2 +m^2\ln \frac{m^2}{M^2} \right)\\
&+\left(\xi-\frac{1}{6}\right)\frac{\dot{H}}{8\pi^2}\left(M^2-m^2+m^2\ln\frac{m^2}{M^2}\right)+...
\end{split}
\end{equation}
Here we have used Eq.\,\eqref{RenVDEexplicit} and neglected the higher order adiabatic terms.  The meaning of $\thickapprox$ is that we have also omitted the first two terms of the mentioned equation, since we know that when we compare the VED at two scales in the same manner as we did  in \eqref{eq:VEDscalesMandM0Final} these terms will exactly cancel each other and only the indicated terms  of \eqref{eq:VacuumPressureCurrentU} will contribute. In fact, when we insert Eq.\,\eqref{eq:VacuumPressureCurrentU} on the \textit{r.h.s.} of  \eqref{eq:DiffH} and solve for $H$, it will all occur  as though the effective vacuum energy density contains not only the $\sim H^2$ dynamical component but also the new one proportional to $\dot{H}$.  We can repeat  a very similar argument to that in Sec.\,\ref{sec:RVMcurrentUniverse}, with the two scales  $M=H$ and $M_0=H_0$,  and we find that the effective expression for the vacuum energy density  in the present universe can be expressed in a generic form as follows:
\begin{equation}\label{eq:RVMgeneralized2}
\rveff(H,\dot{H})=\rvo+\frac{3\nu_{\rm eff}}{8\pi G_N}\,(H^2-H_0^2)+ \frac{3\tilde{\nu}_{\rm eff}}{8\pi G_N}\,(\dot{H}-\dot{H}_0)\,.
\end{equation}
We have normalized this relation such that  $\rveff(H=H_0, \dot{H}=\dot{H}_0)=\rvo$ at the present time, where  $H_0$ and  $\dot{H}_0$ stand  for the respective current  values of  $H$ and $\dot{H}$. Notice also that we have placed two generic coefficients $\nueff$ and $\tilde{\nu}_{\rm eff}$ for each of the two terms of adiabatic order $2$,  $H^2$ and $\dot{H}$, rather than the specific ones  in \eqref{eq:VacuumPressureCurrentU} (which would entail just $\tilde{\nu}_{\rm eff}=(2/3){\nu}_{\rm eff}$  for a single scalar field and no other matter field)  because in general we expect that these coefficients will receive contributions from different sorts of fields, fermions and bosons.  These coefficients are naturally small in magnitude, viz. of order  $\sim m^2/\mpl^2\ll1$, cf. Eq.\,\eqref{eq:nueffAprox}, but not hopelessly small if $m$ is the mass of a GUT particle. In the limit $\nueff,\tilde{\nu}_{\rm eff}\to 0$ we just recover the $\CC$CDM with constant $\rvo=\CC/(8\pi G_N)$, but if they are small though nonvanishing they can impact nontrivially on the phenomenology of the dark energy.  Here we have computed  the effect from a  single scalar field only,  but in general we have to sum over the concomitant contributions from other bosons and  fermions\,\cite{JCS2022}.

The above formula \eqref{eq:RVMgeneralized2} obviously extends the structure of   Eq.\,\eqref{eq:RVM2}.
In this way we have found an alternative justification for the generalized form of the  RVM,  which was already motivated in Ref.\,\cite{CristianJoan2020} from a different perspective.  Basically, one expects an extended form for the vacuum structure \eqref{EMTvacuum} such that it comprises more geometric structures which are not possible in Minkowski  spacetime but are certainly available in curved spacetime. Namely, one may naturally conceive a generalization of the form
 \begin{equation}\label{eq:GeneralVDE1}
 \left\langle T_{\mu\nu}^{\rm vac} \right\rangle= -\rL g_{\mu\nu} +\left\langle T_{\mu\nu}^{\delta\phi} \right\rangle + \alpha_1 R g_{\mu\nu} +\alpha_2 R_{\mu\nu}+\mathcal{O}(R^2)\,.
\end{equation}
In the above expression, $\mathcal{O}(R^2)$ represents possible contributions from geometric tensors of adiabatic order 4, that is $R^2$, $R_{\mu\nu}R^{{\mu\nu}},\dots$,  and $\alpha_i$ are parameters of dimension $+2$ in natural units.  In a more realistic picture,  contributions from all fields (bosons and fermions) are expected, and general covariance leads to a generic form as represented by  \,\eqref{eq:GeneralVDE1}.   In fact, the prospect for new terms in the effective vacuum action has been discussed in various ways in  the literature\,\cite{Maggiore2011,Maggiore2011b,Bilic2011,Bilic2012}. These terms are also expected in the aforementioned stringy version of the RVM, see\,\cite{NickJoan2020,NickJoan2021}.

Even though the vacuum dynamics from cosmological observations will receive contributions from all fields at a time, and in this sense the values of the coefficients  $\nueff$ and $\tilde{\nu}_{\rm eff}$ can only be determined observationally, what matters here  is that the theoretical framework leads  to small values for them, as we have seen.  After all the $\CC$CDM with a rigid cosmological constant works relatively well. Even so we  know that the latter is afflicted with persisting tensions which call for an explanation. The RVM seems to encode the key theoretical features for such an explanation  and  appears phenomenologically preferred as well\,\footnote{The fitting results with different data sets confirm that the coefficients $\nueff$ and $\tilde{\nu}_{\rm eff}$ are of order $10^{-3}$ at most, see e.g. \cite{RVMpheno1,RVMpheno2,PericoTamayo2017,CQG2017,ApJL2019,Mehdi2019,Tsiapi2019,BD2020,Singh2021,Manos2021,Yu2022,deCruzPerez:2023wzd}.  This suffices to have a nontrivial impact on the $\sigma_8$ and $H_0$ tensions\,\cite{EPL2021}. }.

\subsection{EoS of the quantum vacuum in the late universe}\label{sec:EoSLateUniverse}

Despite of the fact that the complete expression for the equation of state (EoS) of the quantum vacuum has been computed in Sec.\,\ref{sec:Quasivacuum}, see Eq.\eqref{eq:VacuumPressureFull}, it is interesting to highlight the specific form that such an EoS adopts in the late universe, which is the most accessible part of the cosmic history.  We therefore set $M=H$ at the corresponding epoch  (cf. Appendix  \ref{sec:appendixAbis1}) and from Eq.\,\eqref{eq:VacuumPressureCurrentU} we find that the current EoS is
\begin{equation}\label{eq:EoSLow1}
\begin{split}
\wv(H)\equiv &\frac{\Pv(H)}{\rv(H)}\simeq -1+\frac{f_2(\dot{H})}{\rv(H)}
\simeq -1+ \left(\xi-\frac{1}{6}\right)\frac{\dot{H} m^2}{8 \pi^2 \rv(H)}\left(1 -\ln \frac{m^2}{H^2} \right)\\
\simeq &  -1- \left(\xi-\frac{1}{6}\right)\frac{\dot{H} m^2}{8 \pi^2 \rvo} \ln \frac{m^2}{H_0^2}= -1-\nueff\, \mpl^2\,\frac{\dot{H}}{4\pi\rvo}\,.
\end{split}
\end{equation}
 In the last equation we have rephrased the result directly  in terms of  the approximately constant parameter $\nueff$ defined in Eq.\,\eqref{eq:nueffAprox}. This is all the more justified if we remain within the recent expansion history.  More details on the exact form of $\nueff(H)$ are given in Appendix \eqref{sec:appendixAbis1}.
Furthermore, we used $\ln \frac{m^{2}}{H^2}\gg1$ and also the fact that,  in linear order in the small parameter  $\nueff={\cal O}\left(10^{-3}\right)$  and for small redshift (which we can estimate in the range $z\lesssim10$)  we can approximate the expression of $\rv(H)$ -- cf. Eq.\,\eqref{eq:RVM2} --  involved in the denominator  of the above EoS formula by the constant value $\rv(H_0)=\rvo$  (the VED at present; and, finally, we have set $H=H_0$ in the log since it makes no significant difference for the cosmic span under consideration.   By simple manipulations from equations \eqref{eq:FriedmanEqs} we can reach  now a beautiful and compact expression for the current EoS of the quantum vacuum, which just depends on $\nueff$ and on the current cosmological parameters $\Omega_m^0$ and $\Omega_\CC^0$, and  can be written  in terms of the cosmological redshift $z$:
\begin{equation}\label{eq:EoSLow2}
\begin{split}
\wv(z)\simeq  -1+ \nueff \frac{\rho_m(a)}{\rvo}\simeq -1+\nueff \frac{\Omega_m^0}{\Omega_{\rm vac}^0}(1+z)^3\,.
\end{split}
\end{equation}
In the second equality  we have set  $\rho_m=\rho_m^0 a^{-3}=\rho_m^0 (1+z)^{3}$, just as in the matter conservation law for the  $\CC$CDM. This  is justified to ${\cal O}(\nueff)$ as we now argue. Recall that in the presence of running vacuum the matter conservation law can be affected in some cases. For example, if there is an exchange between vacuum and cold dark matter (CDM), the matter conservation law takes corrections of the type
 $\rho_{\rm  cdm}^0a^{-3(1-\nueff)}$\,\cite{RVMpheno1,RVMpheno2}.  However, even in these cases it does not modify the leading form of the EoS that we have found, as it only amounts to add a  ${\cal O}(\nueff^2)$ correction to it.  In other situations, such as e.g.  the one that will be studied in Sec.\,\ref{sec:RenormalizedFriedmann} and also the (so-called type-II) scenario addressed \cite{EPL2021}, the gravitational coupling also runs with the cosmic expansion but the running is of the form  $\sim \nueff \ln H$.  This term would induce once more a negligible  ${\cal O}(\nueff^2)$  correction to the equation of state \eqref{eq:EoSLow2}. As it turns out, therefore,  the   EoS   that  ensues  from our QFT approach is pretty universal for the RVM in its various implementations, at least to order ${\cal O}(\nueff)$ and for the indicated  low redshift range.
A more detailed treatment of the EoS in the general regime will be presented elsewhere\cite{Moreno-Pulido:2022upl}.

In summary,   Eq.\,\eqref{eq:EoSLow2} reconfirms what we had already advanced in Sec.\ref{sec:RVMcurrentUniverse}, namely that  for $\nueff>0$  (resp. $\nueff<0$) the evolving VED mimics quintessence (resp. phantom DE). This result is a bit provocative and comes as something of a surprise. The usual picture of the vacuum is that it is a kind of medium with a strict EoS equal to $-1$.  However,  the QFT analysis of vacuum in curved spacetime shows that it is not so.  The so-called quintessence or phantom fields (and in fact DE in general) could  be nothing else but a manifestation of the (dynamical) quantum vacuum, and if so {there is no need of  \textit{ad hoc} fields with particular potentials to explain the DE.  Vacuum fluctuations of quantum matter fields could just make it since upon  proper renormalization they lead to small contributions of order $m^2 H^2$.}   The current fits to $\nueff$ suggest that it should be in the ballpark of  $\sim 10^{-3}$ and positive\,\,\cite{ApJL2015,ApJ2017,RVMphenoOlder1,Elae2015,RVMphenoOlder2,RVMpheno1,RVMpheno2,PericoTamayo2017,CQG2017,ApJL2019,Mehdi2019,Tsiapi2019,BD2020,Singh2021,Manos2021,Yu2022,deCruzPerez:2023wzd,EPL2021}, and hence the favored dynamical DE  mimicked by the quantum vacuum seems to be that of quintessence.  A related result  was already hinted at long ago but on much more phenomenological grounds\,\cite{JoanHrvoje2005}.  Surprisingly, it also  holds in Brans-Dicke theory with a cosmological constant, as this context has recently been shown to mimic the RVM, see \cite{GRF2018,JavierJoan2018} and\,\cite{ApJL2019,BD2020}.  After we have studied the impact of the new pressure terms of ${\cal O}(T^{-2})$ at low energies, in the next section we consider the impact of the ${\cal O}(T^{-6})$ terms at high energies.


\section{Predicting inflation from running vacuum: RVM-inflation}\label{sec:RVMInflation}

Next we move to the opposite end of the cosmic history and consider the possible implications for the  very early universe. It is interesting to note that once the vacuum energy density in cosmological spacetime is renormalized through the  adiabatic procedure a definite prediction for a mechanism of early inflation emerges which is characterized by a short period where $H$=const. This constant must take, of course, a large value which we expect to lie around a characteristic GUT scale. It is nevertheless  totally  unrelated to the ground state  of a scalar field potential and hence does not require  any \textit{ad hoc} inflaton field. Such an alternative form of  inflation,  based on the constancy of $H$ for a short lapse of time,   is called  `RVM-inflation'.  To set off an inflationary phase with this mechanism we need powers of $H$ higher than  $H^2$, see\,\cite{rvmInflationpheno,JSPRev2015,GRF2015,Yu2020} for a phenomenological description.  In the current work,  however,  we can establish  `RVM-inflation'  from first principles.  For this we need to go beyond  ${\cal O}(T^{-2})$ (i.e. beyond second adiabatic order), but in fact we must go even further.  In our previous work \cite{CristianJoan2020}, the calculation reached  up to ${\cal O}(T^{-4})$ ($4th$ adiabatic order) since this was sufficient for discussing the renormalization of the EMT.  However, the ${\cal O}(T^{-4})$  terms that we found turn out to vanish for $H=$const.  since they all depend on  time derivatives,  as it is manifest in  Eq.\,\eqref{RenVDEexplicit}.  The absence of the power $H^4$ in the renormalized EMT hinges on the renormalization prescription \eqref{EMTRenormalized}, in which the divergences are removed by subtracting the EMT up to order adiabatic $4$ at the scale $M$.  However, the  ${\cal O}(T^{-6})$ terms accounted for here take now the lead, see Eq.\,\eqref{renormalizedONSHELL6th}. They originate from finite contributions unrelated to renormalization.  Collecting the relevant terms from our $6th$-order  calculation,  we find
\begin{equation}\label{eq:RVMinflation}
\rv^{\rm inf}(m)=\frac{\langle T_{00}^{\delta \phi}\rangle^{(6)}_{\rm ren}(m)}{a^2}=\frac{\txi}{80\pi^2 m^2}\, H^6+ f(\dot{H}, \ddot{H},\vardot{3}{H}...)\,,
\end{equation}
which we have labeled  with a superindex `inf' because such an effective VED  triggers inflation, as we shall see immediately.   In computing the overall coefficient of $H^6$, we have defined the parameter
\begin{equation}\label{eq:xitilde}
 \txi=\left(\xi-\frac16\right)-\frac{2}{63}-360\left(\xi-\frac16\right)^3\,.
\end{equation}
Notice that in \eqref{eq:RVMinflation} we have only stood out  the  contributions from \eqref{renormalizedONSHELL6th} which can be responsible for fast inflation  in a transient  $H=$const. regime.  The only relevant terms for such an inflationary interval are those  carrying the power $H^6$ with some constant coefficient.  The remaining terms, collected in the function $ f(\dot{H}, \ddot{H},\vardot{3}{H}...)$,  consist of different combinations of powers of $H$ accompanied  in all cases with derivatives of $H$, and hence all these terms vanish for $H=$const. In other words, $f=0$ for $H=$const. in Eq.\,\eqref{eq:RVMinflation}.  As a result,  up to $6th$ adiabatic order the only terms which do not vanish for constant Hubble rate are the isolated powers $H^6$.  The overall coefficient upon collecting all these terms is given by \eqref{eq:xitilde}.

For the current discussion on inflation  we may admit the presence of incoherent matter with density and pressure $(\rho_m,p_m)$ beyond our original field $\phi$, which is of course a most realistic assumption for this consideration.  The primeval, highly energetic, vacuum can then decay into relativistic particles of all species.  In the mentioned phenomenological approach, it is  considered a generalized RVM model of the form
\begin{equation}\label{eq:EffLambda}
\rv(H)=\frac{3}{\kappa^2}\left(c_0+\nu H^2+\tal \frac{H^{2p+2}}{H_I^{2p}}\right)\ \ \ \ \ \ \ \  \ \ \ \  (p=1,2,3,...)\,,
\end{equation}
where $\tal$ is another dimensionless coefficient. For  $\tal=0$  we recover the low-energy form of the RVM, i.e.  Eq.\,\eqref{eq:RVM2}, after we impose the boundary condition $\rv(H=H_0)=\rvo$  to determine the coefficient $c_0$.  At higher energies, however, the presence of higher powers of $H$ beyond $H^2$ can bring about inflation in the early universe, and in this case $H_I$ defines (up to a coefficient) the scale of inflation, as we shall see. The effect of the higher powers of $H$ is negligible for the  current universe.  For $\tal\neq 0$ the primeval vacuum can decay into matter (most likely relativistic) at high energies (when  $H\sim H_I$  very large),  thus  we can set  $\rho_m=\rho_r$ and $p_m=p_r=w_r\rho_r$  (with  $w_m=w_r=1/3$ in this case)  in \eqref{eq:FriedmanEqs}  and  \eqref{eq:DiffH}.  Additionally,  since $f_6=0$ for $H=$const.  we have  $\rveff=(4/3)\rv$ on the \textit{r.h.s.} of \eqref{eq:DiffH} for the inflationary period.  Coefficients $c_0$ and $\nu$ are not important for the early universe, and hence  we may just concentrate on the effect of $H^{2p+2}$  for  the study of the RVM-inflationary mechanism.   Once more the presence of only even powers  $H^{2p+2}\, (p=1,2,3,...)$  is related to the general covariance.   Our QFT calculation has revealed that  $p=2$ is singled out as the lowest possible power  ($\sim H^6$) available for triggering inflation in the present framework.

Let us summarize the main traits of implementing RVM-inflation from our predicted VED  form \eqref{eq:RVMinflation} in the very early universe.  First, we observe that we can  take  $\Pv\simeq-\rv$ for an inflationary regime in which  $H$ remains approximately constant since functions  $f_2,f_4$ and $f_6$  are essentially vanishing in Eq.\,\eqref{eq:VacuumPressureFullsplit}  during such an inflationary phase.  In these conditions,  taking  $p=2$ in \eqref{eq:EffLambda} and neglecting the terms $c_0$ and $\nu H^2$ in front of the higher power $H^6$, we can actually solve for $H$ directly from \eqref{eq:DiffH}:
\begin{eqnarray}\label{Hfunction}
H(a)=\frac{\tHI}{\left[1+\displaystyle{\left(\frac{a}{\astar}\right)^{8}}\right]^{1/4}}\,.
\end{eqnarray}
where we have traded cosmic time for the scale factor variable,  $a$.
Next, using equations \eqref{eq:FriedmanEqs} we can solve for  the explicit form of the radiation and vacuum energy densities\,\footnote{See Appendix B of Ref.\,\cite{Yu2020}, where the analytic solution is given for arbitrary $\nu$, $\tal$ and $p$, but still with $c_0=0$. The analytic solution for $c_0\neq 0$ is only possible for the late universe, where the high power  $H^{2p+2}\, (p\geq 1)$ is negligible\,\cite{Yu2020}. }:
\begin{eqnarray}\label{rhodensities}
\rho_r(a)=\rI\,\frac{\displaystyle{\left(\frac{a}{\astar}\right)^{8}}}{\left[1+\displaystyle{\left(\frac{a}{\astar}\right)^{8}}\right]^{\frac{3}{2}}}\,\,, \ \ \ \ \ \ \ \ \
\rv(a)=\frac{\rI}{\left[1+\displaystyle{\left(\frac{a}{\astar}\right)^{8}}\right]^{\frac{3}{2}}}\,\,.
\end{eqnarray}
We are expressing the above results in terms of the  point $\astar$, which  defines  the transition between vacuum dominance and the radiation era, i.e. the point which satisfies $\rho_r(\astar)=\rv(\astar)$.  It can be estimated as  $\astar\sim 10^{-29}$ within a typical GUT defined at the scale at $M_X\sim 10^{16}$ GeV\,\cite{Yu2020}. Furthermore, we have defined
\begin{equation}\label{eq:defsHIrI}
\tHI=\frac{H_I}{\tal^{1/4}}\,, \ \ \ \ \ \ \ \ \ \ \ \  \rI=\frac{3}{\kappa^2}\,\tHI^2\,.
\end{equation}
Since $H(a=0)=\tHI$, it follows that this is the value of the Hubble rate in the very early universe.  Similarly,  $\rv(0)=\rI$ is the VED at that initial point.  We can see that they are both finite. The model, therefore, presents no early singularities at all.  On comparing\,\eqref{eq:RVMinflation}  with the generic form \eqref{eq:EffLambda} -- with $p=2$ in our case -- and using the above definitions  we can easily identify
\begin{equation}\label{eq:HIeffective}
\tHI=\left(\frac{240\pi^2}{\txi}\right)^{1/4}\sqrt{\MPl\, m}\,.
\end{equation}
Clearly, in order to have inflation  near a typical GUT scale we need   very massive particles with masses $m$ in the neighborhood of that scale. In addition, it  is  imperative, of course,  to have  $\txi>0$. For a single scalar field nonminimally coupled to gravity, we find numerically  from \eqref{eq:xitilde}  that this occurs   for $\xi\lesssim0.1023$.  Such a range excludes $\xi=1/6$, for which $\txi<0$, but it admits the minimal coupling situation $\xi=0$, and the negative values $\xi<0$.  In general,  we can expect to have many fields nonminimally coupled to gravity with different couplings and masses,  especially if we consider the matter content of GUT's.  Therefore, a proper study of inflation must take into account the fact that we have a much wider parameter space than with just one scalar field.  We have checked that, in general, RVM-inflation can be made compatible with nonvanishing running of the VED at low energies,  see Eq.\eqref{eq:RVM2}. However, a detailed account of RVM-inflation in the QFT context requires a  dedicated study and will be presented elsewhere.  In particular, we note that the fermionic contribution might be important as well. Although the generalization of the point-splitting and ARP methods for fermions have been amply discussed in the literature\,\cite{Christensen1978,Landete2013Barbero2018},  its application to compute the scaling evolution of the  VED, and more specifically  within the off-shell ARP procedure that we have been using for scalar fields,  has been considered only very recently\,\cite{JCS2022}.  Overall, theoretical scenarios indeed exist for which RVM-inflation occurs along with  $\nueff>0$  in Eq.\eqref{eq:nueffAprox},  thereby  being consistent with the sign picked up by the current phenomenological  analyses\,\cite{RVMpheno1,RVMpheno2,EPL2021}.

\begin{figure}[t]
\begin{center}
\includegraphics[scale=0.65]{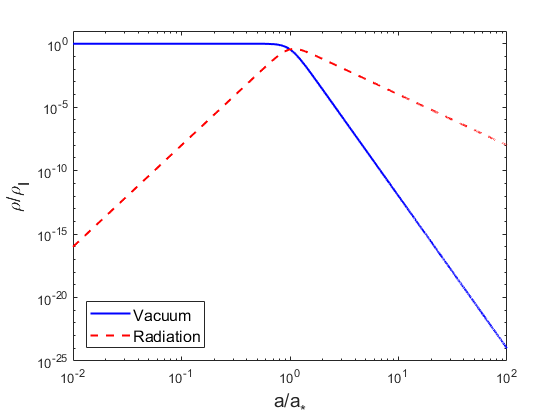}
\end{center}
{\bf Figure  1}.   Inflationary period.  It is shown the evolution of the energy densities \eqref{rhodensities} of vacuum and relativistic matter before and after  the transition point $a_*\sim 10^{-29}$  from inflation to the early radiation epoch (see the text).  The constant vacuum energy density during inflation decays into radiation and the standard FLRW regime starts.
\label{Fig1}
\end{figure}

From equations \eqref{rhodensities} we learn that the VED is initially  constant and large, $ \rv(a)\simeq \rI$  for  $a\ll \astar$, and decreases very fast beyond that point.  On the other hand, the radiation energy is initially zero, $\rho_r(0)=0$,  but  increases very fast in the beginning  ($\rho_r\sim a^8$) owing to the vacuum decay into radiation.   We can appraise this behavior in Fig.\,1, where we can see the transition from the pure vacuum state that brings about the inflationary phase, in which the VED remains approximately constant,  into an incipient radiation epoch, which  soon dominates the evolution of the universe.  For $a\gg\astar$,  the frenzied growing  $\sim a^8$ of the radiation density turns into the dilution law $\rho_r(a)\sim a^{-4}$  and hence we retrieve the standard  behavior of the radiation until our days.  The VED, on the other hand, remains negligible during the radiation epoch as compared to  $\rho_r$ and therefore  it cannot perturb in any significant way the BBN processes which will occur  much later at  $a\sim 10^{-9}$.   In the long run, after entering the matter-dominated epoch,  the VED  will  recover the (much more tranquil)  form \eqref{eq:RVM2}, which is characteristic of the current universe and  evolves just  as a constant plus a mild component $ H^2\sim a^{-3}$ with a small coefficient. Obviously such form will appear in the solution only if we  keep the terms $c_0$ and $\nu$  in solving the equations, but in this section  we focus exclusively on the very early times when the  inflationary regime turns into the primeval radiation epoch.  Needless to say, the above solution is approximate and is in effect only when the terms carrying the time derivatives of $H$ are negligible as compared to $H^6$  during the short inflationary stage set off by $H\simeq$ const.  This estimate  is nonetheless sufficient to exhibit the main features of the early  phase of RVM-inflation.  The  terms with  time derivatives  of $H$ are suppressed and cannot perturb significantly the inflationary period characterized by $H\simeq$const. nor can have any sizeable influence beyond the very early universe  once the $H^2$ terms take over until the constant term $c_0$ becomes eventually dominant around our present time.

The inflationary mechanism based on  a short lapse of  time in which $H\simeq$ const. (unrelated to the ground state value of the potential of a scalar field) is the distinctive hallmark of RVM-inflation as compared to other  related mechanisms. For example,  in the well-known  Starobinsky's $R^2$-inflation\cite{Starobinsky80}  there exists no inflationary phase with  $H\simeq$ const.  Instead,  there exists a short interval in which it is the time derivative of the Hubble rate (rather than the Hubble rate itself)  which remains constant,  i.e. $\dot{H}\simeq$ const.  During such a phase  inflation is prompted as well.  The approximations made in both cases are similar since during the $\dot{H}\simeq$ const. regime  all higher  time derivatives are neglected -- cf. Ref.\,\cite{JSPRev2015} for a  comparative discussion of RVM-inflation versus Starobinsky inflation.

Despite RVM-inflation had been explored phenomenologically in the mentioned previous works (cf.\,\cite{rvmInflationpheno} and \cite{JSPRev2015,GRF2015,Yu2020}),  all of them were based on assuming the \textit{ad hoc} structure \eqref{eq:EffLambda}. Here, in contrast, we have been able to show  that  such a structure indeed emerges from QFT in curved spacetime and that  $p = 2$ is the lowest possible value. In other words, we have proven from an explicit QFT calculation that  RVM-inflation can be  unleashed by the  $\sim H^6$-term in the vacuum energy density.  Thus, the two specific powers $H^2$  and  $H^6$ are the ones which are picked out by the quantum effects of matter fields in the FLRW background.   The former power  affects the dynamics of the current universe, whilst the latter is responsible for inflation in this context.  Remember that only even powers are allowed by general covariance. The conspicuous absence of  the power  $H^4$  is not surprising: it is  a built-in consequence of the subtraction procedure  in the  adiabatic renormalization of the EMT in four dimensions.  At the end of the day,  the RVM description has the ability  to encompass  the entire history of the universe from inflation up to our days from first principles\,\footnote{We remind the reader that a related inflationary mechanism, based on the power $H^4$,  is conceivable in the framework of\,\cite{BMS2020,NickJoan2020,NickJoan2021}. However, such a proposal is not based on the QFT action considered here but on gravitational-anomaly aspects which are specific of the bosonic part of the effective action of string theory.  Because the leading power  in this case is $H^4$ rather than $H^6$, such a stringy RVM-inflation is, in principle,  distinguishable from the QFT one addressed here. These two independent options strengthen the general support for RVM inflation.}.  A thorough account of the RVM mechanism  of inflation will be presented in a devoted study.

\section{Heat-Kernel approach:   renormalization of the effective action}\label{HeatKernel}
The effective action describing the quantum  matter vacuum effects of QFT in curved spacetime, $W$,  is defined through its  relation  to the VEV of the  EMT\,\cite{BirrellDavies82,ParkerToms09,Fulling89}. In our conventions,
\begin{equation}\label{eq:DefW}
\langle T^{\mu\nu}\rangle=\frac{2}{\sqrt{-g}} \,\frac{\delta W}{\delta g_{\mu\nu}}\ \ \ \ \ \Longleftrightarrow\ \ \ \ \ \ \langle T_{\mu\nu}\rangle=-\frac{2}{\sqrt{-g}} \,\frac{\delta W}{\delta g^{\mu\nu}}\,.
\end{equation}
Such an effective  action provides the quantum matter vacuum effects on top of the classical action. These quantum effects can be computed through a loopwise expansion in powers of $\hbar$. Thus, if the expansion is truncated at the one loop level it contains all terms of the complete theory to order $\hbar$.
It is  well-known that at leading (one loop) order the value of $W$ for the free theory  is essentially given by the trace of the logarithm of the  inverse of the Green's function.  More specifically:
\begin{equation}\label{eq:EAW}
\begin{split}
W= &\frac{i\hbar}{2}Tr \ln (-G_F^{-1})=\frac{i\hbar}{2}Tr \ln (-G_F)^{-1}= -\frac{i\hbar}{2}Tr \ln (-G_F)\\
=&  -\frac{i\hbar}{2} \int d^4 x \sqrt{-g}\lim\limits_{x\to x'} \ln\left[ -G_F(x,x')\right]\equiv \int d^4 x \sqrt{-g}\, L_W\,,
\end{split}
\end{equation}
wherein in the second line we have indicated the precise  computational meaning of the trace in the spacetime continuum. The last equality defines the  Lagrangian density $\sqrt{-g}\, L_W$,  and for simplicity we will call the piece $L_W$ the (effective) quantum vacuum Lagrangian, as it accounts for the quantum vacuum  effects from the quantized matter fields (in our case just the scalar field $\phi$).  We retain $\hbar$ in the above expression just to emphasize the aforementioned fact that the above formula describes a pure quantum effect at one loop. From now on, however, we continue with $\hbar=1$  (as it has been done in most of the paper). The above action gives the vacuum effects, i.e. the  effects originating  from the matter  vacuum-to-vacuum (`bubble') diagrams. These  are closed loop diagrams without external tails, thereby rendering the zero-point energy  contributions (ZPE).  Let us note that,  in the case of the free field theory based on the action \eqref{eq:Sphi},  we have no self-interactions of $\phi$ since there is no effective potential.  In this situation we have one single bubble diagram and  the one-loop effective action is the exact result since in the absence of matter interactions we cannot have additional vertices to insert in the bubble diagram to produce higher order loops.

The ZPE is usually discarded in flat spacetime on account of normal  ordering of the operators (in the operator formulation) or by normalizing to one the generating functional of the Green's functions at zero value of the source (in the functional approach to QFT).  In the context of gravity, none of these arbitrary settings is permitted.  Although we have  already performed  the computation of the ZPE by directly computing the VEV of the enery-momentum tensor, i.e. the \textit{l.h.s.} of Eq.\,\eqref{eq:DefW},  here we wish to dwell further on these considerations in the context of the effective action approach  by computing $W$ and then re-deriving the vacuum EMT using  Eq.\,\eqref{eq:DefW}.  While the procedure is well-known \cite{BunchParker1979,BirrellDavies82,ParkerToms09,Fulling89}, we wish to discuss the changes introduced in it when we compute the effective action off-shell, as this is convenient  to better understand the connection with the previous sections, in which we subtracted the EMT off-shell.

In curved spacetime, the Feynman propagator, $G_F$, is the solution to the following distributional differential equation\,\cite{BirrellDavies82,ParkerToms09,Fulling89}:
\begin{equation}\label{KGPropagatorOnShell}
\left(\Box_x-m^2-\xi R(x)\right)G_F(x,x^\prime)=-\left(-g(x)\right)^{-1/2}\delta^{(n)}(x-x^\prime)\,,
\end{equation}
where $\delta^{(n)}$ is the Dirac $\delta$ distribution in $n$ spacetime dimensions. For all practical purposes in our work, $n=4$.  Notwithstanding we can keep $n$ general at the moment since DR will be employed  for regularizing the UV divergences  in the calculations presented in  this section.    We now reformulate the above on-shell equation in an appropriate form as follows:
\begin{equation}\label{KGPropagatorOffShell}
\left(\Box_x-M^2-\Delta^2-\xi R(x)\right)G_F(x,x^\prime)=-\left(-g(x)\right)^{-1/2}\delta^{(n)}(x-x^\prime)\,.
\end{equation}
Here we have introduced a new scale, $M$, and also the important quantity:
\begin{equation}\label{eq:Delta2}
\Delta^2\equiv m^2-M^2\,.
\end{equation}
Although we have already introduced $\Delta^2$  in the context of the WKB expansion (cf. Sec.\,\ref{sec:WKB}),  we now endow it with  a different perspective that may help to better understand its meaning. The strategy behind Eq.\,\eqref{KGPropagatorOffShell} is to delve into the solution to the propagator  equation (and hence of the effective action) for an arbitrary mass scale $M$. We can recover the on-shell case $M=m$ by simply setting $\Delta=0$.  But if the quantity $\Delta^2$ is to be used to explore the off-shell regime it must be dealt with as being of  adiabatic order higher than $M$ (which is of order zero). Hence  $\Delta^2$ must be conceived as being of adiabatic order $2$, which is the next-to-leading order compatible with general covariance.  Taking into account that  the term  $\xi R$  in \eqref{KGPropagatorOffShell} is also of adiabatic order $2$, the combination $\Delta^2+\xi R$  can be treated as a block  of adiabatic order $2$.  This adiabaticity assignment for $\Delta^2$  is consistent with our former considerations in Sec.\,\ref{sec:WKB} and can be regarded as an alternative justification for it.  As long as the adiabaticity order of the terms must be hierarchical respected, the fact  that the mass scale $M$ is of adiabatic order zero whereas the special quantity $\Delta^2$ is of adiabatic order $2$ is precisely what makes the solution to the Green's function equation \eqref{KGPropagatorOffShell} different from the solution to the original (on-shell) equation \eqref{KGPropagatorOnShell}.   The adiabatic expansion of the solution to Eq.\,\eqref{KGPropagatorOffShell} will generate new ($\Delta^2$-dependent) terms which are genuinely distinct as compared to the adiabatic expansion of the solution to \eqref{KGPropagatorOnShell}.   In what follows we work out such an expansion of the Green's function for a scalar field in curved spacetime\,\cite{BirrellDavies82}, with the purpose of identifying  the (extra) $\Delta$-dependent terms characteristic of our off-shell subtraction procedure. See also the approach of \cite{FerreiroNavarroSalas2}, which is however slightly different as we will comment later on.

\subsection{Effective action of QFT in curved spacetime}

The solution to \eqref{KGPropagatorOffShell} can be obtained from the adiabatic expansion of the Green's function. The method is well-known\,\,\cite{BirrellDavies82,ParkerToms09} but is outlined here (with some more technical details disclosed in Appendix \ref{sec:appendixB}) since,  as advertised, we wish to pay special attention to the modification introduced by the presence of the extra terms  $\Delta^2$ on top of  the usual procedure.  The final result allows us to determine the effective Lagrangian defined in Eq.\,\eqref{eq:EAW} and express it in the form of an asymptotic DeWitt-Schwinger expansion\,\cite{DeWitt1975}. Using DR as a regularization procedure in this section, we find after  a considerable amount of calculations (see Appendix \ref{sec:appendixB} for an expanded exposition) the following result\footnote{For convenience we use DR in the context of the effective action, but this is (as we remarked before)  just optional. We have explicitly checked that one can obtain the same results with the subtraction procedure used to renormalize  the EMT  in the previous sections. See Appendix \ref{sec:appendixA2} for some more considerations along these lines.}:
\begin{equation}\label{eq:effLagrangian}
\begin{split}
L_W=&\frac{\mu^{4-n}}{2(4\pi)^{n/2}} \sum_{j=0}^\infty \hat{a}_j (x) \int_0^\infty (is)^{j-1-n/2}e^{-iM^2 s}ids
=\frac{1}{2(4\pi)^{n/2}}\left(\frac{M}{\mu}\right)^{n-4}\sum_{j=0}^\infty \hat{a}_j (x) M^{4-2j}\Gamma \left(j-\frac{n}{2}\right)\\
=&\frac{1}{2(4\pi)^{2+\frac{\varepsilon}{2}}}\left(\frac{M}{\mu}\right)^{\varepsilon}\sum_{j=0}^\infty \hat{a}_j (x) M^{4-2j}\Gamma \left(j-2-\frac{\varepsilon}{2}\right)\,,
\end{split}
\end{equation}
where $\varepsilon\equiv n-4$ and the limit $\varepsilon \rightarrow 0$ is understood;  $\Gamma$ is Euler's gamma function and $\mu$ is 't Hooft's mass unit to keep the effective Lagrangian with natural dimension  $+4$ of energy   in $n$ spacetime dimensions. The final results will not depend on it.  The sum is over   $j=0,1,2,...$ and includes  the even adiabatic orders only.  The modified  DeWitt-Schwinger coefficients resulting from our calculation up to fourth adiabatic order read
\begin{equation}\label{eq:ModifiedDWSchcoeff}
\begin{split}
&\hat{a}_0 (x)=1=a_0 (x),\\
&\hat{a}_1 (x)=a_1(x)-\Delta^2 ,\\
&\hat{a}_2 (x)=a_2(x)+\frac{\Delta^4}{2}+\Delta^2 R \left(\xi-\frac{1}{6}\right)
\end{split}
\end{equation}
and include  the zero, second and fourth adiabatic orders, respectively.
The hatless $a_i(x)$ are the ordinary  DeWitt-Schwinger coefficients for  $\Delta=0$ (on-shell expansion),  which the reader can find in the in Appendix \ref{sec:appendixB}.  The effective Lagrangian \eqref{eq:effLagrangian} and corresponding effective action are UV-divergent quantities since the Euler's $\Gamma$-function is divergent for $j=0,2,4$  in  $n=4$ spacetime dimensions.

Let us now consider some renormalization issues. Our starting point was the Einstein-Hilbert action \eqref{eq:EH}  together with the quantum  matter action \eqref {eq:SrL}.  The former is associated to the Lagrangian
\begin{equation}
L_{EH}=-\rho_\Lambda+\frac{1}{16\pi G} R=-\rho_\Lambda+\frac{1}{2}\MPl^2 R\,,
\end{equation}
which represents the starting  (classical) vacuum action. We have nevertheless observed in our discussion on the EMT renormalization in Sec. \ref{sec:RenormEMT} that, even though we did not start with  higher derivative (HD)  terms in the action,  such as  $R^2(x)$, $R^{\mu\nu}(x)R_{\mu\nu}(x)$, etc.,  these purely geometric structures are generated by the quantum fluctuations of the matter field, which probe the short distances around $x$. Therefore,  renormalizability of  QFT in the FLRW background requires that the more general classical action comprises also these HD geometric structures.  Let us write the extended classical gravitational Lagrangian for the vacuum with all the necessary terms  in two alternative ways as follows:
\begin{equation}\label{eq:LEHHD}
\begin{split}
L_G^{\rm cl.}=L_{EH}+L_{HD}=&-\rho_\Lambda +\frac{1}{2}\MPl^2R+\alpha_Q \frac{{Q^\lambda}_\lambda}{3}+\alpha_2 R^2\\
=&-\rho_\Lambda+\frac{1}{2}\MPl^2  R+\alpha_1 C^2+\alpha_2 R^2+\alpha_3 E+\alpha_4\Box R\,,
\end{split}
\end{equation}
in which the notation in $L_G^{\rm cl.}$  indicates that this is the classical Lagrangian part of the  gravitational field, to which we still have to add the quantum vacuum effects.
Coefficients $\alpha_i$  for  $i=1,3,4$ in the second expression can be easily related with the coefficient $\alpha_Q$ if we take into account that the combined HD structure   ${Q^\lambda}_\lambda$  can be phrased in terms of the square of the Weyl tensor ($C^2$), the Euler density ($E$) and a total derivative term as follows (cf. Appendix \ref{sec:appendixA1}):
\begin{equation}\label{eq:traceQ1}
\frac{1}{3}{Q^\lambda}_\lambda =-\frac{1}{120}C^2+\frac{1}{360}E+\frac{1}{6}\left(\xi-\frac{1}{5}\right)\Box R\,.
\end{equation}
These HD terms did not appear when we renormalized the EMT in Sec.\ref{sec:RenormEMT} since we used a restricted generalization of Einstein's equations, viz. Eq.\,\eqref{eq:MEEs}, which is sufficiently general for the FLRW spacetime. The three terms \eqref{eq:traceQ1} appear in a natural way in the effective action approach since they are involved as part of the DeWitt-Schwinger coefficient $a_2$ (cf. Appendix \ref{sec:appendixB}), so we have just computed them within the natural flow of the effective action procedure, but none of these terms actually plays any role for FLRW spacetime since the latter in conformal to the Minkowski metric and hence the Weyl tensor vanishes identically. The other two are also irrelevant at the level of the action since $E$ leads to  a topological invariant in $n=4$ dimensions, the Gauss-Bonnet term $\GB$  (cf. Appendix \ref{sec:appendixA1}),  and $\Box R$ is a total derivative.  We have carried along these HD terms up to this point just for completeness, but in effect the only HD term which stays in the FLRW background is $R^2$, as we warned in Sec.\ref{sec:RenormEMT}. We will nonetheless still keep these terms in the next section so as to close our discussion on the effective action method in a more complete way.

Starting from $L_W$ and following a procedure similar to our definition of adiabatically renormalized EMT, see Eq.\,\eqref{EMTRenormalized}, we  define now the renormalized  quantum vacuum Lagrangian  at the scale $M$.  It is obtained by subtracting the divergent adiabatic orders at this scale from the on-shell value  $L_W (m)$:
\begin{equation}\label{eq:LWrenormalized}
L_W^{\rm ren}(M )= L_W (m)-L_W^{(0-4)}(M)\equiv  L_W (m)-L_{\rm div} (M)\,,
\end{equation}
where  $L_{\rm div}(M)\equiv  L_W^{(0-4)}(M)$ is the divergent part  of Eq.\,\eqref{eq:effLagrangian}; by this we mean that   $L_{\rm div}$ is that part of  $L_W$ involving only  the terms $j=0,1,2$, i.e. up to fourth adiabatic order.   Of course both  $L_W (m)$  and $L_{\rm div} (M)$ are divergent, but the former is assumed to involve the full DeWitt-Schwinger expansion at the scale $m$, whilst the latter stops the expansion at $j=2$ and is evaluated at a different scale $M$.  This subtraction prescription for the quantum vacuum Lagrangian is the exact analogue of the off-shell ARP that we used for the EMT  and it is sufficient to make  $L_W^{\rm ren}(M)$ a finite quantity.  The above renormalized Lagrangian describes the vacuum effects from the quantum matter (in this case, the scalar field $\phi$) and it must be added up to the classical vacuum Lagrangian so as to form the total vacuum Lagrangian. We do this in the next section.

Upon expanding  Euler's  $\Gamma$-function in the limit $\varepsilon\to 0$ (cf. Appendix \ref{sec:appendixA2})  and using the explicit form of the modified DeWitt-Schwinger coefficients \eqref{eq:ModifiedDWSchcoeff}, we find after a relatively lengthy but straightforward calculation the following result (cf. the current Appendix \ref{sec:appendixB} for  more details):
\begin{equation}\label{eq:LWrenM}
\begin{split}
L_W^{\rm ren}(M)=\delta \rho_\Lambda(M)-\frac{1}{2}\delta\MPl^2(M) R-\delta \alpha_Q(M) \frac{{Q^\lambda}_\lambda}{3}-\delta \alpha_2(M) R^2+\cdots\,,
\end{split}
\end{equation}
where the dots stand for subleading contributions which decouple at large $m$, and
\begin{equation}\label{eq:deltacouplings}
\begin{split}
&\delta\rL(M)=\frac{1}{8\left(4\pi\right)^2}\left(M^4-4m^2M^2+3m^4-2m^4 \ln \frac{m^2}{M^2}\right),\\
&\delta\MPl^2(M) =\frac{\left(\xi-\frac{1}{6}\right)}{(4\pi)^2}\left(M^2-m^2+m^2\ln \frac{m^2}{M^2}\right),\\
&\delta \alpha_Q(M)=-\frac{1}{2(4\pi)^2}\ln\frac{m^2}{M^2},\\
&\delta{\alpha_2}(M)=\frac{\left(\xi-\frac{1}{6}\right)^2}{4(4\pi)^2}\ln\frac{m^2}{M^2}.
\end{split}
\end{equation}
As promised, the dependence on  $\mu$ fully cancelled out  along with the poles at $n=4$.  We have used DR to verify the cancellation of the UV-divergences (similarly to the procedure used  in the Appendix B of Ref.\,\cite{CristianJoan2020}).  We emphasize that the use of  DR is auxiliary here,  it can be done with other regulators, the final result has no memory of this intermediate step. The chief difference here is not so much  about regularization but about renormalization\,\footnote{We emphasize that the subtracted term $L_{\rm div}(M)$ at the scale $M$  in \eqref{eq:LWrenormalized} involves not just the UV-divergences  but the full expression obtained from the sum of the first three terms ($j=0,1,2$)  in the DeWitt-Schwinger expansion\,\eqref{eq:effLagrangian}, including  their finite parts (cf. Eq.\,\eqref{eq:LdivMdef}), hence fully in consonance with the procedure Eq.\,\eqref{EMTRenormalized} utilized for the EMT. This renormalization prescription is, of course,  entirely different from MS renormalization. }.  The quantities  \eqref{eq:deltacouplings}  are finite renormalization effects associated to the quantum vacuum Lagrangian $L_W$.

\subsection{Running couplings}\label{sec:RunningCouplings}

We are now ready to modify the classical or background vacuum Lagrangian \eqref{eq:LEHHD}  by including the quantum matter effects generated in our scalar field model and in this way  to track the shift received by each parameter as a function of the renormalization point $M$.  This will allow us  to derive the running couplings.  The full effective Lagrangian from which we can extract physical information up to one loop (actually the complete result at the quantum level, in the absence of scalar self-interactions)  is obtained by adding the extended classical  Lagrangian of gravity plus the (renormalized) quantum effects, i.e. the sum of equations \eqref{eq:LEHHD} and \eqref{eq:LWrenM}:
\begin{equation}\label{eq:Full-Leff1}
\begin{split}
L_{\rm eff}&=L_G^{\rm cl.}(M)+L_W^{\rm ren}(M)=-\rL(M)+\frac{1}{2}\MPl^2(M)  R+\alpha_1(M) C^2+\alpha_2(M) R^2+\alpha_3(M) E\\
&+\alpha_4(M)\Box R+\delta \rL(M)-\frac{1}{2}\delta\MPl^2(M) R-\delta \alpha_Q(M) \frac{{Q^\lambda}_\lambda}{3}-\delta \alpha_2(M) R^2+\cdots\\
\end{split}
\end{equation}
where the dots represent the subleading finite pieces emerging from the DeWitt-Schwinger expansion \,\eqref{eq:effLagrangian}\footnote{We have not computed these  terms in the effective action formalism (in contrast to the calculation that we have previously performed within the direct EMT approach, where we have reached up to the (finite)  $6th$ adiabatic order. Here we just want to cross-check the core design of the renormalization procedure within the effective action method and confirm that we obtain the same results.}.
Notice that the couplings of the classical part are dependent on the renormalization scale $M$ since the above expression represents the full effective renormalized Lagrangian of the theory. Overall it is independent of $M$ (i.e. RG-invariant), but each coupling `runs' (scales) with $M$ even though there is a net internal compensation among all the scaling dependencies.
It is convenient to rearrange \eqref{eq:Full-Leff1} as follows:
\begin{equation}\label{eq:Full-Leff}
\begin{split}
L_{\rm eff}&=\left[-\rL(M)+\delta\rL(M)\right]+\frac12\left[\MPl^2(M)-\delta\MPl^2(M)\right] R+\left[\alpha_1 (M) +\frac{1}{120} \delta\alpha_Q(M)\right] C^2\\
&+\left[\alpha_3 (M)-\frac{1}{360}\delta\alpha_Q(M)\right] E+\left[\alpha_4 (M)-\frac{1}{6}\left(\xi-\frac15\right) \delta\alpha_Q(M)\right]\Box R+\left[\alpha_2 (M)-\delta\alpha_2(M)\right] R^2+\cdots\\
\end{split}
\end{equation}
where we have used Eq.\,\eqref{eq:traceQ1}.  As previously remarked, the  full effective Lagrangian $L_{\rm eff}$  must be independent of the renormalization point $M$.  It follows that each one of the quantities in the square brackets of \eqref{eq:Full-Leff} must be independent of the scale $M$, and  this allows us to readily compute the $\beta$-functions for each of the couplings:
\begin{equation}\label{eq:BetaFunctionrL}
\beta_{\rL} (M)=\frac{1}{2(4\pi)^2}(M^2-m^2)^2
\end{equation}
\begin{equation}\label{eq:BetaFunctionMPl}
\beta_{\MPl^2} (M)=\frac{\left(\xi-\frac{1}{6}\right)}{8\pi^2} (M^2-m^2)
\end{equation}
and
\begin{equation}\label{eq:BetaFunctions12}
\begin{split}
\beta_{\alpha_1}=-\frac{1}{120(4\pi)^2}\ \ \ \ \ \
 \beta_{\alpha_2}=-\frac{\left(\xi-\frac{1}{6}\right)^2}{2(4\pi)^2}
 \end{split}
\end{equation}
 \begin{equation}\label{eq:BetaFunctions34}
\begin{split}
  \beta_{\alpha_3}=\frac{1}{360 (4\pi)^2}\ \ \ \ \ \ \
   \beta_{\alpha_4}=\frac{\xi-\frac15}{6(4\pi)^2}\,.
\end{split}
\end{equation}
We have used the explicit expressions  \eqref{eq:deltacouplings} for the calculation of the $\beta$-functions through
\begin{equation}\label{eq:BetaFunction}
  \beta_i=M \frac{\partial \lambda_i(M)}{\partial M}
\end{equation}
for each of the couplings  $(\lambda_i=\rL,\MPl^2,\alpha_1,...,\alpha_4)$\,\footnote{Related formulas have been considered in \cite{FerreiroNavarroSalas2}. Let us, however,  note that they differ from ours in that we consider the scale $M$ as the primary off-shell quantity from which to parameterize the quantum effects, rather than the difference $\Delta^2$ (called $-\mu^2$ in their case).}.

Let us note that in our approach the decoupling effects of physical quantities, such as the vacuum energy density itself, satisfy the Appelquist-Carazzone theorem\,\cite{AppelquistCarazzone75}. This is  apparent  in our $6th$-order formulas, in the limit of large $m$, see e.g. Eq.\,\eqref{renormalizedONSHELL6th}.   This is not to be expected for the couplings in general, as they do not have the same level of physical significance.  For example, we know that $\rL(M)$, which satisfies the first renormalization group equation (RGE)  above, is a formal quantity which does not appear in the physical results. Only the EMT has physical meaning, and in particular the VED, so there is no need in general for the couplings to satisfy manifest decoupling.

It is straightforward to  integrate the corresponding RGE's and derive the explicit running of the couplings with the renormalization point $M$, assuming that they are defined at some initial value $M_0$:
\begin{equation}\label{eq:RGEscouplings}
\begin{split}
&\rho_\Lambda(M)=\rho_\Lambda(M_0 )+\frac{1}{8(4\pi)^2}\left(M^4-M_0^4-4m^2(M^2-M_0^2)+2m^4\ln \frac{M^2}{M_0^2}\right),
\\
&\MPl^2(M)=\MPl^2(M_0)+ \frac{\left(\xi-\frac{1}{6}\right)}{(4\pi)^2}\,\left(M^2-M_0^2-m^2\ln \frac{M^2}{M_0^2}\right),\\
&\alpha_1 (M)=\alpha_1 (M_0)-\frac{1}{240(4\pi)^2}\ln \frac{M^2}{M_0^2},
\\
&\alpha_2 (M)=\alpha_2 (M_0)-\frac{\left(\xi-\frac{1}{6}\right)^2}{4(4\pi)^2}\ln \frac{M^2}{M_0^2},\\
&\alpha_3 (M)=\alpha_3 (M_0)+\frac{1}{720(4\pi)^2 }\ln \frac{M^2}{M_0^2},\\
&\alpha_4 (M)=\alpha_4 (M_0)+\frac{\xi-\frac15}{12(4\pi)^2}\ln \frac{M^2}{M_0^2}.
\end{split}
\end{equation}
The  equation for the running (reduced) Planck mass squared  $\MPl^2(M)= 1/\left(8\pi G(M)\right)$  given above can also be cast  in terms of the running Newton's constant:
\begin{equation}\label{eq:RGENewton}
G(M)=\frac{G(M_0)}{1+\frac{\left(\xi-\frac{1}{6}\right)}{2\pi}G(M_0)\left(M^2-M_0^2-m^2\ln \frac{M^2}{M_0^2}\right)}\,.
\end{equation}
The previous equation  can be related to the physical running of the gravitational coupling during the cosmological expansion. In Sec. \ref{sec:RenormalizedFriedmann} we further dwell upon the running of the gravitational coupling in combination with that of the VED, and will come back to Eq.\,\eqref{eq:RGENewton}.

We can see that the first two RGE solutions and the fourth one  in  \eqref{eq:RGEscouplings} are nothing but equations \eqref{SubtractionrL}, \eqref{SubtractionMPl} and \eqref{Subtractionalpha},  respectively (with  $\alpha_2=\alpha/2$) which we found in the process of renormalization of the EMT.  Overall we have met at this point  a rather nontrivial consistency check between the renormalization procedure of the EMT from which we started our calculation in Sec. \ref{sec:RenormEMT},  and the alternative approach based on the renormalization of the effective action (and corresponding effective Lagrangian), which we have undertaken in this section\,\footnote{Although the RGEs for $\alpha_1,\alpha_3$ and $\alpha_4$ were not discussed in  Sec. \ref{sec:RenormEMT}, we have derived them en route in the effective action approach only for completeness.}. In other words, it confirms that the renormalized couplings that we have now computed from the effective Lagrangian method  are indeed the same parameters  which appeared in the renormalized EMT following from the original prescription \eqref{EMTRenormalized} and performing the corresponding subtraction $\delta X(M,M_0)= X(M)-X(M_0)$  in the renormalized Einstein's equations \eqref{eq:EqsVac}.  In a similar way, we can easily check that the relations \eqref{eq:deltacouplings} can be recovered now as a particular case  of the above running solutions for the case $M_0=m$ upon defining $\delta X(M)\equiv\delta X(M,m)= X(M)-X(m)$  for each of the parameters $X=\rL, \MPl^2, \alpha_i$. For instance, using the first relation in \eqref{eq:RGEscouplings} we find
\begin{equation}\label{eq:RGErL}
\delta\rL(M)=\rL(M)-\rL(m)=\frac{1}{8\left(4\pi\right)^2}\left(M^4-4m^2M^2+3m^4-2m^4 \ln \frac{m^2}{M^2}\right)\,,
\end{equation}
which matches the first one of \eqref{eq:deltacouplings}.  Similarly for the other parameters.  The complete formulas are obtained after inserting $M_0=m$ in the various  relations\,\eqref{eq:deltacouplingsB} of the Appendix \ref{sec:appendixB}.

Let us finally pause at this point to observe that there is a long way mediating between these two approaches, namely, the one based on tackling a direct renormalization of  the EMT by means of the adiabatic procedure and the other based  on computing the effective action from the DeWitt-Schwinger expansion. However different they are,  they  appear to be fully consistent.  This fact is, of course, very much welcome as it demonstrates the cogency and congruence of the results obtained in our calculation.  The  touchstone of such a consistency  can be made even more transparent if we compute the functional derivative of the action associated to the  renormalized vacuum effective Lagrangian with respect to the metric, i.e. if we show how to recover the renormalized EMT obtained in previous sections using the effective action itself, Eq.\,\eqref{eq:DefW}.

\subsection{Full consistency between the EMT and effective action results}

We take up the renormalized quantum vacuum Lagrangian defined in the previous section, Eq.\,\eqref{eq:LWrenormalized}.  The effective action associated to  such Lagrangian is
\begin{equation}\label{eq:Wren}
W_{\rm ren}(M)\equiv\int d^4 x \sqrt{-g} \  L_{\rm W}^{\rm ren}(M)=\int d^4 x \sqrt{-g} \ \left(L_W (m)-L_{\rm div} (M)\right)\,.
\end{equation}
Let us now show that with this action we can recompute  the renormalized vacuum EMT that we have previously found in Sec.\,\ref{Sec:RenEMToffshell}.
In fact, on  inserting Eq.\,\eqref{eq:LWrenM} in it we find
\begin{equation}\label{eq:effActionLren}
\begin{split}
W_{\rm ren}(M)=\int d^4 x\sqrt{-g} \left( \delta \rL(M)-\frac{1}{2}\delta\MPl^2(M) R-\delta \alpha_Q(M) \frac{{Q^\lambda}_\lambda}{3}-\delta \alpha_2(M) R^2\right)\,.
\end{split}
\end{equation}
The renormalized vacuum EMT now follows from
\begin{equation}\label{eq:DefWMM0}
\langle T_{\mu\nu}^{\delta \phi}\rangle_{\rm ren}(M)=-\frac{2}{\sqrt{-g}} \,\frac{\delta W_{\rm ren}(M)}{\delta g^{\mu\nu}}\,.
\end{equation}
Using \eqref{eq:effActionLren} on the \textit{r.h.s.} of \eqref{eq:DefWMM0} we may compute the metric functional variation.  In performing the variation of the HD term $\frac13 {Q^\lambda}_\lambda$ as given in Eq.\,\eqref{eq:traceQ1}, we can use some of the formulas quoted in Appendix \ref{sec:appendixA1}.  In particular,  we drop  the contribution  from the Euler density  $E$ (since the metric functional variation of the Gauss-Bonnet term $\GB$ is exactly zero in $n=4$ spacetime dimensions)  and of course that of the total derivative term $\Box R$.  Therefore, using the mentioned appendix,
\begin{equation}\label{eq:VariationQ}
\frac{1}{\sqrt{-g}}\frac{\delta\left( {Q^\lambda}_\lambda/3\right)}{\delta g^{\mu\nu}}=\frac{1}{\sqrt{-g}}\frac{\delta}{\delta g^{\mu\nu}}\left( -\frac{1}{120} C^2\right)=-\frac{1}{60}\left( \leftidx{^{(2)}}{\!H}_{\mu\nu}-\frac13 \leftidx{^{(1)}}{\!H}_{\mu\nu}\right)\,.
\end{equation}
With this proviso, the sought-for metric functional variation can be easily performed and \eqref{eq:DefWMM0} can be written in the compact form
\begin{equation}\label{eq:TrenMm}
\begin{split}
\langle T_{\mu\nu}^{\delta \phi}\rangle_{\rm ren}(M)= \delta\MPl^2(M) G_{\mu\nu}+ \delta \rL(M) g_{\mu\nu}+\delta\alpha(M)\leftidx{^{(1)}}{\!H}_{\mu\nu}-\frac{1}{30}\delta\alpha_Q(M)\left(\leftidx{^{(2)}}{\!H}_{\mu\nu}-\frac13 \leftidx{^{(1)}}{\!H}_{\mu\nu}\right)\,,
 \end{split}
\end{equation}
where we have used the fact that $2\delta\alpha_2=\delta\alpha$ and we recall that the coefficients of the various tensor expressions on the \textit{r.h.s.} of the previous formula are given explicitly  by Eqs.\,\eqref{eq:deltacouplings}.   For conformally flat spacetimes (which comprise, in particular, all the FLRW backgrounds) the two HD tensors $\leftidx{^{(1)}}{\!H}_{\mu\nu}$ and $\leftidx{^{(2)}}{\!H}_{\mu\nu}$ are related in the form $\leftidx{^{(2)}}{\!H}_{\mu\nu}=\frac13 \leftidx{^{(1)}}{\!H}_{\mu\nu}$  -- cf.  Eq.\,\eqref{eq:ConfFlat} of Appendix \ref{sec:appendixA1}\footnote{The notations `(1)'  and `(2)' as upper indices on the left for these HD tensors is standard\cite{BirrellDavies82}, it has nothing to do with adiabatic orders.}.  As  a consequence,  the previous equation simplifies into
\begin{equation}\label{eq:TrenMm0FLRWMm}
\begin{split}
\langle T_{\mu\nu}^{\delta \phi}\rangle_{\rm ren}(M)=
& \delta \rL(M) g_{\mu\nu}+\delta\MPl^2(M) G_{\mu\nu}+\delta\alpha(M)\leftidx{^{(1)}}{\!H}_{\mu\nu}\,.
 \end{split}
\end{equation}
If we  take the $00th$-component of this result and use the formulas given in Appendix \ref{sec:appendixA1}  we find
\begin{equation}\label{RenormalizedfromAction}
\begin{split}
&\langle T_{00}^{\delta \phi}\rangle_{\rm ren}(M)=\delta \rL(M) g_{00}+\delta\MPl^2(M) G_{00}+\delta\alpha(M)\leftidx{^{(1)}}{\!H}_{00}\\
&=\frac{a^2}{128\pi^2 }\left(-M^4+4m^2M^2-3m^4+2m^4 \ln \frac{m^2}{M^2}\right)\\
&-\left(\xi-\frac{1}{6}\right)\frac{3 \mathcal{H}^2 }{16 \pi^2 }\left(m^2-M^2-m^2\ln \frac{m^2}{M^2} \right)+\left(\xi-\frac{1}{6}\right)^2 \frac{9\left(2  \mathcal{H}^{\prime \prime} \mathcal{H}- \mathcal{H}^{\prime 2}- 3  \mathcal{H}^{4}\right)}{16\pi^2 a^2}\ln \frac{m^2}{M^2}+\dots
\end{split}
\end{equation}
The upshot is that we are led once more to Eq.\,\eqref{RenormalizedExplicit2}, as it should be. The corresponding result for the VED obtains now  from \eqref{RenormalizedfromAction} using the relation \eqref{RenVDE}.  In this way we reach again the final result \eqref{RenVDEexplicit} and hence we have demonstrated the perfect consistency between the two renormalization procedures.

We can perform a similar computation to find the scaling evolution of the EMT between the renormalization point $M$ and $M_0$. One option is to  use Eq.\,\eqref{RenormalizedfromAction} to reproduce the results we have already found in Sec.\,\ref{sec:RenormalizedVED}.   But we may also repeat the above procedure ab initio, now using the subtracted effective action at the two mentioned scales:
\begin{equation}\label{eq:effActionExplicit2}
\begin{split}
W_{\rm ren}(M)-W_{\rm ren}(M_0)=&\int d^4 x \sqrt{-g} \ \left(L_{\rm W}^{\rm ren}(M)-L_{\rm W}^{\rm ren}(M_0)\right)\\
=&\int d^4 x \sqrt{-g} \ \left(L_{\rm div} (M_0)-L_{\rm div} (M)\right)\\
=&\int d^4 x\sqrt{-g} \left( \delta \rL(m,M,M_0)-\frac{1}{2}\delta\MPl^2(m,M.M_0) R\right.\\
&\left.-\delta \alpha_Q(M,M_0) \frac{{Q^\lambda}_\lambda}{3}-\delta \alpha_2(M,M_0) R^2\right)\,.
\end{split}
\end{equation}
Here we have used Eq.\,\eqref{eq:divMdivM0B} of the Appendix \ref{sec:appendixB} and we note  that the coefficients of the various tensor expressions on the \textit{r.h.s.} of the previous formula are not the same as in \eqref{eq:effActionLren} but are  given explicitly  in  Eq.\,\eqref{eq:deltacouplingsB} of that appendix .  The difference of  vacuum EMT values at the two scales reads
\begin{equation}\label{eq:DefWMM0bis}
\delta\langle T_{\mu\nu}^{\delta \phi}\rangle\equiv \langle T_{\mu\nu}^{\delta \phi}\rangle_{\rm ren}(M)-\langle T_{\mu\nu}^{\delta \phi}\rangle_{\rm ren}(M_0)=-\frac{2}{\sqrt{-g}} \,\frac{\delta}{\delta g^{\mu\nu}}\left(W_{\rm ren}(M)-W_{\rm ren}(M_0)\right)\,.
\end{equation}
and upon computing the metric functional variation we find
\begin{equation}\label{eq:TrenMm0General}
\begin{split}
\delta\langle T_{\mu\nu}^{\delta \phi}\rangle=&
\,\delta\MPl^2(m,M.M_0) G_{\mu\nu}+ \delta \rL(m,M,M_0) g_{\mu\nu}+\delta\alpha(M,M_0) \leftidx{^{(1)}}{\!H}_{\mu\nu}\\
&-\frac{1}{30}\,\delta\alpha_Q(M,M_0)\left(\leftidx{^{(2)}}{\!H}_{\mu\nu}-\frac13\,\leftidx{^{(1)}}{\!H}_{\mu\nu}\right).
 \end{split}
\end{equation}
 For conformally flat spacetimes we can repeat the same argument as given above and the above result boils down to
\begin{equation}\label{eq:TrenMm0FLRWl}
\begin{split}
\delta\langle T_{\mu\nu}^{\delta \phi}\rangle=
& \,\delta\MPl^2(m,M,M_0) G_{\mu\nu}+ \delta \rL(m,M,M_0) g_{\mu\nu}+\delta\alpha(M,M_0)\leftidx{^{(1)}}{\!H}_{\mu\nu}\,.
 \end{split}
\end{equation}
The obtained expression is just the subtracted form of Eq.\,\eqref{eq:EqsVac} at the two scales $M$ and $M_0$. Thus, if we  take the $00th$-component of this result and use the formulas given in Appendix \ref{sec:appendixA1} to perform the identifications on both sides and the definition of VED we encounter once more  the important Eq.\,\eqref{eq:VEDscalesMandM0Final} which gives the smooth evolution of the VED between the two scales with the total absence of quartic mass contributions.  This corroborates the perfect consistency between the two approaches. Having found the very same renormalization results with  the effective action formalism,  all of the discussions made in Sec.\,\ref{sec:RenormalizedVED}  can be iterated exactly as they are there.

\subsection{Renormalization group equation for the VED}\label{Sec:RGE-VED}

To compute the RGE for $\rv(M)$ we have to take into account that only the adiabatic orders up to ${\cal O}(T^{-4})$ carry $M$-dependence since the higher orders are finite and hence need not be subtracted. It follows that the exact $\beta$-function for the VED can be obtained from Eq.\,\eqref{RenVDEexplicit}  as follows:
\begin{equation*}
\begin{split}
\beta_{\rv}=&M\frac{\partial\rv(M)}{\partial M}=\beta_{\rL}+\frac{1}{128\pi^2}\left(-4M^4+8 m^2 M^2- 4 m^4\right)\\
&-\left(\xi-\frac{1}{6}\right)\frac{3 \mathcal{H}^2 }{16 \pi^2 a^2}\left(-2M^2+2m^2\right)+\left(\xi-\frac{1}{6}\right)^2 \frac{9\left(2  \mathcal{H}^{\prime \prime} \mathcal{H}- \mathcal{H}^{\prime 2}- 3  \mathcal{H}^{4}\right)}{16\pi^2 a^4}(-2)\\
\end{split}
\end{equation*}
\begin{equation}\label{eq:RGEVED1}
=\left(\xi-\frac{1}{6}\right)\frac{3 \mathcal{H}^2 }{8 \pi^2 a^2}\left(M^2-m^2\right)+\left(\xi-\frac{1}{6}\right)^2 \frac{9\left(\mathcal{H}^{\prime 2}-2\mathcal{H}^{\prime \prime}\mathcal{H}+3 \mathcal{H}^4\right)}{8\pi^2 a^4}\,.
\end{equation}
In the first line we have used the $\beta$-function for $\rL(M)$ that we have just obtained  in \eqref{eq:BetaFunctionrL}. It is seen that $\beta_{\rL}$  exactly cancels  against the contribution from the second term on the \textit{r.h.s.} of the above equation. This cancellation is most welcome, as it  leaves the $\beta$-function of the VED completely  free from quartic mass contributions.  It follows that the running of $\rv(M)$  rests only on the presence of quadratic mass scales in the final result.  Integrating the above  RGE we find
\begin{equation}\label{eq:VEDintegrated}
\begin{split}
&\rv(M)=\rv(M_0)+\left(\xi-\frac16\right)\frac{3\cH^2}{16\pi^2 a^2}\,\left(M^2 - M_0^{2} -m^2\ln \frac{M^{2}}{M_0^2}\right)\\
&\phantom{XXXXXXXXXXXXXXXX}+\left(\xi-\frac16\right)^2\frac{9}{16 \pi^2 a^4}\left(\mathcal{H}^{\prime 2}-2\mathcal{H}^{\prime \prime}\mathcal{H}+3 \mathcal{H}^4 \right)\ln \frac{M^2}{M_0^{2}}\,.\\
\end{split}
\end{equation}
Thus we have recovered the expected result \eqref{eq:VEDscalesMandM0Final}, which gives the evolution of the VED with the scale $M$, starting from another scale $M_0$, and  such relation involves only quadratic mass scales which in leading order are highly tempered by the presence of quadratic powers of the Hubble rate. In other words, rather than the hard $\sim m^4$ behavior we obtain the much softened one $\sim m^2H^2$.

The following observation is now in order.  As we warned in Sec.\,\ref{sec:RenormalizedVED}, the renormalization of the EMT and in particular of the VED involves the renormalization of formal quantities which do not ever play a role in the physical interpretation of the VED.  Such is the case of the quantity \eqref{SubtractionrL}, which carries the quartic powers of the masses. This quantity cancels exactly in the important expression \eqref{eq:VEDscalesMandM0Final},  which physically relates the VED at the two scales $M$ and $M_0$ and hence no dependence is left of the unwanted terms $\sim m^4$. As we know, this is the clue  to avoid the need for fine-tuning in our renormalization procedure. Now we can see that  the primary reason for that stems from the soft behavior of the VED $\beta$-function \eqref{eq:RGEVED1}.

In contrast, the $\beta$-function for  the parameter $\rL(M)$ is proportional to the quartic power of the particle mass, as indicated in Eq.\,\eqref{eq:BetaFunctionrL}, and for this reason the solution to the corresponding RGE,  as  represented by the first equation in \eqref{eq:RGEscouplings},  is also proportional to those quartic terms.  However,  the running  of $\rL$ with $M$ has  no physical implication since these terms  exactly cancel out in the VED, as we have just seen.  This situation can be compared to the  running of $\rL$ with the unphysical mass unit $\mu$ in the MS approach to the VED, as discussed  in Sec. \ref{sec:VEDMSS}.  Yet there is an important difference: in the MS case one usually interprets that the  renormalized VED is given by Eq.\eqref{eq:VEDMSS}\footnote{Recall the footnote on p. 35.}.  If so,  then, as a  (purported) physical quantity one is  enforced to fine tune $\rL(\mu)$  against the large  $\sim m^4$ contribution (represented by the second term in that expression).  The counterpart of these formulas in our calculation is just given by a single piece of our Lagrangian  \eqref{eq:Full-Leff} (since all the others are geometric contributions from curved spacetime), to wit:  it is  just  (minus) the first term, or  $\rL(M)-\delta\rL(M)$.  The last equation indeed contains, among others,  the terms involved in Eq. \eqref{eq:VEDMSS}. This  can be checked from  Eq.\,\eqref{eq:RGErL}, conveniently rewritten as
\begin{equation}\label{eq:deltarhocommon}
-\delta\rL(M)=\frac{m^4}{64\pi^2}\left(\ln\frac{m^2}{M^2}-\frac32-\frac{M^4}{2m^4}+\frac{2M^2}{m^2}\right)\,,
\end{equation}
upon replacing $M$ with $\mu$ and neglecting the last two terms since we consider  $M^2=H^2\ll m^2$.
The RG-invariant expression  $\rL(M)-\delta\rL(M)=\rL(m)$   is not at all the VED.   In Minkowski spacetime, we saw  that the correctly renormalized VED is zero in our framework (cf. Sec.\,\ref{sec:VEDMinkowski}).  What is more, on comparing the VED at two different scales the effect of $\rL(M)$  always  cancels against the quartic terms emerging from the renormalized ZPE, and the net result is free  from the influence of the quartic masses, cf. Eq.\,\eqref{eq:VEDintegrated}.  Thanks to this crucial fact the observable running of the VED  in curved spacetime  depends only on the quadratic mass scales  times the Hubble rate square, i.e.  $\sim m^2H^2$,  as shown in that expression. The presence of $H^2$ makes the running rate much more temperate:  it just follows the evolution of the cosmic flow itself.  In fact, this is nothing but the  characteristic running law of the RVM\,\cite{JSPRev2013,JSPRev2022}.

\section{Friedmann's equations and conservation laws with running vacuum}\label{sec:RenormalizedFriedmann}

Friedmann's equations  in the presence of running vacuum are a bit more complicated than usual.  They have been dealt with previously in \eqref{eq:FriedmanEqs} assuming the generic RVM form \eqref{eq:EffLambda}, which boils down to \eqref{eq:RVM2} at low energy.  This is sufficient for an effective treatment of the RVM since they offer the possibility to confront the predictions with the data and put bounds to the $\nueff$ parameter.  This has led to a fruitful phenomenology, cf.\,\cite{ApJL2015,ApJ2017,RVMphenoOlder1,Elae2015,RVMphenoOlder2,RVMpheno1,RVMpheno2,PericoTamayo2017,CQG2017,ApJL2019,Mehdi2019,Tsiapi2019,BD2020,Singh2021,Manos2021,Yu2022,EPL2021}, for instance.  The phenomenological approach is very useful because there may be many  QFT models (even string models\,\cite{NickJoan2020,NickJoan2021}) whose effective behaviour leads to a vacuum energy density of the form \eqref{eq:EffLambda}.  However, it is also interesting to study the exact form of Friedmann's equations of the RVM using directly the field variables involved in the QFT model under consideration, which in the present instance is based on a nonminimally coupled scalar field with action \eqref{eq:Sphi}.  As can be expected, this part is more cumbersome but  it reveals some new clues on the internal consistency of our calculation.   Assuming an  FLRW background, the starting point is  Einstein's equations in renormalized form, see  \eqref{eq:EqsVac2}, which we have to combine with  the explicit formulas that we have derived in the previous sections for the vacuum energy density and pressure in our adiabatically renormalization approach. The fact that the quantum vacuum satisfies a quasi-vacuum equation of state means that there is a  relationship between vacuum density and pressure which is not the naive one we usually have in mind, see Sec.\ref{sec:Quasivacuum}.

\subsection{Field equations and matter conservation law}

In the context of the model \eqref{eq:Sphi}, we have to distinguish between the background field density and pressure and their fluctuating or vacuum components (cf. Sec. \ref{sec:AdiabaticVacuum}).  We denote by $(\rho_\phi, P_\phi)$  the background components. The fluctuating parts of these quantities have been object of devoted study in the previous sections and are represented by the quantities $(\rv, \Pv)$,  which have been computed up to $6th$ adiabatic order.  The generalized Friedmann's equation emerging from the $00th$-component of Eq.\,\eqref{eq:EqsVac2}  can be written  as follows,
\begin{equation}\label{Friedmann_density}
H^2=\frac{8\pi}{3}G(M)\left[\rv(M)+\rho_\phi+\rho_X(M)\right],
\end{equation}
where  the running gravitational coupling  $G(M) $ is related to the parameter  $\MPl^2(M)$ frequently used in the previous sections  through Eq.\,\eqref{eq:RenCouplings}.  Needless to say, if this equation were to apply to the current universe we would need to add baryons and CDM, but here we just want to illustrate the interplay between the field $\phi$ and the vacuum without introducing more elements. In fact, the main actor here are the quantum vacuum effects produced by $\phi$.  Its background part is not the main focus, but we include it for completeness and self-consistency.
In this context,  we have got also the  gravitational contribution from  the HD tensor $\leftidx{^{(1)}}{\!H}_{\mu\nu}$ in the generalized Einstein's equations, which contributes the term $\rho_X$ in the above  Friedmann's equation as follows:
\begin{equation}\label{eq:rhoX}
\rho_X \equiv -\alpha(M)\frac{\leftidx{^{(1)}}{\!H}_{00}}{a^2}= 18\alpha (M) (\dot{H}^2-2H\ddot{H}-6H^2\dot{H})\,.
\end{equation}
This  effective energy density  (acting as an effective fluid $X$)  stems  from the $00th$-component of the mentioned HD tensor  (cf.  Appendix \ref{sec:appendixA1}).
Similarly, the generalized pressure equation within the Friedmann's pair can be written as
\begin{equation}\label{Friedmann_pressure}
3H^2+2\dot{H}=-8\pi G (M)\left[P_{\rm vac}(M)+P_\phi+P_X\right],
\end{equation}
where
\begin{equation}\label{eq:PX}
P_X \equiv  -\alpha(M)\frac{\leftidx{^{(1)}}{\!H}_{11}}{a^2}=  \alpha(M)\left(108H^2\dot{H}+54\dot{H}^2+72H\ddot{H}+12 \vardot{3}{H} \right)\,.
\end{equation}
Combining the two generalized Friedmann's equations given above we find
\begin{equation}\label{eq:dotHgen}
\dot{H}=-4\pi G(M) \left[P_{\rm vac}(M)+\rv(M)+P_\phi+\rho_\phi+P_X+\rho_X\right]\,.
\end{equation}
The conservation equation for  the fluid X  reads
\begin{equation}\label{ConservationOfX}
\dot{\rho}_X+3H\left(\rho_X+P_X\right)=18\dot{\alpha}\left(\dot{H}^2-2H\ddot{H}-6H^2\dot{H}\right)=\frac{\dot{\alpha}}{\alpha}\rho_X.
\end{equation}
The local conservation law for the background field $\phi_b$ (which we have just denoted $\phi$ here for simplicity)  ensues from the fact that $\nabla^\mu T_{\mu\nu}^\phi=0 $.  Indeed, using  the explicit form of the classical EMT, Eq.\,\eqref{EMTScalarField}, we find
\begin{equation}
\begin{split}\label{MatterCovConservation}
&\nabla^\mu T_{\mu\nu}^\phi=\left(1-2\xi\right)\left(\Box \phi\right) \left( \nabla_\nu \phi\right)+\left(1-2\xi\right)\left(\nabla_\mu \phi \right)\left(\nabla^\mu \nabla_\nu \phi\right)+\left(2\xi-\frac{1}{2}\right)\nabla_\nu \left(\nabla^\alpha \phi \nabla_\alpha \phi\right)\\
&-2\xi\left(\nabla^\mu \phi\right)\left( \nabla_\mu \nabla_\nu \phi\right)-2\xi\phi \nabla^\mu \nabla_\mu \nabla_\nu  \phi+2\xi  \nabla_\nu \left(\phi \Box \phi\right)+\xi \left(\nabla^\mu G_{\mu\nu}\right)\phi^2+\xi G_{\mu\nu}\nabla^\mu \phi^2-\frac{1}{2}m^2\nabla_\nu \phi^2\\
&=\left(\Box-m^2\right)\phi \nabla_\nu \phi+2\xi \phi \left(G_{\mu\nu}\nabla^\mu \phi+\nabla_\nu \Box \phi-\Box\nabla_\nu \phi\right)\\
&=\xi R\phi \nabla_\nu \phi+2\xi \phi \left(G_{\mu\nu}\nabla^\mu \phi-R_{\mu\nu}\nabla^\mu \phi\right)=0\,.
\end{split}
\end{equation}

In the above derivation we have used the Klein Gordon equation \eqref{eq:KG} and the Bianchi identity $\nabla^\mu G_{\mu\nu}=0$.
At the same time we have made use of the formula \eqref{eq:CommutationNablaBox} in the Appendix \ref{sec:appendixA1} in  order to commute the covariant box operator $\Box$ and $\nabla_\nu$.
Equation \eqref{MatterCovConservation} for $\nu=0$ can be rephrased in terms of the energy density and pressure:
\begin{equation}\label{MatterConservation1}
\nabla^\mu T_{\mu\nu}^\phi=g^{\mu\alpha}\left(\partial_\alpha T_{\mu\nu}-\Gamma^\sigma_{\alpha\mu}T_{\sigma\nu}-\Gamma^\sigma_{\alpha\nu}T_{\mu \sigma}\right)=-\rho_\phi^\prime-3\mathcal{H}\left(\rho_\phi+P_\phi\right)=0\,,
\end{equation}
which, if rewritten in cosmic time differentiation (using $d/d\tau=a (d/dt)$),  implies that
\begin{equation}\label{MatterConservation}
\dot{\rho}_{\phi}+3H\left(\rho_\phi+P_\phi\right)=0\,,
\end{equation}
with
\begin{equation}
\begin{split}
\rho_\phi\equiv\frac{T_{00}^{\phi_b}}{a^2}=\frac{1}{2}\dot{\phi}^2+\frac{1}{2}m^2\phi^2+3\xi\left(2H\phi\dot{\phi}+H^2\phi^2\right)
\end{split}
\end{equation}
and
\begin{equation}
\begin{split}
P_\phi\equiv\frac{T_{11}^{\phi_b}}{a^2}=\frac{1}{2}\dot{\phi}^2-\frac{1}{2}m^2\phi^2-\xi\left(2\dot{\phi}^2+4H\phi\dot{\phi}+2\phi\ddot{\phi}+3H^2\phi^2+2\dot{H}\phi^2\right)\,.
\end{split}
\end{equation}
The ratio $w_\phi=P_\phi/\rho_\phi$ from the last two equations defines the EoS of the nonminimally coupled  ($\xi\neq 0$) scalar field $\phi$, which is seen to be nontrivial.  All in all, we have found that the background matter field $\phi$ does not interact with the vacuum, and hence  its energy density  is covariantly self-conserved during the expansion, cf. \eqref{MatterConservation}.

It is easy to see that the local conservation law \eqref{MatterConservation} is just another way to write the Klein-Gordon equation for the  background field $\phi$:
\begin{equation}
\ddot{\phi}+3H\dot{\phi}+(m^2+\xi R)\phi=\ddot{\phi}+3H\dot{\phi}+m^2\phi+\xi\left(12H^2+6\dot{H}\right)\phi=0\,.
\end{equation}
This equation is, of course,  the same as Eq.\,\eqref{eq:KGexplicit}, but written in terms of the cosmic time and after having neglected the term $\nabla^2\phi$ owing to homogeneity and isotropy for the background field  $\phi$ -- which, as advertised, corresponds to  $\phi_b(t)$ in Eq.\,\eqref{ExpansionField}.

\subsection{Conservation equation for the quantum vacuum}

The vacuum, however, does not obey the same conservation equation as matter in general.  In point of fact, it is not generally conserved. We find
\begin{equation}\label{eq:NonConserVED1}
\begin{split}
\dot{\rho}_{\rm vac}+3H\left(\rv+P_{\rm vac}\right)=\frac{3\dot{M}}{8\pi^2 M}\left[\left(\xi-\frac{1}{6}\right)H^2(M^2-m^2)
+3\left(\xi-\frac{1}{6}\right)^2\left(\dot{H}^2-2H\ddot{H}-6H^2\dot{H}\right)\right]\,.
\end{split}
\end{equation}
It is remarkable that this equation can be written very succinctly in terms of the $\beta$-function of the running vacuum obtained in \eqref{eq:RGEVED1}:
\begin{equation}\label{eq:NonConserVED2}
\dot{\rho}_{\rm vac}+3H\left(\rv+P_{\rm vac}\right)=\frac{\dot{M}}{M}\,\beta_{\rv}\,.
\end{equation}
Here we have taken into account that the scale $M$ in cosmology is associated to a dynamical variable ($H$ in our case, although we do not implement any particular choice at this point), and hence it evolves with the cosmic time,  $\dot{M}\neq0$.  The compact form \eqref{eq:NonConserVED2} illustrates the fact that the non-conservation of the VED is due to both the running of $\rv$ with $M$  (i.e. the fact that $\beta_{\rv}\neq 0$)  and to the cosmic time dependence of $M$. This feature is in contradistinction to ordinary gauge theories of strong and electroweak interactions\,\cite{FritzschSola}, and allows us to probe the effect of the time-dependence of $M$ in the running couplings and in particular in the VED. This is possible and even necessary in cosmology since the scale $M$ should be linked with cosmological variables changing with the cosmic time.  When one studies situations where the ordinary gauge couplings participate in cosmological problems, it is perfectly possible to find out that they run both with the (time-independent) 't Hooft's mass unit $\mu$ and  also with the cosmic (time-dependent) scale $M$, which is associated to $H$\, \cite{FritzschSola}.  In Appendix \ref{sec:appendixAbisbis1} we use Eq.\,\eqref{eq:NonConserVED2} to further investigate the time evolution of the VED.  We show that the result is consistent with Eq.\,\eqref{eq:RVM2}, as it should.

The following comments are pertinent at this point. We remark that equation \,\eqref{eq:NonConserVED1},  or  equivalently \eqref{eq:NonConserVED2}, is exact, that is to say, fulfilled to all adiabatic orders. This must be so since the scale dependence of the running quantities stops at order four.  The higher order adiabatic terms do not bring additional $M$-dependent terms.  However, let us not forget that  the time-dependence is carried by all orders through the powers of $H(t)$ and its derivatives.  We have used this fact to explicitly check that  even in the presence of the complicated $6th$ order contributions in the structure of $\rv$ and $\Pv$ -- see Sections \ref{sec:ZPE6th} and \ref{sec:Quasivacuum} -- Eq.\,\eqref{eq:NonConserVED1} is exactly satisfied and the $6th$ order effects in it  just cancel out precisely.  The calculation has been performed in the following way. The quantity  $\rv(M)$ depends on time implicitly through $M$ but also through the many terms which depend on the Hubble rate and its time derivatives: $H,\dot{H}, \ddot{H}\dots \vardot{5}{H}$. Let's separate the first term on the \textit{l.h.s.} of \eqref{eq:NonConserVED1} as follows :
\begin{equation}\label{eq:chainruletM}
\dot{\rho}_{\rm vac}+3H\left(\rv+P_{\rm vac}\right)=\dot{M}\frac{\partial\rv}{\partial M}+ \frac{\partial\rv}{\partial t}\Bigg|_{M}+3H\left(\rv+P_{\rm vac}\right)\,,
\end{equation}
where $\dot{M}=dM/dt$ and  $|_M$ in the second term  on the \textit{r.h.s.}  is to emphasize that we keep $M$ constant in time when performing such a  differentiation. The  first term on the \textit{r.h.s.},  $\dot{M}\frac{\partial \rv}{\partial M}$, is entirely responsible for the result  \eqref{eq:NonConserVED1}.
The last two terms on the \textit{r.h.s.} of \eqref{eq:chainruletM}  can be shown to yield an identically vanishing result:
\begin{equation}
\frac{\partial \rho_{\rm vac}}{\partial t}\Bigg|_{M}+3H\left(\rho_{\rm vac}+P_{\rm vac}\right)=0\,.
\end{equation}
We have verified the exactness of this equation.  Carrying out the check explicitly is a bit ponderous as it  implies using the full structure (up to $6th$ adiabatic order) of the expressions for the vacuum density and pressure,  i.e. equations \eqref{renormalizedONSHELL6th} and \eqref{eq:VacuumPressureFull}. For this reason we have performed it  with the help of Mathematica\,\cite{Mathematica}.  We believe it constitutes  a pretty robust consistency check of our formulas. The net outcome  is just the expression on the RHS of \eqref{eq:NonConserVED1}, which involves effects up to $4th$ adiabatic order.  All the remaining contributions from higher order cancel identically.

As emphasized above, the  VED is not locally conserved since the scale  $M$ evolves with the cosmic time and the VED runs with $M$.  The integration of  Eq.\,\eqref{eq:NonConserVED1} yields, of course,  the characteristic RVM evolution law \eqref{RenVDEexplicit}.  The full local conservation equation containing all the ingredients is more complicated.  Let us find it.  We first extend the generalized Einstein's equations \eqref{eq:EqsVac}  by including also the background EMT contribution from the scalar field (i.e. by inserting the term $T_{\mu\nu}^{\phi}$ on its \textit{r.h.s.}), as in this way we take into account all of the components exchanging energy in the system:
\begin{equation}\label{eq:EqsVacExt}
\MPl^2(M) G_{\mu \nu}+\rho_\Lambda (M) g_{\mu \nu}+\alpha(M) \leftidx{^{(1)}}{\!H}_{\mu\nu}= \langle T_{\mu\nu}^{\delta \phi}\rangle_{\rm ren}(M)+T_{\mu\nu}^{\phi}\,.
\end{equation}
We  multiply next this equation by $8\pi G(M)=1/\MPl^2(M)$ and take the covariant divergence on both sides (i.e. we  apply the operator $\nabla^\mu$ on each term). Taking into account that $G_{\mu\nu}$ is a conserved tensor (i.e. we have the Bianchi identity $\nabla^\mu G_{\mu \nu}=0$) and that the HD tensor $ \leftidx{^{(1)}}{\!H}_{\mu\nu}$ is also conserved ($\nabla^\mu\, \leftidx{^{(1)}}{\!H}_{\mu\nu}=0$, see Appendix \ref{sec:appendixA1}), we are left with a reduced expression where $\nabla^\mu$ acts now only on the running parameters and on the vacuum part of the  EMT:
\begin{equation}\label{eq:EqsVacExt2}
\nabla^\mu\left(G(M)\rho_\Lambda (M)\right) g_{\mu \nu}+\nabla^\mu\left(G(M)\alpha(M)\right) \leftidx{^{(1)}}{\!H}_{\mu\nu}= \nabla^\mu\left(G(M)\langle T_{\mu\nu}^{\delta \phi}\rangle_{\rm ren}(M)\right)+\nabla^\mu\left(G(M)\right) T_{\mu\nu}^{\phi}\,.
\end{equation}
We have used $\nabla^\mu T_{\mu\nu}^\phi=0 $ as well ---cf. Eqs.\,\eqref{MatterConservation1}-\eqref{MatterConservation}.   Performing the remaining derivatives and writing down the $\nu=0$ component of the final result,  one finds after some  calculations the following expression:
\begin{equation}
\begin{split}\label{MixedConservation}
&\dot{G}(M)\left(\rho_\phi+\rv\right)+G(M)\dot{\rho}_{\rm vac}+3HG(M)\left(\rv(M)+P_{\rm vac}(M)\right)\\
&=\left( \alpha(M)\dot{G}(M)+G(M)\dot{\alpha}(M)\right) \frac{\leftidx{^{(1)}}{\!H}_{00}}{a^2}\,.
\end{split}
\end{equation}
With this result it is straightforward to show that the generalized Friedmann's equations and local conservation laws given above lead to the following overall conservation law involving all of the ingredients entering our quantum matter system nonminimally coupled to gravity:
\begin{equation}\label{BianchiIdentity_all}
\frac{d}{dt}\left(G(M)\left(\rho_\phi+\rv(M)+\rho_X\right)\right)+3HG(M)\left(\rho_\phi+\rv(M)+\rho_X+P_\phi+P_{\rm vac}(M)+P_X\right)=0.
\end{equation}

\subsection{Running gravitational coupling}\label{sec:runningG}

If we neglect the effect of the HD term in the current universe, the \textit{r.h.s.} of \eqref{MixedConservation} can be set to zero and we are left with
\begin{equation}\label{MixedConservationApprox1}
\dot{G}(M)\left(\rho_\phi+\rv(M)\right)+G(M)\dot{\rho}_{\rm vac}(M)+3HG(M)\left(\rv(M)+P_{\rm vac}(M)\right)=0\,,
\end{equation}
where we have used the conservation law  for the background component of the scalar field,  Eq.\,\eqref{MatterConservation}.
A further simplification can be obtained if we assume that the EoS of the quantum vacuum  is exactly $\Pv=-\rv$.  We shall check right next what is the effect of the correction we have found in Sec.\,\ref{sec:EoSvacuum}. In the meantime, if we just take the standard EoS of vacuum it  allows us to dispense with the last term in the above equation.  Because the two  terms left involve derivatives with respect to the cosmic time we can write down \eqref{MixedConservationApprox1} as a simple differential form:
\begin{equation}\label{MixedConservationApprox2}
\left(\rho_\phi+\rv\right) dG+G d\rv=0\,.
\end{equation}
This equation can be used together with Friedmann's equation \eqref{Friedmann_density} in the same approximation (i.e. neglecting the HD terms and hence ignoring the  $\rho_X$ component in it).  The two equations can be easily combined  in the nicely separable form,
\begin{equation}\label{eq:diffeqG}
\frac{3H^2}{8\pi G} dG+Gd\rv=\frac{3H^2}{8\pi G} dG+G\frac{3\nueff}{4\pi} \mpl^2{H} dH=0 \,,
\end{equation}
in which the sum $\rho_\phi+\rv$ has been replaced with $3H^2/(8\pi G)$ thanks to Friedmann equation,  and $d\rv$ has been computed from \eqref{eq:RVM2} within the approximation of constant $\nueff$  -- cf.\,Eq.\,\eqref{eq:nueffAprox}.  Notice that  $G$ varies with $H$ and our aim is to find this function in this approximation.  Dividing out the above equation by $G$ and  upon identifying  $G(H_0)=G_N=1/\mpl^2$ with the current local value measured in Cavendish-type experiments,  we can solve for  the function $G(H)$, with the result
\begin{equation}\label{eq:GH}
G(H)=\frac{G_N}{1+\nueff \ln\frac{H^2}{H_0^2}}\,,
\end{equation}
where $\nueff$ here is given by the constant coefficient \eqref{eq:nueffAprox}. This equation was previously met in \cite{FritzschSola} within a more simplified theoretical context and used to study the potential variation of the fundamental constants of Nature.  However, equation \eqref{eq:GH} is only approximate in our context.  Remember that we assumed that $\nueff$ is constant in its derivation, but we know from Appendix  \ref{sec:appendixAbis1} that it is not a strict constant.

 It is natural to compare the above formula  with Eq.\,\eqref{eq:RGENewton} at this point. The latter stems  from the existence of running couplings, which  is of course a direct reflex of the RG invariance of the effective action.  In it, $M$ can be arbitrary since the effective action is independent of $M$.
Moreover, our assumption that  $\Pv=-\rv$, which we also used in the above derivation, is not a sufficiently good approximation since we know from Sec. \ref{sec:EoSvacuum}  that there is indeed a departure of the quantum vacuum EoS from $-1$, see Eq.\,\eqref{eq:VacuumPressureFull}.  Admittedly the last equation is a bit cumbersome, but if we consider only the contributions that can be relevant for the current universe, the departure is given by the term $f_2(M,\dot{H})$ on the \textit{r.h.s.} of  Eq.\,\eqref{eq:VacuumPressureFullsplit}. This is the same approximation used in our discussion of the EoS of the quantum vacuum for the current universe, see Sec.\,\ref{sec:EoSLateUniverse}. Thus,  from Eq.\,\eqref{eq:f2} and setting $M=H$, according to our usual prescription (cf. Appendix \ref{sec:appendixAbis1} for details), we find
\begin{equation}\label{eq:EoSaprox}
\Pv(H)+\rv(H)\simeq \frac{\left(\xi-\frac{1}{6}\right)}{8\pi^2}\dot{H}\left(m^2-H^2-m^2\ln\frac{m^2}{H^2}\right)\simeq \frac{\left(\xi-\frac{1}{6}\right)}{8\pi^2}\dot{H} m^2\left(1-\ln\frac{m^2}{H^2}\right)\,,
\end{equation}
where we have neglected a term of ${\cal O}(\dot{H}H^2)={\cal O}(H^4)$  but we have  kept   the terms proportional to $m^2\dot{H}$  as they are not necessarily negligible in the present universe. The above expression gives the leading deviation of the quantum vacuum EoS from $-1$ at low energy,  being such a deviation of the same order of magnitude as the $\sim H^2$ term involved in $\rv(H)$.  The above correction to the quantum vacuum EoS genuinely originates from our calculation of the vacuum pressure in  Sec.\,\ref{sec:EoSvacuum}.  Therefore,  it must be considered on equal footing with the dynamical term of  $\rv(H)$  in the correct calculation.

By duly taking into account Eq.\,\eqref{eq:EoSaprox}  in the calculation of $G(H)$  and using the exact function  $\nueff(H)$ given in the Appendix  \ref{sec:appendixAbis1} rather than just inserting the approximate constant result \eqref{eq:nueffAprox}, we find after some calculations the following expression for the running of the gravitational coupling (see the details in Appendix  \ref{sec:appendixAbis2}):
\begin{equation}\label{eq:GNHfinal}
G(H)=\frac{G_N}{1-\frac{\left(\xi-\frac{1}{6}\right)}{2\pi}\frac{m^2}{m^2_{\rm Pl}}\ln \frac{H^2}{H_0^2}}\,.
\end{equation}
This formula is not only more rigorous than Eq.\,\eqref{eq:GH} in our context, but in  contrast to the latter  it is entirely consistent with the running coupling formula \eqref{eq:RGENewton}  when we set $M=H$ and $M_0=H_0$ in it and use the fact that $H^2-H_0^2$ is fully negligible versus $m^2\ln\left(H^2/H_0^2\right)$ for all $H$ (in post-inflationary times).  In fact,  for $H$ around the current value $H_0$, this follows from
\begin{equation}\label{eq:ratio}
  \frac{H^2-H_0^2}{m^2\ln\frac{H^2}{H_0^2}}=\frac{H_0^2}{m^2}\left(1+\frac12 x+{\cal O}(x^2)\right)\ll1 \ \ \ \  \  \ \ \  \left( 0\leq |x|<1\right)\,,
\end{equation}
with $x\equiv (H^2-H_0^2)/H_0^2$ and $H_0^2/m^2\ll1$ for any known particle.  On the other hand, for large values of $H$ we also have $H^2/m^2\ll1$ since $m$ is assumed to be a mass of a typical GUT particle.  Notice that the running of $G(H)$ from Eq.\eqref{eq:GNHfinal} is very mild, not only because it is a logarithmic running but also because the coefficient of the log is of order  {$m^2/\mpl^2\ll 1$, which holds good even for $m$ in the GUT range, i.e. $m\sim M_X\sim 10^{16}$ GeV, assuming that $\xi$ is not very big.

We should emphasize that  the setting $M=H$   used above to study the running of $G(H)$  is the same one employed to infer the running vacuum formula (Appendix \ref{sec:appendixAbis1}).  The fully consistent derivation of \eqref{eq:GNHfinal} from  two diverse roads; namely one (more physical) relying  on the overall  local conservation law \eqref{MixedConservationApprox1}, and the other (more formal) based on  the running coupling formula  \eqref{eq:RGENewton} -- and ultimately on the RG-invariance of the effective action --   is a most remarkable feature. At the end of the day, the scale setting prescription $M=H$ proves to be the clue for exploring the physical consequences of our renormalization framework.  Overall the obtained results speak up of the full mathematical and physical consistency of our approach.

Finally, we should note that although our discussion in this section has focused on the background and  vacuum effects from the single quantum matter field $\phi$, this does not exclude the possibility that additional contributions from  incoherent dust matter and radiation become involved.  In the discussion of Appendix \ref{sec:appendixAbis2} we show that the presence of ordinary matter does not alter  at all the results presented above, provided one assumes that such an ordinary matter is conserved.  Apart from that, there is also the possibility that matter components  interact with the running vacuum.  The safest possibility would be to assume potential interactions between CDM and vacuum, as in this way  the most sensitive and well  known components  of the  universe (baryons and photons) are unaffected.  These new types of interaction between DM and vacuum can certainly be  important but  are model dependent, as they rely on introducing new  parameters in the theory.

%
\begin{figure}[t]
\begin{center}
\includegraphics[scale=0.55]{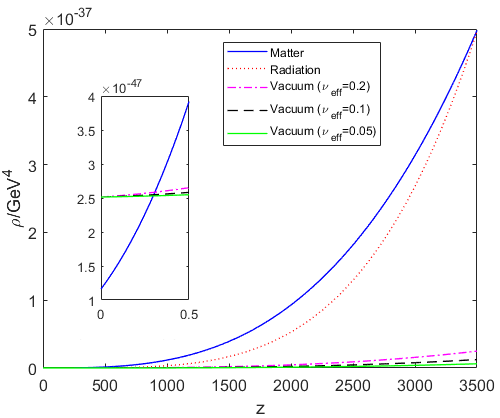}\
\includegraphics[scale=0.55]{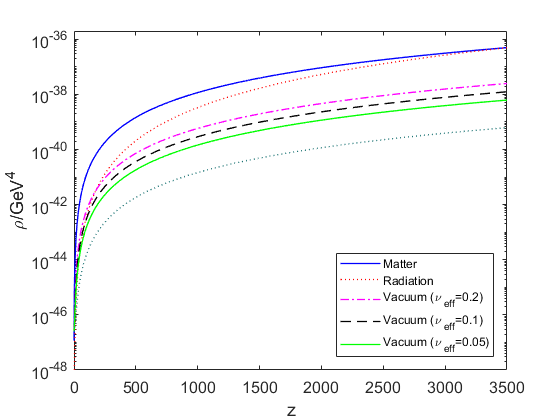}\
\end{center}
{\bf Figure  2}.  Evolution of the different energy densities with the expansion in the RVM context. The plot on the right provides a complementary view in a (vertical) logarithmic scale. The VED  exhibits a very mild dynamics up to the early radiation dominated epoch. Parameters $H_0$ and $\Omega_m^0$ are taken from the best-fit values of\cite{PlanckCollab}, see the text. The vacuum evolution is very mild and to make it more apparent it is shown for different values of   $\nueff$ . The additional (green-dotted) line on the right plot corresponds to $\nu=0.005$. Since it could not be appreciated on the left plot (which uses a linear scale), it has not been marked there.
\protect\label{Fig2}
\end{figure}

\section{Running vacuum: some phenomenological implications}\label{sec:PhenoImplications}

We have devoted most of this work to  put  the foundations of the running vacuum model (RVM) on a sound theoretical basis within  QFT in curved spacetime. In this last section, which precedes the conclusions,  we would like to put forward some phenomenological considerations in order to illustrate the possible physical impact of the quantum running vacuum in the light of the observational data.  Although a more detailed phenomenological analysis will be presented elsewhere\,\cite{CosmoTeam2021}, here we just  highlight a few potentially significant implications.  We have already mentioned that the RVM has been tested in the past in  a variety of phenomenological contexts, where the basic parameter $\nueff$ has been fitted to different sets of  cosmological data \cite{ApJL2015,ApJ2017,RVMphenoOlder1,Elae2015,RVMphenoOlder2,RVMpheno1,RVMpheno2,PericoTamayo2017,CQG2017,ApJL2019,Mehdi2019,Tsiapi2019,BD2020,Singh2021,Manos2021,Yu2022,deCruzPerez:2023wzd,EPL2021}.  The fact that we have now been able to account for the structure of the RVM on  QFT grounds, it obviously strengthens its position.   To test the RVM, we are going to use the generalized expression for the VED  which we have predicted in Sec.\,\ref{sec:Generalized RVM}.  It  includes the two low energy terms  $H^2$ and  $\dot{H}$, each one with independent coefficients. More specifically, we consider a generic RVM structure of the form \eqref{eq:RVMgeneralized2}.  From the QFT perspective, the two coefficients $\nueff$  and $\tilde{\nu}_{\rm eff}$  depend both on the number of bosons and fermions in a way which can be computed theoretically. Although  in this work we have presented the calculation for the scalar contribution only,  the yield from the fermionic part will be reported in a separate study\,\cite{JCS2022}.  Let us however note that despite of the fact that  the vacuum running is theoretically computable in QFT,  which is perhaps the most remarkable observation of our work,  in practice we ignore  the details of the underlying GUT which ultimately accounts  for such a vacuum dynamics. Therefore, at present  we cannot predict the precise quantitative evolution of the VED.  But this does not preclude us, of course,   from  testing the phenomenological performance of the model.   While the  general VED form \eqref{eq:RVMgeneralized2} was analyzed on pure phenomenological grounds in \cite{ApJL2015,ApJ2017}, here we will minimize the number of parameters  and shall consider the convenient situation $\tilde{\nu}_{\rm eff}=\nueff/2$, as in this way  the VED adopts the form
\begin{equation}\label{eq:RRVM}
 \rv(H) = \frac{3}{8\pi G_N}\left(c_{0} + \nu_{\rm eff}\,H^2+ \tilde{\nu}_{\rm eff}\,\dot{H}\right)= \frac{3}{8\pi G_N}\left(c_{0} + \frac{\nueff}{12} {R}\right)\,,
 \end{equation}
with  ${R} = 12H^2 + 6\dot{H}$ the curvature scalar.  This scenario was analyzed in \cite{EPL2021} under different hypotheses, in particular it was assumed that the vacuum was interacting with cold dark matter.  We  restrict the number of  assumptions to the minimum and  just adapt to the precise situation that we have encountered in the QFT analysis presented in this work,  in which  matter is locally conserved (as in the standard $\CC$CDM) and  the VED and  gravitational coupling $G$ possess a mild cosmic evolution (as studied in the previous section).   Interactions between matter and vacuum are not considered. This scenario  is particularly  well-behaved in  the radiation dominated era since the relativistic matter density is not altered as compared to the standard model  and hence cannot  impinge on the BBN physics\,\cite{ApJL2015,ApJ2017}. This is all the more true if we take into account that the vacuum energy density \eqref{eq:RRVM} remains also subdominant in the radiation epoch since  $R\simeq 0$ in it.

\begin{figure}[t]
\begin{center}
\includegraphics[scale=0.55]{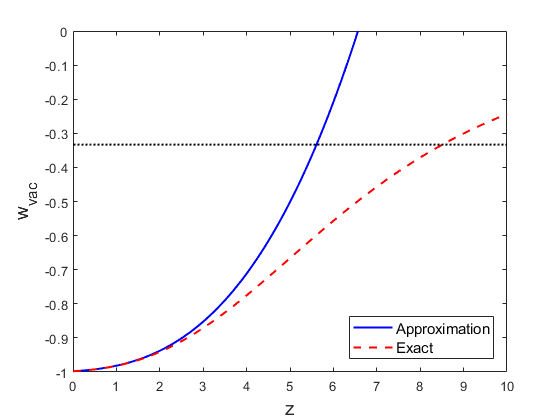}
\end{center}
{\bf Figure  3}.  Vacuum EoS at low redshifts including the quantum effects computed in this work.  We show both the curve corresponding to the approximate formula \eqref{eq:EoSLow2} and the exact curve computed numerically for  $\nueff=0.005$. Because the fitting analyses generally  favour the sign  $\nueff>0$\,\cite{EPL2021}, the EoS deviates slightly  from $\wv=-1$ already at $z=0$ and  mimics quintessence.
\label{Fig3}
\end{figure}
\begin{figure}[t]
\begin{center}
\includegraphics[scale=0.6]{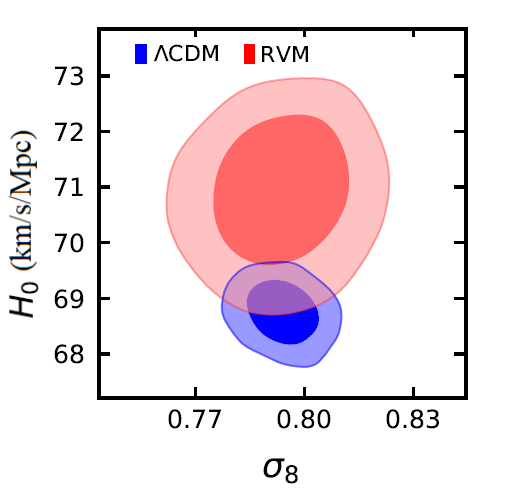}
\end{center}
{\bf Figure  4}.   Comparative contours  at $1\sigma$ and $2\sigma$ c.l.  in the ($\sigma_8$-$H_0$)-plane corresponding to the RVM and the $\CC$CDM  for the data sets mentioned in the text.  It can be seen that the RVM is quite effective in alleviating the $H_0$ tension and at the same time it reduces  the $\sigma_8$ one.
\label{Fig4}
\end{figure}

As previously indicated,  the RVM under consideration preserves local matter conservation.  However, the VED evolves together with $G$ such that the Bianchi identity can be satisfied, just as explained in Sec.\ref{sec:RenormalizedFriedmann}.  On solving explicitly the model (details will be provided elsewhere)   we find  the following evolution for the VED in linear order in the small parameter $\nueff$:
\begin{equation}\label{eq:VDEm}
\rv(a)=\left(\frac{\Omega_{\rm  vac}^0}{\Omega_m^0}-\frac14\,\nueff\right) \rho_m^0+\frac14\nueff\rho_m^{0}a^{-3}+\mathcal{O}(\nueff^2)\,,
\end{equation}
where the current cosmological parameters $\Omega_i^0=(8\pi G_N) \rho_i^0/(3 H_0^2)$ satisfy  $\Omega_{\rm vac}^0+\Omega_m^0=1$.
For $\nueff=0$ the VED is constant and  $\rv=\frac{3 H_0^2}{8\pi G_N} \Omega_{\rm vac}^0=\rvo=$const.  (i.e. we recover  the $\CC$CDM behavior, as it should be),  but for nonvanishing $\nueff$ the VED  has  a moderate dynamics since this parameter is small.  In Fig. 2 we plot the various energy densities for matter, radiation and vacuum using the  best-fit values from Planck TT, TE, EE + low E + lensing data\,\cite{PlanckCollab}.  In addition,  in Fig. 3 we display the EoS of the quantum vacuum at low energy, which is important for the late time universe.  As discussed in Sec.\ref{sec:EoSLateUniverse},  the EoS of the vacuum departs from the classical value $\wv=-1$ when the quantum effects are taken into account.  In Fig. 3 we plot the approximate formula \eqref{eq:EoSLow2}  as well as the exact result (for $\nueff=0.005$).  They remain close for low redshift only.  Because $\nueff>0$ (as it follows from different phenomenological studies mentioned above), the EoS of the quantum vacuum satisfies $\wv>-1$ even at $z=0$. Hence it mimics quintessence without need of invoking \textit{ad hoc} scalar fields.  This is quite revealing, as it suggests that such an effective quintessence behaviour may emerge from a fundamental QFT origin.

Finally, in Fig.\,4 we illustrate an important phenomenological implication of the present framework, namely the possibility that the RVM could help in solving or at least alleviating the persistent $H_0$ and $\sigma_8$ tensions mentioned in the introduction. These tensions are still the main  focus of  interest of many cosmologists \cite{TensionsLCDM,IntertwinedH0,PerivoSkara2021,ValentinoReviewTensions,Intertwined8,TensionsJSP2018,WhitePaper2022,Riess2019}.  In that figure we show  the  $1\sigma$ and $2\sigma$ c.l. contours  in the $(\sigma_8, H_0)$ plane both for the RVM and the $\CC$CDM.  The used data sets in this analysis involve type Ia supernovae, baryonic acoustic oscillations, cosmic chronometers, large scale structure formation data (on $f\sigma_8$) and the cosmic microwave background observations from Planck 2018 (i.e. the data string SnIa+BAO+$H(z)$+LSS+CMB).  We refer the reader to the previous studies \cite{BD2020,EPL2021}   for a detailed definition and description of these data  and the methodology used in the fitting analysis.
In the light of the results presented in  Fig. 4,  we can say that the comparative performances  between  the  RVM  and  $\CC$CDM clearly show that the RVM may alleviate the two tensions $\sigma_8$ and $H_0$  at a time, which is remarkable.  The value of $H_0$ tends to be higher in the case of the RVM as compared to the $\CC$CDM and as a result the current  $5\sigma$ tension\cite{Riess2022}  between the measurements of the local value of $H_0$ and from the early universe (CMB)  is rendered at the residual level of $\sim 1.6\sigma$.  Similarly, the   $\sigma_8$ value is also  reduced and the corresponding tension is brought  to the inconspicuous level of $\sim 1.3\sigma$.  A more detailed and exhaustive presentation  including additional data sources (e.g. from BICEP2/Keck Array  CMB polarization experiments\,\cite{KeckBICEP2} as well as  updated observations  from the data string mentioned above)  will be undertaken in a future phenomenological study.  However, we believe that even from the short considerations highlighted in this section the reader can already get  a flavor of the real potential of the RVM  for improving the description of the cosmological data, i.e.  when the  cosmological vacuum becomes dynamical, and more specifically when it runs according to the QFT description presented in this paper.

\section{Discussion and conclusions}\label{sec:conclusions}

The focus point of  this paper has been  the vacuum energy density (VED) in the context of quantum field theory (QFT) in  FLRW spacetime and the possible connection with the cosmological term in Einstein's equations.  As we know, the difficulties in reconciling QFT (and string theory) predictions with cosmological observations in connection to this subject are at the basis of the so-called Cosmological Constant Problem (CCP)\,\cite{Weinberg89,Witten2000}. Time and again this is one of the main problems of theoretical physics and cosmology that demands an urgent explanation at a fundamental level. The methods to deal with QFT in curved spacetime  are well-known since long \,\cite{BirrellDavies82,ParkerToms09,Fulling89,MukhanovWinitzki07}, and yet some of the most pressing problems of modern cosmology still remain unaccounted.  The CCP  is certainly a focus issue for any formal theory of the cosmic evolution.  It  is a must to be addressed in this kind of field theoretical studies since the physical interpretation of the cosmological term $\CC$ has traditionally been linked to the current value of the  VED, $\rvo$,  through Lema\^\i tre's formula $\rvo=\CC/(8\pi G_N)$.   As we have mentioned in the introduction, the discrepancy existing between the measured value of $\rvo$ and the generic prediction made  in field theories of the fundamental interactions (e.g. the standard model of particle physics)  is utterly disproportionate.  Such an appalling clash of theoretical concurring ideas versus direct astrophysical observations is at the root of the CCP. Furthermore,  irrespective of the fact that there are many sources of vacuum energy in QFT (which are,  in principle, uncorrelated), each one of them is very large as compared to $\rvo\sim 10^{-47}$ GeV$^4$, and hence the possible compensation among these sources leads to hopeless fine-tuning among the parameters of the theory. If that is not enough, the adjustment  must be redone order by order in perturbation theory.   Such an unending process of  tuning and  retuning makes the CCP even harsher, in fact unacceptable as a natural solution\,\cite{JSPRev2013,JSPRev2022}.  When tackling big theoretical hurdles of this sort one may expect to get some helping hand from the concepts,  methods and tools underlying quantum gravity and string theory\,\cite{Witten2000}. The bare truth, however,  is that neither one of them has succeeded in improving significantly the situation  for the time being.  Quantum gravity (QG) does not  exist as a consistent theory yet; and string theory  somehow abhors de Sitter space, as  `swampland' conjectures preclude the construction of  metastable de Sitter vacua in the string framework\,\cite{Vafa}. While we remain agnostic about these problems a lot of exciting QG phenomenology is still possible with the advent of the multi-messenger era, characterized by a steep increase in the quantity and quality of experimental data that are being obtained from
the detection of the various cosmic messengers (photons, neutrinos, cosmic rays and gravitational waves) from numerous origins\,\cite{CostAction}.  At the same time we expect that the more pedestrian methods of the semiclassical QFT approach may still shed some  light on those pending issues in the cosmological arena, in particular on the vacuum energy and its renormalization.  This has been the main aim guiding our task here.

The pivotal message which we can extract  from our  investigation should be made clear:  if the VED of the expanding universe is to be tackled  from  the fundamental point of view of QFT in curved spacetime, the vacuum energy should be a mild dynamical quantity.  The reason is that, in the  QFT context,  the  VED appears initially as a UV-divergent quantity, and hence it is subject to renormalization.  Intrinsic to the process of renormalization is the appearance of a scale $M$ (which plays the role of renormalization point).  Scale dependencies (explicit and implicit)  obviously cancel out in the full effective action of the theory,  which is  renormalization group (RG)-invariant, and involves  in particular the vacuum action as well as  many other scale-dependent terms, e.g. from the classical matter Lagrangian (such as masses, couplings and fields, all of them implicitly scale-dependent).  Now except for  the effective action itself,  which is  a rather  formal object, in cosmology we cannot play with  RG-invariant quantities which are more common in particle physics (such as scattering cross-sections, decay rates  etc.  and in general  different kinds of  Green's functions related to observable quantities).  Thus, we must content ourselves with using different parts of the full effective action that remain scale dependent.   The VED is one of these parts and hence it appears as one of the scale-dependent quantities upon renormalization.  To explore the cosmic evolution of the VED we find that the setting $M=H$ is most appropriate and it also leads to a mildly logarithmic running of the gravitational coupling  $G$.

Even though a full-blown calculation  of the VED in QFT cannot be faced at this point, in this work  we have focused on the simplest, and yet nontrivial, QFT model interacting with the FLRW spacetime background that we can think of,  namely  a scalar field $\phi$  nonminimally coupled to gravity.  For the sake of a more simple presentation and, therefore, to avert  `not seeing the forest for the trees', we have assumed that $\phi$ has no self-interactions and hence no spontaneous symmetry breaking. This assumption allows us not only to avoid dealing with the renormalization of the corresponding effective potential but also to concentrate on the computation of the zero-point energy (ZPE) part, which is a pure quantum effect and thereby constitutes the most genuine  quantum vacuum piece within the whole  VED structure. With these basic assumptions, we have undertaken the renormalization of the corresponding energy-momentum tensor (EMT)  using the  adiabatic regularization and  renormalization method.  A key point in our approach has been to implement an appropriate renormalization of the  EMT by performing a subtraction of its on-shell value  at an arbitrary renormalization point $M$.  The presence of this floating scale brings  into play the renormalization group  flow. Since the renormalized EMT  becomes a function of   $M$, we can compare the renormalized result at different epochs of the cosmic history characterized by different energy scales, which we  have tested with the value of $H$ (the Hubble rate) at each epoch. This is certainly along the lines of the original RG approach\,\cite{JSPRev2013,JSPRev2022} but goes well beyond it since it provides a more formal and explicit QFT calculation.  The renormalized VED is composed not only of the  ZPE (involving the quantum fluctuations of the scalar field) but also of $\rL(M)$,  the renormalized value of  the parameter $\rL$ in the bare vacuum action. We find it remarkable that  when we compute the evolution of the VED from one scale to another  within our renormalization framework, the result  is free from quartic contributions  $\sim m^4$. This is in stark contrast with other renormalization schemes in which the $\sim m^4$ effects  are responsible for the outrageously  large contributions to the VED and hence are badly in need of extreme  fine-tuning arrangements.  In our opinion, even if being well aware  of the many other difficulties ahead of us,  with the absence of these terms in our framework  we might be  inching into  an eventual  solution of the CCP.

Let us, however, not forget that the CCP is a tough polyhedric problem with many faces.  One thing is to glimpse at some light at the end of the fine-tuning `tunnel', and quite another is to solve all aspects of the cosmological constant problem.  As we have emphasized from the very beginning,  renormalization theory stops at this point.  While we can use input data at present to predict (within the current renormalization framework) a well-defined value and a smooth evolution of the VED at any time in the past (and even in the future),  which is already something,   let us not be mistaken:  we cannot predict its value at present.  Renormalization theory uses data at one point to make a prediction at another point, but it does not aim at self-predicting the initial data, of course.  Therefore, the so-called `old cosmological constant problem'\,\cite{Weinberg89} (namely, the problem about explaining the value itself of the cosmological constant), persists.  Notwithstanding, it may have been  freed of the unbearable `shakes and jerks' impinged on the result by the seeming need of continuously having to do unnatural  fine-tuning  among the parameters of the theory.  Instead, in our case we find a mild time evolution of the VED which is proportional to $H^2$ in the present universe, and hence it is small and changes smoothly and slowly  with the   flow of the cosmic expansion.

The renormalized results obtained  in our analysis  are robust. Indeed, in order  to strengthen our conclusion, which we have first presented on the basis of the WKB expansion of the Fourier modes of the field, we have subsequently corroborated it from the perspective of the  effective action formalism.  It means that we have solved the curved spacetime  Feynman propagator of  the nonminimally coupled scalar field to gravity using the adiabatic method and computed the effective action using the heat-kernel expansion. Since that expansion has also been performed off-shell (i.e. at the arbitrary scale $M$ rather than at the physical mass $m$), it was necessary to compute the corresponding corrections induced on the DeWitt-Schwinger coefficients.  With the help of the effective action  we have rederived the renormalized EMT and obtained the same result as with the original WKB method.  As a bonus we have extracted the renormalization group equations (RGE's) for the couplings and also for the VED itself.  For this we have had to find out the explicit form of $\beta$-function for the VED running,  Eq.\,\eqref{eq:RGEVED1}. The latter appears to be free from quartic mass scales, which otherwise would recreate the usual (unfathomable) fine-tuning problem which we wanted to eschew.  The smoothly behaving RGE  for the VED that we have found  was long suspected from semi-qualitative RG arguments, see \cite{JSPRev2013,JSPRev2022} and references therein, but  only here (and previously in the shorter presentation\,\cite{CristianJoan2020})  this  result  has been demonstrated for the first time in the literature in a full-fledged QFT context. Besides, we provide the RGE  for the gravitational coupling.  From the dynamical  interplay with the VED we find that it evolves very mildly as a logarithmic function of the Hubble rate, $G=G(\ln H)$.

The above accomplishments for the low energy behavior of the VED  can be summarized by saying   that the renormalized VED  obtained in our calculation for the accessible expansion history of the universe adopts essentially the canonical  form of the running vacuum model (RVM), in which $\rv=\rv(H)$ consists of a dominant `rigid' term (playing the role of cosmological constant) plus a series of powers of $H$  and its time derivatives.  This form was originally  motivated from general  renormalization group arguments.  Finding this result is most reassuring since it shows that the long predicted form of the RVM structure for the VED can be attained  from direct calculations of QFT in the FLRW spacetime.   The structure of $\rv(H)$ is such that  all of the involved terms carry an even number of time derivatives of the scale factor, which is mandatory to preserve the general covariance of the effective action.  The lowest order terms are just $H^2$ and $\dot{H}$ both of second adiabatic order.  In the original RVM formulation, only the  $H^2$ was taken into account.  In the present calculation, we have shown that its coefficient $\nueff$ can be computed in QFT in curved spacetime and is naturally predicted to be small ($|\nueff|\ll1$).  The detailed structure of $\nueff$ depends on all the quantum matter fields involved  in the calculation.  In this sense, its value  must ultimately be determined experimentally by confronting the model to the cosmological data.  The $H^2$ term is nevertheless sufficient to describe the dynamics of the vacuum in the current universe, while the higher order components can play a role in the early universe, and in particular for describing inflation.

As indicated, our findings here reconfirm and improve the results of \cite{CristianJoan2020}, where the $4th$ order adiabatic solution was first presented. Extending the solution into the next ($6th$) order is a  computationally demanding task, which has allowed us to determine the on-shell renormalized form of the EMT and derive the equation of state (EoS) of the quantum vacuum.  To the best of our knowledge these are brand-new results in the literature. We have found that the EoS  is slightly different from $-1$ and hence that the quantum vacuum is really a quasi-vacuum state.  As a result,  this may have consequences for the present universe (see below), but at the same time these effects  can have nontrivial  implications for the very early universe. In fact, we  predict a new mechanism for inflation which is triggered by the aforementioned  $\sim H^6$ terms.  In contrast to the ${\cal O}(H^4)$ terms, which vanish for $H=$const. (as they all depend on time derivatives), the term  $\sim H^6$ can perfectly bring about a short but fast period  of $H=$const. inflation.

We should stress, however,  that the RVM inflationary mechanism is distinct from Starobinsky's inflation\,\cite{Starobinsky80}, for which it is the time derivative of the Hubble rate which remains constant for a short stretch of time (viz. $\dot{H}=$ const.)   Noteworthy, there exists a stringy version of the RVM inflationary mechanism which can operate with $H^4$ terms,  i.e. for which these terms appear  without time derivatives and hence  do not vanish for $H=$const.\,\cite{NickJoan2020,NickJoan2021}. This is in contrast to the QFT version that we have described here,  being characterized  by  $H^6$-inflation.  This means that the stringy and non-stringy (QFT) mechanisms of RVM-inflation can, in principle, be distinguished.  In  both cases  inflation is unleashed during a short time interval of the early universe where $H=$const. and therefore requires that the effective behavior of the VED carries a higher  (even) power $H^{N}\, (N=4,6,...)$  (beyond $H^2$).  The existence of  a high power of the Hubbe rate in the VED  is the characteristic trademark of RVM-inflation. In the QFT case, it appears as a fundamental mechanism of inflation solely induced by the pure quantum matter effects on a classical gravitational background.

Even if the RG approach was actually  the first qualitative idea behind  the RVM (see \cite{JSPRev2013,JSPRev2022} and references therein), with the present QFT calculations and those of \cite{CristianJoan2020}  we have provided for the first time a solid foundation of the  RVM, in which the dynamical structure of the VED is seen to ensue from first principles, namely from the quantum effects associated with the proper renormalization of the EMT.   Let us  also clarify that despite we have illustrated our results employing one single (real)  scalar field, further ongoing investigations show that the generalization of  these results for  multiple fields,  involving bosons as well as  fermionic components, lead also  to the generic RVM structure mentioned here (up to nontrivial computational details\cite{JCS2022}).

The unsatisfactory status of the $m^4$ terms in cosmology  is alike to the hierarchy problem linked to the role played by the  $m^2$ terms  in ordinary gauge theories\,\cite{Veltman1981}, although much worse in magnitude.  In  the approach we have outlined in this work we  need not call for special cancellations  (fine-tuning) among the  $m^4$ contributions, such as e.g. when using the Pauli sum rules\,\cite{Pauli}  to insure the (fine-tuned) cancellation of  quartic, quadratic and logarithmic contributions from bosons against fermions\,\cite{Visser2018}, nor to invoke the existence of emergent scales or very small dimensionless parameters suppressing the undesired effects, or treating the standard model of particle physics as a low energy effective theory, see e.g.\,\cite{Bjorken2011,Jegerlehner2014,Bass2020,Elahe2016} for a variety of contexts of this sort.  Instead, the problem is dealt with here by suitable renormalization leading to more physical results.

The upshot and final summary of this investigation is that from calculations of QFT in  FLRW spacetime  we find that the VED  takes the form of an expansion in powers of $H$ (and its time derivatives)  of even adiabatic order. This is essential to preserve the general covariance of the effective action.   This form of  $\rho_{\rm vac}(H)$ is just as predicted by the RVM.   At low energies, it is particularly simple and consists of an additive constant  together with a small dynamical component  $\sim \nueff H^2$, in which the dimensionless parameter $\nueff$  is predicted to be small ($|\nueff|\ll1$). The latter is proportional to the coefficient of the $\beta$-function of the running VED, although they have opposite signs. This is a reflex of the fact that the two leading effects on the late time dynamics of the VED, namely one from the scaling evolution with $M$ (before we fix its value) and the other from the expansion rate $H$,  are actually opposed. However, the second one is dominant and this fact implies that  the net sign of the  slope  of the VED  is fixed by the sign of $\nueff$.  So for $\nueff>0$  (resp. $\nueff<0$) the evolving VED mimics quintessence (resp. phantom DE).   Such a behavior  is also manifest at the level of the equation of state  of the quantum vacuum  in the late universe, which we have also determined explicitly
(cf. Sec.\,\ref{sec:EoSLateUniverse}) and is found to be redshift dependent. This is in contradistinction  to the usual preconceptions on the vacuum state, in which the vacuum EoS is assumed to have the rigid value $-1$ since one is  ignoring the renormalization effects on it. The dynamical VED, in fact, turns out to have a dynamical EoS too.

The physical  outcome is that even today's cosmic vacuum  should be mildly dynamical and the same holds for  the gravitational coupling , although in the latter case is much milder ($\sim \ln H$).  The corrections are small for the current universe, but they are not necessarily negligible, especially for the VED.  In practice, the value of $\nueff$ can only be determined  upon fitting the RVM to the overall cosmological data.  In previous works, the RVM  has been phenomenologically fitted to a large wealth of cosmological data and the value of $\nueff$  has been found to be positive and of order $\sim 10^{-3}$\, cf.\cite{ApJL2015,ApJ2017,RVMphenoOlder1,Elae2015,RVMphenoOlder2,RVMpheno1,RVMpheno2}. This has been reconfirmed in the most recent studies\,\cite{EPL2021,CosmoTeam2021}.  In the present work,   we have also highlighted (cf.  Sec. \ref{sec:PhenoImplications}) some of these important phenomenological applications of the RVM, which may help to improve the description of the overall cosmological data, and in particular to alleviate the $H_0$ and $\sigma_8$ tensions\cite{WhitePaper2022}.

Being $\nueff>0$  preferred by the fitting results,  the VED  decreases (slowly) with the expansion and hence the RVM mimics quintessence. {From this more fundamental perspective,  the effects that are usually attributed to the hypothetical existence of quintessence or phantom fields (and other forms of DE in general) might  be nothing else but a manifestation of the quantum effects of matter fields in curved spacetime.  If  so there would be no need of appealing to the existence of \textit{ad hoc} scalar fields with particular effective potentials capable  of describing the DE in the current universe.  Such a description is only possible at the (huge)  price of arranging  severe fine tuning among the parameters.  Instead, the observed DE might just be a property of the quantum vacuum dynamics, which is triggered by the quantum matter fields  through pure virtual vacuum-to-vacuum  fluctuations, and hence being sensitive to all mass scales available.  In our framework no particular potential was used and the renormalized vacuum energy density evolves smoothly without fine tuning. }  All that said, we cannot exclude, of course, that quantum gravity effects may also be involved and help in further shaping the structure and properties of the quantum vacuum,  especially once a consistent theory of quantum gravity will  (hopefully)  be available in the future, but these additional  vacuum contributions fall out of the scope of the present study.

 Finally, we should mention that despite of the fact that our QFT calculation focused on the ZPE,  we conjecture that a similar dynamical  structure  should emerge from the  VED in the general case since the expansion of the full effective action in powers of momenta in the context of FLRW spacetime should result in an even power series  of the Hubble rate (owing once more to general covariance).

\vspace{0.5cm}

 \subsection*{Acknowledgements}
 We are funded by projects  PID2019-105614GB-C21 and FPA2016-76005-C2-1-P (MINECO, Spain), 2017-SGR-929 (Generalitat de Catalunya) and CEX2019-000918-M (ICCUB).  CMP is also partially supported  by  the fellowship 2019 FI$_{-}$B 00351. The work of JSP is  also partially supported by the COST Association Action CA18108  ``Quantum Gravity Phenomenology in the Multimessenger Approach  (QG-MM)''. We are  very  grateful  to A.  G\' omez-Valent and  J. de Cruz P\'erez  for  the fruitful collaboration in the task of understanding the RVM and its manyfold phenomenological implications.

\vspace{0.5cm}

\appendix
\section{Conventions and useful formulas}
\subsection{Geometric quantities}\label{sec:appendixA1}
We use natural units, therefore $\hbar=c=1$ and $G_N=1/\mpl^2$, where $\mpl\simeq 1.22\times 10^{19}$ GeV is the Planck mass. As for the conventions on geometrical quantities used throughout this work, they read as follows: signature of the metric  $g_{\mu\nu}$, $(-, +,+,+ )$; Riemann tensor,
$R^\lambda_{\,\,\,\,\mu \nu \sigma} = \partial_\nu \, \Gamma^\lambda_{\,\,\mu\sigma} + \Gamma^\rho_{\,\, \mu\sigma} \, \Gamma^\lambda_{\,\, \rho\nu} - (\nu \leftrightarrow \sigma)$; Ricci tensor, $R_{\mu\nu} = R^\lambda_{\,\,\,\,\mu \lambda \nu}$; and Ricci scalar,  $R = g^{\mu\nu} R_{\mu\nu}$.  Overall, these correspond to the $(+, +, +)$ conventions in the classification by Misner-Thorn-Wheeler\,\cite{MTW}.  The Einstein field equations read $G_{\mu\nu}+\CC g_{\mu\nu}=8\pi G\, T_{\mu\nu}$,  where   $G_{\mu\nu}=R_{\mu\nu}-\frac12\,R g_{\mu\nu}$  is the Einstein tensor.  We assume spatially flat three-dimensional geometry. The nonvanishing Christoffel symbols corresponding to the conformally flat metric $ds^2=a^2(\tau)\eta_{\mu\nu}dx^\mu dx^\nu$, with $\eta_{\mu\nu}={\rm diag} (-1, +1, +1, +1)$, are the following:
\begin{equation}
\Gamma_{00}^{0}=\mathcal{H},\qquad \Gamma_{ij}^0=\mathcal{H}\delta_{ij}, \qquad \Gamma_{j0}^i=\mathcal{H}\delta_j^i\,.
\end{equation}
We denote with primes the derivatives with respect to conformal time ($\tau$) and with dots the derivatives with respect to  cosmic time  ($t$).  Thus,  ${\cal H}=a H$,    $a'=a \mathcal{H}=a^2 H$ and  $a''=a^3(2 H^2+\dot{H})$. We can also derive the following useful relations to convert the derivatives of the Hubble rate with respect to conformal time  into derivatives with respect to cosmic time, which are repeatedly used in the calculations quoted in the main text:
\begin{equation}
\begin{split}
\mathcal{H}^\prime=&a^2(H^2+\dot{H}),\\
\mathcal{H}^{\prime\prime}=&a^3\left(2H^3+4 H\dot{H}+\ddot{H}\right),\\
\mathcal{H}^{\prime\prime\prime}=&a^4\left(6H^4+18 H^2\dot{H}+ 4\dot{H}^2+ 7 H\ddot{H}+\vardot{3}{H}\right),\\
\mathcal{H}^{\prime\prime\prime\prime}=&a^5\left(24H^5+96 H^3\dot{H}+ 52 H\dot{H}^2+ 46 H^2\ddot{H}+15\dot{H}\ddot{H}\ddot{H}+11 H\vardot{3}{H}+\vardot{4}{H}\right),\\
\mathcal{H}^{\prime\prime\prime\prime\prime}=&a^6 \left( 120 H^6 + 600 H^4 \dot{H} + 548 H^2 \dot{H}^2 + 52 \dot{H}^3 +
326 H^3 \ddot{H} + 271 H \dot{H} \ddot{H} + 15 \ddot{H}^2 \right.\\
&\left. +101 H^2 \vardot{3}{H} + 26 \dot{H} \vardot{3}{H} + 16 H \vardot{4}{H} + \vardot{5}{H} \right).
\end{split}
\end{equation}
For convenience we quote the Ricci scalar and the nonvanishing components of the curvature tensors in alternative forms:
\begin{equation}\label{eq:R}
R={6}\frac{a^{\prime\prime}}{a^3}=\frac{6}{a^2}\,(\mathcal{H}^\prime+\mathcal{H}^2)=6\,\left(\frac{\dot{a}^2}{a^2}+\frac{\ddot{a}}{a}\right)=6(2H^2+\dot{H})
\end{equation}
and
\begin{equation}\label{eq:R00G00}
 \ \ \ R_{00}=-3\mathcal{H}^\prime=-3a^2(H^2+\dot{H})\,,\qquad G_{00}=3\mathcal{H}^2=3a^2H^2\,.
\end{equation}

For reference we  also quote  the  well-known definitions of Euler's density $E$ and the square of the Weyl tensor ($C^2$):
\begin{equation}\label{eq:EC2}
E= R^{\alpha\beta\gamma\delta}R_{\alpha\beta\gamma\delta}-4R^{\alpha\beta}R_{\alpha\beta}+R^2, \qquad \ \
C^2=R^{\alpha\beta\gamma\delta}R_{\alpha\beta\gamma\delta}-2R^{\alpha\beta}R_{\alpha\beta}+\frac{1}{3}R^2.
\end{equation}
It follows that
\begin{equation}\label{eq:R2R4}
R^{\alpha\beta\gamma\delta}R_{\alpha\beta\gamma\delta}= 2 C^2-E+\frac13\,R^2, \qquad \ \
R^{\alpha\beta}R_{\alpha\beta}=\frac12\, (C^2-E)+\frac13\,R^2\,.
\end{equation}
From the density  $E$  one defines the  Gauss-Bonnet term,
\begin{equation}\label{eq:GaussBonnet}
 \GB= \int d^nx  \sqrt{-g} E\,,
\end{equation}
which is a topological invariant in $n=4$ (not so in other dimensions). Such a topological invariance implies that the metric functional variation of $\GB$ vanishes identically in four dimensions:
\begin{equation}\label{eq:Topological}
  \frac{\delta \GB}{\delta g^{\mu\nu}}=0\ \ \ \ \ \ \ (n=4)\,.
\end{equation}

From the basic HD terms one may construct the higher derivative (HD) part of the vacuum action, henceforth  $n=4$:
\begin{equation}\label{eq:HDaction}
S_{\rm HD}= \int d^4x  \sqrt{-g} \left (\alpha_1 C^2+\alpha_2  R^2+\alpha_3  E+\alpha_4 \Box R\right)\equiv \int d^4x  \sqrt{-g} L_{\rm HD} \,.
\end{equation}
The purely geometric terms in \eqref{eq:HDaction} are generated by the quantum matter contributions, and hence these HD terms are necessary for the renormalization procedure.  The bare couplings $\alpha_i$ become renormalized couplings  $\alpha_i$(M) which run with the renormalization scale $M$.
That HD gravitational action (with effective Lagrangian $  L_{\rm HD}$) is to be added to the EH action plus matter, cf. Eq.\,\eqref{eq:EH}, in order to have a well-defined and renormalizable  semiclassical theory of quantum fields  in curved spacetime.  The total action of gravity plus matter therefore reads
\begin{equation}\label{eq:Totalaction}
S_{\rm tot}= S_{\rm EH}+S_{\rm HD}+S_{\rm m} \,.
\end{equation}
In it, the total vacuum action is the sum of the first two pieces, whereas  the last piece is the matter action.
By functionally differentiating the $R^2$ and $R_{\alpha\beta}R^{\alpha\beta}$ terms with respect to the metric, we obtain two (conserved)  higher order curvature tensors  (of adiabatic order $4$), namely
\begin{equation}\label{eq:H1munu}
 \leftidx{^{(1)}}{\!H}_{\mu\nu}=\frac{1}{\sqrt{-g}}  \frac{\delta}{\delta g^{\mu\nu}}  \int d^4x  \sqrt{-g} R^2=-2\nabla_\mu\nabla_\nu R + 2 g_{\mu\nu} \Box R - \frac12 g_{\mu\nu} R^2 +2 R R_{\mu\nu}
\end{equation}
and
\begin{equation}
\begin{split}
\leftidx{^{(2)}}{\!H}_{\mu\nu}&=\frac{1}{\sqrt{-g}}\frac{\delta}{\delta g^{\mu\nu}}\int dx^4 \sqrt{-g}R_{\alpha\beta}R^{\alpha\beta}\\
&=2{R^\alpha}_\mu R_{\alpha \nu}-2g_{\mu\beta}\nabla_\alpha\nabla_\nu R^{\alpha\beta}+\Box R_{\mu\nu}+\frac{1}{2}g_{\mu\nu}\Box R-\frac{1}{2}g_{\mu\nu}R^{\alpha\beta}R_{\alpha \beta}\,.
\end{split}
\end{equation}
One can also define $H_{\mu\nu}=\frac{1}{\sqrt{-g}}\frac{\delta}{\delta g^{\mu\nu}}\int dx^4 \sqrt{-g}R^{\alpha\beta\gamma\delta}R_{\alpha\beta\gamma\delta}$. However, because of the topological property \eqref{eq:Topological} in $n=4$, one can easily show that the new  HD tensor can be written in terms of the previously defined ones as follows:  $H_{\mu\nu}=4\leftidx{^{(2)}}{\!H}_{\mu\nu}-\leftidx{^{(1)}}{\!H}_{\mu\nu}$. Using this property to compute  the functional derivative of the Weyl tensor squared defined in \eqref{eq:EC2} we find
\begin{equation}\label{eq:deltaC2}
\frac{1}{\sqrt{-g}}\frac{\delta C^2}{\delta g^{\mu\nu}}=H_{\mu\nu}-2\leftidx{^{(2)}}{\!H}_{\mu\nu}+\frac13 \leftidx{^{(1)}}{\!H}_{\mu\nu}= 2 \leftidx{^{(2)}}{\!H}_{\mu\nu}-\frac23 \leftidx{^{(1)}}{\!H}_{\mu\nu}\,.
\end{equation}
The previous relation implies that for conformally flat spacetimes (like FLRW), for which the Weyl tensor vanishes identically,  the basic two HD tensors $ \leftidx{^{(2)}}{\!H}_{\mu\nu}$ and $ \leftidx{^{(1)}}{\!H}_{\mu\nu}$ are not independent:
\begin{equation}\label{eq:ConfFlat}
\leftidx{^{(2)}}{\!H}_{\mu\nu}=\frac13\,\leftidx{^{(1)}}{\!H}_{\mu\nu}\,.
\end{equation}
We  remark that the two HD tensors $,\leftidx{^{(1)}}{\!H}_{\mu\nu}$ and $,\leftidx{^{(2)}}{\!H}_{\mu\nu}$ are conserved tensors,  namely they satisfy the local  conservation laws
\begin{equation}\label{eq:conservH1H2}
\nabla^\mu\, \leftidx{^{(1)}}{\!H}_{\mu\nu}=0\,, \ \ \ \ \ \ \ \ \ \  \ \ \ \ \  \nabla^\mu\, \leftidx{^{(2)}}{\!H}_{\mu\nu}=0\,.
\end{equation}
These laws are fulfilled identically and  independently of each other,  even if the background geometry is non-conformally flat and the relation \eqref{eq:ConfFlat} is not satisfied. This should not be surprising  for the following reason.  Tensors $\leftidx{^{(1)}}{\!H}_{\mu\nu}$ and $\leftidx{^{(2)}}{\!H}_{\mu\nu}$ represent the most general  modification of the \textit{l.h.s.} of Einstein's equations in the presence of HD terms. In fact, the metric variation of the total action \eqref{eq:Totalaction} produces the generalized Einstein's equations:
\begin{equation}\label{eq:MostGenEinstEqs}
 G_{\mu \nu}+b_1(M) \leftidx{^{(1)}}{\!H}_{\mu\nu}+ b_2(M) \leftidx{^{(2)}}{\!H}_{\mu\nu}=8\pi G(M) \langle T^{\rm tot}_{\mu\nu}\rangle_{\rm ren}(M)\,.
\end{equation}
One would expects that  $\leftidx{^{(1)}}{\!H}_{\mu\nu}$ and $\leftidx{^{(2)}}{\!H}_{\mu\nu}$ should not perturb the consistency between the Bianchi identity $\nabla^{\mu}G_{\mu\nu}=0$ satisfied by the Einstein tensor and the local conservation law $\nabla^{\mu}T^{\rm tot}_{\mu\nu}=0$  (where the EMT  $T^{\rm tot}_{\mu\nu}$ involves all forms of energy, matter and vacuum, whether interacting or not).  One can  verify, of course, by explicit  calculations from the above definitions  that the two local conservation laws \eqref{eq:conservH1H2}  are indeed  satisfied. This fact insures that acting with  $\nabla^\mu$ on both sides of \eqref{eq:MostGenEinstEqs} gives consistently zero. For the explicit derivation of the relations \eqref{eq:conservH1H2}, the following standard relation can be used:
\begin{equation}\label{eq:CommutationNablas}
 \left(\nabla_\nu \nabla_\mu-\nabla_\mu \nabla_\nu\right)v_\alpha =R^\sigma_{\textrm{	}\alpha \mu \nu}v_\sigma\,,
\end{equation}
which holds for any covariant vector field $v_\alpha$. In particular, for $v_\alpha=\nabla_\alpha\phi$ we find
\begin{equation}\label{eq:CommutationNablas2}
\left(\nabla_\nu\nabla_\mu-\nabla_\mu\nabla_\nu\right)\nabla_\alpha \phi=R^\sigma_{\
\alpha\mu\nu}\nabla_\sigma \phi\,.
\end{equation}
It shows that in curved spacetime the successive action of three $\nabla_\mu$ operators cannot be performed by commuting the last two being applied, while of course $\nabla_\nu\nabla_\mu\phi=\nabla_\mu\nabla_\nu\phi$ (because the Christoffel symbols are symmetric).  The relation \eqref{eq:CommutationNablas2} can be used to  derive the rule for commuting the  nabla and box operators, which we need as well in the text:
\begin{equation}\label{eq:CommutationNablaBox}
\nabla_\mu \Box \phi-\Box\nabla_\mu \phi=-R_{\mu\nu}\nabla^\nu \phi\,.
\end{equation}

Additional formulas which are used in the main text  involving the above HD tensors  in the specific context of the FLRW metric are the following.
The   $00th$ and $11th$-components of the $\leftidx{^{(1)}}{\!H}_{\mu\nu}$ tensor  in the conformally flat metric reads
\begin{equation}\label{eq:H100}
\leftidx{^{(1)}}{\!H}_{00}=\frac{-18}{a^2}\left(\mathcal{H}^{\prime 2}-2\mathcal{H}^{\prime \prime}\mathcal{H}+3 \mathcal{H}^4 \right)= -18 a^2\left(\dot{H}^2-2H\ddot{H}-6H^2\dot{H}\right)\,,
\end{equation}
\begin{equation}\label{eq:H111}
\leftidx{^{(1)}}{\!H}_{11}=-a ^2\left(108  H^2 \dot{H}+54\dot{H}^2+72\dot{H}\ddot{H}+12\vardot{3}{H}\right)\,.
\end{equation}
We also need
\begin{equation}\label{eq:RmunusquareBoxR}
 R^{\mu\nu} R_{\mu\nu}=\frac{12}{a^4}\left({\cH^\prime}^2+\cH^\prime\cH^2+\cH^4\right)\,,\qquad\Box R=-\frac{6}{a^4}\left(\cH^{\prime\prime\prime}-6\cH'\cH^2\right)\,.
\end{equation}
As warned, these formulas assume vanishing three-dimensional curvature.

\subsection{Master integral}\label{sec:appendixA2}
Integrals over $3$-dimensional momentum appear quite often in our calculations. For our purposes it will suffice to  focus on integrals of the form
\begin{equation}\label{eq:MasterIntegral}
I_3(p,Q)\equiv\int \frac{d^3 k}{(2\pi)^3}\frac{1}{\omega^p_k(Q)}=\frac{1}{2\pi^2}\int dk k^2 \frac{1}{\omega^p_k(Q)}=\frac{1}{2\pi^2}\int dk k^2\frac{1}{(k^2+Q^2)^{p/2}}\,,
\end{equation}
where  $k\equiv|\bk|$, $\omega_k(Q)=\sqrt{k^2+Q^2}$ and $Q$ is an arbitrary scale.  In $n-1$ spatial dimensions,
\begin{equation}
\begin{split}
I_{n-1}(p,Q)&\equiv \int \frac{\mu^{3-(n-1)}d^{n-1} k}{(2\pi)^{(n-1)}}\frac{1}{(k^2+Q^2)^{p/2}}=\frac{\mu^{3-(n-1)}}{(4\pi)^{(n-1)/2}}\frac{\Gamma\left(\frac{p-(n-1)}{2} \right)}{\Gamma\left( \frac{p}{2}\right)}\left(Q^2\right)^{\frac{(n-1)-p}{2}}\\
&=\frac{1}{(4\pi)^{3/2}}\frac{\Gamma\left(\frac{p-3}{2} +\epsilon\right)}{\Gamma\left( \frac{p}{2}\right)}\left(Q^2\right)^{\frac{3-p}{2}}\left(\frac{Q^2}{4\pi \mu^2}\right)^{-\epsilon}\,. \label{DRFormula}
\end{split}
\end{equation}
Here  $\Gamma(x)$ is Euler's $\Gamma$ function, which satisfies the functional relation  $\Gamma(x+1)= x\,\Gamma (x)$. The scale  $\mu$ (with natural dimension one) has been  introduced such that the new  integration measure  $d^{n-1} k\to \mu^{2\epsilon} d^{n-1} k$  has the same dimension as $d^3k$, where  $\epsilon\equiv\frac {3-(n-1)}{2}=\frac {4-n}{2}$. Of course, the limit  $\epsilon\to 0$ (i.e. $n-1\to 3$)  at the end of the calculation is understood.  Such a  limit is trivial for $p> 3$,  but not so for $p\leq 3$  since in the last case poles $\sim\frac{1}{\epsilon}$ appear  in the result of (\ref{DRFormula}),  which can be used to regularize the UV-divergent terms appearing in many of the integrals appearing in our calculation, see e.g.  Eq.\eqref{EMTFluctuations}. The limit $\epsilon\to 0$  also generates finite parts which must be carefully included. Despite the fact that the adiabatic subtraction procedure   provides overall UV-convergent integrals, as explained in detail in the main text,  one can also use dimensionally regularized integrals to track the poles found in intermediate results.   The following properties of the $\Gamma$ function are useful:
\begin{equation}\label{eq:Gammaepsilon}
\Gamma(\epsilon)=\frac{1}{\epsilon} - \gamma_E  +{\cal O}(\epsilon)\,,\ \ \ \ \  \Gamma(-1+\epsilon)=-\frac{1}{\epsilon} - 1+\gamma_E  +{\cal O}(\epsilon)\,,
\end{equation}
where $\gamma_E$ is Euler's constant.  Using the  functional definition of $\Gamma$ mentioned above, one can easily extend these formulas  to parameterize the divergent behavior of $\Gamma$ around any negative integer.

The following observation is in order at this point. It is important to clarify  that, in our renormalization scheme, the auxiliary  't Hooft's mass unit $\mu$ used in the above formulae plays no role and cancels out completely at the level of the final results. This is so in all the computations presented in this work. The appearance of $\mu$ in intermediate steps is related to have used (optionally)  dimensional regularization  in some parts of our calculation.  Use of DR, however,  is not essential at all and  it can be totally circumvented. This was shown e.g. in the calculations given in Appendix B of  Ref.\,\cite{CristianJoan2020}, where the regularization of the EMT was performed using DR after the results had already been obtained using the subtraction prescription in the main text.   Similarly,  use of DR in the effective action approach of Sec.\,\ref{HeatKernel} is only for convenience, we have rederived the same results using the scale subtraction procedure, i.e.  the one we have employed in  Sections \ref{sec:RenormEMT} and \ref{sec:RenormalizedVED} of the actual study.  We do not deem necessary to provide more details here after we have already  illustrated the perfect matching of the final results using the two alternative methods for the renormalization of the EMT\,\,\cite{CristianJoan2020}.  Use of one procedure over the other  is ultimately a matter of convenience.  We emphasize, however,  that we did not use  the MS scheme  of renormalization  at any point in our study of the VED, although this is of course independent of using DR as an intermediate regularization technique, if desired.  In contrast, the subtracting scale $M$ remains always in our results as it is inherent to our renormalization method, no matter whether we decide to use DR in intermediate steps  for regularization or just proceed to rearrange the terms of the integrands of our subtracted integrals to show by explicit calculation that the result is overall convergent. See e.g. the current Appendix \ref{sec:appendixC} for another example.

\section{Running  vacuum and gravitational coupling in the RVM}\label{sec:appendixAbis}

In this appendix,  we provide calculational details on the formulas for the running vacuum and gravitational coupling introduced in the main text,  and their interrelationship. Recall that $ \rv(M)$ is an abridged notation for  $\rv=\rv(M,H, \dot{H},\ddot{H},...)$, i.e.  the vacuum energy density (VED), which is a function not only of the scale  $M$ but also of the Hubble rate and its time derivatives.   The value of  $H=H(t)$ defines an expansion history time $t$. When compared with our current cosmic time, $t_0$, the difference $t_0-t$ defines our lookback time to the events occurring around the expansion history epoch $H$.  It is advisable to make the original shorthanded notation  a bit more explicit for the kind of discussion in  this appendix. Rather than denoting  the renormalized value of the  VED  at the scale $M$  for a fixed expansion rate $H$ (and corresponding time derivatives) by just $\rv(M)$ ,  we will use  $\rv(M,H)$. The second argument denotes generically all the dependency in  $H,\dot{H},\ddot{H},\dots$  The values of $M$ and $H$ are independent, of course, but a selected choice of the renormalization point  $M$ near $H$ corresponds to choose the RG scale around the characteristic energy scale of  FLRW spacetime at a given moment, and hence it should have more physical significance.  This is actually in analogy with the standard practice in ordinary gauge theories, where the choice of  RG scale  is usually  made near the typical energy of the process. For the FLRW universe, the natural choice for the process of expansion is  $M=H$ and we will see it is consistent.  In what follows we derive the `low energy'  form  of the VED  along these lines.  Subsequently we will focus on the running gravitational coupling $G(M)$ and its relation with the running $\rv(M)$.

\subsection{Running VED}\label{sec:appendixAbis1}

The expression for  VED  at the scale $M$  for a given expansion history time $H$   is provided  by our  renormalization procedure and it is  given by Eq.\,\eqref{RenVDEexplicit}.  This expression contains the contributions from all the possible adiabatic orders up to the limit of the asymptotic expansion. Suppose, however, that  we consider  the renormalized VED  at a given expansion history time $H$ for different values of the renormalization scale, say $M$ and $M_0$.  The difference of  renormalized VED values at these scales at a fixed $H$  can be computed in an exact way,  see  Eq.\,\eqref{eq:VEDscalesMandM0Final}.  The exactness  of such formula stems from the fact that the renormalization scale dependence of the EMT (i.e. the $M$-dependence) can only be carried by the terms that are originally divergent (those up to $4th$ adiabatic order). The renormalized  EMT at the scales $M$ and $M_0$ at fixed cosmic time (hence at fixed $H$)  is obtained upon subtracting  the corresponding on-shell value at these respective scales, as explained in Sec.\ref{sec:RenormEMT}. Therefore, the difference of renormalized VED values at $M$ and $M_0$ is free from all of the finite contributions from $6th$ adiabatic order and higher.  Only ${\cal O}\left(H^2\right)$  and ${\cal O}\left(H^4\right)$ (i.e. second and fourth adiabatic orders, respectively)  remain, as it is manifest in  Eq.\,\eqref{eq:VEDscalesMandM0Final}.  However, despite of the fact that such result is exact, we wish to focus on lookback times accessible to observations, hence with  values of $H$ which are moderate enough for the ${\cal O}\left(H^4\right)$ terms to be negligible.  The desired difference between  $\rv(M,H)$ and $\rv(M_0,H)$ within  our lookback observational range therefore reads
\begin{equation}\label{RenVDECurrentUniv}
\begin{split}
&\rv(M,H)-\rv(M_0,H)= \rho_\Lambda (M)- \rho_\Lambda (M_0)\\
&+\frac{1}{128\pi^2 }\left(-M^4+M_0^4+4m^2(M^2-M_0^2)-2m^4 \ln \frac{M^2}{M_0^2}\right)\\
&+\left(\xi-\frac{1}{6}\right)\frac{3 \mathcal{H}^2 }{16 \pi^2 a^2}\left(M^2-M_0^2-m^2\ln \frac{M^2}{M_0^2} \right)+\cdots\\
&=  \left(\xi-\frac{1}{6}\right)\frac{3 {H}^2 }{16 \pi^2}\left(M^2-M_0^2-m^2\ln \frac{M^2}{M_0^2} \right)+\cdots
\end{split}
\end{equation}
where the dots denote the kind of neglected contributions mentioned above.  As we know, the first two terms in the above formula  cancel  against each other thanks to the relation \eqref{SubtractionrL}. The obtained result is given, of course, by the ${\cal O}\left(H^2\right)$  part of {Eq.\,\eqref{eq:VEDscalesMandM0Final}.

Similarly,  from Eq.\,\eqref{RenVDEexplicit} we may also find the difference between the values of the VED  corresponding  to two accessible lookback times for  a given renormalization point $M$:
\begin{equation}\label{DiffHH0Mfix}
\begin{split}
\rv(M,H)-\rv(M,H_0)=3\frac{\left(\xi-\frac{1}{6}\right)}{16\pi^2}(H^2-H_0^2)\left(M^2-m^2+m^2\ln \frac{m^2}{M^2}\right)+\cdots
\end{split}
\end{equation}
In this expression we have disregarded not only the ${\cal O}\left(H^4\right)$  terms but also the higher order ones (which are indeed present here,  in contrast to Eq.\,\eqref{RenVDECurrentUniv}), as they entail no significant contribution at present. They can be important  only for the early universe, e.g. during the inflationary regime, see Sec.\,\ref{sec:RVMInflation}.

We may as well compute the scaling evolution of the VED when we change both the cosmic times and the renormalization points.  Keeping our focus on cosmic epochs $H$ and $H_0$ accessible to our observations,  the result  immediately  follows from Eq.\,\eqref{RenVDEexplicit} upon  neglecting the ${\cal O}\left(H^4\right)$ terms and higher:
\begin{equation}\label{DiffHH0MM0}
\begin{split}
\rv(M,H)-\rv(M_0,H_0)&
=\frac{3\left(\xi-\frac{1}{6}\right)}{16\pi^2}\left[H^2\left(M^2-m^2+m^2\ln\frac{m^2}{M^2}\right)\right.\\
&\left.-H_0^2\left(M_0^2-m^2+m^2\ln\frac{m^2}{M_0^2}\right)\right]+\cdots\,,
\end{split}
\end{equation}
where again the first two terms on the \textit{r.h.s.} of Eq.\,\eqref{RenVDECurrentUniv} are involved here,  but cancel each other for the aforementioned reasons.
Finally, let us consider what should be the physical (measurable) difference between the VED values at different epochs of the cosmic evolution within our observational range.  According to our prescription,  choosing the renormalization point  $M$ near $H$  (and hence bringing  the RG scale near the characteristic energy scale  of the FLRW spacetime at the  given epoch)  ought to  be the most suited physical choice in consonance with the usual practice based on selecting the RG scale choice near the typical energy of the process in particle physics.   As indicated, in our case  the `process'  is nothing but the cosmic expansion of the universe at a given epoch.  Thus, to  compute the scaling evolution of the VED in the span mediating in between the two cosmic epochs  $H$ and $H_0$,  follows directly from Eq.\,\eqref{DiffHH0MM0}  upon  picking out the renormalization points $M$ and $M_0$  at precisely the values of the Hubble rate in those epochs, respectively: $M=H$ and $M_0=H_0$.  Defining for convenience  $\rv(H)\equiv\rv(M=H,H)$  and similarly  $\rv(H_0)\equiv \rv(H_0,H_0)$,  and neglecting as always  the higher order terms ${\cal O}\left(H^4\right)$ , we find
\begin{equation}\label{DiffVEDphys}
\begin{split}
\rv(H)-\rv(H_0)&=\frac{3\left(\xi-\frac{1}{6}\right)}{16\pi^2}\left[H^2\left(H^2-m^2+m^2\ln\frac{m^2}{H^2}\right)-H_0^2\left(H_0^2-m^2+m^2\ln\frac{m^2}{H_0^2}\right)\right]+\cdots\\
\simeq& \frac{3\left(\xi-\frac16\right)m^2 }{16\pi^2}\left[-\left(H^2-H_0^2\right)+H^2\ln\frac{m^2}{H^2}-H_0^2\ln\frac{m^2}{H_0^2}\right]\\
&=\frac{3\left(\xi-\frac16\right)m^2 }{16\pi^2}\left[-1+\ln \frac{m^2}{H^2}-\frac{H_0^2}{H^2-H_0^2}\ln \frac{H^2}{H_0^2}\right] \left(H^2-H_0^2\right)\,.
\end{split}
\end{equation}
The previous formula shows that there is in effect  a `running'  or change of the VED from $H_0$ to $H$.  Notice that if $m$ is an ordinary particle mass (e.g. within the standard model of particle physics) the running would be very small. Suppose, however, that $m$ is a particle mass near some GUT scale, then  it is natural to measure its value in units of the Planck mass $\mpl$ and factor out the ratio $m/\mpl$.  We do this in defining the effective running parameter
\begin{equation}\label{eq:nueff2}
\nueff(H)\equiv\frac{1}{2\pi}\,\left(\xi-\frac16\right)\,\frac{m^2}{\mpl^2}\left(-1+\ln \frac{m^2}{H^{2}}-\frac{H_0^2}{H^2-H_0^2}\ln \frac{H^2}{H_0^2}\right)\,.
\end{equation}
The running VED  formula \eqref{DiffVEDphys} can now be written in a rather compact form as follows:
\begin{equation}\label{eq:RVMcanonical}
\rv(H)\simeq \rvo+\frac{3\nueff(H)}{8\pi}\,(H^2-H_0^2)\,\mpl^2=\rvo+\frac{3\nueff(H)}{8\pi G_N}\,(H^2-H_0^2)\,,
\end{equation}
where $\rv(H_0)$ is identified with today's VED value, $\rvo$, and $G_N$ is assumed to be the currently  measured value of the gravitational constant.
As a matter of fact, $\nueff(H)$ in \eqref{eq:nueff2} is not a parameter, of course, since it is a function of $H$. However, it varies very slowly with the Hubble rate. The last term of \eqref{eq:nueff2} is logarithmic and becomes quickly suppressed for increasingly large values of $H$ above $H_0$, whereas the second term  furnishes (on account of  $\ln \frac{m^{2}}{H^2}\gg1$) the dominant contribution  to the effective running parameter:
\begin{equation}\label{eq:nueffAprox2}
\nueff(H)\simeq\frac{1}{2\pi}\,\left(\xi-\frac{1}{6}\right)\,\frac{m^2}{\mpl^2}\ln\frac{m^2}{H^2}\,.
\end{equation}
In the approximation $H=H_0$  (valid to within a  few percent in the accessible part of the expansion history for large $m$), it  just renders  Eq.\,\eqref{eq:nueffAprox} in the main text, and Eq.\,\eqref{eq:RVMcanonical} is nothing but  the canonical form of the VED for the running vacuum model (RVM), as given in the main text in Eq.\,\eqref{eq:RVM2}.  In it, the running parameter  is treated essentially as a constant.  In actual fact,  for a large stretch of the recent universe we can just set $H=H_0$ in Eq.\,\eqref{eq:nueffAprox2} since it  differs  less than $7\%$  through the entire period from now up to the decoupling time. Even if $\xi$ is not known,  the ratio $m^2/\mpl^2\ll1$ in the prefactor  of \eqref{eq:nueff2} is  very small, so we can expect that  $\nueff$ is essentially a tiny quantity with a very mild variation with $H$.  From the foregoing,  it follows that it can be treated  to a good approximation  as a small parameter within the observable universe.  Notice, however,  that for   $m$ large, say of  order of a GUT scale $M_X\sim 10^{16}$ GeV, we have $m^2/\mpl^2\sim 10^{-6}$,  which is not hopelessly small. Since $\xi$ is,  in principle, arbitrary  and we have in general  a large multiplicity of heavy scalar particles in a typical  GUT, the effective value of $\nueff$ can not be excluded to be in the small  but sizeable  range $10^{-4}-10^{-3}$\,\cite{Fossil2008}. This theoretical expectation  is actually corroborated by the phenomenological analysis. The RVM has  been fitted to the data  and the obtained results for $\nueff $ lie in  expected ballpark of $\sim 10^{-3}$,   see e.g. \cite{RVMpheno1,RVMpheno2}  and  \cite{EPL2021}.

\subsection{Time versus scaling evolution of the VED}\label{sec:appendixAbisbis1}

To further illustrate the meaning and consistency of the above formulas for the running VED, it is interesting to compare  the scaling versus  time evolution laws  of the VED at the differential level.  The former is, of course, determined by the $\beta$-function \eqref{eq:RGEVED1} of the VED,  whereas the latter can be computed as follows.  For two given expansion epochs $H$ and $H_0$,  the time evolution  is  determined by Eq.\,\eqref{DiffVEDphys}, or equivalently by  \eqref{eq:nueff2} and \eqref{eq:RVMcanonical}. However, we would like this result for an infinitesimal change of $H$ around $H=H_0$,  which means to compute the derivative of $\rv(H)$ with respect to $H$ at $H=H_0$, or,  for convenience,  the logarithmic derivative $d\rv(H)/d\ln H=H d\rv(H)/dH$.  The result follows from Eq.\,\eqref{DiffVEDphys}.
We find
\begin{equation}\label{eq:DiffLimiting}
\begin{split}
H_0\frac{d\rv(H_0)}{dH}=
\left(\xi-\frac{1}{6}\right)\frac{3 H_0^2 m^2}{8 \pi^2 }\left(-2+\ln\frac{m^2}{H_0^2}\right)\,.
\end{split}
\end{equation}
This equation can be written in an approximate way as follows:
\begin{equation}\label{eq:ApproxDiffLimiting}
\begin{split}
H_0\frac{d\rv(H_0)}{dH}\simeq \left(\xi-\frac{1}{6}\right)\frac{3 H_0^2 m^2}{8 \pi^2 }\ln\frac{m^2}{H_0^2}=\frac{3\nueff }{4\pi} \,\mpl^2 H_0^2\,.
\end{split}
\end{equation}
where in the last step we used  $\ln\frac{m^2}{H^2}\gg 1$ and took the approximate expression \eqref{eq:nueffAprox} for the coefficient $\nueff$.
It is also instructive to obtain the same result using the chain rule to compute the total derivative with respect to $M$ and set  $M=H_0$ at the end of the calculation, as in this way the role of the $\beta$-function for the VED, $\beta_{\rv}$,  becomes manifest:
\begin{equation}\label{eq:DerivTotal}
\begin{split}
M\frac{d\rv(M,H)}{dM}=&M\left(\frac{\partial\rv}{\partial M}+\frac{\partial\rv}{\partial H}\frac{\partial H}{\partial M}\right)=\beta_{\rv}+M\frac{\partial\rv}{\partial H}\frac{\partial H}{\partial M}\,,
\end{split}
\end{equation}
or, more explicitly, for $H=H_0$:
\begin{equation}\label{eq:DerivTotal2}
\begin{split}
M\frac{d\rv(M,H_0)}{dM}=&\left(\xi-\frac{1}{6}\right)\frac{3 H_0^2 }{8 \pi^2 }\left(M^2-m^2\right)+M\left(\xi-\frac16\right)\frac{3H_0}{8\pi^2}\left(M^2-m^2+m^2\ln\frac{m^2}{M^2}\right)\frac{\partial H_0}{\partial M}\, ,
\end{split}
\end{equation}
where both for  the VED expression \eqref{RenVDEexplicit} and for  its $\beta$-function \eqref{eq:RGEVED1}  we used the relevant ${\cal O}(H^2)$ terms only, and of course we borrowed the $\beta$-function for the renormalized coupling $\rL(M)$,  as given in Eq.\,\eqref{eq:BetaFunctionrL}. Setting at this point  $M=H_0$ and dropping the terms higher than  ${\cal O}(H^2)$ we strike once more  Eq.\,\eqref{eq:DiffLimiting}.  From the latter we can see that the sign of the total variation of the VED is given by the sign of $\xi-\frac16$ and hence also by the sign of $\nueff$ , see  \eqref{eq:ApproxDiffLimiting}. This is in full accordance with the result \eqref{eq:RVM2}. In particular, we can see from the last derivation that the total evolution of the VED is dominated by the variation of $\rv$ with $H$ rather than with the scale $M$ (before setting $M=H$). When we set $M=H$ at the end,  the dominant term is that one carrying  $\partial\rv/\partial H$ in \eqref{eq:DerivTotal}, whose sign is the same as that of $\nueff$.  It is therefore the total, rather than just the partial, derivative with respect to $M$ what matters for the study of the physical evolution of the VED. This is a consequence of the time dependence of $M$ in cosmology, in contrast to the situation in ordinary gauge theories.  Writing the leading term of Eq.\,\eqref{eq:RGEVED1}   at low energies as
\begin{equation}\label{eq:BetaFuncoeff}
\begin{split}
\beta_{\rv}=\frac{3\,\bv}{4\pi}\, H_0^2 m^2\,\ \ \ \ \ \ \ \ \ \ \ \ \ \  \bv\equiv -\frac{1}{2\pi}\left(\xi-\frac16\right)\,,
\end{split}
\end{equation}
we find from \eqref{eq:nueffAprox2} the approximate formula relating $\nueff$ with  the coefficient $ \bv$ of the $\beta$-function for the running VED:
\begin{equation}\label{eq:DTotalversusBeta}
\nueff=- \bv\,\frac{m^2}{\mpl^2}\,\ln\frac{m^2}{H_0^2}\,.
\end{equation}
The sign between $\nueff$ and $\bv$ reflects the mentioned antagonism between  the variations of the VED with $H$ and with $M$  before the latter  is set equal to the former.  Recall that $H_0$ here may be the current value of the Hubble rate, but for that matter it can represent any point of the expansion history at low energies. As noted in the previous section,  $\nueff$ remains essentially constant since the change of  $\ln\frac{m^2}{H^2}$ is bound within a few percent from now until recombination.

Finally, it is also instructive to derive once more the above result using a third alternative procedure, as in this way we may crosscheck our formulas from different perspectives. In particular, let us remember that Eq.\,\eqref{eq:NonConserVED1}  provides direct and precious information about  the time evolution of the VED and it involves the influence from  the vacuum pressure, which as we know is not exactly equal to minus the vacuum density in this QFT framework (i.e. the EoS of the quantum vacuum is not exactly equal to $-1$, see Sec.\,\ref{sec:EoSvacuum}). We want to show that we  can test the consistency of this feature as well.   Upon inserting \eqref{eq:EoSaprox} into the term $\rv+\Pv$  of Eq.\,\eqref{eq:NonConserVED2} we find
\begin{equation}\label{eq:ThirdTest}
\begin{split}
\dot{\rho}_{\rm vac}=\frac{\dot{M}}{M}\beta_{\rho_{vac}}-3H(\rho_{vac}+P_{vac})=\frac{\dot{M}}{M}\beta_{\rho_{vac}}-3H\frac{\xi-\frac16}{8\pi^2}\dot{H}m^2\left(1-\ln\frac{m^2}{H^2}\right)+\cdots
\end{split}
\end{equation}
where the dots  just denote that we are not considering terms  higher than  ${\cal O}(H^2)$, which is our usual assumption for the present considerations.  Using once more the  $\beta$-function \eqref{eq:RGEVED1} of the VED to the same consistent order and setting $M=H_0$, we find
\begin{equation}\label{eq:ThirdTest2}
\begin{split}
\dot{\rho}_{\rm vac}(t_0)=\left(\xi-\frac16\right)\frac{3m^2}{8\pi^2}H_0\dot{H_0}\left(-2+\ln\frac{m^2}{H_0^2}\right)\,.
\end{split}
\end{equation}
This equation gives the rate of change of the VED at $t=t_0$ (corresponding to  $H=H_0$), which may refer to the present time or for that matter  any other cosmic time.  For  $\nueff>0$ (respectively for  $\nueff<0$)  and taking into account that  $\dot{H}<0$ at all (post-inflationary) times,  together with the fact that  $\ln \frac{m^{2}}{H^2}\gg1$,  it is not difficult to see that the VED increases (resp. decreases) towards the past and decreases (resp. increases)  towards the future.  We can actually show that the above equation coincides with Eq.\,\eqref{eq:DiffLimiting}.  Indeed, bearing in mind  that $\dot{\rho}_{\rm vac}(t)=\dot{H}\frac{d\rv}{d H}$ we find
\begin{equation}\label{eq:ThirdTest3}
\begin{split}
H \frac{d\rv}{d H}=\frac{H}{\dot{H}} \dot{\rho}_{\rm vac}(t)=\left(\xi-\frac{1}{6}\right)\frac{3 H^2 m^2}{8 \pi^2 }\left(-2+\ln\frac{m^2}{H_0^2}\right)\,,
\end{split}
\end{equation}
which for $t=t_0$ exactly matches Eq.\,\eqref{eq:DiffLimiting}  (q.e.d.)  Thus the three approaches do converge to the same result, which can be summarized as follows:  for $\nueff>0$ the VED is larger in the past and behaves effectively as  quintessence, whereas for $\nueff<0$ the VED is smaller in the past (equivalently, it increases towards the future)  and hence it behaves effectively as phantom DE.   This is exactly the same message encoded in Eq.\,\eqref{eq:RVM2}.  The fact that  the quantum vacuum can mimic both quintessence and phantom DE shows that it may not be necessary to introduce \textit{ad hoc} scalar fields in the classical action to generate dynamical DE, since the latter could just be caused by the fact that the quantum vacuum is in permanent cosmic evolution!

\subsection{Running $G$}\label{sec:appendixAbis2}
Here we elaborate further on the derivation of Eq.\,\eqref{eq:GNHfinal}  in the main text.  We take up the discussion from Eq.\,\eqref{MixedConservationApprox1}, in which we have disregarded the HD contributions present in Eq.\,\eqref{MixedConservation}, as they are negligible for  the present universe (and for that matter, also at all times after inflation). We can admit the concurrence of relativistic and  nonrelativistic ordinary  matter components  $(\rho_m,p_m)$ apart from the background scalar field $(\rho_\phi,p_\phi)$. If the former are  locally conserved ($\dot{\rho}_m+3H(\rho_m+p_m)=0$) it is not difficult to see that the structure of \eqref{MixedConservationApprox1} remains unaltered:
\begin{equation}\label{ModifiedConsLaw}
\dot{G}(M)\left(\rho_m+\rho_\phi+\rv(H)\right)+G(H)\dot{\rho}_{\rm vac}(H)+3HG(H)\left(\rv(H)+P_{\rm vac}(H)\right)=0\,,
\end{equation}
where we have set $M=H$, according to the prescription discussed in the previous section.
Using the Friedmann equation we can get rid of the total energy density in the above expression no matter the number of components involved in it:  $\rho_m+\rho_\phi+\rv=3H^2/(8\pi G)$.  In addition,  we insert the expression $\rv(H)$ from \eqref{eq:RVM2} in the above equation,  and also $\rv(H)+P_{\rm vac}(H)$ from \eqref{eq:EoSaprox}.  As always we neglect of course the higher order terms ${\cal O}\left(H^4\right)$  generated in intermediate calculations, which are irrelevant after the inflationary epoch.  All in all, Eq.\,\eqref{ModifiedConsLaw} can be rewritten

\begin{equation}\label{ModifiedConsLaw2}
\begin{split}
& \frac{3H^2 \dot{G}}{8\pi G}+3 G\frac{d}{dt}\left[ \rvo+\frac{1}{\kappa^2}\nueff (H)(H^2-H_0^2)\right]+3HG\frac{\left(\xi-\frac{1}{6}\right)}{8\pi^2}\dot{H}m^2\left(1-\ln \frac{m^2}{H^2}\right)\\
&=\frac{3H^2 \dot{G}}{8\pi G}+3\frac{G}{\kappa^2}\left(\dot{\nu}_{\rm eff}(H) (H^2-H_0^2)+2H\dot{H}\nueff (H)+H\frac{\left(\xi-\frac{1}{6}\right)}{\pi}\dot{H}\frac{m^2}{m^2_{\rm Pl}}\left(1-\ln \frac{m^2}{H^2}\right)\right)=0\,,
\end{split}
\end{equation}
where we recall that $\kappa^2= 8\pi G_N$  is constant, whereas $G=G(H(t))$  is the function that we wish to determine by solving the above differential equation.
Notice that to compute $\dot{\nu}_{\rm eff}(H)=\frac{d}{dt}{\nu}_{\rm eff}(H)$ we use the exact expression \eqref{eq:nueff2}  rather than just a constant approximation.  At this point it is important to keep all terms since,  in general,  expressions that are neglected are not necessarily negligible after being differentiated.  We find that there is  a partial cancellation between the last two terms of \eqref{ModifiedConsLaw2} in the second line, which can be pinpointed  if we use the explicit form of \eqref{eq:nueff2}.  The  intermediate result at this point, prior to calculating the time derivative of the last term  of \eqref{eq:nueff2},  reads as follows:
\begin{equation}\label{MixedConservationAppox2}
\begin{split}
&\frac{3H^2 \dot{G}}{8\pi G}+3\frac{G}{\kappa^2}\left[\frac{\left(\xi-\frac{1}{6}\right)}{2\pi}\frac{m^2}{m_{\rm Pl}^2}(H^2-H_0^2)\left(-2\frac{\dot{H}}{H}-H_0^2\frac{d}{dt}\left(\frac{\ln\frac{H^2}{H_0^2}}{H^2-H_0^2}\right)\right)\right.\\
&\left. -\frac{H\dot{H}\left(\xi-\frac{1}{6}\right)}{\pi}\frac{m^2}{m_{\rm Pl}^2}H_0^2\frac{\ln\frac{H^2}{H_0^2}}{H^2-H_0^2}\right]=0\,.\\
\end{split}
\end{equation}
Computing the pending derivative it is possible to produce further simplifications among the various terms  until reaching the following beautifully simple and compact expression:
\begin{equation}\label{MixedConservationApprox3}
\frac{d{G}}{G^2}=\frac{\left(\xi-\frac{1}{6}\right)}{\pi}m^2\frac{dH}{H}\,,
\end{equation}
in which we have replaced the time derivatives $\dot{G}=dG/dt$ and $\dot{H}=dH/dt$  by just the differentials $dG$ and $dH$  since $dt$ cancels on both sides.
Finally, integrating by simple quadrature Eq.\,\eqref{MixedConservationApprox3}  from the present time $(H_0,G(H_0)=G_N)$ up to an arbitrary moment around the present $(H,G(H))$  we  meet after some elementary algebra the final result
\begin{equation}\label{eq:runGH}
G(H)=\frac{G_N}{1-\frac{\left(\xi-\frac{1}{6}\right)}{2\pi}\frac{m^2}{m^2_{\rm Pl}}\ln \frac{H^2}{H_0^2}}=\frac{G_N}{1+\bv\frac{m^2}{m^2_{\rm Pl}}\ln \frac{H^2}{H_0^2}}\,,
\end{equation}
in which  $G_N$ defines the local gravity value usually associated to the inverse Planck mass squared:  $ G(H_0)=G_N=1/m^2_{\rm Pl}$ (in natural units).  In this way we have proven \eqref{eq:GNHfinal} (q.e.d.)  The obtained formula is our QFT prediction for the physical running of the gravitational coupling with the cosmic expansion. The presence of the coefficient $\bv$  in it -- cf.  Eq.\,\eqref{eq:BetaFuncoeff} -- denotes its connection with the  scaling evolution  of the VED.  See Sec.\,\ref{sec:runningG} for  more discussions on Eq. \eqref{eq:runGH}.

\section{DeWitt-Schwinger expansion in the off-shell formulation}\label{sec:appendixB}
To solve the  Green's function equation in curved spacetime as  given in Sec.\,\ref{HeatKernel}, namely
\begin{equation}\label{KGPropagatorB}
\left(\Box_x-M^2-\Delta^2-\xi R(x)\right)G_F(x,x^\prime)=-\left(-g(x)\right)^{-1/2}\delta^{(n)}(x-x^\prime)\,,
\end{equation}
 is not such a simple task  as to find the corresponding solution in the flat spacetime case.  Here we summarize the well-known procedure \cite{BunchParker1979,BirrellDavies82,ParkerToms09,Fulling89} putting  special emphasis on  highlighting the differences introduced by the parameter $\Delta^2\equiv m^2-M^2$, Eq.\,\eqref{eq:Delta2}, which is crucial in our off-shell approach, see also \cite{FerreiroNavarroSalas} for a related formulation.

 \subsection{Computing the effective action from the heat-kernel}\label{sec:HeatKernel}

 A traditional method to circumvent the difficulty of dealing with a curved spacetime manifold has been to expand the metric  around Minkowski space.  A suitable implementation of this idea  is to make a local expansion of the metric in Riemann normal coordinates, up to four derivatives of the metric (hence  up to fourth adiabatic order).
In these coordinates, the metric admits the following expansion up to $4th$ order\cite{BunchParker1979}:
\begin{equation}
\begin{split}\label{NormalCoordExp}
&g_{\mu\nu}(y)=\eta_{\mu \nu}-\frac{1}{3}R_{\mu \alpha \nu \beta}y^\alpha y^\beta-\frac{1}{6}R_{\mu \alpha \nu \beta;\gamma}y^\alpha y^\beta y^\gamma+\left[-\frac{1}{20}R_{\mu\alpha\nu\beta ; \gamma \delta}+\frac{2}{45}R_{\alpha\mu\beta\lambda}R^\lambda_{\ \gamma\nu\delta}\right]y^\alpha y^\beta y^\gamma y^\delta+\dots
\end{split}
\end{equation}
Here $y$ stands for the difference between the spacetime coordinate $x$ and the source point $x^\prime$ taken as a reference point in normal coordinates, i.e. $y=x-x^\prime$. The different curvature tensors and its derivatives (for instance $R_{\mu\alpha\nu\beta}$) that appear in the expansion above are assumed to be computed at the source point  $x'$ (i.e. at $y=0$). The same is true for the expansion of the determinant and the inverse of the metric.
For simplicity it is easier to define
\begin{equation}\label{Redefine}
\mathcal{G}_F(x,x^\prime)=\left(-g(x)\right)^{1/4}G_F(x,x^\prime)\,.
\end{equation}
We can operate using the standard definition of curved spacetime box operator:
\begin{equation}\label{boxoperator}
\begin{split}
&\Box_x G_F(x,x^\prime)=\Box_x \left((-g(x))^{-1/4}\mathcal{G}_F(x,x^\prime)\right)=\frac{1}{(-g(x))^{1/2}}\partial_\mu\left((-g(x))^{1/2}\partial^\mu \left((-g(x))^{-1/4}\mathcal{G}_F(x,x^\prime)\right)\right)\\
=&(-g(x))^{1/4}\left[\frac{3}{16}\mathcal{G}_F(x,x^\prime)\frac{\partial_\mu (-g(x))\partial^\mu (-g(x))}{(-g(x))^2}-\frac{\partial_\mu \left(\mathcal{G}_F(x,x^\prime)\right)\partial^\mu (-g(x))}{4(-g(x))}-\frac{\mathcal{G}_F(x,x^\prime)\partial_\mu\partial^\mu (-g(x))}{4(-g(x))}\right.\\
&+\left. \frac{\partial^\mu \left(\mathcal{G}_F(x,x^\prime)\right)\partial_\mu (-g(x))}{4(-g(x))}+\partial_\mu \partial^\mu \left(\mathcal{G}_F(x,x^\prime)\right)\right]\,.
\end{split}
\end{equation}
In order to continue, we need to know the expansion of the determinant of the metric as well its inverse. For convenience,  we define
\begin{equation}
g_{\mu\nu}(y)=\eta_{\mu \nu}+h_{\mu \nu}\,,
\end{equation}
where the  deviation $h_{\mu \nu}$ from flat spacetime is written  in powers of the normal coordinate $y$ according to  \eqref{NormalCoordExp}.  We denote it as follows:
\begin{equation}\label{eq:hmunu}
h_{\mu \nu}=h^{(1)}_{\mu \nu}+h^{(2)}_{\mu \nu}+h^{(3)}_{\mu \nu}+h^{(4)}_{\mu \nu}+\cdots= h^{(2)}_{\mu \nu}+h^{(3)}_{\mu \nu}+h^{(4)}_{\mu \nu}+\cdots
\end{equation}
where $h^{(i)}_{\mu\nu}$ stands for the $ith$-term in the indicated order of  \eqref{NormalCoordExp}.
The missing term in the second equality is because \eqref{NormalCoordExp} tells us that   $h^{(1)}_{\mu \nu}=0$.  Such linear term  is missing because the expansion refers to a local inertial (Lorentz) frame, which  is the tangent Lorentz frame at the point $x'$ of the curved spacetime manifold. This demands not only $g_{\mu\nu}(0)=\eta_{\nu\nu}$ but also $\partial_{\alpha}g_{\mu\nu}(0)\equiv\partial g_{\mu\nu}/\partial y^{\alpha}(0)=0$,  both being satisfied at  $x'$ (i.e. at $y=0$).
From the above expansion of the metric we can Taylor expand the corresponding determinant of it, $g(y)$:
\begin{equation}
g(y)=g(0)+\frac{\partial g}{\partial g_{\mu \nu}}\Bigg|_{y=0} \,h_{\mu \nu}
+\frac{1}{2!}\,\frac{\partial^2 g}{\partial g_{\gamma \lambda}\partial g_{\mu \nu}}\Bigg|_{y=0}\,h_{\mu \nu}\,h_{\gamma \lambda}+\cdots\,,
\end{equation}
with $g(0)=-1$ and  $h_{\mu \nu}$ given by \eqref{eq:hmunu}.
The derivatives of the determinant can be computed as follows:
\begin{equation}
\frac{\partial g}{\partial g_{\mu \nu}}=g(y) g^{\mu \nu}(y), \qquad
\frac{\partial^2 g}{\partial g_{\gamma \lambda}\partial g_{\mu \nu}}=g(y) g^{\gamma\lambda}(y)g^{\mu \nu}(y)-g(y) g^{\mu \gamma}(y) g^{\lambda \nu}(y).
\end{equation}
Furthermore, the expansion of the inverse of the metric in powers of the normal coordinate reads
\begin{equation}
g^{ab}=\eta^{ab}-\eta^{a\mu}\eta^{b\nu}\,h_{\mu \nu}+\frac{1}{2!}\left(\eta^{a\lambda}\eta^{\mu \rho}\eta^{b\nu}+\eta^{a\mu}\eta^{b\lambda}\eta^{\nu \rho}\right)\,h_{\mu \nu}\,h_{\rho \lambda}+\cdots
\end{equation}
Thus, the previous calculations can be expanded up to fourth order as follows:
\begin{equation}
\begin{split}\label{NormalCoordExpDet}
g(y)=&g(0)+h_{\mu\nu}\frac{\partial g}{\partial g_{\mu \nu}}\Bigg|_{y=0}+\frac{1}{2!}h^{(2)}_{\mu \nu}h^{(2)}_{\rho \lambda}\frac{\partial^2 g}{\partial g_{\rho \lambda}\partial g_{\mu \nu}}\Bigg|_{y=0}+\cdots\\
=&g(0)-\left( h^{(2)}_{\mu \nu}+h^{(3)}_{\mu \nu}+h^{(4)}_{\mu \nu}+\cdots\right) \eta^{\mu\nu}-\frac{1}{2!}h^{(2)}_{\mu \nu}h^{(2)}_{\rho \lambda}\left(\eta^{\rho\lambda}\eta^{\mu \nu}-\eta^{\mu \rho}\eta^{\lambda \nu}\right)+\cdots
\end{split}
\end{equation}
and
\begin{equation}
\begin{split}
&g^{ab}=\eta^{ab}-\eta^{a\mu}\eta^{b\nu}\left(h_{\mu \nu}^{(2)}+h_{\mu \nu}^{(3)}+h_{\mu \nu}^{(4)}\right)+\frac{1}{2!}\left(\eta^{a\lambda}\eta^{\mu \rho}\eta^{b\nu}+\eta^{a\mu}\eta^{b\lambda}\eta^{\nu \rho}\right) h_{\mu \nu}^{(2)}h_{\rho \lambda}^{(2)}+\cdots
\end{split}
\end{equation}
Using \eqref{NormalCoordExp} and the previous results, we find after some calculations:
\begin{equation}
\begin{split}
g(&y)=-1+\frac{1}{3}R_{\alpha\beta}y^\alpha y^\beta+\frac{1}{6}R_{\alpha\beta;\gamma}y^\alpha y^\beta y^\gamma
+\left[\frac{1}{20}R_{\alpha\beta ; \gamma \delta}-\frac{1}{18}R_{\alpha \beta}R_{\gamma \delta}+\frac{1}{90}R^\mu_{\ \alpha\beta\lambda} R^{\lambda}_{\ \gamma\delta \mu}\right]y^\alpha y^\beta y^\gamma y^\delta+\cdots
\end{split}
\end{equation}
and
\begin{equation}
\begin{split}
&g^{ab}=\eta^{ab}+\frac{1}{3}\eta^{a\mu}\eta^{b\nu}R_{\mu\alpha\nu\beta}y^\alpha y^\beta+\frac{1}{6}\eta^{a\mu}\eta^{b\nu}R_{\mu\alpha\nu\beta;\gamma}y^\alpha y^\beta y^\gamma\\
&+\left[\frac{1}{20}\eta^{a\mu}\eta^{b\nu}R_{\mu\alpha\nu\beta;\gamma\delta}-\frac{2}{45}\eta^{a\mu}\eta^{b\nu}R_{\alpha\mu\beta\lambda}R^\lambda_{\ \gamma\nu\delta}+\frac{1}{18}\left(\eta^{b\nu}\eta^{ak}\eta^{\lambda\mu}+\eta^{a\mu}\eta^{bk}\eta^{\lambda \nu}\right)R_{\mu\alpha\nu\beta}R_{k\gamma\lambda\delta}\right]y^\alpha y^\beta y^\gamma y^\delta+\cdots
\end{split}
\end{equation}
It will be useful to consider the Fourier integrals and transforms
\begin{equation}\label{FourierTransform1}
\mathcal{G}_F(x,x^\prime)=\frac{1}{(2\pi)^n}\int d^n k e^{iky}\mathcal{G}_F(k),
\end{equation}
\begin{equation}\label{FourierTransform2}
i\eta^{\alpha \beta}y_\beta\mathcal{G}_F(x,x^\prime)=\frac{1}{(2\pi)^n}\int d^n k e^{iky}\frac{\partial}{\partial k_\alpha}\mathcal{G}_F(k)\,,
\end{equation}
with $ky\equiv \eta^{\alpha\beta}y_\alpha k_\beta$. Note that in normal coordinates we can raise and lower indices with the Minkowskian metric, as it can be easily shown from \eqref{NormalCoordExp}.
We organize the solution in adiabatic orders, i.e.  counting the number of time derivatives of the metric:
\begin{equation}
\mathcal{G}_F(k)=\mathcal{G}_F^{(0)}(k)+\mathcal{G}_F^{(1)}(k)+\mathcal{G}_F^{(2)}(k)+\mathcal{G}_F^{(3)}(k)+\mathcal{G}_F^{(4)}(k)+\cdots
\end{equation}
Introducing this expansion in the propagator equation \eqref{KGPropagatorB} one can generate a solution of it in terms of an adiabatic series. The results, up to  $4th$-order, are
\begin{equation}
\begin{split}
&\mathcal{G}_F^{(0)}(k)=\frac{1}{k^2+M^2}, \qquad \mathcal{G}_F^{(1)}(k)=0, \qquad \mathcal{G}_F^{(2)}(k)=-\frac{1}{(k^2+M^2)^2}\left(\left(\xi-\frac{1}{6}\right)R+\Delta^2\right),\\
&\mathcal{G}_F^{(3)}=-\frac{i}{2}\left(\xi-\frac{1}{6}\right)R_{;\alpha}\frac{\partial}{\partial k_\alpha}\left(\frac{1}{(k^2+M^2)^2}\right),\\
&\mathcal{G}_F^{(4)}=\frac{1}{3}Q_{\alpha \beta}\frac{\partial^2}{\partial k_\alpha \partial k_\beta}\left(\frac{1}{(k^2+M^2)^2}\right)+\left[\left(\xi-\frac{1}{6}\right)^2R^2+\Delta^4+2\Delta^2R \left(\xi-\frac{1}{6}\right) -\frac{2}{3}{Q^\lambda}_\lambda\right]\frac{1}{(k^2+M^2)^3}\,,
\end{split}
\end{equation}
where  we have defined
\begin{equation}\label{eq:Qalphabeta}
Q_{\alpha\beta}\equiv\frac{1}{2}\left(\xi-\frac{1}{6}\right)R_{;\alpha\beta}+\frac{1}{120}R_{;\alpha\beta}-\frac{1}{40}{R_{\alpha\beta;\lambda}}^\lambda+\frac{1}{30}{R_{\alpha}}^\lambda R_{\lambda \beta}-\frac{1}{60}{{{R^\kappa}_\alpha}^\lambda}_\beta R_{\kappa \lambda}-\frac{1}{60}{R^{\lambda \mu \kappa}}_\alpha R_{\lambda \mu \kappa \beta}.
\end{equation}
As we know, of the two parameters  $M^2$  and  $\Delta^2=m^2-M^2$ entering the propagator equation  \eqref{KGPropagatorB}, the former  is of adiabatic order $0$ whereas the latter is of adiabatic order $2$, see the main text.  One can easily recognize that the  terms $\mathcal{G}_F^{(i)}(k)$  are of adiabatic orders  $i=0,1, 2,3,4$, respectively, and represent successive corrections to the propagator solution up to $4th$-order.

The obtained solution represents an adiabatic  expansion of the propagator in momentum space.  Using  Fourier integral formulas such as \eqref{FourierTransform1}-\eqref{FourierTransform2} we can transfer  the solution to position space. Integrating by parts  and neglecting the boundary terms, we find:
\begin{equation}\label{eq:FourierGF}
\begin{split}
\mathcal{G}_F (x,x^\prime)=\frac{1}{(2\pi)^n}\int d^n k e^{iky}&\left\{ \hat{a}_0 (x,x^\prime)+\hat{a}_1 (x,x^\prime)\left(-\frac{\partial}{\partial M^2}\right)\right.\\
&\left. +\hat{a}_2(x,x^\prime)\left(-\frac{\partial}{\partial M^2}\right)^2+\cdots\right\}\left(\frac{1}{k^2+M^2}\right),
\end{split}
\end{equation}
with
\begin{equation}\label{eq:bilocalWScoeff}
\begin{split}
&\hat{a}_0 (x,x^\prime)=1,\\
&\hat{a}_1 (x,x^\prime)=-\left(\xi-\frac{1}{6}\right)R-\Delta^2-\frac{1}{2}\left(\xi-\frac{1}{6}\right)R_{;\alpha}y^\alpha-\frac{1}{3}Q_{\alpha\beta}y^\alpha y^\beta ,\\
&\hat{a}_2 (x,x^\prime)=\frac{1}{2}\left(\xi-\frac{1}{6}\right)^2 R^2+\frac{\Delta^4}{2}+\Delta^2 R \left(\xi-\frac{1}{6}\right)-\frac{1}{3}{Q^\lambda}_\lambda\,.
\end{split}
\end{equation}
As we can see, these bilocal coefficients receive $\Delta^2$-dependent corrections in our case.
The quantity  ${Q^\lambda}_\lambda$  in the last expression can be found explicitly by taking the trace of \eqref{eq:Qalphabeta}:
\begin{equation}\label{eq:traceQ}
{Q^{\lambda}}_\lambda= -\frac{1}{60} R^{\alpha\beta\gamma\delta}R_{\alpha\beta\gamma\delta}+\frac{1}{60} R^{\alpha\beta}R_{\alpha\beta}+\frac12 \left(\xi-\frac15\right)\Box R\,.
\end{equation}
Using the Euler's density $E$ and the square of the Weyl tensor ($C^2$) -- see  Appendix  \ref{sec:appendixA1} --
we can rewrite \eqref{eq:traceQ}   as follows:
\begin{equation}\label{eq:traceQ2}
\frac{1}{3}{Q^\lambda}_\lambda =-\frac{1}{120}C^2+\frac{1}{360}E+\frac{1}{6}\left(\xi-\frac{1}{5}\right)\Box R\,,
\end{equation}
We  use this expression in Sec.\,\ref{HeatKernel} and below.
The pole in \eqref{eq:FourierGF} must be shifted $M^2\rightarrow M^2-i\epsilon$ in order to have a time ordered product.  In addition, we employ Schwinger's proper time representation\,\cite{ProperTime} of the zeroth order propagator through the following identity and corresponding derivatives with respect to the scale $M$:
\begin{equation}\label{eq:propertime}
\begin{split}
&(k^2+M^2-i\epsilon)^{-1}=i\int_0^\infty ds e^{-is(k^2+M^2-i\epsilon)},\\
&\left(-\frac{\partial}{\partial M^2}\right)^j (k^2+M^2-i\epsilon )^{-1}=i\int_0^\infty (is)^j e^{-is(k^2+M^2-i\epsilon )}ds.
\end{split}
\end{equation}
This is the basis for subsequently obtaining the  DeWitt-Schwinger representation of the sought-for Green's function in curved spacetime\,\cite{DeWitt1975}, originally derived by DeWitt\,\cite{DeWitt1965} following the work of Schwinger\,\cite{ProperTime}.
Using the integral representations  \eqref{eq:propertime} in  the expression \eqref{eq:FourierGF} we can interchange the order of integration and  perform first the following Gaussian integral in momentum space
\begin{equation}
\int d^nk e^{iky-isk^2}=i\left(\frac{\pi}{is}\right)^{n/2}e^{-\sigma(x,x^\prime)/(2is)}\,,
\end{equation}
where the characteristic function $\sigma(x,x^\prime)$ (sometimes called the world function\,\cite{Fulling89}) is one-half of the square of the geodesic distance between $x$ and $x^\prime$:  $\sigma(x,x^\prime )= \frac{1}{2}y_\alpha y^\alpha\equiv \frac{1}{2}\,(x-x^\prime)^2$.
In this way the desired final form for the proper time representation of the Green's function \eqref{Redefine} ensues:
\begin{equation}\label{eq:HK}
G_F(x,x^\prime)=\frac{i{\cal D}^{1/2}(x,x^\prime)}{(4\pi)^{n/2}}\int_0^\infty i ds \frac{e^{-iM^2 s-\sigma (x,x^\prime)/(2is)}}{(is)^{n/2}}\left[a_0(x,x^\prime)+is a_1 (x,x^\prime)+(is)^2 a_2 (x,x^\prime)+\cdots\right],
\end{equation}
where ${\cal D} (x,x^\prime)\equiv -\frac{det\left(-\partial_\mu \partial_{\nu^\prime} \sigma (x,x^\prime) \right)}{\sqrt{g(x)g(x^\prime)}}$ is the general expression for the Van Vleck-Morette determinant, which reduces to  ${\cal D}(x,x^\prime)=(-g(x))^{-1/2}$ for the case of normal coordinates. This, of course, agrees with the redefinition we made in \eqref{Redefine}.  One can easily recognize in \eqref{eq:HK} a generalized form of the fundamental solution of the heat (or diffusion) equation, i.e. its integral kernel.
Once we have the proper time representation of the propagator we may compute the effective Lagrangian   $L_W$  associated to the quantum vacuum effective action,
\begin{equation}
W=-\frac{i}{2}Tr \ln (-G_F)=\int d^4 x \sqrt{-g} L_W\,.
\end{equation}
The trace in this expression is to be computed as specified in Eq.\,\eqref{eq:EAW}. Proceeding now  in the standard way\,\cite{BirrellDavies82}  the effective Lagrangian in $n$ spacetime dimensions can finally  be put in the form of a DeWitt-Schwinger expansion at the arbitrary scale $M$:
\begin{equation}\label{eq:LWDR}
L_W(M)=\frac{\mu^{4-n}}{2(4\pi)^{n/2}}\sum_{j=0}^\infty \hat{a}_j (x) \int_0^\infty (is)^{j-1-n/2}e^{-iM^2 s}ids,
\end{equation}
where $\mu$ is 't Hooft's mass unit  introduced by dimensional purposes (viz. in this case to maintain  $L_W$ with natural dimension $4$ in $n$ spacetime dimensions) and  $\hat{a}_j(x)\equiv\hat{ a}_j(x,x)$ are the corresponding DeWitt-Schwinger coefficients, which appear after computing the coincidence limits $x\to x'$ (i.e. $y\to 0$) of the bilocal coefficients \eqref{eq:bilocalWScoeff}. Upon implementing this limit, the final DeWitt-Schwinger coefficients carry $\Delta^2$-dependent correction terms, as follows:
\begin{equation}\label{eq:ModifDWScoeff}
\begin{split}
&\hat{a}_0 (x)=1=a_0 (x),\\
&\hat{a}_1 (x)=-\left(\xi-\frac{1}{6}\right)R-\Delta^2=a_1(x)-\Delta^2 ,\\
&\hat{a}_2 (x)=\frac{1}{2}\left(\xi-\frac{1}{6}\right)^2R^2+\frac{\Delta^4}{2}+\Delta^2 R \left(\xi-\frac{1}{6}\right)-\frac{1}{3}Q^\lambda_{\ \lambda}=a_2(x)+\frac{\Delta^4}{2}+\Delta^2 R \left(\xi-\frac{1}{6}\right)\,,
\end{split}
\end{equation}
where the hatless $a_j(x)$ represent the ordinary  DeWitt-Schwinger coefficients when $\Delta=0$ (on-shell expansion),
\begin{equation}\label{eq:ClassicDWScoeff}
\begin{split}
&a_0 (x)=1,\\
&a_1(x)=-\left(\xi-\frac{1}{6}\right)R ,\\
&a_2(x)=\frac{1}{2}\left(\xi-\frac{1}{6}\right)^2R^2-\frac{1}{3}{Q^\lambda}_\lambda\,.
\end{split}
\end{equation}
Expressed in this way we can more clearly see what is the effect of performing the expansion off-shell.
Computing the integral involved in \eqref{eq:LWDR}  with the help  of the Euler $ \Gamma$ function, we find
\begin{equation}\label{eq:LWDR2}
L_W(M)=\lim\limits_{\varepsilon \rightarrow 0}\frac{1}{2(4\pi)^{2+\frac{\varepsilon}{2}}}\left(\frac{M}{\mu}\right)^{\varepsilon}\sum_{j=0}^\infty \hat{a}_j (x) M^{4-2j}\Gamma \left(j-2-\frac{\varepsilon}{2}\right),
\end{equation}
where  $\varepsilon\equiv n-4$  is the departure of the spacetime dimension from $4$ in DR.

\subsection{Renormalizing the effective action}\label{sec:RenW}

The effective vacuum action $W$  and corresponding Lagrangian $L_W$ obtained in the previous section  shows up in the form of a DeWitt-Schwinger expansion. However, it is divergent
since the first terms $j=0,1,2$ are UV-divergent and include the contributions up to $4th$ adiabatic order. This so-called divergent part of  $L_W$  at the scale $M$ is defined through
\begin{equation}\label{eq:LdivMdef}
L_{\rm div}(M)\equiv L_W^{(0-4)}(M)=\lim\limits_{\varepsilon \rightarrow 0}\frac{1}{2(4\pi)^{2+\varepsilon}}\left(\frac{M}{\mu}\right)^{\varepsilon}\sum_{j=0}^2 \hat{a}_j (x) M^{4-2j}\Gamma \left(j-2-\frac{\varepsilon}{2}\right)\,.
\end{equation}
The divergent character is apparent since the $\Gamma$ function has poles for $j\leq 2 $ in the limit $\epsilon\to 0$.  Therefore it requires renormalization.  Since we are tracking the poles through DR, it is convenient to  expand $L_{\rm div}$  for $\epsilon\to 0$. We find
\begin{equation*}
\begin{split}
&L_{\rm div}(M)=\frac{1}{2\left(4\pi\right)^2}\left(1+ \frac{\epsilon}{2}\ln\frac{M^2}{4\pi\mu^2}+\mathcal{O}\left(\epsilon^2\right)\right)\Bigg[\hat{a}_0 (x)M^4\left(-\frac{1}{\epsilon}-\frac{\gamma_E}{2}+\frac{3}{4}+\mathcal{O}\left(\epsilon\right)\right)\\
&+\hat{a}_1(x)M^2\left(\frac{2}{\epsilon}+\gamma_E-1+\mathcal{O}\left(\epsilon\right)\right)+\hat{a}_2(x)\left(-\frac{2}{\epsilon}-\gamma_E+\mathcal{O}\left(\epsilon\right)\right)\Bigg]\\
\end{split}
\end{equation*}
\begin{equation}\label{eq:LdivM}
\begin{split}
&=\frac{1}{2\left(4\pi\right)^2}\Bigg[\frac{1}{\epsilon}\left(-\hat{a}_0(x)M^4+2\hat{a}_1(x)M^2-2\hat{a}_2(x)\right)+\gamma_E\left(-\frac{1}{2}\hat{a}_0(x)M^4+\hat{a}_1(x)M^2-\hat{a}_2(x)\right)\\
&+\hat{a}_0(x)M^4\left(\frac{3}{4}-\frac{1}{2}\ln\frac{M^2}{4\pi\mu^2}\right)+\hat{a}_1(x)M^2\left(-1+\ln\frac{M^2}{4\pi\mu^2}\right)-\hat{a}_2(x)\ln\frac{M^2}{4\pi\mu^2}\Bigg]\,.
\end{split}
\end{equation}
 To perform the renormalization, we could generate UV-divergent counterterms by splitting the parameters of the extended classical Lagrangian (including the HD terms) into a renormalized parameter plus an UV-divergent counterterm -- cf.  Eq.\,\eqref{eq:splitcounters} --  and then cancel the divergences of $L_W$ leaving some arbitrary finite parts. However, we  do not want to use this procedure (MS scheme  or variations thereof) since it  does not produce acceptable results in this context.  Instead, we wish to renormalize the effective action $W$ and corresponding effective Lagrangian  in the same way as we did with the EMT, namely by performing a subtraction at another scale.  Thus, we define the renormalized vacuum  effective  Lagrangian $L_W$ at the scale $M$  through the subtraction prescription \eqref{eq:LWrenormalized}:
$L_W^{\rm ren}(M)=   L_W (m)-L_{\rm div} (M)$, with  $L_{\rm div} (M)$  the divergent part of $L_W$, as defined in \eqref{eq:LdivMdef}-\eqref{eq:LdivM}.  The latter involves terms only up to adiabatic order $4$, precisely as in the case of the definition of the renormalized  EMT -- see Eq.\,\eqref{EMTRenormalized}. Thus, $L_W^{\rm ren}(M)$ is a finite quantity.  Notice that $L_W^{\rm ren}(m )= L_W (m)-L_{\rm div} (m)$ is also finite, of course: it is zero if  $L_W(m)$ is evaluated up to $j=2$   but  is nonvanishing if $L_W(m)$ is evaluated beyond $j=2$ (i.e. beyond $4th$ adiabatic order).

To exhibit the finiteness of the renormalized Lagrangian, let us compute $L_W^{\rm ren}(M)$ explicitly.  Notice that
\begin{equation}\label{eq:LWRenaprox}
L_W^{\rm ren}(M)=   L_W (m)-L_{\rm div} (M)= L_{\rm div}  (m)-L_{\rm div} (M)+\cdots
\end{equation}
The dots in this expression represent finite subleading terms (viz. higher than $4th$ adiabatic order) emerging from the  DeWitt-Schwinger expansion\,\eqref{eq:effLagrangian} of $L_W (m)$.  These subleading terms decouple for large values of the mass $m$ of the scalar field, as can be easily seen from Eq.\,\eqref{eq:LWDR2} for $j>2$ and $M=m$.  Thus, if we are just interested in tracking the cancellation of divergences and the finite parts left in the process, it is enough to compute $L_{\rm div}  (m)-L_{\rm div} (M)$.  For the sake of convenience concerning other formulas used in the main text, it will be more useful to first perform the subtraction between  two arbitrary scales $M$ and $M_0$:
\begin{equation}\label{eq:Lren}
L_W^{\rm ren}(M)-L_W^{\rm ren}(M_0)=   L_{\rm div} (M_0)-L_{\rm div} (M)\,.
\end{equation}
Although the calculation of this quantity is straightforward it is a bit laborious, as there are many terms. In particular,  one has to use the explicit form of the modified DeWitt-Schwinger coefficients \eqref{eq:ModifDWScoeff}. Notice that these coefficients  depend on the quantity $\Delta^2(M)=m^2-M^2$   when we perform the calculation of $L_{\rm div}(M)$,  whereas they depend  on $\Delta^2(M_0)=m^2-M_0^2$   when we compute  $L_{\rm div}(M_0)$.  All these terms must be tracked carefully, as they are responsible for the precise expressions \eqref{eq:deltacouplingsB} quoted in the final result given below. Most important, one has to check that the poles cancel in the subtraction and that no trace is left of the arbitrary mass unit $\mu$ either.
Some terms can  be shown immediately to vanish in this subtraction, e.g. it is easy to check  that the overall coefficient of $\gamma_E$ in \eqref{eq:LdivM}, namely $-\frac{1}{2}\hat{a}_0(x)M^4+\hat{a}_1(x)M^2-\hat{a}_2(x)$,  does not depend on $M$ and therefore this term will automatically cancel in the subtraction.  Other terms require more work and one has to go through all the details.  After some tedious algebra one finds that the poles which appear in the limit $\epsilon\to 0$   indeed cancel along with the dependence on the arbitrary mass unit $\mu$, and the final result can be cast in the compact form
\begin{equation}\label{eq:divMdivM0B}
\begin{split}
  L_{\rm div}(M_0)-L_{\rm div}(M)=&\delta\rL(m,M,M_0)-\frac{1}{2}\delta\MPl^2(m,M,M_0) R\\
  &-\delta \alpha_Q(M,M_0) \frac{{Q^\lambda}_\lambda}{3}- \delta{\alpha_2}(M,M_0)R^2\,,
\end{split}
\end{equation}
where  the various contributions read  as follows:
\begin{equation}\label{eq:deltacouplingsB}
\begin{split}
&\delta\rL(m,M,M_0)=\frac{1}{8\left(4\pi\right)^2}\left(M^4-M_0^4-4m^2M^2+4m^2M_0^2+2m^4\ln\frac{M^2}{M_0^2}\right),\\
&\delta\MPl^2(m,M,M_0) =\frac{\left(\xi-\frac{1}{6}\right)}{(4\pi)^2}\left(M^2-M_0^2-m^2\ln \frac{M^2}{M_0^2}\right),\\
&\delta \alpha_Q(M,M_0)=\frac{1}{2(4\pi)^2}\ln\frac{M^2}{M_0^2},\\
&\delta{\alpha_2}(M,M_0)=-\frac{\left(\xi-\frac{1}{6}\right)^2}{4(4\pi)^2}\ln\frac{M^2}{M_0^2}.
\end{split}
\end{equation}
In the light of these general subtraction formulas we may now evaluate the leading terms involved in our original expression \eqref{eq:LWRenaprox} by just setting $M_0=m$ in the above equations. We can easily check that the result is precisely given by the equations  \eqref{eq:LWrenM} and \eqref{eq:deltacouplings}  quoted  in the main text.  Note that the relation between the parameters in the aforementioned equations with those of  \eqref{eq:deltacouplingsB} reads as follows:  $\delta\rL(M)=\delta\rL(m,M,m), \, \delta\MPl^2(M)=\delta\MPl^2(m,M,m)$, \, $\delta \alpha_Q(M)=\delta \alpha_Q(M,m)$ and $\delta{\alpha_2}(M)=\delta{\alpha_2}(M,m)$.

 In the original renormalization approach  to the EMT some of these  quantum effects appeared as parameter differences computed  at the two scales under consideration, ie.
$\delta X\equiv X(M)-X(M_0)$, for the various couplings  $X$ and using the renormalized form of Einstein's equations \eqref{eq:EqsVac}, see Sec. \ref{sec:RenormalizedZPE}.  This is because we renormalized the EMT following the subtraction prescription defined in  Eq.\,\eqref{EMTRenormalized}. We can indeed recognize the first two expressions in \eqref{eq:deltacouplingsB} as being identical to the parameter subtractions \eqref{SubtractionrL}-\eqref{SubtractionMPl}. The third and fourth expressions  in \eqref{eq:deltacouplingsB}  are related to the coefficients of  the HD terms  $Q^\lambda_{\ \lambda}$  and $R^2$.  In particular, $\delta{\alpha_2}(M,M_0)$ is just one half of $\delta\alpha$ given in Eq.\,\eqref{Subtractionalpha}. The factor of $1/2$ is because the parameter $\alpha_2$ in the Lagrangian \eqref{eq:LEHHD}  is related to the parameter $\alpha$ in the generalized Einstein's equations \eqref{eq:MEEs} through $\alpha_2=\alpha/2$. Recall that $\alpha$ is the coefficient of  the HD tensor $\leftidx{^{(1)}}{\!H}_{\mu \nu}$ in these equations, and that tensor is given by  the functional derivative of $R^2$ with respect to the metric, see Appendix \ref{sec:appendixA1}.

At this point we have fully justified the important Eq.\,\eqref{eq:LWrenM}, which gives the renormalized effective Lagrangian of vacuum. From here we may construct the full effective Lagrangian \eqref{eq:Full-Leff}   and reproduce  the remaining considerations. In particular, we can obtain the coefficients of the $\beta$-functions for the various couplings \eqref{eq:BetaFunctionrL}-\eqref{eq:BetaFunctions34} and solve the corresponding renormalization group equations,  with the result \eqref{eq:RGEscouplings}.

\section{Calculation of the vacuum trace of the EMT}\label{sec:appendixC}

\subsection{Full trace up to  $6th$ adiabatic order}\label{sec:Full6thOrder}
The following expression is the expanded form of Eq.\,\eqref{eq:TraceVEV} in the main text\footnote{Extensive use of Mathematica\,\cite{Mathematica} is made in performing these lengthy computations. }:
\begin{equation*}
\begin{split}
& \langle T^{\delta \phi}\rangle =\frac{1}{4\pi^2 a^4}\int dk k^2 \Bigg\{ -\frac{a^2 M^2}{\omega_k}-\frac{a^4 M^4}{4 \omega_k^5}\left(2 \mathcal{H}^2+\mathcal{H}^\prime  \right)+\frac{a^4 M^4}{16\omega_k^7}\left(8\mathcal{H}^4+24\mathcal{H}^2 \mathcal{H}^\prime+6(\mathcal{H}^\prime)^2+8\mathcal{H}\mathcal{H}^{\prime \prime}+\mathcal{H}^{\prime \prime \prime} \right)\\
& +\frac{5 a^6 M^6}{8\omega_k^7} \mathcal{H}^2-\frac{7a^6 M^6}{32\omega_k^9}\left(28 \mathcal{H}^4+36\mathcal{H}^2 \mathcal{H}^\prime+3 (\mathcal{H}^\prime)^2+4\mathcal{H} \mathcal{H}^{\prime \prime} \right)+\\
&-\frac{a^4M^4}{64\omega_k^9}\bigg(32\mathcal{H}^6+240\mathcal{H}^4\mathcal{H}^\prime+60\left(\mathcal{H}^{\prime}\right)^3+160\mathcal{H}^3\mathcal{H}^{\prime \prime}+20\left(\mathcal{H}^{\prime\prime}\right)^2+30\mathcal{H}^\prime \mathcal{H}^{\prime \prime  \prime}+360\mathcal{H}^2 \left(\mathcal{H}^\prime\right)^2\\
&+60\mathcal{H}^2\mathcal{H}^{\prime \prime  \prime}+240\mathcal{H}\mathcal{H}^\prime\mathcal{H}^{\prime\prime}+12\mathcal{H}\mathcal{H}^{\prime \prime\prime\prime}+\mathcal{H}^{\prime\prime\prime\prime\prime}\bigg)\\
&+\frac{3a^6 M^6}{128\omega_k^{11}}\left(1264 \mathcal{H}^6+1512 \mathcal{H}^3\mathcal{H}^{\prime \prime}+23\left(\mathcal{H}^{\prime\prime}\right)^2+228\left(\mathcal{H}^\prime\right)^3+38\mathcal{H}^\prime\mathcal{H}^{\prime\prime\prime}+940\mathcal{H}\mathcal{H}^{\prime}\mathcal{H}^{\prime\prime}+18\mathcal{H}\mathcal{H}^{\prime\prime\prime\prime}\right.\\
&\left.+4672\mathcal{H}^4\mathcal{H}^\prime+3276\mathcal{H}^2\left(\mathcal{H}^{\prime}\right)^2+256\mathcal{H}^2\mathcal{H}^{\prime\prime\prime}\right)+\frac{231 a^8 M^8}{32 \omega_k^{11}}\left(2\mathcal{H}^4+\mathcal{H}^\prime \mathcal{H}^2 \right)-\frac{1155 a^{10} M^{10} \mathcal{H}^4}{128 \omega_k^{13}}\\
\end{split}
\end{equation*}
\begin{equation*}
\begin{split}
&-\frac{11a^8 M^8}{128\omega_k^{13}}\left(3152\mathcal{H}^6+61\left(\mathcal{H}^\prime\right)^3+1116\mathcal{H}^3\mathcal{H}^{\prime\prime}+258\mathcal{H}\mathcal{H}^\prime\mathcal{H}^{\prime\prime}+2364\mathcal{H}^2\left(\mathcal{H}^\prime\right)^2+75\mathcal{H}^2\mathcal{H}^{\prime\prime\prime}+6660\mathcal{H}^4\mathcal{H}^\prime\right)\\
&+\frac{429a^{10}M^{10}\mathcal{H}^2}{256\omega_k^{15}}\left(492\mathcal{H}^4+572\mathcal{H}^2\mathcal{H}^\prime+83\left(\mathcal{H}^\prime\right)^2+40\mathcal{H}\mathcal{H}^{\prime \prime}\right)-\frac{255255a^{12}M^{12} \mathcal{H}^4}{512 \omega_k^{17}}\left(2\mathcal{H}^2+\mathcal{H}^\prime\right)\\
&+\frac{425425a^{14}M^{14}\mathcal{H}^6}{1024\omega_k^{19}}+\Delta^2\bigg(-\frac{a^2}{\omega_k}+\frac{a^4 M^2}{2\omega_k^3}-\frac{a^4 M^2 \left(2\mathcal{H}^2+\mathcal{H}^\prime\right)}{2\omega_k^5}+\frac{5a^6 M^4}{8\omega_k^7}\left(5\mathcal{H}^2+\mathcal{H}^\prime \right)-\frac{35a^8M^6}{16\omega_k^9}\mathcal{H}^2\bigg)\\
&+\Delta^4\left(\frac{a^4}{2\omega_k^3}-\frac{3a^6M^2}{8\omega_k^5}\right)\Bigg\}\\
\end{split}
\end{equation*}
\begin{equation*}
\begin{split}
&+\frac{\left(\xi-\frac{1}{6} \right)}{4\pi^2 a^4}\int dk k^2 \Bigg\{ \frac{6 \mathcal{H}^\prime}{\omega_k}+\frac{3 a^2 M^2}{\omega_k^3} \left(\mathcal{H}^2 +2\mathcal{H}^\prime\right)-\frac{9a^4 M^4 \mathcal{H}^2}{ \omega_k^5}-\frac{a^2 M^2}{2 \omega_k^5}\left(3\mathcal{H}^{\prime \prime \prime}+12\mathcal{H} \mathcal{H}^{\prime \prime}+9(\mathcal{H}^\prime )^2+12 \mathcal{H}^2 \mathcal{H}^\prime \right)\\
&+\frac{3a^2M^2}{8\omega_k^7}\bigg(32\mathcal{H}^4\mathcal{H}^\prime+96\mathcal{H}^2\left(\mathcal{H}^\prime\right)^2+24\left(\mathcal{H}^\prime\right)^3+40\mathcal{H}^3\mathcal{H}^{\prime\prime}+92\mathcal{H}\mathcal{H}^\prime\mathcal{H}^{\prime\prime}+13\left(\mathcal{H}^{\prime\prime}\right)^2+20\mathcal{H}^2\mathcal{H}^{\prime\prime\prime}\\
&+18\mathcal{H}^\prime \mathcal{H}^{\prime \prime\prime}+6\mathcal{H}\mathcal{H}^{\prime\prime\prime\prime}+\mathcal{H}^{\prime\prime\prime\prime\prime}\bigg)+\frac{a^4 M^4}{4\omega_k^7}\left( 210 \mathcal{H}^4+390 \mathcal{H}^2 \mathcal{H}^\prime+45(\mathcal{H}^\prime )^2+60 \mathcal{H} \mathcal{H}^{\prime \prime}\right)\\
&-\frac{a^6 M^6}{8 \omega_k^9}\left(1365 \mathcal{H}^4+840\mathcal{H}^2 \mathcal{H}^\prime \right)-\frac{21a^4M^4}{16\omega_k^9}\bigg(152\mathcal{H}^6+57\left(\mathcal{H}^\prime\right)^3+288\mathcal{H}^3\mathcal{H}^{\prime \prime}+7\left(\mathcal{H}^{\prime \prime}\right)^2\\
&+12\mathcal{H}^\prime\mathcal{H}^{\prime\prime\prime}+232\mathcal{H}\mathcal{H}^{\prime}\mathcal{H}^{\prime\prime}+6\mathcal{H}\mathcal{H}^{\prime\prime\prime}+724\mathcal{H}^4\mathcal{H}^\prime+642\mathcal{H}^2\left(\mathcal{H}^\prime\right)^2+61\mathcal{H}^2\mathcal{H}^{\prime\prime\prime}\bigg)+\frac{945a^8 M^8\mathcal{H}^4}{8\omega_k^{11}}\\
\end{split}
\end{equation*}
\begin{equation*}
\begin{split}
&+\frac{63a^6 M^6}{32\omega_k^{11}}\left(1332\mathcal{H}^6+40\left(\mathcal{H}^\prime\right)^3+636\mathcal{H}^3\mathcal{H}^{\prime\prime}+172\mathcal{H}\mathcal{H}^\prime\mathcal{H}^{\prime\prime}+1359\mathcal{H}^2\left(\mathcal{H}^\prime\right)^2+52\mathcal{H}^2\mathcal{H}^{\prime\prime\prime}+3276\mathcal{H}^4\mathcal{H}^\prime\right)\\
&-\frac{693a^8 M^8\mathcal{H}^2}{64\omega_k^{13}}\left(860\mathcal{H}^4+1117\mathcal{H}^2 \mathcal{H}^\prime+180\left(\mathcal{H}^\prime\right)^2+88\mathcal{H}\mathcal{H}^{\prime\prime}\right)+\frac{9009 a^{10}M^{10}\mathcal{H}^4}{128\omega_k^{15}}\left(173\mathcal{H}^2+94\mathcal{H}^\prime\right)\\
&-\frac{675675a^{12}M^{12}\mathcal{H}^6}{128\omega_k^{17}}+\Delta^2\bigg(\frac{3a^2}{\omega_k^3}\left(\mathcal{H}^2+\mathcal{H}^\prime\right)-\frac{9a^4M^2}{2\omega_k^5}\left(5\mathcal{H}^2+2\mathcal{H}^\prime\right)+\frac{45a^6 M^4}{2\omega_k^5}\mathcal{H}^2 +\frac{45a^6 M^4 \mathcal{H}^2}{2\omega_k^7}\bigg)\Bigg\}
\end{split}
\end{equation*}
\begin{equation*}\label{EqTrace2}
\begin{split}
&+\frac{\left(\xi-\frac{1}{6} \right)^2}{4\pi^2 a^4}\int dk k^2 \Bigg\{ \frac{9}{\omega_k^3}\left(-6\mathcal{H}^2\mathcal{H}^\prime+\mathcal{H}^{\prime\prime\prime}\right)-\frac{27 a^2 M^2}{2\omega_k^5}\left(\mathcal{H}^4 +12\mathcal{H}^2\mathcal{H}^\prime+3\left(\mathcal{H}^\prime\right)^2+4\mathcal{H}\mathcal{H}^{\prime\prime}\right)\\
&+\frac{9}{4\omega_k^5}\left(4\left(\mathcal{H}^\prime\right)^3+16\mathcal{H}\mathcal{H}^\prime\mathcal{H}^{\prime\prime}-6\left(\mathcal{H}^{\prime\prime}\right)^2+4\mathcal{H}^2\mathcal{H}^{\prime\prime\prime}-6\mathcal{H}^\prime\mathcal{H}^{\prime\prime\prime}-\mathcal{H}^{\prime\prime\prime\prime\prime}\right)\\
&+\frac{45a^2 M^2}{8\omega_k^7}\left(28\left(\mathcal{H}^\prime\right)^3+40\mathcal{H}^3\mathcal{H}^{\prime\prime}+7\left(\mathcal{H}^{\prime\prime}\right)^2+12\mathcal{H}^{\prime}\mathcal{H}^{\prime\prime\prime}+6\mathcal{H}\left(18\mathcal{H}^\prime\mathcal{H}^{\prime\prime}+\mathcal{H}^{\prime\prime\prime\prime}\right)\right.\\
&\left.+32\mathcal{H}^4\mathcal{H}^\prime+24\mathcal{H}^2\left(4\left(\mathcal{H}^\prime\right)^2+\mathcal{H}^{\prime\prime\prime}\right)\right)+\frac{135a^4M^4}{\omega_k^7}\left(\mathcal{H}^4+\mathcal{H}^2\mathcal{H}^\prime\right)\\
&-\frac{315a^4M^4}{8\omega_k^9}\left(26\mathcal{H}^6+119\mathcal{H}^4\mathcal{H}^\prime+5\left(\mathcal{H}^\prime\right)^3+44\mathcal{H}^3\mathcal{H}^{\prime\prime}+22\mathcal{H}\mathcal{H}^\prime\mathcal{H}^{\prime\prime}+\mathcal{H}^2 \left(96\left(\mathcal{H}^\prime\right)^2+7\mathcal{H}^{\prime\prime\prime}\right)\right)\\
&\left.+\frac{2835a^6M^6}{16\omega_k^{11}}\mathcal{H}^2\left(25\mathcal{H}^4+56\mathcal{H}^2\mathcal{H}^\prime+15\left(\mathcal{H}^\prime\right)^2+8\mathcal{H}\mathcal{H}^{\prime\prime}\right)-\frac{31185a^8 M^8}{8\omega_k^{13}}\mathcal{H}^4\left(\mathcal{H}^2+\mathcal{H}^\prime\right)\right)\Bigg\}\\
\end{split}
\end{equation*}
\begin{equation}
\begin{split}
&+\frac{\left(\xi-\frac{1}{6} \right)^3}{4\pi^2 a^4}\int dk k^2 \Bigg\{ \frac{81}{\omega_k^5}\left(5\mathcal{H}^4\mathcal{H}^\prime-\left(\mathcal{H}^\prime\right)^3-4\mathcal{H}\mathcal{H}^\prime\mathcal{H}^{\prime\prime}-\left(\mathcal{H}^{\prime\prime}\right)^2-\mathcal{H}^{\prime\prime\prime}\left(\mathcal{H}^2+\mathcal{H}^\prime\right)\right)\\
&+\frac{135a^2M^2}{2\omega_k^7}\left(\mathcal{H}^2+\mathcal{H}^\prime \right) \left(\mathcal{H}^4+29\mathcal{H}^2\mathcal{H}^\prime+4 \left(\mathcal{H}^\prime\right)^2+12\mathcal{H}\mathcal{H}^{\prime\prime}\right)-\frac{2835a^4M^4\mathcal{H}^2}{2\omega_k^9}\left(\mathcal{H}^2+\mathcal{H}^\prime\right)^2 \Bigg\}\\
&+{\cal O}\left(\Delta^2\right){\cal O}\left(T^{-4}\right)\,.
\end{split}
\end{equation}
The notation ${\cal O}\left(\Delta^2\right){\cal O}\left(T^{-4}\right)$  at the end of the above formula denotes the  collection of all the contributions of adiabatic order 6 or higher, i.e.  ${\cal O}\left(T^{-6}\right)$, which  are constructed from the product of terms  proportional to  $\Delta^2$ times other contributions of adiabatic order $4th$ (or higher) made out of more powers of $\Delta^2$, the Hubble rate and its derivatives: $\mathcal{H}, \mathcal{H}^\prime,  \mathcal{H}^{\prime\prime}, \mathcal{H}^{\prime\prime\prime}...$. We do not include these  ${\cal O}\left(T^{-6}\right)$ terms  in our analysis because they vanish on-shell,  thereby playing no role in the renormalization of $\langle T^{\delta \phi}\rangle$ according to the subtraction prescription given below -- cf. Eq.\,\eqref{eq:TraceEMTsubtracted}.

The above expression is UV-divergent. It is easy to identify the few divergent terms in it, all of them of 4th adiabatic order:
\begin{equation}\label{eq:UVdivergentTrace}
\begin{split}
&\langle T^{\delta \phi}\rangle_{\rm Div}\equiv \frac{1}{4\pi^2 a^4}\int dk k^2 \Bigg\{ -\frac{a^2 M^2}{\omega_k}+\left(\xi-\frac{1}{6}\right) \left(\frac{6}{\omega_k}\mathcal{H}^\prime  +\frac{3 a^2 M^2}{\omega_k^3}\left(\mathcal{H}^2+2\mathcal{H}^\prime \right)\right)\\
& + \left( \xi-\frac{1}{6}\right)^2\left(\frac{1}{\omega_k^3}\left(-54 \mathcal{H}^2 \mathcal{H}^\prime+9\mathcal{H}^{\prime \prime \prime} \right) \right)-\frac{a^2 \Delta ^2}{\omega_k}+\frac{a^4}{2\omega_k^3}\left(M^2 \Delta^2+\Delta^4 \right) \\
& +\left(\xi-\frac{1}{6}\right)\frac{3a^2 \Delta^2}{\omega_k^3}\left(\mathcal{H}^2+\mathcal{H}^\prime \right) \Bigg\}\,.
\end{split}
\end{equation}
The remaining  terms, given of course by $\langle T^{\delta \phi}\rangle_{\rm Non-Div}\equiv\langle T^{\delta \phi}\rangle-\langle T^{\delta \phi}\rangle_{\rm Div}$,  are finite.  In order to meet a well defined renormalized expression within the ARP we must perform the subtraction of the trace of the EMT up to the $4th$ adiabatic order at an arbitrary scale $M$:
\begin{equation}\label{eq:TraceEMTsubtracted}
\langle T^{\delta \phi} \rangle_{\rm ren}(M)=\langle T^{\delta \phi} \rangle (m)-\langle T^{\delta \phi} \rangle^{(0-4)}  (M)\,.
\end{equation}
Let us emphasize that we are  following the same prescription as for the $00th$-component of the EMT (cf. Sec. \ref{sec:RenormEMT}), namely  the subtraction is performed in this case over all of the terms of $\langle T^{\delta \phi} \rangle_{\rm ren}(M)$, whether UV-divergent or UV-convergent.  This overall subtraction  is crucial to insure the consistency of the procedure, namely to avoid that the net finite part that remains in the subtractions turns out to be ambiguous.
After the subtraction the integrations left in that expression are finite and can be performed with the help of the formulas of Appendix \ref{sec:appendixA1}.  Albeit the overall integration left is indeed convergent it is not evident if one considers the isolated pieces.  Following the method of Appendix B of \cite{CristianJoan2020} one can either proceed by explicitly exhibiting the divergent parts of these isolated pieces (for example through  the poles in DR) and showing that they cancel out altogether, or one can rearrange that  expression to show that the integrals can be put in a convergent form.  If one wishes to follow the last procedure, the following relations prove useful to show that upon performing the subtraction \eqref{eq:TraceEMTsubtracted} on the integral \eqref{eq:UVdivergentTrace} one finds manifestly convergent expressions:
\begin{equation}
\begin{split}
&-\frac{a^2 m^2}{\omega_k (m)}+\frac{a^2 M^2}{\omega_k (M)}+\frac{a^2 \Delta^2}{\omega_k (M)}-\frac{a^4}{2\omega_k^3 (M)}\left(M^2 \Delta^2+\Delta^4 \right) \\
&=-\frac{a^6 m^2 \Delta^4 }{2\omega_k^3(M) \omega_k(m)\left(\omega_k (m)+\omega_k(M)\right)}\left(1+\frac{\omega_k (M)}{\omega_k (M)+\omega_k (m)}\right),
\end{split}
\end{equation}
and
\begin{equation}
\begin{split}
&\frac{6 \mathcal{H}^\prime}{\omega_k (m)}+\frac{3a^2 m^2}{\omega_k^3 (m)}(\mathcal{H}^2+2\mathcal{H}^\prime)-\frac{6\mathcal{H}^\prime}{\omega_k (M)}-\frac{3a^2 M^2}{\omega_k^3 (M)}(\mathcal{H}^2+2\mathcal{H}^\prime)-\frac{3a^2 \Delta^2}{\omega_k^3 (M)}(\mathcal{H}^2+\mathcal{H}^\prime)\\
&=-\mathcal{H}^\prime\left[\frac{3a^4\Delta^4}{\omega_k^2 (m)\omega_k (M)(\omega_k (m)+\omega_k (M))}\left(\frac{1}{\omega_k (m)}+\frac{1}{\omega_k (m)+\omega_k(M)}\right)\right.\\
&\left.-\frac{3a^4(m^4-M^4)}{\omega_k^2 (M)\omega_k (m)}\left(\frac{1}{\omega_k^2 (m)}+\frac{1}{\omega_k (M)(\omega_k (m)+\omega_k (M))}\right)\right]\\
&-\mathcal{H}^2\frac{3a^4 m^2\Delta^2}{\omega_k^2 (M)\omega_k (m)}\left(\frac{1}{\omega_k^2 (m)}+\frac{1}{\omega_k (M)(\omega_k (m)+\omega_k (M))}\right).\\
\end{split}
\end{equation}
As we can see after these rearrangements,  the integration $\int dk k^2$  of these expressions leads to convergent  integrals.  They all behave as  $\sim \int dk k^2/k^5\sim \int dk /k^3$  in the UV region, similarly as in the situation indicated in the mentioned footnote on p. 26.
The remaining task is to compute these converging integrals, which is not completely trivial.  With the help of Mathematica\,\cite{Mathematica}  the final  result can be expressed as in Eq.\,\eqref{eq:TraceIntegrated} of the main text.  Alternatively, if one performs the calculation in DR one may account for all the integrals (whether  UV-divergent or convergent) on using  the master formula \eqref{DRFormula} of Sec.\,\ref{sec:appendixA2}.  The result is the same.

\subsection{Trace anomaly: explicit computation}\label{sec:TraceAnomalyComp}
As a nontrivial check of our calculations, let us use our results to reproduce the famous trace or conformal anomaly of the energy-momentum tensor in curved spacetime.  In our case it is obtained upon computing Eq.\,\eqref{eq:TraceVEVconformal1} in the limit $m\to 0$.  While this is not the interesting limit for our purposes, it serves nevertheless as a nontrivial calculational check.
The explicit form of the result remains nonvanishing for $m\to 0$ and can be easily extracted from Eq.\,\eqref{EqTrace2} above by setting  $\xi=1/6$ and $M=m$, and then picking out the terms of $4th$ adiabatic order which are independent of $m$ under an appropriate change of integration variable (see below).  These are the following:
\begin{equation}\label{eq:TraceVEVconformal2}
\begin{split}
 \left.\lim\limits_{m\to 0} \langle T^{\delta \phi}\rangle\right|_{(\xi=1/6,M=m )} &=\frac{1}{4\pi^2 a^4}\int dk k^2 \Bigg\{\frac{a^4 m^4}{16\omega_k^7}\left(8\mathcal{H}^4+24\mathcal{H}^2 \mathcal{H}^\prime+6(\mathcal{H}^\prime)^2+8\mathcal{H}\mathcal{H}^{\prime \prime}+\mathcal{H}^{\prime \prime \prime} \right)\\
& -\frac{7a^6 m^6}{32\omega_k^9}\left(28 \mathcal{H}^4+36\mathcal{H}^2 \mathcal{H}^\prime+3 (\mathcal{H}^\prime)^2+4\mathcal{H} \mathcal{H}^{\prime \prime} \right)\\
&+\frac{231 a^8 m^8}{32 \omega_k^{11}}\left(2\mathcal{H}^4+\mathcal{H}^\prime \mathcal{H}^2 \right)-\frac{1155 a^{10} m^{10} \mathcal{H}^4}{128 \omega_k^{13}}\Bigg\}\,,\\
\end{split}
\end{equation}
where the limit $m\to 0$ remains implicit on the \textit{r.h.s.} of the above expression.  However, it is not necessary since the involved  integrals  are actually independent of $m$.  To see that, let us make the change of variable $k= am x$.   Then, $\omega_k=\sqrt{k^2+a^2m^2}=am\sqrt{1+x^2}$,  and we realize that the powers of $m$ in the numerator  of all the above terms (plus the three added ones from $dk k^2 $)  exactly cancel against those in the denominator, so the integrals do not actually depend on $m$ and hence the expression  \eqref{eq:TraceVEVconformal2} cannot vanish for $m\to 0$.  The result after some computations  (with the help of the master integral formula in Appendix \ref{sec:appendixA2}) reads
\begin{equation}\label{eq:TraceVEVconformal3}
\begin{split}
 \left.\lim\limits_{m\to 0} \langle T^{\delta \phi}\rangle\right|_{(\xi=1/6,M=m )}
&=\frac{1}{4\pi^2 a^4}\int_0^\infty dx x^2\left\{\frac{8\mathcal{H}^4+24\mathcal{H}^2 \mathcal{H}^\prime +6\left(\mathcal{H}^\prime\right)^2 +8\mathcal{H} \mathcal{H}^{\prime\prime}+\mathcal{H}^{\prime\prime\prime}}{16\left(1+x^2\right)^{7/2}} \right.\\
&\left. -\frac{7(28\mathcal{H}^4+36\mathcal{H}^2\mathcal{H}^\prime+3\left(\mathcal{H}^\prime\right)^2+4\mathcal{H}\mathcal{H}^{\prime\prime})}{32\left(1+x^2\right)^{9/2}}+\frac{231\left(2\mathcal{H}^4+\mathcal{H}^\prime\mathcal{H}^2\right)}{32\left(1+x^2\right)^{11/2}}-\frac{1155\mathcal{H}^4}{128\left(1+x^2\right)^{13/2}}\right\}\\
&=\frac{1}{480\pi^2 a^4}\left(-4\mathcal{H}^2\mathcal{H}^\prime+\mathcal{H}^{\prime\prime\prime}\right)=\frac{1}{480\pi^2}\left(\frac{\ddot{a}^2}{a^2}+\frac{\vardot{4}{a}}{a}+3\frac{\dot{a}\vardot{3}{a}}{a^2}-3\frac{\dot{a}^2\ddot{a}}{a^3}\right).
\end{split}
\end{equation}
The trace anomaly can also be elucidated in the framework of the effective action, $W$\,\cite{BirrellDavies82}.  The latter is related to the VEV of the EMT,  as noted in \eqref{eq:DefW}.
The effective action is purely geometric and involves the quantum effects of $\phi$ in our case. Since the UV-divergences are inherent to short-distance effects, they all involve the behavior of  geometric tensors $R^2, R_{\mu\nu}  R^{\mu\nu},\dots$ at short distances (cf. Appendix \ref{sec:appendixB}).  The nontrivial  local behavior of the curved spacetime at the level of the effective action is the counterpart to the UV behavior of the field modes in the EMT.   The two languages lead to the same answer.  Thus,  although one can use $W$  and the UV-divergences associated to these geometric terms to derive the trace anomaly\,\cite{BirrellDavies82},  here we have used directly the VEV of the EMT corresponding to the quantum matter field $\phi$.  The divergences of $W$ are of course the same as those of the vacuum  EMT. So, if we write $W=W_{\rm div}+W_{\rm ren}$, the divergent and renormalized  parts of the vacuum EMT must correspond respectively to $W_{\rm div}$ and $W_{\rm ren}$.
This means that the  obtained expression \eqref{eq:TraceVEVconformal3} can be identified with  the vacuum trace  emerging from the divergent part of the  effective action, which in the massless conformal limit turns out to be finite. Thus,
\begin{equation}\label{eq:EMTWdiv}
\left.\lim\limits_{m\to 0} \langle T^{\delta \phi}\rangle\right|_{(\xi=1/6,M=m )}=\frac{2}{\sqrt{-g}}\, g_{\mu\nu}\,\frac{\delta W_{\rm div}}{\delta g_{\mu\nu}}\,.
\end{equation}
Because the vacuum trace of the total EMT derived from the effective action must vanish in the massless conformally coupled limit\,\cite{BirrellDavies82}, the trace associated to $W_{\rm ren}$ (the so-called  renormalized part of the effective action) is given by minus the previous result \eqref{eq:EMTWdiv}, and this defines the trace anomaly:
\begin{equation}\label{eq:FormalTraceAnomaly}
\left.\lim\limits_{m\to 0} \langle T^{\delta \phi}\rangle\right|_{(\xi=1/6,M=m )}^{\rm anomaly} =\frac{2}{\sqrt{-g}}\, g_{\mu\nu}\,\frac{\delta W_{\rm ren}}{\delta g_{\mu\nu}}=-\frac{2}{\sqrt{-g}}\, g_{\mu\nu}\,\frac{\delta W_{\rm div}}{\delta g_{\mu\nu}}=- \left.\lim\limits_{m\to 0} \langle T^{\delta \phi}\rangle\right|_{(\xi=1/6,M=m )}\,.
\end{equation}
It can be expressed in an invariant form, showing that the anomaly is a general coordinate scalar. With the help of the geometric relations given  in Appendix  \ref{sec:appendixA1} one can readily show that the obtained expression can be written in a covariant way as follows:
\begin{equation}\label{eq:TraceInvariant}
\begin{split}
\left.\lim\limits_{m\to 0} \langle T^{\delta \phi}\rangle\right|_{(\xi=1/6,M=m )}^{\rm anomaly} &=-\left.\lim\limits_{m\to 0} \langle T^{\delta \phi}\rangle\right|_{(\xi=1/6,M=m )}=-\frac{1}{480\pi^2 a^4}\left(-4\mathcal{H}^2\mathcal{H}^\prime+\mathcal{H}^{\prime\prime\prime}\right)\\
&=+\frac{1}{2880\pi^2}\left[R^{\mu\nu}R_{\mu\nu}-\frac{1}{3} R^2+\Box R\right]\,.
\end{split}
\end{equation}
It is well known that there is no contribution from the square of the Weyl tensor  $C^2=C^{\alpha\beta\gamma\delta}C_{\alpha\beta\gamma\delta}$  for conformally flat spacetimes since that  tensor vanishes identically for them.
The above expression is the form which we have quoted  in the main text, see  Eq.\,\eqref{eq:TraceAnomaly}.  In general the conformal anomaly can also be written in a very succinct way in terms of the DeWitt-Schwinger coefficient of adiabatic order $4$ -- cf.  Appendix\,\ref{sec:HeatKernel} and \cite{BirrellDavies82}.  Borrowing equations \eqref{eq:traceQ2} and \eqref{eq:ModifDWScoeff},  one finds
\begin{equation}\label{eq:TraceInvariant2}
\left.\lim\limits_{m\to 0} \langle T^{\delta \phi}\rangle\right|_{(\xi=1/6,M=m )}^{\rm anomaly} =+\left.\frac{a_2}{16\pi^2}\right|_{\xi=1/6}=-\left.\frac{1}{48\pi^2} {Q^\lambda}_\lambda\right|_{\xi=1/6}= +\frac{1}{1920\pi^2}\left[C^{2}-\frac13 E+\frac{2}{3}\Box R\right]\,.
\end{equation}
 Using   \eqref{eq:R2R4},  with $C_{\alpha\beta\gamma\delta}=0$ for FLRW spacetime, the previous expression boils down to the particular form \eqref{eq:TraceInvariant}.  Notice that  $\hat{a}_2=a_2$ in this case, since $M=m$ and hence $\Delta=0$ in  \eqref{eq:ModifDWScoeff}.


\end{document}